\documentclass[11pt]{article}
\usepackage[margin=1.05in]{geometry}

\usepackage{amsmath,amssymb,amsthm,mathtools,stmaryrd}
\usepackage{microtype}
\usepackage{booktabs,array,longtable}
\usepackage{enumitem}
\usepackage{needspace}
\usepackage[section]{placeins}
\usepackage[dvipsnames]{xcolor}
\usepackage{tikz}
\usetikzlibrary{arrows.meta,decorations.pathreplacing,calc}
\IfFormatAtLeastTF{2026-06-01}{}{\usepackage{aliascnt}}
\usepackage[colorlinks=true,linkcolor=blue!50!black,citecolor=blue!50!black,urlcolor=blue!50!black]{hyperref}
\usepackage[capitalise,nameinlink]{cleveref}

\newcolumntype{R}[1]{>{\raggedright\arraybackslash}p{#1}}

\newtheorem{theorem}{Theorem}[section]
\makeatletter
\IfFormatAtLeastTF{2026-06-01}
  {\newcommand{\sharedtheorem}[2]{%
     \newtheorem{#1}[theorem]{#2}%
     \expandafter\g@addto@macro\csname cl@section\endcsname{\@elt{#1}}}}
  {\newcommand{\sharedtheorem}[2]{%
     \newaliascnt{#1}{theorem}%
     \newtheorem{#1}[#1]{#2}%
     \expandafter\xdef\csname the#1\endcsname{\noexpand\thetheorem}}}
\makeatother
\sharedtheorem{lemma}{Lemma}
\sharedtheorem{proposition}{Proposition}
\sharedtheorem{corollary}{Corollary}
\sharedtheorem{observation}{Observation}
\theoremstyle{definition}
\sharedtheorem{definition}{Definition}
\sharedtheorem{example}{Example}
\sharedtheorem{question}{Question}
\theoremstyle{remark}
\sharedtheorem{remark}{Remark}
\crefname{theorem}{Theorem}{Theorems}
\crefname{lemma}{Lemma}{Lemmas}
\crefname{proposition}{Proposition}{Propositions}
\crefname{corollary}{Corollary}{Corollaries}
\crefname{observation}{Observation}{Observations}
\crefname{definition}{Definition}{Definitions}
\crefname{example}{Example}{Examples}
\crefname{question}{Question}{Questions}
\crefname{remark}{Remark}{Remarks}

\newenvironment{proofof}[1]{\renewcommand{\proofname}{Proof of \cref{#1}}}{}

\makeatletter
\def\@fnsymbol#1{\ensuremath{\ifcase#1\or *\or \dagger\or \ddagger\or
  \mathsection\or \mathparagraph\or \|\or **\or \dagger\dagger\or
  \ddagger\ddagger\else\@ctrerr\fi}}
\makeatother

\setlist{itemsep=2pt,topsep=4pt}

\newcommand{\CO}{\textup{\textbf{\small[COMPUTATIONAL]}}}
\newcommand{\KN}{\textup{\textbf{\small[KNOWN]}}}

\newcommand{\eps}{\varepsilon}
\newcommand{\elad}{\eps_{\mathrm{lad}}}
\newcommand{\EI}{E_I^\ast}

\newcommand{\BOX}{\mathrm{BOX}}
\newcommand{\Bot}{\mathrm{bot}}
\newcommand{\Top}{\mathrm{top}}
\newcommand{\omax}{o_{\max}}
\newcommand{\xp}{x_+}
\newcommand{\xm}{x_-}
\newcommand{\tmin}{\tau_{\min}}
\newcommand{\WSI}{\mbox{\textsf{WS-I}}}
\newcommand{\WSW}{\mbox{\textsf{WS-W}}}
\newcommand{\OWW}{\mbox{\textsf{1W-W}}}
\newcommand{\stP}{\mathsf{P}}
\newcommand{\stN}{\mathsf{N}}
\newcommand{\stp}{\mathsf{p}}
\newcommand{\stn}{\mathsf{n}}

\newcommand{\cl}[1]{\llbracket #1\rrbracket}
\newcommand{\PCirc}{\textup{\textsc{Pure-Circuit}}}
\newcommand{\Select}{\textup{\textsc{Select}}}

\title{Approximate clearing in financial networks\\ with credit default swaps:\\ hardness thresholds, a box algorithm, barriers,\\ and exact solvency at bounded degree}
\author{Michal \v{C}ert\'ik\\[2pt] \small The World bank Group\\ \small Washington, DC, USA\\ \small\texttt{michal.certik@gmail.com}\\ \small ORCID: \href{https://orcid.org/0009-0008-6880-4896}{0009-0008-6880-4896}}
\date{30 October 2026}

\hypersetup{pdfauthor={Michal Čertík},pdftitle={Approximate clearing in financial networks with credit default swaps: hardness thresholds, a box algorithm, barriers, and exact solvency at bounded degree}}

\begin{document}
\maketitle

\begin{abstract}
We study the complexity of computing approximate clearing vectors in financial networks with credit default swaps (CDSs). For the weak approximation notions (W) and (I), we prove PPAD-hardness for every rational error below $E_I^\ast\approx0.2530$. On networks whose CDS writers are solvent ex ante we prove it below $\alpha=(13-\sqrt{153})/4\approx0.1577$, even with a single CDS writer (for (W) through calibrated copies of the whole circuit). A simple interval algorithm solves both notions in polynomial time for every rational error above $\eps_{\mathrm{lad}}\approx0.2834$ in general and above $3-2\sqrt2\approx0.1716$ on writer-solvent networks, and both thresholds are sharp for algorithms of bounded radius (for (I) on writer-solvent networks, and for (I) and (W) in general). On non-degenerate networks, general or writer-solvent, (I) and (W) have the same supremum of PPAD-hard errors and the same infimum of polynomially solvable errors. For the exact-solvency notion (D) of Dohn, Hansen and Klinkby, PPAD-hardness holds on non-degenerate networks for every rational error below $1$; without non-degeneracy this notion is not total, while (W), (I) and an exempted variant of it remain in PPAD. On a restricted class in which core banks hold CDSs of notional $1$, we classify (D) by the largest number $d$ of contracts held by a core bank, counted with multiplicity: for every rational $E\in[1/2,1)$ it is solvable in polynomial time for $d<(\lceil\rho\rceil+1)\rho$ and PPAD-complete for $d\ge(\lceil\rho\rceil+1)\rho$, where $\rho=(1+E)/(1-E)$; at $E=1/3$ it is solvable in polynomial time for $d\le5$ and PPAD-complete for $d\ge6$; and for $E<1/9$ the problem is PPAD-complete already at $d=2$, while $d=1$ is exactly solvable. Assuming that PPAD is not contained in FP, this determines the threshold error $p_d$ for $d\in\{6,7\}$ and all $d\ge12$, and $\sqrt d\,(1-p_d)\to2$. These degree results are specific to (D) and leave the windows for (I) and (W) unchanged. Reductions from \PCirc\ whose gadgets are certified by interval propagation cannot prove hardness of writer-solvent (I) at any error in $[\alpha,1)$ unless PPAD is contained in FP; this holds for gadgets of any topology with disjoint input wires, outputs on the gadget's own receivers, auxiliary banks whose boxes do not depend on the inputs, and arbitrary fixed pairs of disjoint low and high intervals on each rail. A finite-precision version of the interval algorithm reduces writer-solvent (I) to selecting a point in a near-stalled box, which we do in polynomial time when the residual matrix is symmetric or monotone on each strongly connected component. The open windows are $[E_I^\ast,\eps_{\mathrm{lad}}]\approx[0.2530,0.2834]$ and $[\alpha,3-2\sqrt2]\approx[0.1577,0.1716]$.
\end{abstract}

\setcounter{tocdepth}{2}
{\small\tableofcontents}

\section{Introduction}\label{sec:intro}

In a financial network, banks owe each other money, and a bank that cannot pay in full pays a fraction of its obligations, its \emph{recovery rate}. A \emph{clearing vector} is a vector of recovery rates that is consistent with itself: every bank pays out what it receives, up to its obligations. For networks of ordinary debts, Eisenberg and Noe~\cite{EN01} showed that clearing vectors exist and that the greatest one can be computed in polynomial time. Credit default swaps change this. A CDS obliges its writer to pay the holder an amount that grows as a third bank, the reference, recovers less, so a payment can depend on another bank's default. Schuldenzucker, Seuken and Battiston~\cite{SSB16,SSB20} showed that such networks can have several clearing vectors, or none once default costs are present, and that finding an approximate clearing vector is PPAD-complete~\cite{SSB17}. Ioannidis, de Keijzer and Ventre showed that clearing vectors can be irrational, that exact clearing is FIXP-complete and that its strong approximation is $\mathrm{FIXP}_a$-complete~\cite{IKV22}, and that approximate clearing in the sense of the notion we call (I) is PPAD-hard for small constant errors~\cite{IKV23,IKV26b}. Dohn, Hansen and Klinkby~\cite{DHK} introduced an approximation notion with an exact solvency clause and proved PPAD-hardness for every error up to $5-2\sqrt6\approx0.101$.

Approximation is measured by a constant error $E$: a recovery vector is acceptable if every bank's recovery is within $E$ of what the clearing map assigns to it, possibly with further clauses. For $E\ge1/2$ the weak problems are trivial: the vector with recovery $1/2$ at every bank that is not solvent ex ante qualifies. The published hardness constants were $(2-\sqrt3)/4\approx0.067$ for (I)~\cite{IKV26b} and $5-2\sqrt6\approx0.101$ for (D)~\cite{DHK} (see \cref{sec:literature}), and we are not aware of a polynomial algorithm for general networks at any constant error below $1/2$ in the literature. Two questions therefore remained: where above these constants the complexity of each notion changes, and whether membership in PPAD survives without the non-degeneracy assumption under which it had been proved; the second was raised explicitly in~\cite[Appendix~A]{DHK}. This paper answers the second question and narrows the first to two short windows. For the exact-solvency notion it also addresses a structural question: the hardness constructions for (D) use CDSs with non-unit notionals, which in \cref{thm:D} grow as the error approaches $1$, and Kanellopoulos and Melissourgos~\cite{KM} recently gave a polynomial algorithm for (D) at error $1/5$ on a restricted class in which every core bank holds at most two CDSs of notional $1$. How many unit contracts must a bank hold before (D) becomes hard? On a unit-notional normal form within that class we prove polynomial algorithms below, and PPAD-completeness from, an explicit held degree for every error in $[1/2,1)$ and for error $1/3$, and PPAD-completeness at held degree $2$ for every error below $1/9$; assuming that PPAD is not contained in FP, this answers the question exactly at these errors.

\subsection{Results}

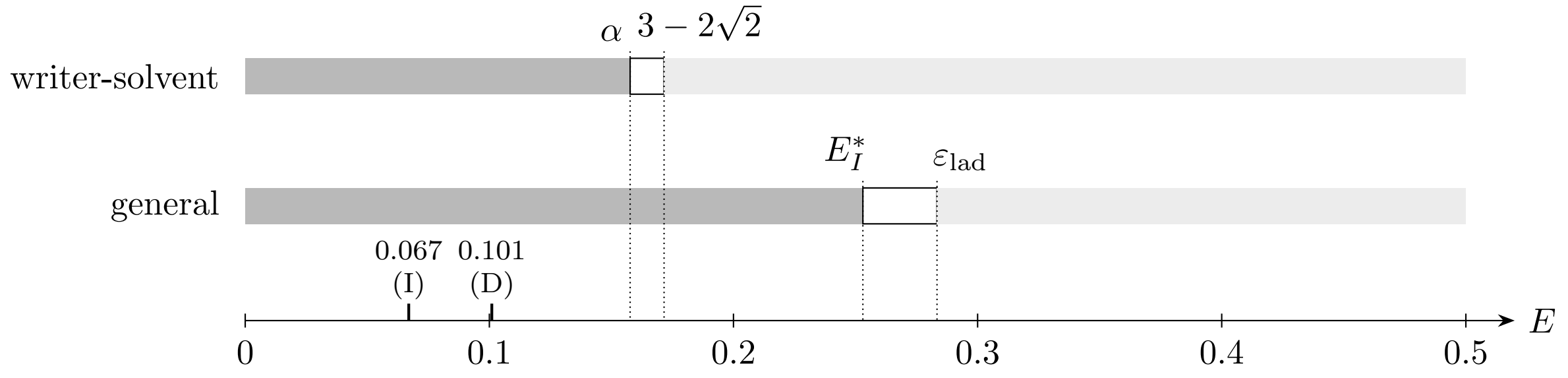
\begin{figure}[t]
\centering
\begin{tikzpicture}[x=24.5cm,y=1cm,font=\small]
\def\yW{1.3} \def\yG{0}
% writer-solvent row
\fill[gray!55] (0,\yW-0.18) rectangle (0.15767,\yW+0.18);
\fill[white] (0.15767,\yW-0.18) rectangle (0.17157,\yW+0.18);
\draw (0.15767,\yW-0.18) rectangle (0.17157,\yW+0.18);
\fill[gray!15] (0.17157,\yW-0.18) rectangle (0.5,\yW+0.18);
\node[anchor=east] at (-0.005,\yW) {writer-solvent};
% general row
\fill[gray!55] (0,\yG-0.18) rectangle (0.25300,\yG+0.18);
\draw (0.25300,\yG-0.18) rectangle (0.28337,\yG+0.18);
\fill[gray!15] (0.28337,\yG-0.18) rectangle (0.5,\yG+0.18);
\node[anchor=east] at (-0.005,\yG) {general};
% axis
\draw[-{Stealth}] (0,-1.15) -- (0.52,-1.15) node[right] {$E$};
\foreach \x/\l in {0/0,0.1/0.1,0.2/0.2,0.3/0.3,0.4/0.4,0.5/0.5}{\draw (\x,-1.08)--(\x,-1.22); \node[below] at (\x,-1.22) {\l};}
\draw[densely dotted] (0.15767,\yW+0.25) -- (0.15767,-1.15); \node[above] at (0.150,\yW+0.22) {$\alpha$};
\draw[densely dotted] (0.17157,\yW+0.25) -- (0.17157,-1.15); \node[above] at (0.186,\yW+0.22) {$3-2\sqrt2$};
\draw[densely dotted] (0.25300,\yG+0.25) -- (0.25300,-1.15); \node[above] at (0.246,\yG+0.22) {$\EI$};
\draw[densely dotted] (0.28337,\yG+0.25) -- (0.28337,-1.15); \node[above] at (0.293,\yG+0.22) {$\elad$};
\draw[thick] (0.06699,-1.15)--(0.06699,-0.98); \node[above,align=center,inner sep=1pt,font=\scriptsize] at (0.06699,-0.98) {$0.067$\\(I)};
\draw[thick] (0.10102,-1.15)--(0.10102,-0.98); \node[above,align=center,inner sep=1pt,font=\scriptsize] at (0.10102,-0.98) {$0.101$\\(D)};
\end{tikzpicture}
\caption{Thresholds for (I) and (W). Dark: PPAD-hard for every rational error in the range. Light: polynomial time. White: open window. The ticks mark the previously published hardness constants $(2-\sqrt3)/4\approx0.067$ for (I)~\cite{IKV26b} and $5-2\sqrt6\approx0.101$ for (D)~\cite{DHK}; the second is for the exact-solvency notion (D), for which \cref{thm:D} gives hardness below $1$, and is shown for comparison only.}\label{fig:landscape}
\end{figure}

We work with five approximation notions (\cref{def:notions}): the weak notion (W), which bounds every residual by $E$; the notion (I) of~\cite{IKV23}, which also fixes ex-ante solvent banks at recovery $1$; the notion (D) of~\cite{DHK}, which adds the clause that a bank with assets at least $(1+E)$ times its liabilities recovers exactly $1$; its variant (D$^\circ$), which exempts banks without liabilities; and the clipped-ratio notion (S), modelled on~\cite{SSB17}. A network is \emph{writer-solvent} if every CDS writer is solvent ex ante, and \emph{one-writer} if a single, ex-ante solvent bank writes all CDSs. \Cref{tab:landscape} and \cref{fig:landscape} summarize the results.

\paragraph{Two principal parts.} The first part concerns (I) and (W): hardness thresholds on general, writer-solvent and one-writer networks, a box algorithm whose thresholds are sharp for algorithms of bounded radius, and barriers for interval-certified reductions; two short windows remain open. The second part concerns the exact-solvency notion (D) on the restricted unit-notional class $\mathcal R(d)$, which we classify by held degree. It concerns a different notion and a restricted class, and it does not move the windows for (I) and (W). Results on membership without non-degeneracy, on totality and on exact existence complete the picture. The main text states the results with their key constructions; every proof that is not in the main text is in the appendices.

\paragraph{General networks: hardness below $\EI$, algorithms above $\elad$.} (I)-$E$ and (W)-$E$ are PPAD-complete for every rational $E<\EI=0.25299586\ldots$, the root in $[1/5,27/100]$ of an explicit polynomial of degree $11$ (\cref{thm:general}). The construction uses ladders of bank pairs that clean signals step by step. It works exactly while the interval $(\tmin(E),T(E))$ of admissible thresholds for its PURIFY gadget is nonempty, that is, for $E<\EI$, and in the ideal limit $\EI$ is the exact endpoint of this design family (\cref{rem:EIendpoint}). The box algorithm below solves both problems in polynomial time for every rational $E>\elad=0.28336553\ldots$, the root in $(0,1/2)$ of $5E^5-45E^4+52E^3-32E^2+20E-4$ (\cref{thm:alggen}). The remaining window is $[0.25299586,0.28336554]$.

\paragraph{Writer-solvent networks: hardness below $\alpha$, algorithms above $3-2\sqrt2$.} (I)-$E$ is PPAD-complete on writer-solvent networks for every rational $E<\alpha=(13-\sqrt{153})/4=0.15767078\ldots$, with one writer and, at every NAND gate, one seed receiver after two input buffers (\cref{thm:wsI}). The box algorithm solves (I) and (W) on these networks for every rational $E>3-2\sqrt2=0.17157287\ldots$ (\cref{thm:algws}). The remaining window is $[0.15767078,0.17157288]$.

\paragraph{One writer under (W): calibration copies.} Under (W) the single writer is only approximately solvent, and a gadget that must work for every value of its recovery loses accuracy. We avoid this by building many copies of the whole circuit, each calibrated for a different value of the writer's recovery, and decoding from the copy that matches (\cref{thm:calib}). The match is made by the decoder and not by the network, because no bank can be forced to react to the writer's recovery (\cref{rem:obstruction}); so each copy is a complete circuit, and the decoder reads every gate from the same copy. This gives PPAD-completeness of one-writer (W) for every rational $E<\alpha$ (\cref{cor:oneW}). At $E=157/1000$ the reduction uses $157$ copies; the explicit instance has $17743$ banks and $17898$ CDSs (\cref{rem:cert157}).

\paragraph{Equal thresholds.} Writer-solvent (I), writer-solvent (W) and one-writer (W) have the same supremum of PPAD-hard rational errors and the same infimum of polynomially solvable ones (\cref{thm:equal}); so do (I) and (W) on non-degenerate networks (\cref{cor:equalgen}). These are statements about suprema and infima only; they say nothing about an individual error.

\paragraph{$\alpha$ is the exact endpoint of the chain-and-seed family.} Within the family of NAND gadgets made of input chains, one seed receiver and an output chain, certified with three statuses, uniform gadgets exist exactly for $E<\alpha$ (\cref{cor:Fendpoint}). In writer-solvent networks the dual-rail wires of the general construction do not help interval-certified gadgets, because all assets there are separable affine.

\paragraph{The box algorithm.} Every bank carries an interval, initially $[E,1-E]$; each round computes an interval enclosing every value of the bank's map on the current boxes; a bank whose enclosing interval has length at most $2E$ settles at the point of its box nearest to the midpoint, and every other bank shrinks its box to the enclosing interval narrowed by $E$ on each side. Every output is an (I)-$E$ vector (\cref{lem:sound}), although the boxes need not contain an exact clearing vector (\cref{ex:stallfail}). Ratio recursions show termination after a number of rounds that depends only on $E$, above $3-2\sqrt2$ for writer-solvent networks and above $\elad$ in general (\cref{thm:algws,thm:alggen}). Both thresholds coincide with lower bounds for algorithms of bounded radius (\cref{thm:locallower}), for (I) on writer-solvent networks and for (I) and (W) on general networks, so within the locality model they are sharp, and $R$ rounds of the algorithm are an optimal $R$-local algorithm for writer-solvent (I) (\cref{cor:local}). This optimality is a statement about bounded-radius algorithms only, not about polynomial-time algorithms in general. We also record that the writer-solvent local lower bound needs its hypothesis on every iterate, not only on the last; a counterexample at $E=3/10$, $R=3$ shows that the weaker hypothesis does not suffice (\cref{rem:L7}). Rounding the boxes outward to a grid of step $\delta$ makes the algorithm stop after at most $nM+1$ rounds with endpoints of polynomial size, at the price of an error $E_0+\delta/2$ for settled banks (\cref{thm:roundedbox}); every writer-solvent instance is thereby reduced to selecting a point of a near-stalled box (\cref{thm:transfer}). On networks whose CDS writers lie in $F$ and whose signed residual matrix is symmetric or monotone on each strongly connected component, such points can be selected in polynomial time, which gives a polynomial approximation scheme for (I) and (W) on that class (\cref{cor:structured}), for instance at $E=1/6$ inside the writer-solvent window; this does not move the window for unrestricted networks.

\paragraph{Barriers.} The barriers concern the gadget model of \cref{def:gadget,def:unary}: input wires are disjoint, output rails are receivers of the gadget (no output is an alias of an input rail), and every other bank enters as a fixed bank whose box does not depend on the condition. Within this model, for $\alpha\le E<1/2$, no writer-solvent NAND gadget for (I) certified by interval propagation, of any topology (feedback, mixing receivers, rereads of raw inputs, intermediate values), has an output bound at most the smaller of its input bounds (\cref{thm:barrier}). From an input bound at least the lower fixed point $\lambda_-$ of the inverter map, a unary gadget never produces an output bound below $\lambda_-$; below $\lambda_-$ it never lowers the bound; and outside $(\lambda_-,\lambda_+)$, and for every input bound $\gamma$ if $E>3-2\sqrt2$, every output bound is at least an uncapped transfer bound $\psi(\gamma)$, which is the bound of the best single receiver when it lies below $1-E$ and leaves no output wire at all otherwise (\cref{thm:unary}). No NAND gadget takes one input with bound at least $\lambda_+$ and the other with bound at least $\lambda_-$ (\cref{thm:asym}), and \PCirc\ instances over any of the PPAD-complete gate sets, realized by interval-certified gadgets with arbitrary bounds (one per wire), are solvable in polynomial time (\cref{thm:nonuniform}). Unless PPAD is contained in FP, every escape route is capped at $3-2\sqrt2$ by the algorithm (\cref{cor:ceiling}). With the bound of a wire replaced by an effective bound computed from the intervals of its rails, these statements extend to wire interfaces in which each rail has an arbitrary fixed pair of disjoint low and high intervals, possibly non-saturated and different for the two rails of a wire (\cref{prop:ifgates,thm:interfaces}). Gadgets certified on exact solution sets, correlation-aware certificates, alias outputs and other source problems remain open (\cref{sec:routes}).

\paragraph{(D) for every error below $1$.} (D)-$E$ is PPAD-complete on non-degenerate networks for every rational $E\in(0,1)$, already with one writer (\cref{thm:D}); for $E\ge1$ it is trivial. The solvency clause forces exact values, so one receiver per NAND gate suffices and no error accumulates.

\paragraph{(D) at bounded held degree.} On the restricted unit-notional class $\mathcal R(d)$ (\cref{def:Rd}), in which every core bank has no external assets, owes a unit debt and holds at most $d$ CDSs of notional $1$, counted with multiplicity, we classify (D) by $d$ (\cref{thm:degmain}). Let $\rho=(1+E)/(1-E)$. For every rational $E\in[1/2,1)$, (D)-$E$ is solvable in polynomial time for $d<(\lceil\rho\rceil+1)\rho$ and PPAD-complete for $d\ge(\lceil\rho\rceil+1)\rho$; at $E=1/3$ it is solvable in polynomial time for $d\le5$ and PPAD-complete for $d\ge6$. It is PPAD-complete at held degree $4$ for $E\le3/13$, at held degree $3$ for $E<1/7$ and at held degree $2$ for $E<1/9$, while held degree $1$ admits an exact clearing vector in linear time. Assuming that PPAD is not contained in FP, the threshold error $p_d$ is determined for $d\in\{6,7\}$ and all $d\ge12$, it is itself a PPAD-hard error, and $\sqrt d\,(1-p_d)\to2$; \cref{tab:phase} gives the full phase diagram and the degrees that remain open. Hardness holds also for (D$^\circ$) and (S). The lower bounds use the sparse NAND and two-arm PURIFY topology, which appears with weighted contracts in~\cite{DHK} and in \cref{thm:D}, with a new placement of unit multiplicities (\cref{thm:dH}), and add restoring chains at low degree (\cref{thm:lowdeg}); the upper bounds refine the supported propagation of~\cite{KM} (\cref{thm:Pk,thm:A2,thm:A3}). Each method is tight at its own boundary, for its own decoder (\cref{app:limits}). These results concern (D) only: on $\mathcal R(d)$, (I) and (W) are solvable in polynomial time for every error above $3-2\sqrt2$ at every degree (by the box algorithm below $1/2$ and trivially from $1/2$ on), so the windows for (I) and (W) are unchanged.

\paragraph{Membership without non-degeneracy.} (W)-$E$ and (I)-$E$ are in PPAD on all networks, with $E$ given in binary (\cref{thm:member}): capping recoveries at $1-E/2$ keeps every nonzero liability bounded away from zero, at the cost of an additive $E/2$. In particular (W)-$E$ vectors exist on every network for every $E>0$.

\paragraph{The exempted notion.} (D$^\circ$)-$E$ and (S)-$E$ are in PPAD on all networks, by a rounding that snaps recoveries to $1$ across an empty interval among $n+1$ geometric scales (\cref{thm:gap}), and they are PPAD-complete for every fixed rational $E\in(0,1)$ (\cref{cor:Dcirc}).

\paragraph{(D) is not total.} Without non-degeneracy, (D)-$E$ vectors need not exist: a network with three banks has none, for every $E\in(0,1)$ and every convention for $0/0$ (\cref{prop:Dfail}). So the question of~\cite[Appendix~A]{DHK} has a negative answer for (D) as written and a positive one for (D$^\circ$).

\paragraph{Exact existence and stalled boxes.} Writer-solvent networks always have exact clearing vectors (\cref{prop:wsexact}); for non-degenerate networks this is known~\cite{DHK,IKV22}, and the assumption cannot be dropped (\cref{prop:ndexact}). Where the box algorithm stalls on writer-solvent networks of inverters, the stall determines an affine system whose solution, when it lies in the stalled box, is exact on the unsettled banks and, with the settled points, forms an (I)-$E$ vector, which is an exact clearing vector of the whole network if, in addition, no bank with $e_i<L_i^{\max}$ settled before the stall (\cref{prop:stall,prop:affinerule}). The solution does lie in the box when two banks are unsettled and for symmetric cycles of any length (\cref{prop:stallcases}), and it did in all $28$ stalls of a computer search reported in~\cite[record~Lsolo]{Supp} (\cref{obs:stalls}); but it does not in general: an example with three unsettled banks inside the writer-solvent window has no exact clearing vector in its stalled box (\cref{ex:stallfail}). A global rule for stalled configurations remains open (\cref{q:stall}). At exact inverter stalls the selectors are exactly the points of finitely many polyhedra, one for each set of saturated banks held at the top of their boxes (a point of such a polyhedron may have fewer clipped banks, \cref{prop:clipsets}); this family need not be closed under intersection (\cref{prop:nointersection}), an exact run can converge without ever stalling (\cref{prop:nostall}), and even an affine residual band can be infeasible on a reached stalled box (\cref{prop:affineband}).

\begin{table}[t]
\centering\small
\begin{tabular}{@{}llllc@{}}
\toprule
networks & notion & PPAD-hard for $E<$ & polynomial for $E>$ & see\\
\midrule
non-degenerate & (D) & $1$ & trivial for $E\ge1$ & \ref{thm:D}\\
all & (D$^\circ$), (S) & $1$ & trivial for $E\ge1$ & \ref{cor:Dcirc}\\
all & (I), (W) & $\EI=0.25299586\ldots$ & $\elad=0.28336553\ldots$ & \ref{thm:general}, \ref{thm:alggen}\\
writer-solvent & (I), (W) & $\alpha=0.15767078\ldots$ & $3-2\sqrt2=0.17157287\ldots$ & \ref{thm:wsI}, \ref{cor:oneW}, \ref{thm:algws}\\
one writer & (W) & $\alpha=0.15767078\ldots$ & $3-2\sqrt2=0.17157287\ldots$ & \ref{cor:oneW}, \ref{thm:algws}\\
\midrule
$\mathcal R(d)$, $d\in\{6,7\}$ or $d\ge12$ & (D) & $\bar p_d$ (inclusive) & $\bar p_d$ & \ref{thm:degmain}\\
\bottomrule
\end{tabular}
\caption{The landscape, for rational errors $E$; the last column gives the theorem numbers. All problems are in PPAD on all networks (\cref{thm:member,thm:gap}), except (D), which is in PPAD on non-degenerate networks~\cite{DHK} and is not total in general (\cref{prop:Dfail}). In each of the rows for (I) and (W) the hard and polynomial ranges are separated by a window that is open. In the last row, $\bar p_d$ is the explicit rational of \cref{sec:degclass} ($1/3$ for $d=6,7$): (D)-$\bar p_d$ is PPAD-hard and (D)-$E$ is polynomial for every rational $E>\bar p_d$, unconditionally (\cref{thm:degmain}(c)), and if PPAD is not contained in FP, then $\bar p_d$ equals the infimum $p_d$ of the polynomially solvable errors. The restricted class has its own phase diagram, \cref{tab:phase}.}\label{tab:landscape}
\end{table}

\subsection{Conventions of this paper}

\paragraph{Status labels.} Statements marked \KN\ are from the literature. Every other theorem, proposition, lemma and corollary has a complete proof in this paper, or, where explicitly indicated, a precise pointer to the research record in which the proof is written out. Statements marked \CO\ report exact finite computations or numerical diagnostics, as stated in each case; they support the proofs but are never used in place of them.

\paragraph{Provenance and availability.} The research records~\cite[records~L1, L3, L7, L11, L13, L14, L15, L16, L17, L18, L19, L21]{Supp} provide auxiliary proof details and computational certificates cited in this paper. The results of \cref{sec:Rclass,sec:rounded,sec:interfaces,sec:degree,sec:selection} and \cref{app:chains,app:limits,app:weighted,app:stalls} were developed and checked in a later series of research records~\cite[records~S1, S2, S3, S4, S5]{Supp}; all their proofs are given in full in this paper. The public support bundle of this version~\cite{Supp} contains a scientific-only derivative of the earlier curated reproducibility package, with its mathematics preserved and historical machine paths normalized, the scientific notes of the later records, as byte-identical or curated copies with a map to their sources, the exact-arithmetic programs with their expected outputs, and a replay script. \Cref{app:verification} distinguishes proofs given here, external auxiliary arguments and unavailable historical inputs, and records the scope of the literature checks. Separate implementations are identified by their technical independence from the producing code; this is not a claim of independent human verification. The general claim of the source note~\cite[record~Lsolo]{Supp} is refuted in \cref{sec:stalls}. The results and qualifications stated in this paper take precedence over superseded claims in the research records.

\paragraph{Organization.} \Cref{sec:prelim} fixes the model, the notions and the restricted class $\mathcal R(d)$. \Cref{sec:general,sec:onewriter,sec:coincidence} give the hardness results for (I), (W) and (D) on general networks, the hardness results on writer-solvent and one-writer networks with the calibration copies, and the coincidence of thresholds with the barrier at $\alpha$. \Cref{sec:box} gives the box algorithm, its optimality among local algorithms and its finite-precision version, and \cref{sec:barriers} the remaining barriers and their scope. \Cref{sec:degree} classifies (D) on $\mathcal R(d)$ by held degree. \Cref{sec:membership,sec:existence,sec:selection} treat membership without non-degeneracy, exact existence and stalled boxes, and selection at near-stalls on structured networks. \Cref{sec:literature} discusses the literature and \cref{sec:open} open problems. The appendices follow the order of the sections: \cref{app:ladder,app:hardnessaux} belong to \cref{sec:general}, \cref{app:onewriter} to \cref{sec:onewriter,sec:coincidence}, \cref{app:box,app:localaux} to \cref{sec:box}, \cref{app:routes} to \cref{sec:barriers}, \cref{app:degree,app:chains,app:limits,app:weighted} to \cref{sec:degree}, \cref{app:membership,app:stalls} to \cref{sec:membership,sec:existence}, and \cref{app:selection} to \cref{sec:selection}. A proof given in an appendix is headed by the name of the statement it proves. \Cref{app:verification} records the dependencies on the research records and the scope of the literature checks.

\section{Preliminaries}\label{sec:prelim}

\subsection{Networks and the clearing map}\label{sec:model}

A \emph{network} has a finite set $N=\{1,\dots,n\}$ of banks and nonnegative rational data: external assets $e_i\ge0$, ordinary debts $c_{ij}\ge0$ with $c_{ii}=0$, and a finite list of credit default swaps. A CDS $(i,j,k;c)$ has a \emph{writer} $i$, a \emph{holder} $j$, a \emph{reference} $k$ and a notional $c>0$; the writer owes the holder $c(1-r_k)$, where $r_k$ is the recovery rate of the reference. Throughout the paper the three roles of every CDS are pairwise distinct banks. We write $c^k_{ij}$ for the total notional of the CDSs with writer $i$, holder $j$ and reference $k$. There are no default costs.

For a recovery vector $r\in[0,1]^n$ put
\[
\ell_{ij}(r)=c_{ij}+\sum_kc^k_{ij}(1-r_k),\qquad \ell_i(r)=\sum_j\ell_{ij}(r),\qquad A_i(r)=e_i+\sum_jr_j\,\ell_{ji}(r),
\]
the liability of $i$ towards $j$, the total liability of $i$, and the assets of $i$. The \emph{clearing map} is
\[
\Phi_i(r)=\begin{cases}\min\bigl(1,A_i(r)/\ell_i(r)\bigr),&\ell_i(r)>0,\\ 1,&\ell_i(r)=0.\end{cases}
\]
The convention at $\ell_i(r)=0$ is that of~\cite[Definition~1]{IKV23} and~\cite{SSB20}. Since $c_{ii}=0$ and the roles of every CDS are distinct, $\Phi_i(r)$ does not depend on $r_i$. A \emph{clearing vector} is a fixed point $r=\Phi(r)$. On networks that are not non-degenerate (\cref{def:classes}) clearing vectors need not exist (\cref{sec:existence}).

Let $L_i^{\max}=\sum_jc_{ij}+\sum_{j,k}c^k_{ij}$ be the largest possible liability of $i$, and
\[
T=\{i:\ e_i>L_i^{\max}\},\qquad F=\{i:\ e_i\ge L_i^{\max}\}\supseteq T .
\]
Banks of $T$ are \emph{ex-ante solvent}. If $i\in F$ then $\Phi_i\equiv1$: either $\ell_i(r)=0$, or $A_i(r)\ge e_i\ge L_i^{\max}\ge\ell_i(r)>0$.

\subsection{Approximate clearing}\label{sec:notions}

For $x\in\mathbb R$ let $\cl x=\min(1,\max(0,x))$.

\needspace{4\baselineskip}
\begin{definition}[approximation notions]\label{def:notions}
Let $E\ge0$ and $r\in[0,1]^n$.
\begin{itemize}
\item[(W)] $r$ is a \emph{(W)-$E$ vector} if $|r_i-\Phi_i(r)|\le E$ for every bank $i$.
\item[(I)] $r$ is an \emph{(I)-$E$ vector} if it is a (W)-$E$ vector and $r_i=1$ for every $i\in T$.
\item[(D)] $r$ is a \emph{(D)-$E$ vector} if it is a (W)-$E$ vector and $A_i(r)\ge(1+E)\ell_i(r)$ implies $r_i=1$, for every bank $i$.
\item[(D$^\circ$)] $r$ is a \emph{(D$^\circ$)-$E$ vector} if both requirements of (D) hold for every bank $i$ with $\ell_i(r)>0$; banks with $\ell_i(r)=0$ are exempt.
\item[(S)] $r$ is an \emph{(S)-$E$ vector} if $\cl{A_i(r)/\ell_i(r)-E}\le r_i\le\cl{A_i(r)/\ell_i(r)+E}$ for every bank $i$ with $\ell_i(r)>0$; banks with $\ell_i(r)=0$ are exempt.
\end{itemize}
\end{definition}

The notion (W) is the weak approximation of~\cite{IKV22,DHK}; \cite[Section~3]{DHK} relate it to the notion of~\cite{SSB19}. The notion (I) is the definition of~\cite[Definition~4]{IKV23}, whose first clause fixes the ex-ante solvent banks; the word ``ex-ante'' is ours. The notion (D) is~\cite[Definition~3]{DHK}, the notion for which~\cite{DHK} define the problem $\eps$-CDS-CLEARING. The exempted notion (D$^\circ$) and the clipped-ratio notion (S), the latter modelled on the $\eps$-solutions of~\cite[Definition~4.1]{SSB17}, are introduced here to separate the role of zero-liability banks (\cref{sec:membership}). The paper~\cite{DHK} writes the map as $f_i=a_i/\max(a_i,\ell_i)$; it agrees with $\Phi_i$ whenever $\max(a_i,\ell_i)>0$, and in particular everywhere on non-degenerate networks.

\begin{lemma}\label{lem:relations}
Let $0\le E'\le E$.
\begin{enumerate}[label=(\alph*)]
\item Every (I)-$E$ vector is a (W)-$E$ vector, every (D)-$E$ vector is a (D$^\circ$)-$E$ vector, and every (S)-$E$ vector is a (D$^\circ$)-$E$ vector.
\item For each of the five notions, every $E'$-vector is an $E$-vector.
\item If $E\ge1$, the all-ones vector is an $E$-vector for all five notions. If $E\ge1/2$, the vector with $r_i=1$ on $T$ and $r_i=1/2$ elsewhere is an (I)-$E$ vector.
\end{enumerate}
\end{lemma}
\begin{proof}
(a) The first two claims are immediate. Let $r$ be an (S)-$E$ vector and $\ell_i(r)>0$, and write $\rho=A_i(r)/\ell_i(r)$. If $\rho\ge1+E$, then $\cl{\rho-E}=1$, so $r_i=1$. If $\rho\le1$, then $r_i\in[\rho-E,\rho+E]$, so $|r_i-\Phi_i(r)|\le E$. If $\rho>1$, then $r_i\ge\min(1,\rho-E)\ge1-E$, so $|r_i-1|\le E$.
(b) The residual condition and the interval in (S) only get weaker as $E$ grows, and the hypothesis $A_i\ge(1+E)\ell_i$ of the (D) clause implies $A_i\ge(1+E')\ell_i$.
(c) Every residual is at most $1$, the clauses of (D) and (D$^\circ$) and the condition $r_T=\mathbf 1$ hold trivially, and $\cl{\rho+E}=1$ for $\rho\ge0$, $E\ge1$. For the second vector, $|1/2-\Phi_i|\le1/2$, and $\Phi_i\equiv1$ on $T$.
\end{proof}

By (c) we restrict attention to $0<E<1$, and for (I) and (W) to $0<E<1/2$. By (b), PPAD-hardness at an error $E$ implies hardness at every $E'<E$, and polynomial solvability at $E$ implies it at every $E'>E$. In our reductions $E$ is a fixed rational; in the membership results of \cref{sec:membership} it may be part of the input, written in binary.

\subsection{Network classes}

\begin{definition}[network classes]\label{def:classes}
A network is \emph{non-degenerate} if every bank has $e_i>0$ or some $c_{ij}>0$~\cite[Definition~2]{DHK}. It is \emph{writer-solvent} if every CDS writer lies in $T$, and \emph{one-writer} if exactly one bank writes CDSs and that bank lies in $T$.
\end{definition}

On a non-degenerate network every bank with $\ell_i(r)=0$ has $e_i>0$, so the convention at $\ell_i=0$ plays no role there. Other papers use other non-degeneracy conditions:~\cite[Definition~4.2]{SSB17} asks only that every CDS writer owe a debt or have positive external assets, and~\cite[Definition~2]{IKV23} adds that every reference owes a debt. For \cref{def:classes} the following problem names are used throughout:
\begin{itemize}
\item \WSI-$E$ and \WSW-$E$: find an (I)-$E$, respectively (W)-$E$, vector of a non-degenerate writer-solvent network;
\item \OWW-$E$: find a (W)-$E$ vector of a non-degenerate one-writer network.
\end{itemize}
The algorithms of \cref{sec:box} need neither non-degeneracy nor the strict inequality in $T$: their writer-solvent bound holds whenever every CDS writer lies in $F$.

\subsection{Search problems, PPAD and Pure-Circuit}\label{sec:pc}

For a notion $X\in\{\mathrm W,\mathrm I,\mathrm D,\mathrm D^\circ,\mathrm S\}$ and a class of networks, the search problem $X$-$E$ asks for an $X$-$E$ vector of a given network of the class. All data are rationals in binary. A \emph{polynomial-time reduction} from a search problem $\Pi_1$ to $\Pi_2$ consists of two polynomial-time maps: one sends an instance $I$ of $\Pi_1$ to an instance $f(I)$ of $\Pi_2$, the other sends $I$ and any solution of $f(I)$ to a solution of $I$. PPAD~\cite{Pap94} is closed under such reductions.

Our hardness results reduce from \PCirc~\cite{DFHM22}. An instance consists of a set $V$ of \emph{nodes} and a set of \emph{gates}. Each gate has a type and a tuple of pairwise distinct nodes, and every node is the output of exactly one gate~\cite[Definition~1]{DFHM22}. A solution is an assignment $x:V\to\{0,1,\bot\}$ that satisfies the robust constraint of every gate:
\begin{center}\small
\begin{tabular}{@{}ll@{}}
\toprule
gate & constraint\\
\midrule
$\mathrm{NOT}(u;w)$ & $x_u=0\Rightarrow x_w=1$;\quad $x_u=1\Rightarrow x_w=0$\\
$\mathrm{COPY}(u;w)$ & $x_u\in\{0,1\}\Rightarrow x_w=x_u$\\
$\mathrm{AND}(u,v;w)$ & $x_u=x_v=1\Rightarrow x_w=1$;\quad $0\in\{x_u,x_v\}\Rightarrow x_w=0$\\
$\mathrm{OR}(u,v;w)$ & $x_u=x_v=0\Rightarrow x_w=0$;\quad $1\in\{x_u,x_v\}\Rightarrow x_w=1$\\
$\mathrm{NAND}(u,v;w)$ & $x_u=x_v=1\Rightarrow x_w=0$;\quad $0\in\{x_u,x_v\}\Rightarrow x_w=1$\\
$\mathrm{NOR}(u,v;w)$ & $x_u=x_v=0\Rightarrow x_w=1$;\quad $1\in\{x_u,x_v\}\Rightarrow x_w=0$\\
$\mathrm{PURIFY}(u;v,w)$ & $x_u\in\{0,1\}\Rightarrow x_v=x_w=x_u$;\quad $\{x_v,x_w\}\cap\{0,1\}\ne\emptyset$\\
\bottomrule
\end{tabular}
\end{center}

\begin{theorem}[\cite{DFHM22}, Theorem~2.1 and Corollary~2.2]\label{thm:pc}\KN\
\PCirc\ is PPAD-complete for each of the gate sets $\{\mathrm{PURIFY},\mathrm{NAND}\}$, $\{\mathrm{PURIFY},\mathrm{NOR}\}$, $\{\mathrm{PURIFY},\mathrm{NOT},\mathrm{OR}\}$ and $\{\mathrm{PURIFY},\mathrm{NOT},\mathrm{AND}\}$.
\end{theorem}

We list the COPY gate as well because the barriers of \cref{sec:barriers} cover it.

\subsection{Constants}\label{sec:constants}

\Cref{tab:constants} lists the constants that recur in the paper, with their defining polynomials. Every digit was computed by the program \texttt{code/constants.py}, which isolates each root by exact real-root isolation over $\mathbb Q$ and evaluates it to $60$ significant digits. The programs of the support bundle~\cite{Supp} also check the polynomial identities and the finite facts used in our proofs. The two thresholds that matter most are $\EI$, the unique root in $[1/5,27/100]$ of
\begin{multline*}
P_I(E)=32E^{11}-352E^{10}+1728E^9-4944E^8+9128E^7-11352E^6\\
+9650E^5-5572E^4+2128E^3-510E^2+69E-4,
\end{multline*}
and $\elad$, the unique root in $(0,1/2)$ of $Q_5(E)=5E^5-45E^4+52E^3-32E^2+20E-4$.

\begin{table}[ht]
\centering\small
\begin{tabular}{@{}lll@{}}
\toprule
constant & defining condition & value\\
\midrule
$\alpha=(13-\sqrt{153})/4$ & $2E^2-13E+2=0$ & $0.15767078078675458763394260801944223113960058\ldots$\\
$3-2\sqrt2$ & $E^2-6E+1=0$ & $0.17157287525380990239662255158060384286065624\ldots$\\
$\EI$ & $P_I(E)=0$ & $0.25299586217212414967585134658539148534836894\ldots$\\
$\elad$ & $Q_5(E)=0$ & $0.28336553249330853976327847784983398726912758\ldots$\\
$1-1/\sqrt2$ & $2E^2-4E+1=0$ & $0.29289321881345247559915563789515096071516406\ldots$\\
$(3-\sqrt5)/2$ & $E^2-3E+1=0$ & $0.38196601125010515179541316563436188227969082\ldots$\\
\midrule
$\sqrt5-2$ & $E^2+4E-1=0$ & $0.23606797749978969640917366873127623544061835\ldots$\\
$5-2\sqrt6$ & $E^2-10E+1=0$ & $0.10102051443364380360543185058821721606810503\ldots$\\
\bottomrule
\end{tabular}
\caption{Recurring constants. The first block lists the four thresholds that delimit our main results and the one-round thresholds $1-1/\sqrt2$ and $(3-\sqrt5)/2$; $5-2\sqrt6$ is the hardness constant of~\cite{DHK}.}\label{tab:constants}
\end{table}

\subsection{The restricted unit-notional class}\label{sec:Rclass}

\Cref{sec:degree} studies (D) on networks of the following normal form, in which every contract that carries information is a CDS of notional $1$ held by a bank of a core of banks with unit debts. Held contracts are counted with multiplicity.

\begin{definition}[restricted unit-notional class]\label{def:Rd}
Let $d\ge1$ be an integer. A network of $\mathcal R(d)$ consists of a finite set $V$ of \emph{core} banks, one \emph{creditor} $z$, and \emph{writers}. Every core bank $b$ has external assets $e_b=0$, owes an ordinary debt $1$ to $z$ and no other debt, writes no CDS, and holds a finite list $N_b$ of CDSs of notional $1$, each written by a writer and referencing a core bank other than $b$. Repetitions of a reference in $N_b$ are allowed and counted. The \emph{held degree} of $b$ is $|N_b|$, counted with multiplicity, and it is at most $d$. Writers hold no contract, owe no debt, and satisfy $e_w\ge L^{\max}_w$, so they lie in $F$. The creditor has $e_z>0$ and no liabilities, and holds no CDS; it receives the ordinary debt claims of the core banks. Every writer writes at least one contract, and every CDS has three distinct parties.
\end{definition}

A bank without contracts and liabilities can be removed from a network and given recovery $1$ without affecting any other bank or any clause of the five notions, so the requirement that every writer writes a contract is no restriction. Every network of $\mathcal R(d)$ is non-degenerate (\cref{def:classes}): every core bank owes a debt, $e_z>0$, and every writer has $e_w\ge L^{\max}_w\ge1$. The class is a normal form within the restricted financial class of~\cite[Definition~F.6]{KM}: there, too, core banks have zero external assets and unit debts to a creditor, receive only CDSs of notional $1$, and repeated references are counted with multiplicity, and the funded writers are assigned recovery $1$. The restrictions on the writers stated above (no holdings, no ordinary debts) are part of \cref{def:Rd}; we do not assert that it coincides with every choice of peripheral banks allowed in~\cite{KM}.

For a recovery vector $r$ put $x_b=1-r_b$ for $b\in V$ (the \emph{default} of $b$), and
\[
S_b(x)=\sum_{k\in N_b}x_k,\qquad T_b(x)=\max\{0,\,1-S_b(x)\},
\]
where the sum runs over the list $N_b$, with multiplicity. From here on, in this subsection, in \cref{sec:degree} and in \cref{app:degree,app:chains,app:limits}, we write $A=1+E$. In default coordinates the assets of a core bank are denoted by $S_b$; the notation $A_i(r)$ of \cref{sec:model} for assets appears there only in \cref{lem:core} and where the notions (D$^\circ$) and (S) are checked, always with an index and an argument, so it is not confused with the constant $A$.

\begin{lemma}[core equations]\label{lem:core}
Consider a network of $\mathcal R(d)$ and let $E\ge0$.
\begin{enumerate}[label=(\alph*)]
\item If $r_w=1$ for every writer $w$ with $\ell_w(r)>0$, then every core bank has $\ell_b(r)=1$ and $A_b(r)=S_b(x)$, so $\Phi_b(r)=1-T_b(x)$.
\item A vector $r$ with $r=1$ on all writers and on $z$ is a (D)-$E$ vector if and only if every core bank $b$ satisfies
\begin{equation}\label{eq:core}
|x_b-T_b(x)|\le E\qquad\text{and}\qquad\bigl(S_b(x)\ge1+E\ \Longrightarrow\ x_b=0\bigr).
\end{equation}
\item If every writer has $e_w\ge(1+E)L^{\max}_w$, then every (D)-$E$ vector has $r=1$ on all writers and on $z$, and every (D$^\circ$)-$E$ vector has $r_w=1$ on every writer with $\ell_w(r)>0$. In both cases (a) applies and \eqref{eq:core} holds at every core bank.
\end{enumerate}
\end{lemma}
\begin{proof}
(a) A core bank owes exactly $1$ and writes no CDS, so $\ell_b=1$. Its assets are $e_b=0$ plus the payments $r_w(1-r_k)$ of its held unit CDSs $(w,b,k)$. If $r_k<1$ then $\ell_w(r)\ge1-r_k>0$, so $r_w=1$; if $r_k=1$ the payment is $0$ whatever $r_w$ is. So $A_b=\sum_{k\in N_b}(1-r_k)=S_b(x)$ and $\Phi_b=\min(1,S_b)=1-T_b$.

(b) Writers lie in $F$, so $\Phi_w\equiv1$ (\cref{sec:model}), and $\ell_z\equiv0$ gives $\Phi_z=1$. With $r=1$ at these banks both requirements of (D) hold there. At a core bank $|r_b-\Phi_b|=|x_b-T_b|$ by (a), and the (D) clause $A_b\ge(1+E)\ell_b\Rightarrow r_b=1$ is the second part of~\eqref{eq:core}.

(c) If $\ell_w(r)>0$, then $A_w\ge e_w\ge(1+E)L^{\max}_w\ge(1+E)\ell_w$, so the (D) clause gives $r_w=1$. If $\ell_w(r)=0$, the premise $A_w\ge(1+E)\cdot0$ holds as well, so (D) gives $r_w=1$; the same applies to $z$. Under (D$^\circ$) banks with $\ell=0$ are exempt, but (a) only needs the writers with $\ell_w>0$. Core banks have $\ell_b=1>0$, so they are never exempt, and~\eqref{eq:core} follows from (a) as in (b).
\end{proof}

A writer with $L^{\max}_w\le e_w<(1+E)L^{\max}_w$ need not be forced to recovery $1$ by (D) (it is forced whenever $\ell_w(r)\le e_w/(1+E)$); on such networks a (D)-$E$ vector need not satisfy~\eqref{eq:core}, since core assets are then scaled by $r_w\in[1-E,1)$. So (b) is an identity for vectors whose outside recoveries are chosen to be $1$, while (c) gives the identity for all solutions under the stronger funding. Nothing below depends on more than this: the algorithms of \cref{sec:degree} output $r_w=1$, which (b) allows, and the hardness instances use $e_w=3>(1+E)L^{\max}_w=1+E$.

\begin{lemma}[monotonicity]\label{lem:Rmono}
Let $X$ be one of the notions (D), (D$^\circ$), (S). If $X$-$E$ on $\mathcal R(d)$ is PPAD-hard, then so is $X$-$E'$ on $\mathcal R(d')$ for all rationals $E'\le E$ and integers $d'\ge d$. If $X$-$E$ on $\mathcal R(d)$ is in FP, then so is $X$-$E'$ on $\mathcal R(d')$ for all rationals $E'\ge E$ and integers $d'\le d$.
\end{lemma}
\begin{proof}
$\mathcal R(d)\subseteq\mathcal R(d')$ for $d\le d'$, and every $X$-$E'$ vector is an $X$-$E$ vector when $E'\le E$ (\cref{lem:relations}(b)).
\end{proof}

In default coordinates \eqref{eq:core} reads: if $S_b\ge A$ then $x_b=0$; if $1\le S_b<A$ then $0\le x_b\le E$; if $S_b<1$ then $\max(0,1-E-S_b)\le x_b\le\min(1,1+E-S_b)$. We use two consequences throughout:
\begin{itemize}
\item[(Sol)] $S_b\ge A$ forces $x_b=0$;
\item[(Res)] $x_b\ge1-E-S_b$ always (for $S_b<1$ this is the residual bound; for $S_b\ge1$ the right side is negative).
\end{itemize}
For $0\le E<1$ and $S\ge0$ put $\mathrm{lo}(S)=\max\{0,1-E-S\}$ and
\[
\mathrm{up}(S)=\begin{cases}1,&S<E,\\1+E-S,&E\le S<1,\\E,&1\le S<A,\\0,&S\ge A.\end{cases}
\]
Both functions are nonincreasing, $\mathrm{lo}$ is continuous, and $\mathrm{lo}\le\mathrm{up}$. Then~\eqref{eq:core} at a core bank with assets $S$ says exactly that its default lies in $\mathrm{Adm}(S):=[\mathrm{lo}(S),\mathrm{up}(S)]$, which is $\{0\}$ when $S\ge A$. Moreover, for $0<E<1$, $\mathrm{up}(S)\ge E>0$ whenever $S<A$, and $\mathrm{lo}(S)<1-E$ whenever $S>0$.

\section{Hardness on general networks}\label{sec:general}

This section presents the two hardness results that hold without restrictions on who writes CDSs. For the exact-solvency notion (D) one receiver per NAND gate suffices, because the solvency clause forces exact values; this gives hardness for every error below~$1$ (\cref{sec:D}). For (I) and (W) no value is forced exactly, and the gadgets must restore signals that degrade by $E$ at every bank. We do this with ladders of bank pairs, and the construction works for every rational $E<\EI$ (\cref{thm:general}); \cref{sec:ladderoverview} outlines it, and \cref{app:ladder} gives the construction and the proof.

\subsection{The exact-solvency notion (D)}\label{sec:D}

\begin{theorem}\label{thm:D}
For every rational $0<E<1$, (D)-$E$ is PPAD-complete on non-degenerate networks. Hardness holds already for one-writer networks in which every bank owes a positive ordinary debt; on such networks every liability is positive, so (D), (D$^\circ$) and (S) are decoded by the same reduction. For $E\ge1$ the all-ones vector is a (D)-$E$ vector.
\end{theorem}

\paragraph{Construction.} Fix a rational $E\in(0,1)$ and put
\[
\gamma=\frac{1+E}2,\qquad C=\frac{2(1+E)}{1-E},\qquad C_A=C^2,\qquad \theta=1-\frac{1+E}{C_A}.
\]
These satisfy
\begin{equation}\label{eq:Did}
C(1-\gamma)=1+E,\qquad C_A(1-\theta)=1+E,\qquad C(1-\theta)=\frac{1-E}2,\qquad C(1-E)=2(1+E),
\end{equation}
and $E<\gamma<\theta<1$. A recovery $r$ is decoded as $1$ if $r=1$, as $0$ if $r\le\gamma$, and as $\bot$ otherwise. Given an instance of \PCirc\ with gates NAND and PURIFY, build a network with a source $s$, a sink $z$ and \emph{receivers}. Every receiver has $e=0$, owes an ordinary debt $1$ to $z$, and writes no CDS. There is a receiver $b_x$ for every node $x$, and two further receivers $a_g,b_g$ for every PURIFY gate $g$. All CDSs are written by $s$:
\begin{itemize}
\item for $\mathrm{NAND}(u,v;w)$, the receiver $b_w$ holds a CDS of notional $C$ on $b_u$ and one of notional $C$ on $b_v$;
\item for $g=\mathrm{PURIFY}(u;v,w)$, the receiver $a_g$ holds a CDS of notional $C_A$ on $b_u$, the receiver $b_g$ holds one of notional $C$ on $b_u$, the receiver $b_v$ holds one of notional $C$ on $a_g$, and $b_w$ holds one of notional $C$ on $b_g$.
\end{itemize}
Finally $s$ owes $1$ to $z$, $z$ owes $1$ to $s$, $e_z=3$, and $e_s=2L_s^{\max}+1$. The nodes of a gate are pairwise distinct, so every CDS has three distinct parties. Every bank owes a positive ordinary debt, so the network is non-degenerate, and $s$ is its only CDS writer and lies in $T$.

\begin{lemma}\label{lem:Dgates}
Let $r$ be a (D$^\circ$)-$E$ vector of the network. Then $r_s=r_z=1$, and:
\begin{enumerate}[label=(\alph*)]
\item for $\mathrm{NAND}(u,v;w)$: if $r_{b_u}\le\gamma$ or $r_{b_v}\le\gamma$ then $r_{b_w}=1$; if $r_{b_u}=r_{b_v}=1$ then $r_{b_w}\le E$;
\item for $g=\mathrm{PURIFY}(u;v,w)$ and $\rho=r_{b_u}$: if $\rho\le\gamma$ then $r_{b_v},r_{b_w}\le E$; if $\rho=1$ then $r_{b_v}=r_{b_w}=1$; if $\rho\le\theta$ then $r_{b_v}\le E$; if $\rho\ge\theta$ then $r_{b_w}=1$.
\end{enumerate}
\end{lemma}
\begin{proof}
Every liability is positive, so both requirements of (D) hold at every bank. We have $A_s\ge e_s>2L_s^{\max}\ge(1+E)\ell_s$ and $A_z\ge3>(1+E)\ell_z$, so the clause gives $r_s=r_z=1$. A receiver $b$ therefore has $\ell_b=1$ and assets $A_b=\sum c(1-r_k)$ over the CDSs it holds, and the clause says that $A_b\ge1+E$ forces $r_b=1$.

(a) If $r_{b_u}\le\gamma$, then $A_{b_w}\ge C(1-\gamma)=1+E$ by~\eqref{eq:Did}, so $r_{b_w}=1$; likewise for $v$. If $r_{b_u}=r_{b_v}=1$, then $A_{b_w}=0$ and $r_{b_w}\le\Phi_{b_w}+E=E$.

(b) If $\rho\le\gamma$, then $A_{a_g}\ge C_A(1-\gamma)\ge C(1-\gamma)=1+E$ and $A_{b_g}\ge1+E$, so $r_{a_g}=r_{b_g}=1$; hence $A_{b_v}=A_{b_w}=0$ and both outputs are at most $E$. If $\rho=1$, then $A_{a_g}=A_{b_g}=0$, so $r_{a_g},r_{b_g}\le E$, and $A_{b_v},A_{b_w}\ge C(1-E)=2(1+E)$, so both outputs equal $1$. If $\rho\le\theta$, then $A_{a_g}\ge C_A(1-\theta)=1+E$, so $r_{a_g}=1$ and $r_{b_v}\le E$. If $\rho\ge\theta$, then $A_{b_g}\le C(1-\theta)=(1-E)/2$, so $r_{b_g}\le(1-E)/2+E=\gamma$, and $A_{b_w}\ge C(1-\gamma)=1+E$ gives $r_{b_w}=1$.
\end{proof}

\begin{proof}[Proof of \cref{thm:D}]
Let $r$ be a (D$^\circ$)-$E$ vector, in particular any (D)-$E$ or (S)-$E$ vector (\cref{lem:relations}), and let $x_y$ be the decoded value of $r_{b_y}$. For $\mathrm{NAND}(u,v;w)$: if $x_u=0$ or $x_v=0$, then $x_w=1$; if $x_u=x_v=1$, then $r_{b_w}\le E<\gamma$ and $x_w=0$. For $\mathrm{PURIFY}(u;v,w)$: a pure input is copied to both outputs, and for every $\rho$ either $\rho\le\theta$ and $x_v=0$, or $\rho\ge\theta$ and $x_w=1$. So $x$ solves the \PCirc\ instance, which is PPAD-hard for these gates (\cref{thm:pc}). The construction and the decoding take polynomial time for fixed $E$. Membership of (D)-$E$ on non-degenerate networks in PPAD is~\cite[Appendix~B, Proposition~3]{DHK}. The case $E\ge1$ is \cref{lem:relations}(c).
\end{proof}

The decoding reads bit $1$ as the single value $1$. This is legitimate because every receiver that must carry a $1$ is forced to exactly $1$ by the solvency clause; a CDS on such a receiver then pays exactly nothing, so no error accumulates and one bank per NAND suffices for every $E<1$. Under (I) and (W) nothing outside $T$ is forced exactly, and \cref{sec:ladderoverview,app:ladder} deal with that.

\subsection{Ladders of bank pairs for (I) and (W)}\label{sec:ladderoverview}

Fix $0<E<1/2$ and write $H=1-E$. The reduction for (I) and (W) uses one source, which writes CDSs and is ex-ante solvent, one sink, and \emph{gadget banks} with zero external assets. A bit is carried by a \emph{pair} $(p,n)$ of gadget banks with \emph{bounds} $(g_p,g_n)$: bit $0$ means $r_p\le g_p$ and $r_n\ge H$, and bit $1$ means $r_p\ge H$ and $r_n\le g_n$. The basic step is a one-coordinate refinement (\cref{lem:step}): a new bank $P$ holds a CDS written by $p$ on $n$, owes a debt and writes a CDS on $p$, and then $(P,n)$ is again a pair. In the ideal limit the new bound is $f(g_p,g_n)$, where
\[
f(x,o)=\min\Bigl(1,\ E+\frac{E^2x}{H(1-x)(1-o)}\Bigr).
\]
For $E\le27/100$, words of such steps clean a pair towards the fixed point $(y,y)$ of $x\mapsto f(x,x)$ exactly on an explicit basin $B_E$ (\cref{prop:basin}). A PURIFY gadget splits a cleaned pair into two seeds with a common writer and a common reference, followed by source-funded inverses with a threshold $\tau$ and by two arms; a NAND gadget has a seed of two banks. For $1/5\le E\le27/100$ the proof chooses $\tau$ in the window
\[
\tmin(E)=E+\sqrt{y-E}\ <\ \tau\ <\ T(E)=\frac{1-3E}{1-2E},
\]
which is nonempty exactly for $E<\EI$ (\cref{lem:window}); smaller errors follow from $E=1/5$. In the ideal limit and on $[\EI,27/100]$, every valid PURIFY of this design family needs a threshold in the same window, which is empty there, so that $\EI$ is the exact endpoint of the family (\cref{rem:EIendpoint}).

\begin{theorem}\label{thm:general}
For every rational $0<E<\EI=0.25299586217212414967\ldots$, (I)-$E$ and (W)-$E$ are PPAD-complete on non-degenerate networks, and hence on all networks. The reduction uses one source.
\end{theorem}

The construction, the ideal maps and the basin, the PURIFY window and the proof of \cref{thm:general} are given in \cref{app:ladder}, together with a direct construction for (W), the endpoint of the design family (\cref{rem:EIendpoint}) and explicit certificates near $\EI$ (\cref{rem:EIcert}); the historical comparisons for this family are proved in \cref{app:hardnessaux}.

\section{Writer-solvent networks and one writer: hardness up to \texorpdfstring{$\alpha$}{alpha}}\label{sec:onewriter}

In a writer-solvent network every CDS writer is ex-ante solvent. Under (I) the writers then recover exactly $1$, and every other bank has constant liabilities and assets that are affine in the recovery vector. This section shows that (I) is nevertheless PPAD-hard below $\alpha=(13-\sqrt{153})/4$ with a single writer (\cref{thm:wsI}), and that a calibration argument transfers this to (W), still with a single writer (\cref{thm:calib,cor:oneW}). The same calibration gives (W) hardness on general networks from (I) hardness (\cref{cor:generalW}).

\subsection{Receivers and inverter chains}

Throughout this subsection every CDS is written by one source $s$ with $e_s=2L_s^{\max}+1$, so $s\in T$, and there is one sink $z$ with $e_z=1$ and no liabilities. Every other bank is a \emph{receiver}: it has $e_b=0$, owes an ordinary debt $1$ to $z$, and holds some CDSs written by $s$. Such a network is non-degenerate and one-writer. In every (I)-$E$ vector $r_s=1$, so a receiver has $\ell_b=1$ and
\[
\Phi_b(r)=\min\bigl(1,A_b(r)\bigr),\qquad A_b(r)=\sum_kc^k_{sb}(1-r_k).
\]
Assume $E\le3-2\sqrt2$, and let
\[
h(g)=E+\frac E{1-g}\quad(0\le g<1),\qquad \lambda_\pm=\tfrac12\Bigl(1+E\pm\sqrt{(1-E)^2-4E}\Bigr),\qquad t=\lambda_+-E=1-\lambda_- .
\]
Then $\lambda_\pm$ are the roots of $x^2-(1+E)x+2E$, equivalently of $(x-E)(1-x)=E$, and $t$ is the larger root of $t^2-(1-E)t+E=0$.

\begin{lemma}\label{lem:hmap}
The map $h$ is increasing, $h(\lambda_\pm)=\lambda_\pm$, and $\lambda_-<h(g)<g$ for $g\in(\lambda_-,\lambda_+)$. Hence for every $g_0<\lambda_+$ and every $\tau\in(\lambda_-,\lambda_+)$ there is $m_0$ with $h^m(g_0)\le\tau$ for all $m\ge m_0$. Moreover $\lambda_+<1-E$ for $E<1/3$.
\end{lemma}
\begin{proof}
$h(g)<g$ if and only if $E<(g-E)(1-g)$, that is, $g\in(\lambda_-,\lambda_+)$. If $g_0\le\lambda_-$ then $h^m(g_0)\le\lambda_-<\tau$ by monotonicity. If $g_0\in(\lambda_-,\lambda_+)$ the iterates decrease and stay above $\lambda_-$, so they converge to a fixed point in $[\lambda_-,g_0)$, which is $\lambda_-$; once below $\tau$ they stay below, since $h(\tau)<\tau$. Finally $\lambda_+<1-E$ if and only if $\sqrt{(1-E)^2-4E}<1-3E$, which holds because $(1-3E)^2-(1-E)^2+4E=8E^2>0$.
\end{proof}

An \emph{inverter} $I(u,g)$ is a receiver that holds one CDS on $u$ of notional $1/(1-g)$. If $r_u\le g$, then $\Phi\ge1$ and its recovery is at least $1-E$; if $r_u\ge1-E$, then $\Phi\le E/(1-g)$ and its recovery is at most $h(g)$. Say that $r_x$ \emph{encodes} $0$ \emph{in code} $g$ if $r_x\le g$, and encodes $1$ if $r_x\ge1-E$. A \emph{chain of length $m$ from $x$ with start code $g_0$} consists of receivers $b_1,\dots,b_m$ with $b_j=I(b_{j-1},g_{j-1})$, $b_0=x$ and $g_j=h(g_{j-1})$. By induction, if $r_x$ encodes $\beta$ in code $g_0<\lambda_+$, then $r_{b_j}$ encodes $\beta\oplus(j\bmod2)$ in code $g_j<\lambda_+<1$; in particular all notionals are finite.

\subsection{Hardness of writer-solvent (I) below \texorpdfstring{$\alpha$}{alpha}}

\begin{theorem}\label{thm:wsI}
For every rational $0<E<\alpha=0.15767078078675458763\ldots$, \WSI-$E$ is PPAD-complete. Hardness holds on one-writer networks in which every NAND gate is realized by two input buffers (chains of inverters of even length) followed by one seed receiver holding two CDSs.
\end{theorem}

The proof is in \cref{app:onewriter}. Every NAND gate is realized by two input buffers, chains of inverters of even length, and one seed receiver holding a CDS on the end of each buffer; every PURIFY gate by an input buffer, two receivers holding CDSs of different notionals on its end, and an odd chain behind each of them. With $t=\lambda_+-E$, the construction works whenever $t^2>2E$ and $t^3>E$. On $(0,3-2\sqrt2]$ the first condition holds exactly for $E<\alpha$, and the second for every $E$ below $0.16071\ldots>\alpha$ (see the proof and \cref{rem:leak}).

\subsection{Calibration copies}\label{sec:calib}

The source of a one-writer network is only $E$-approximately solvent under (W), and a single gadget that must work for every source recovery $r_s\in[1-E,1]$ loses accuracy. Calibration copies avoid this: the decoder reads $r_s$ and uses the copy of the whole circuit that was built for that value. The decoder chooses a complete copy of the normalized input network using the observed value of $r_s$, and returns the recoveries from that copy. Different copies interact only through the shared source and sink. The approximation estimate applies within one selected copy and does not justify combining outputs from different copies. In the one-writer setting, this selection is performed by the decoder and does not require the banks themselves to select or react to the calibration, consistently with \cref{rem:obstruction}.

\begin{theorem}[calibration copies]\label{thm:calib}
Let $0<\eps<\eta<1$ be fixed rationals. There is a polynomial-time reduction from finding an (I)-$\eta$ vector of a non-degenerate network to finding a (W)-$\eps$ vector of a non-degenerate network, which adds one source. If every CDS writer of the input lies in $T$, the output is a one-writer network.
\end{theorem}

The normalization $\mathcal N'$ of the input network and the proof of \cref{thm:calib} are given in \cref{app:onewriter}.

The number of copies is $K+1=\lceil\eps/(2(\eta-\eps)(1-\eps))\rceil+1$, so the reduction is polynomial for each fixed pair $\eps<\eta$, not uniformly as $\eta-\eps\to0$. This is all that the following corollaries need.

\begin{corollary}\label{cor:oneW}
For every rational $0<E<\alpha$, \OWW-$E$ is PPAD-complete, and so is \WSW-$E$.
\end{corollary}
\begin{proof}
Choose a rational $\eta\in(E,\alpha)$. By \cref{thm:wsI}, \WSI-$\eta$ is PPAD-hard on one-writer networks, and \cref{thm:calib} reduces it to \OWW-$E$. One-writer networks are writer-solvent, so \WSW-$E$ is hard as well. Membership is \cref{thm:member}.
\end{proof}

\begin{corollary}\label{cor:generalW}
For every rational $0<E<\EI$, (W)-$E$ is PPAD-hard on non-degenerate networks.
\end{corollary}
\begin{proof}
Choose a rational $\eta\in(E,\EI)$. (I)-$\eta$ is PPAD-hard by the (I) part of the proof of \cref{thm:general}, and \cref{thm:calib} reduces it to (W)-$E$.
\end{proof}

\begin{remark}[why calibration does not contradict the one-writer obstruction]\label{rem:obstruction}
In a one-writer network with writer $s$, let $r^\ast$ be an exact clearing vector. For every $t\in[1-E,1]$, the vector that equals $r^\ast$ except for $r_s=t$ is a (W)-$E$ vector, because every payment of $s$ scales with $t$ and no CDS references $s$~\cite[record~L13; Section~3]{Supp}. So no bank's recovery can be forced to respond to the writer's recovery, and a gadget whose income must grow as $r_s$ falls cannot be built with one writer. This excludes one mechanism, not one-writer hardness. Calibration copies need no such response: in \cref{thm:calib} each copy has to work only for the values of $t$ that select it, and the other copies, which may decode incorrectly, are ignored.
\end{remark}

\section{Equal thresholds, and the barrier at \texorpdfstring{$\alpha$}{alpha}}\label{sec:coincidence}

\subsection{The problems in each class share their thresholds}

For a family $X$ of search problems indexed by the error, put
\begin{align*}
h(X)&=\sup\{E\in(0,1)\cap\mathbb Q:\ X\text{-}E\text{ is PPAD-hard}\},\\
p(X)&=\inf\{E\in(0,1]\cap\mathbb Q:\ X\text{-}E\text{ is solvable in polynomial time}\},
\end{align*}
with $\sup\emptyset=0$ and $\inf\emptyset=1$. By \cref{lem:relations}(b) the first set is closed downwards and the second upwards.

\begin{theorem}\label{thm:equal}
$h(\WSI)=h(\WSW)=h(\OWW)$ and $p(\WSI)=p(\WSW)=p(\OWW)$.
\end{theorem}
\begin{proof}
We use three reductions: (R1) \OWW-$E$ reduces to \WSW-$E$ by the identity, since one-writer networks are writer-solvent; (R2) \WSW-$E$ reduces to \WSI-$E$ by the identity, since an (I)-$E$ vector is a (W)-$E$ vector; (R3) for rationals $0<\eps<\eta<1$, \WSI-$\eta$ reduces to \OWW-$\eps$ by \cref{thm:calib}.

Hardness. By (R1) and (R2), hardness of \OWW-$E$ implies hardness of \WSW-$E$ and of \WSI-$E$, so $h(\OWW)\le h(\WSW)\le h(\WSI)$. Conversely, if \WSI-$\eta$ is PPAD-hard, then by (R3) \OWW-$\eps$ is PPAD-hard for every rational $\eps<\eta$, so $h(\OWW)\ge\eta$; hence $h(\OWW)\ge h(\WSI)$.

Polynomial time. By (R1) and (R2), $p(\OWW)\le p(\WSW)\le p(\WSI)$. Conversely, let $p(\OWW)<E_0\le1$. Some rational $\eps_1<E_0$ has \OWW-$\eps_1$ solvable in polynomial time. Pick a rational $\eta\in(\eps_1,E_0)$; by (R3), \WSI-$\eta$ is solvable in polynomial time, so $p(\WSI)\le\eta<E_0$. Hence $p(\WSI)\le p(\OWW)$.
\end{proof}

\Cref{thm:calib} is stated for arbitrary non-degenerate networks, so the same argument without (R1) gives the general case.

\begin{corollary}\label{cor:equalgen}
On non-degenerate networks, (I) and (W) have the same supremum of PPAD-hard rational errors and the same infimum of polynomially solvable rational errors.
\end{corollary}

These statements compare suprema and infima only; they say nothing about any individual rational error. Their use is to reduce the determination of $h$ and $p$ to the notion (I). By Theorem~\ref{thm:wsI}, Corollary~\ref{cor:oneW} and Theorem~\ref{thm:algws}, the common values satisfy $\alpha\le h$ and $p\le3-2\sqrt2$ for the writer-solvent problems, and by Theorems~\ref{thm:general} and~\ref{thm:alggen}, $\EI\le h$ and $p\le\elad$ for the general ones. Since the hard errors form a down-set and the polynomial ones an up-set, $h\le p$ unless PPAD is contained in FP. Any proof of writer-solvent (W) hardness above $\alpha$, for instance with several ex-ante solvent writers, would give writer-solvent (I) hardness at the same error by (R2), and then one-writer (W) hardness at every smaller error by (R3).

\subsection{Receivers in writer-solvent networks}

Work under (I)-$E$ in a writer-solvent network, $0<E<1$, $H=1-E$; the results below that need $E<1/2$ say so. A \emph{receiver} is a bank outside $T$. It writes no CDS, so its liability is a constant; if that constant is positive we normalize it to $1$, and then $\Phi_b=\min(1,A_b)$ with
\[
A_b(r)=a_0+\sum_k\bigl(a_kr_k+a'_k(1-r_k)\bigr),\qquad a_0,a_k,a'_k\ge0,
\]
where $k$ runs over receivers: $a_0$ collects external assets and the payments of banks of $T$, which recover exactly $1$ (a CDS that references a bank of $T$ pays $0$), $a_k$ is the ordinary debt owed by receiver $k$, and $a'_k$ the total notional of CDSs on $k$. A receiver with zero liability has $\Phi_b\equiv1$, and we put $A_b\equiv1$ for it. We call such maps \emph{separable affine}. By distinctness of roles, $A_b$ does not involve $r_b$. The products $r_w(1-r_k)$ of \cref{sec:general} do not occur, because every writer recovers exactly $1$.

For $0\le x<H$ put
\[
\kappa(x)=\min\Bigl(\frac E{1-x},\frac xH\Bigr),\qquad \psi(x)=E+\min(1,\kappa(x)),\qquad
\Lambda(x)=\min\Bigl(1,\frac{1+x}{2H},\frac{1+E}{2(1-x)}\Bigr).
\]
All three are nondecreasing. For $E\le3-2\sqrt2$ let $\lambda_-\le\lambda_+$ be the roots of $x^2-(1+E)x+2E$, as in \cref{sec:onewriter}.

\subsection{Interval-certified NAND gadgets}\label{sec:gadgets}

\begin{definition}[NAND gadget]\label{def:gadget}
A \emph{rail} is a bank with a polarity and a bound $\gamma\in(0,H)$: for a $+$rail bit $0$ is the box $[0,\gamma]$ and bit $1$ the box $[H,1]$; for a $-$rail the two boxes are exchanged. A \emph{wire} is one rail or two rails. A NAND gadget has input wires $u,v$ with bounds $\gamma_u,\gamma_v$, whose rails are pairwise distinct banks; a finite set $\mathcal R$ of receivers, containing no input rail, whose maps $A_b$ depend only on the input rails, on receivers of $\mathcal R$, on constants, and on \emph{fixed banks}, which are neither input rails nor receivers of $\mathcal R$ and are each given one nonempty closed interval inside $[0,1]$ as their box in all conditions; and an output wire made of receivers of $\mathcal R$, with bound $\gamma_{\rm out}$. The three \emph{conditions} give each input rail a box: in $\sigma_1$, $u$ carries $0$ and the rails of $v$ get $[0,1]$; in $\sigma_2$, the rails of $u$ get $[0,1]$ and $v$ carries $0$; in $\sigma_3$, both carry $1$.
\end{definition}

Single-rail and dual-rail wires of either polarity are included, the receivers may form any directed graph (feedback is allowed), and any receiver may read any input rail. The input wires must be disjoint. This matters: if $u$ and $v$ were the same rail, a single receiver reading $1-r_u$ with notional $1/(1-\gamma)$ would be certified (\cref{def:cert}) with $\gamma_{\rm out}=E+E/(1-\gamma)=\psi(\gamma)<\gamma$ for $\lambda_-<\gamma<\lambda_+$. It is no restriction for reductions, because the nodes of a \PCirc\ gate are distinct.

\begin{definition}[interval certificates]\label{def:cert}
For a receiver $b$ and a condition $j$ let $\Pi_b^j$ be the product of the current boxes of the banks that $A_b$ depends on (input rails get their condition boxes, fixed banks their fixed boxes). Put $\alpha_b^j=\min_{\Pi_b^j}A_b$, $\beta_b^j=\max_{\Pi_b^j}A_b$ (the bare symbol $\alpha$ always denotes $(13-\sqrt{153})/4$) and
\[
P_b^j=\bigl[\max(0,\min(1,\alpha_b^j)-E),\ \min(1,\beta_b^j+E)\bigr].
\]
An \emph{update} of $b$ replaces its boxes $B_b^j$ by $P_b^j$ (for all three $j$). An \emph{interval certificate} is the family of boxes obtained from $B_b^j=[0,1]$ for all $b\in\mathcal R$ by a finite sequence of updates. The gadget is \emph{certified correct} if, in some interval certificate, every output rail has (a subset of) its bit-$1$ box in $\sigma_1$ and $\sigma_2$, and a subset of its bit-$0$ box with bound $\gamma_{\rm out}$ in $\sigma_3$.
\end{definition}

Since $A_b$ is affine in each coordinate, its extremes over $\Pi_b^j$ are attained at endpoints. If $b$ reads no bank with both signs ($a_ka'_k=0$ for all $k$), then $\alpha_b^j=a_0+\sum_k(a_k\Bot_k^j+a'_k(1-\Top_k^j))$ and $\beta_b^j=a_0+\sum_k(a_k\Top_k^j+a'_k(1-\Bot_k^j))$, where $[\Bot_k^j,\Top_k^j]$ is the box of $k$. Using these termwise formulas in all cases gives coarser boxes. The impossibility results below hold for such coarser certificates as well, because their proofs use only $\alpha_b^j\le A_b\le\beta_b^j$ on $\Pi_b^j$ and the monotonicity of the update. The positive statements (\cref{lem:threestatus} and the existence part of \cref{cor:Fendpoint}) use the exact extremes; termwise extremes can destroy a status $\mathrm H$. Boxes need not be of the form $[0,\lambda]$, $[H,1]$ or $[0,1]$, so intermediate (multi-level) values inside a gadget are allowed. For a feedforward gadget, updating every receiver once in topological order gives the tightest certificate, which is exact interval propagation.

\begin{lemma}\label{lem:valid}
Updates only shrink boxes, and every interval certificate is valid: every (I)-$E$ vector whose input rails lie in their condition boxes and whose fixed banks lie in their fixed boxes has $r_b\in B_b^j$ for all $b\in\mathcal R$.
\end{lemma}
\begin{proof}
$P_b^j$ is monotone in the boxes of the inputs of $b$, and the initial boxes are $[0,1]$, so by induction every update produces a box inside the previous box of $b$. If $r$ lies in the current boxes, then $\alpha_b^j\le A_b(r)\le\beta_b^j$, so $r_b\in[\Phi_b(r)-E,\Phi_b(r)+E]\cap[0,1]\subseteq P_b^j$.
\end{proof}

\subsection{The barrier theorem}

\begin{theorem}\label{thm:barrier}
Let $\alpha\le E<1/2$, and let a NAND gadget (\cref{def:gadget}) be certified correct by an interval certificate. Put $\gamma_{\rm in}=\min(\gamma_u,\gamma_v)$. Then $\gamma_{\rm out}>\gamma_{\rm in}$. If moreover $E\le3-2\sqrt2$ and $\gamma_{\rm in}\ge\lambda_-$, then $\gamma_{\rm out}\ge\lambda_+$.
\end{theorem}

The proof, in \cref{app:barrierproof}, attaches to every bank four comparison inequalities (P1)--(P4) between its boxes in the three conditions, with parameters $c_1,s,c_\ast\in[0,1]$. Under the scalar conditions (C1)--(C3) and the input-base conditions of \cref{lem:base}(b), these inequalities hold at the input rails and at every bank before its first update, and every update preserves them (\cref{lem:K,lem:closure,lem:base}, which rest on the comparison lemma for separable affine maps, \cref{lem:comparison}). At an output rail they force $\gamma_{\rm out}\ge E+\min(s,c_\ast)$, and the parameter choices give $\gamma_{\rm out}>\gamma_{\rm in}$ and, if $E\le3-2\sqrt2$ and $\gamma_{\rm in}\ge\lambda_-$, also $\gamma_{\rm out}\ge\lambda_+$. The key scalar inequality is $\lambda_++E\le2\lambda_-$; for $0<E\le3-2\sqrt2$ it holds if and only if $E\ge\alpha$ (\cref{lem:psi}(iv) and its proof). \Cref{app:barrierproof} also proves that for $\alpha\le E<1$ no interval-certified NAND gadget has $\gamma_{\rm out}\le\gamma_{\rm in}$, in particular none has a uniform interface (\cref{cor:uniform}), and that $\alpha$ is the exact endpoint of the family $\mathfrak F$ of chain-and-seed gadgets certified with three statuses, to which the gadget of \cref{thm:wsI} belongs (\cref{cor:Fendpoint}).

\begin{remark}[what the barrier covers, and dual rails]\label{rem:barrierscope}
\Cref{thm:barrier} places no restriction on the topology of the gadget: chains and seeds of any size, several mixing receivers, receivers that reread the raw input rails at any depth, dual-rail wires, feedback loops, and arbitrary intermediate values inside the gadget are all allowed. The restrictions are: separable affine receivers (writer-solvent networks under (I)), disjoint input wires, output rails that are receivers of the gadget (no output is an alias of an input rail), other banks entering only as fixed banks whose boxes do not depend on the condition, the wire interface with states $[0,\gamma]$ and $[H,1]$ (\cref{thm:interfaces} removes this restriction), and correctness certified by interval propagation. In particular the dual-rail machinery of \cref{sec:general} does not help in writer-solvent networks: its gain came from refinement steps whose new bank holds a CDS written by a rail bank, with assets $k\,r_p(1-r_n)$, and in a writer-solvent network these product terms do not exist. What the theorem does not cover is listed in \cref{sec:routes}.
\end{remark}

\section{The box algorithm and the polynomial thresholds}\label{sec:box}

This section gives a simple polynomial-time algorithm for (I), and hence (W), for every rational error above $3-2\sqrt2$ on writer-solvent networks and above $\elad$ on all networks. The algorithm shrinks an interval for every bank and fixes a bank's value once its interval is short. Both thresholds coincide with lower bounds for algorithms of bounded radius (\cref{sec:local}), for (I) on writer-solvent networks and for (I) and (W) on general networks, so within that model they are sharp.

\subsection{The algorithm}

\paragraph{Algorithm $\BOX(E)$, for $0<E<1/2$.} Every bank $i$ carries a box $I_i=[lo_i,hi_i]\subseteq[0,1]$ and is \emph{settled} or \emph{unsettled}. Initially $I_i=\{1\}$ (settled) for $i\in F$ and $I_i=[E,1-E]$ (unsettled) for $i\notin F$. A \emph{round} does the following. For every unsettled bank $i$ compute, from the current boxes,
\begin{align*}
A_i^-&=e_i+\sum_jc_{ji}\,lo_j+\sum_{j,k}c^k_{ji}\,lo_j(1-hi_k), &
A_i^+&=e_i+\sum_jc_{ji}\,hi_j+\sum_{j,k}c^k_{ji}\,hi_j(1-lo_k),\\
\ell_i^-&=\sum_jc_{ij}+\sum_{j,k}c^k_{ij}(1-hi_k), &
\ell_i^+&=\sum_jc_{ij}+\sum_{j,k}c^k_{ij}(1-lo_k),
\end{align*}
and put $(\lambda_i,\mu_i)=(1,1)$ if $\ell_i^+=0$, and $\lambda_i=\min(1,A_i^-/\ell_i^+)$, $\mu_i=\min(1,A_i^+/\ell_i^-)$ otherwise. Then, for all unsettled $i$ simultaneously: if $\mu_i-\lambda_i\le2E$, settle $i$ at the point $p_i$ of $I_i$ nearest to $(\lambda_i+\mu_i)/2$ (so $I_i:=\{p_i\}$); otherwise put $I_i:=[\lambda_i+E,\mu_i-E]$, the set of points whose $E$-neighbourhood lies inside $[\lambda_i,\mu_i]$. The algorithm stops when every bank is settled and outputs $r_i=p_i$ (with $p_i=1$ on $F$).

The interval $[\lambda_i,\mu_i]$ encloses the values of the map $\Phi_i$ on the current product of boxes; the box $I_i$ itself need not contain these values. A first round with the rule ``settle everything'' is the one-pass algorithm of~\cite[record~L1; Section~2]{Supp}, which achieves $(3-\sqrt5)/2$ in general and $1-1/\sqrt2$ on writer-solvent networks.

The boxes need not contain an exact clearing vector: each update cuts $E$ off both ends of the enclosure, and in \cref{ex:stallfail} the network has exact clearing vectors but none in the box at which $\BOX(4/25)$ stalls. Soundness does not use one: the boxes are nested, so the output lies in the boxes of every round; for a bank settled in some round, its map value at the output therefore lies in the enclosure of that round, and its settled value is within $E$ of every point of that enclosure (\cref{lem:sound}).

\begin{lemma}[soundness]\label{lem:sound}
In every round: (a) if $\ell_i^+>0$ then $\ell_i^->0$; (b) $\lambda_i\le\Phi_i(r)\le\mu_i$ for every $r$ in the current product of boxes; (c) the new boxes lie inside the old ones, and $\mu_i-E\le p_i\le\lambda_i+E$ for every bank settled in the round. Consequently, if the algorithm stops, its output satisfies $r_i=1$ for $i\in F$ and $|r_i-\Phi_i(r)|\le E$ for every $i$; so it is an (I)-$E$ vector, and also a (W)-$E$ vector.
\end{lemma}

The proof is in \cref{app:box}.

\subsection{Termination}

Let $U$ be the set of banks that are unsettled at the start of a round, and put $\rho_r=\max_{k\in U}hi_k/lo_k$ and $\rho_n=\max_{k\in U}(1-lo_k)/(1-hi_k)$. In the round every unsettled bank has $\mu_i\le\min(1,K\lambda_i)$ with $K=\rho_r\rho_n^2$, and with $K=\max(\rho_r,\rho_n)$ if no bank outside $F$ writes a CDS (\cref{lem:ratio}). A bank that stays unsettled gets a new box whose two ratios are at most $R_1(K)=(1-E-1/K)/E$, the ratio of its endpoints being at most $R_2(K)=(1-E)/(1/K+E)$ (\cref{lem:newbox}). Iterating $K_{s+1}=R_1(K_s)$ from $K_0=(1-E)/E$ when every writer lies in $F$, and $K_{s+1}=R_2(K_s)R_1(K_s)^2$ from $K_0=((1-E)/E)^3$ in general, gives the following bounds on the number of rounds.

\begin{theorem}[writer-solvent networks]\label{thm:algws}
Let $E$ be rational with $3-2\sqrt2<E<1/2$, and put $t=(1-E)/E$, $\Delta=1-(1-E)^2/(4E)>0$ and $h_E(g)=E+E/(1-g)$. On every network in which every CDS writer lies in $F$, $\BOX(E)$ stops after at most
\[
R_{\rm ws}(E)=1+\max\Bigl(0,\Bigl\lceil(1-E)\Bigl(t-\frac1{1-2E}\Bigr)\Big/\Delta\Bigr\rceil\Bigr)
\]
rounds and outputs an (I)-$E$ vector. More precisely, there is a least integer $R_E\ge1$ with $h_E^{R_E}(E)\ge1-E$; the iterates satisfy $E\le h_E^s(E)<1-E$ for $0\le s<R_E$; and $\BOX(E)$ stops after at most $R_E$ rounds. Hence \WSI-$E$, \WSW-$E$ and \OWW-$E$ are solvable in polynomial time for every rational $E>3-2\sqrt2=0.17157287525380990239\ldots$ (and for $E\ge1/2$ by \cref{lem:relations}(c)).
\end{theorem}

\begin{theorem}[all networks]\label{thm:alggen}
Let $E$ be rational with $\elad<E<1/2$, put $t=(1-E)/E$, and
\[
c(E)=\max_{0\le u\le1-2E}\frac{(1-E)\,u\,(1-E-u)^2}{E^2(u+E)}.
\]
Then $c(E)<1$, and on every network (non-degeneracy is not needed) $\BOX(E)$ stops after at most $R_{\rm gen}(E)=1+\max\bigl(0,\lceil\log(t^3(1-2E))/\log(1/c(E))\rceil\bigr)$ rounds and outputs an (I)-$E$ vector. Hence (W)-$E$ and (I)-$E$ are solvable in polynomial time on all networks for every rational $E>\elad=0.28336553249330853976\ldots$.
\end{theorem}

The proofs, and the bound on the encoding length of the box endpoints that makes $\BOX(E)$ polynomial for each fixed rational $E$ above the threshold (\cref{lem:bits}), are given in \cref{app:box}. The bounds are not uniform as $E$ approaches the threshold.

\subsection{Optimality among local algorithms}\label{sec:local}

We use the following locality model~\cite[record~L3; Section~7]{Supp}. For a bank $i$ let $\mathcal S_i$ consist of $i$, its debtors, the writers of the CDSs it holds, and the references of the CDSs it holds or writes. The \emph{dependency graph} joins two banks if they lie together in some $\mathcal S_k$. The \emph{view} $V_R(i)$ consists of the banks at distance at most $R$ from $i$ in this graph, with their identifiers and external assets and every contract in which one of them participates. A deterministic algorithm is \emph{$R$-local} if its output at $i$ is a function of $E$ and $V_R(i)$, and it has \emph{bounded radius} if it is $R$-local for some $R$.

\begin{theorem}[local lower bounds]\label{thm:locallower}
Let $0<E<1/2$.
\begin{enumerate}[label=(\alph*)]
\item \cite[record~L7; Theorem~6.1]{Supp}, with the hypothesis corrected in \cref{rem:L7}. If $h_E^s(E)<1-E$ for all $1\le s\le R$, then no $R$-local algorithm outputs an (I)-$E$ vector for every non-degenerate writer-solvent network.
\item \cite[record~L3; Theorem~7.1(b)]{Supp}. If $E<\elad$, then for every $R$ no $R$-local algorithm outputs a (W)-$E$ vector for every non-degenerate network; the same holds for (I).
\end{enumerate}
\end{theorem}

\begin{corollary}[bounded-radius thresholds]\label{cor:local}
Let $E\in(0,1/2)$ be rational. Within the locality model above:
\begin{enumerate}[label=(\alph*)]
\item On non-degenerate writer-solvent networks, an $R$-local algorithm for (I) exists if and only if $h_E^s(E)\ge1-E$ for some $1\le s\le R$, and then $R$ rounds of $\BOX(E)$ form one. Algorithms of bounded radius for (I) exist if and only if $E>3-2\sqrt2$.
\item On non-degenerate networks, algorithms of bounded radius for (W) and for (I) exist if and only if $E>\elad$.
\end{enumerate}
\end{corollary}

This corollary is a statement about algorithms of bounded radius in the model just defined. It does not say anything about polynomial-time algorithms that look at the whole network, and in the windows $(\alpha,3-2\sqrt2]$ and $(\EI,\elad]$ it only shows that a polynomial algorithm for (I), and on general networks also for (W), cannot have bounded radius. For (W) on writer-solvent networks the best known bounded-radius lower bound is $0.1609$~\cite[record~L11; Proposition~5.1]{Supp}, so there the bounded-radius threshold is only known to lie in $[0.1609,3-2\sqrt2]$.

The proofs of \cref{thm:locallower,cor:local}, the correction of the hypothesis of~\cite[record~L7; Theorem~6.1]{Supp} (\cref{rem:L7}), the round counts for small radii (\cref{rem:rounds}) and the tests of the algorithm (\cref{rem:boxtests}) are given in \cref{app:box}; the historical $1$-local bound and the bound $0.1609$ are proved in \cref{app:localaux}.

\subsection{Finite precision and near-stalls}\label{sec:rounded}

Below the thresholds the exact algorithm can run forever without settling, and its endpoints can grow in size with every round: in the run of \cref{prop:nostall} the lower endpoints after $t$ rounds have reduced denominator $100^{t+1}$, and \cref{lem:bits} bounds the growth only in terms of the number of rounds. Rounding the boxes outward to a fixed grid removes both problems, at the price of a small loss in the error of settled banks. Neither non-degeneracy nor writer solvency is assumed in this subsection unless stated.

\paragraph{Algorithm $\BOX(E_0,M)$, for rational $0<E_0<1/2$ and an integer $M\ge1$.} Put $\delta=(1-2E_0)/M$ and let $G=\{E_0+j\delta:\ 0\le j\le M\}$. For $y\in[E_0,1-E_0]$ let $\lfloor y\rfloor_G$ and $\lceil y\rceil_G$ be the largest point of $G$ that is at most $y$ and the least point of $G$ that is at least $y$. Banks in $F$ are settled at $1$, and every other bank starts unsettled with the box $[E_0,1-E_0]$. A round computes the enclosures $[\lambda_i,\mu_i]$ of \cref{sec:box} from the current boxes and then updates all unsettled banks simultaneously:
\begin{itemize}
\item if $\mu_i-\lambda_i\le2E_0$, clamp the midpoint $(\lambda_i+\mu_i)/2$ into the current box $[l_i,h_i]$, replace it by a nearest point of $G$ in $[l_i,h_i]$ (ties broken downwards) and settle $i$ there;
\item otherwise replace $[l_i,h_i]$ by $[\max(l_i,\lfloor\lambda_i+E_0\rfloor_G),\ \min(h_i,\lceil\mu_i-E_0\rceil_G)]$.
\end{itemize}
The algorithm stops when every bank is settled, and also when a round settles no bank and changes no box; in the latter case it returns the remaining product $\mathcal B=\prod_{i\in U}[l_i,h_i]$ of the unsettled set $U$, together with the settled points. It does not claim to have solved the clearing problem in that case. For a continuing bank both arguments of the rounding lie in $[E_0,1-E_0]$, because $\mu_i-\lambda_i>2E_0$ and $0\le\lambda_i\le\mu_i\le1$.

\begin{theorem}[finite-precision BOX]\label{thm:roundedbox}
On every network, $\BOX(E_0,M)$ has the following properties.
\begin{enumerate}[label=(\arabic*)]
\item All boxes are nonempty and nested; every unsettled box has positive width and endpoints in $G$. Every enclosure contains the values of $\Phi_i$ on the current product of boxes, also on degenerate networks.
\item Every bank settled in some round has residual at most $E_0+\delta/2$ at every point of every later product of boxes. Banks of $F$ have residual $0$.
\item It stops after at most $nM+1$ rounds. All box endpoints and settled points have encoding length polynomial in the encoding length of $E_0$ and in $\log M$; every round takes time polynomial in the input size and $\log M$, and the total running time is polynomial in the input size and $M$.
\item If it stops with a nonempty unsettled set $U$, the enclosures computed on the final product satisfy, for every $i\in U$,
\begin{equation}\label{eq:nearstall}
\mu_i-\lambda_i>2E_0,\qquad 0\le\lambda_i+E_0-l_i<\delta,\qquad 0\le h_i-(\mu_i-E_0)<\delta .
\end{equation}
\item Every product of boxes reached by the algorithm contains an (I)-$(E_0+\delta/2)$ vector $r$ with $r_F=\mathbf1$ and with the settled banks at their points. This is a statement of existence; it does not give a way to compute such a vector.
\end{enumerate}
\end{theorem}

The proof is in \cref{app:box}. By the same argument with $\delta=0$, every product of boxes reached by the exact algorithm $\BOX(E)$ contains an (I)-$E$ vector (\cref{rem:exactbox}), although it need not contain an exact clearing vector (\cref{ex:stallfail}).

\paragraph{Precision accounting.} Given a rational target $E$ and a rational working accuracy $0<E_0<E<1/2$, the choice
\[
M\ge\max\Bigl(1,\Bigl\lceil\frac{1-2E_0}{2(E-E_0)}\Bigr\rceil\Bigr)
\]
gives $E_0+\delta/2\le E$. If $E-E_0$ is at least an inverse polynomial in the input length, the algorithm runs in polynomial time; for a fixed positive gap it uses $O(n)$ rounds with a constant depending on the gap. The bound on the running time is polynomial in $M$, not in $\log M$, so this bound is not polynomial in $\log(1/(E-E_0))$.

\paragraph{Near-stall selectors.} Let $\mathcal C$ be a class of networks. A \emph{near-stall selector} for $\mathcal C$ at target $E$ receives a network of $\mathcal C$, a working accuracy $E_0$ and grid step $\delta$ with $E_0+\delta/2\le E$, and a terminal configuration of $\BOX(E_0,M)$ on that network (the product $\mathcal B$, the settled points and the set $F$) satisfying~\eqref{eq:nearstall}; it returns a point of $\mathcal B$ at which every unsettled bank has residual at most $E$.

\begin{theorem}[reduction to near-stall selection]\label{thm:transfer}
Let $E_0$ and $M$ satisfy $E_0+\delta/2\le E$, where $\delta=(1-2E_0)/M$. If a near-stall selector for $\mathcal C$ at target $E$ runs in time polynomial in the sizes of the network and of the terminal configuration, then running $\BOX(E_0,M)$ and then the selector solves (I)-$E$ on $\mathcal C$ in time polynomial in the input size and $M$. Banks settled by $\BOX(E_0,M)$ keep residual at most $E$, without being reopened.
\end{theorem}
\begin{proof}
By \cref{thm:roundedbox}, $\BOX(E_0,M)$ either returns a vector all of whose banks are settled, which is an (I)-$(E_0+\delta/2)$ vector by (2) and hence an (I)-$E$ vector, or returns in the stated time a terminal configuration of polynomial size satisfying~\eqref{eq:nearstall}. The selector returns a point of the same product, so (2) certifies every previously settled bank, its own guarantee certifies every unsettled bank, and the banks of $F$ stay at $1$.
\end{proof}

This is a reduction to a uniform near-stall problem, not a solution of it. An oracle that resolves only \emph{exact} stalls cannot be applied directly, because the terminal inequalities~\eqref{eq:nearstall} have defects below $\delta$ rather than equalities, and we do not know how to use one: we do not know a way to realize an arbitrary near-stall as an exact stall of a nearby network. \Cref{sec:selection,app:selection} construct selectors on structured classes.

\section{Escape routes from \texorpdfstring{$\alpha$}{alpha} and their exact scope}\label{sec:barriers}

\Cref{thm:barrier} rules out interval-certified NAND gadgets with a uniform wire bound for $\alpha\le E<1/2$. A reduction could still try to escape with other gate types, with wires of different bounds, or with gadgets whose inputs have different bounds. This section closes these routes for interval-certified gadgets (\cref{thm:unary,thm:asym,thm:nonuniform}), shows that every route is capped by the algorithm of \cref{sec:box} (\cref{cor:ceiling}), and lists what remains open (\cref{sec:routes}). Throughout we work under (I)-$E$ in writer-solvent networks, with the receivers and notation of \cref{sec:coincidence}.

\subsection{Unary gadgets}

\begin{definition}[unary gadget]\label{def:unary}
A \emph{unary gadget} has one input wire $u$ with bound $\gamma$ (with distinct rails), a finite set $\mathcal R$ of receivers, containing no rail of $u$, whose maps depend only on the rails of $u$, on receivers of $\mathcal R$, on constants and on fixed banks (as in \cref{def:gadget}), and output wires made of receivers of $\mathcal R$, each with its own bound. There are two conditions: in $\tau_0$ the wire $u$ carries $0$, and in $\tau_1$ it carries $1$. Interval certificates are defined as in \cref{def:cert}, with the conditions $\tau_0,\tau_1$. An output wire is \emph{copying} if it should carry $0$ in $\tau_0$ and $1$ in $\tau_1$, and \emph{inverting} if the reverse. The gadget is \emph{certified correct} if, in some interval certificate, every output rail has a subset of its bit-$1$ box in the condition in which its wire should carry $1$, and a subset of its bit-$0$ box, with the bound of its wire, in the other condition.
\end{definition}

PURIFY and COPY gadgets are unary gadgets with two, respectively one, copying output wires, and a NOT gadget is one with an inverting output wire.

\begin{theorem}[unary gadgets]\label{thm:unary}
Let $0<E<1/2$ and let a unary gadget with input bound $\gamma$ be certified correct. If $c\in[0,1]$ satisfies $c\le\kappa(\gamma)$ and $c(1-E-c)\le E$, then every output wire has bound at least $E+c$. Consequently every output bound $\gamma_o$ satisfies:
\begin{enumerate}[label=(\alph*)]
\item $\gamma_o\ge\psi(\gamma)$ if $E>3-2\sqrt2$, or if $E\le3-2\sqrt2$ and $\gamma\notin(\lambda_-,\lambda_+)$;
\item $\gamma_o\ge\lambda_-$ if $E\le3-2\sqrt2$ and $\lambda_-<\gamma<\lambda_+$.
\end{enumerate}
In particular $\gamma_o>\gamma$ if $E>3-2\sqrt2$ or $\gamma<\lambda_-$, and $\gamma_o\ge\gamma$ if $\gamma\ge\lambda_+$.
\end{theorem}

The proof is in \cref{app:routes}, which also shows that in case~(a) the best single receiver attains the bound $\psi(\gamma)$ whenever $\psi(\gamma)<H$, and that no unary gadget with input bound $\gamma$ and an output wire is certified correct when $\psi(\gamma)\ge H$.

\subsection{Asymmetric inputs}

\begin{theorem}[asymmetric inputs]\label{thm:asym}
Let $0<E\le3-2\sqrt2$. No NAND gadget with $\gamma_u\ge\lambda_+$ and $\gamma_v\ge\lambda_-$ is certified correct by an interval certificate; by symmetry, neither is one with $\gamma_v\ge\lambda_+$ and $\gamma_u\ge\lambda_-$.
\end{theorem}

The proof, in \cref{app:routes}, uses weighted versions of the comparison inequalities of the barrier theorem.

\subsection{Nonuniform bounds and all standard gate types}

We first record that circuits made only of unary gates are easy.

\begin{lemma}[unary circuits~{\cite[record~L18; Theorem~10.1]{Supp}}]\label{lem:unarycircuits}
Every \PCirc\ instance whose gates are of types NOT, COPY and PURIFY has a solution, computable in linear time. If there is no NOT gate, the all-zero assignment is a solution.
\end{lemma}

\begin{theorem}[nonuniform wire bounds and other gates]\label{thm:nonuniform}
Let $\alpha\le E<1/2$. Consider a \PCirc\ instance whose gates are binary gates of types AND, OR, NAND, NOR and unary gates of types NOT, COPY, PURIFY, in which the nodes of every gate are distinct and every node is the output of exactly one gate. Suppose it is realized in a writer-solvent network as follows. Every node $x$ is a wire (one rail, or two distinct rails) with a declared bound $\gamma_x\in(0,H)$. Every binary gate is a NAND gadget, after the relabeling described below, with input wires its two input nodes, which must have disjoint rails, and output wire its output node. Every unary gate is a unary gadget with input wire its input node and output wires its output nodes, inverting for NOT and copying for COPY and PURIFY. Every gadget is certified correct by an interval certificate with the declared boxes. Then the instance has a solution, which can be computed in polynomial time; if it has no NOT gate, the solution can be taken Boolean.
\end{theorem}

\emph{Relabeling.} Exchanging the polarities of all rails of a wire exchanges the meaning of its bits $0$ and $1$ and changes no box and no bound. An AND gadget must output $0$ when some input carries $0$ and $1$ when both carry $1$; after relabeling its output wire this is the requirement of a NAND gadget. An OR gadget must output $1$ when some input carries $1$ and $0$ when both carry $0$; relabeling both input wires gives the NAND requirement. A NOR gadget becomes a NAND gadget after relabeling all three wires. By \cref{thm:pc} the theorem covers every PPAD-complete gate set of~\cite[Corollary~2.2]{DFHM22}.

The proofs of \cref{lem:unarycircuits,thm:nonuniform} are in \cref{app:routes}. For $E>3-2\sqrt2$ the proof of the theorem shows that the instance has no gates. For $\alpha\le E\le3-2\sqrt2$ it shows that every bound is at least $\lambda_-$ and that every node reachable from the output of a binary gate has bound at least $\lambda_+$, so that by \cref{thm:asym} no input of a binary gate is reachable from such an output; it then solves the unary part by \cref{lem:unarycircuits} and assigns the remaining nodes in order of their distance from the binary outputs.

\begin{remark}[scope of \cref{thm:nonuniform}]\label{rem:nonuniformscope}
The gadgets may have any topology, with feedback, mixing receivers, rereads of the input rails and intermediate values. The theorem is stated for wires whose rails share one bound; wires with a separate bound for each rail are not covered by the statement or its proof, but they are covered, together with non-saturated intervals, by \cref{thm:interfaces} below. Below $\alpha$ the theorem fails, as it must: uniform interval-certified NAND gadgets exist (\cref{cor:Fendpoint}), and by \cref{thm:asym} their common bound $\gamma$ is below $\lambda_+$, so Step~1 or Step~2 of the proof in \cref{app:routes} fails at such a gadget. An analogous obstruction for the family $\mathfrak F$ and for unary feedforward receiver graphs certified with three statuses was proved in~\cite[record~L18; Theorem~8.1]{Supp}.
\end{remark}

\subsection{Arbitrary two-interval interfaces}\label{sec:interfaces}

The results above use the saturated wire interface: a rail carries one bit in $[H,1]$ and the other in $[0,\gamma]$, with one bound $\gamma$ per wire. This subsection and \cref{app:routes} show that nothing changes if every rail has its own pair of disjoint intervals, which may be non-saturated, may have a positive low endpoint or a high endpoint below $1$, and may differ between the two rails of a wire and between wires. The gadget model of \cref{def:gadget,def:unary} and the interval certificates of \cref{def:cert} are otherwise unchanged.

\begin{definition}[two-interval interfaces]\label{def:interfaces}
In a \emph{two-interval interface} every rail $q$ has a polarity and two fixed nonempty disjoint closed intervals
\[
L_q=[b_q,l_q],\qquad U_q=[h_q,t_q],\qquad 0\le b_q\le l_q<h_q\le t_q\le1 .
\]
A $+$rail represents bit $0$ by $L_q$ and bit $1$ by $U_q$; a $-$rail represents bit $0$ by $U_q$ and bit $1$ by $L_q$. A wire has one rail or two distinct rails, and the four endpoints may differ between wires and between the rails of one wire, but they are fixed for each rail across all its uses. In a condition in which an input wire is unspecified, its rails get the box $[0,1]$. A NAND or unary gadget with such interfaces is \emph{certified correct} if, in some interval certificate, every output rail has a box inside the interval of the bit its wire must carry, in every condition in which that bit is prescribed.
\end{definition}

The interface of \cref{def:gadget} is the case $b_q=0$, $l_q=\gamma$, $h_q=H$, $t_q=1$ for all rails of a wire.

\begin{theorem}[barrier for arbitrary two-interval interfaces]\label{thm:interfaces}
Let $\alpha\le E<1$. Consider a \PCirc\ instance and a writer-solvent network that realize it as in \cref{thm:nonuniform}, except that every node is a wire with a two-interval interface (\cref{def:interfaces}) instead of a wire with one bound. If every gadget is certified correct, then:
\begin{enumerate}[label=(\alph*)]
\item if $E>3-2\sqrt2$, the instance has no gates;
\item if $\alpha\le E\le3-2\sqrt2$, no input of a binary gate is reachable from a binary output;
\item in both cases the instance has a solution, which can be computed in polynomial time, and which can be taken Boolean if there is no NOT gate.
\end{enumerate}
Consequently, unless PPAD is contained in FP, \PCirc\ reductions for writer-solvent (I) whose gates are realized by interval-certified gadgets with two-interval interfaces cannot prove PPAD-hardness at any $E\in[\alpha,1)$.
\end{theorem}

The theorem keeps every other restriction of the model: separable affine receivers (writer-solvent networks under (I)), disjoint input wires, outputs that are receivers of the gadget, fixed banks whose boxes do not depend on the condition, and rectangular (interval) certificates. It says nothing about correlated certificates or about algorithms. For separate \emph{saturated} bounds on the two rails of a wire, the least-bound argument of the proof was already noted in the research record~\cite[record~L19]{Supp}; the new step is the normalization of non-saturated intervals.

The proof, in \cref{app:routes}, assigns to every rail $q$ with intervals $[b_q,l_q]$ and $[h_q,t_q]$ the effective bound $\gamma_q=E+(l_q-E)/(h_q+E)$, shows that every produced rail satisfies $E\le l_q<h_q\le H$, and proves that the comparison inequalities of the saturated case hold at a rail as soon as they hold for the saturated interface with bound $\gamma_q$ (\cref{lem:basecost}); the gate inequalities of \cref{prop:ifgates} then reduce the theorem to the argument for \cref{thm:nonuniform}.

\subsection{A universal ceiling}

\begin{corollary}\label{cor:ceiling}
If for some rational $E>3-2\sqrt2$ any of \WSI-$E$, \WSW-$E$, \OWW-$E$ is PPAD-hard under polynomial-time reductions, then every problem in PPAD is solvable in polynomial time. The same holds for (I)-$E$ and (W)-$E$ on general networks with $E>\elad$.
\end{corollary}
\begin{proof}
\Cref{thm:algws} and \cref{thm:alggen} solve these problems in polynomial time for $E<1/2$, and \cref{lem:relations}(c) for $E\ge1/2$.
\end{proof}

So, unless PPAD is contained in FP, every escape from $\alpha$, by any method, must stay inside the window $(\alpha,3-2\sqrt2]=(0.15767078\ldots,0.17157287\ldots]$, of width $0.0139$, and every improvement of $\EI$ inside $(\EI,\elad]=(0.25299586\ldots,0.28336553\ldots]$. Inside these windows, \cref{cor:local} shows that a polynomial algorithm for (I) on writer-solvent networks, and for (I) and (W) on general networks, cannot have bounded radius. On the hardness side, \cref{thm:barrier,thm:nonuniform} show that a \PCirc\ reduction for writer-solvent (I) whose gates are realized by interval-certified gadgets as in \cref{def:gadget,def:unary}, with the standard wire interface, cannot prove PPAD-hardness for any $E\in[\alpha,1)$ unless PPAD is contained in FP: for $E<1/2$ the circuits it handles are solvable in polynomial time, and for $E\ge1/2$ no NAND gadget is certified at all (\cref{cor:uniform}). By \cref{thm:interfaces} the same holds for arbitrary two-interval interfaces, which may differ between rails and need not be saturated.

\subsection{What remains open}\label{sec:routes}

The barriers above are statements about a delimited class of constructions: writer-solvent networks under (I), receivers with separable affine maps, disjoint input wires, output rails that are receivers of the gadget, wire interfaces given by a fixed pair of disjoint intervals for each rail (\cref{thm:interfaces}; the saturated interface with states $[0,\gamma]$ and $[H,1]$ and one bound per wire is a special case), other banks entering as fixed banks, and correctness certified by interval propagation. The following routes lie outside that class and remain open.

\begin{center}\small
\begin{tabular}{@{}R{3.9cm}R{5.0cm}R{6.4cm}@{}}
\toprule
route & status & what remains open\\
\midrule
Multi-level (interval) signals inside gadgets & closed for interval certificates, any topology (\cref{thm:barrier}) & certificates finer than intervals\\
Non-saturated or rail-dependent wire interfaces & closed for interval certificates, any topology, all standard gates (\cref{thm:interfaces}) & the same interfaces with certificates finer than intervals\\
Larger, mutually dependent seeds; helpers that read raw rails; non-adjacent rereads & closed for interval certificates, feedback allowed (\cref{thm:barrier}); a two-receiver mutual-inhibition template is closed even for exact solution sets (\cref{prop:tworeceiver}, from~\cite[record~L18; Theorem~9.1]{Supp}) & two or more receivers outside that template, when correctness is proved on exact internal solution sets\\
Correlations between receivers & open & certificates that use the joint solution set rather than products of intervals\\
Nonuniform wire bounds & closed for interval-certified gadgets of any topology (\cref{thm:nonuniform}), also with separate bounds or intervals for the two rails of a wire (\cref{thm:interfaces}) & outputs that are aliases of input rails; nonuniform bounds with non-interval certificates\\
Other source problems & unary circuits are easy (\cref{lem:unarycircuits}); every PPAD-complete gate set of~\cite[Corollary~2.2]{DFHM22} is closed for interval certificates (\cref{thm:nonuniform}) & source problems other than \PCirc; outputs that are aliases of input rails (for example a NOT gate realized only by relabeling polarities)\\
Writers that are not ex-ante solvent & outside the writer-solvent model & this is the general-network question, with its own window $(\EI,\elad]$\\
All routes & capped at $3-2\sqrt2$ unless PPAD $\subseteq$ FP (\cref{cor:ceiling}) & the window $(\alpha,3-2\sqrt2]$\\
\bottomrule
\end{tabular}
\end{center}

\section{Exact solvency at bounded held degree}\label{sec:degree}

The reduction of \cref{thm:D} uses CDS notionals $C=2\rho$ and $C^2$, where $\rho=(1+E)/(1-E)$; written with contracts of notional $1$, its receivers hold of the order of $\rho^2$ contracts. This section studies how many unit contracts a bank must hold for (D) to be hard, on the restricted class $\mathcal R(d)$ of \cref{sec:Rclass}. For every rational $E\in[1/2,1)$, and for $E=1/3$, the problem is solvable in polynomial time below an explicit held degree and PPAD-complete from that degree on, and for $E<1/9$ it is PPAD-complete already at held degree $2$ (\cref{thm:degmain}); assuming that PPAD is not contained in FP, these are therefore exactly the held degrees at which (D) becomes hard. Two methods meet. On the hardness side we keep the sparse two-arm PURIFY topology, which appears with weighted contracts in~\cite{DHK} and in \cref{thm:D}, place unit multiplicities differently and decode bit $0$ as default at least $1-E$ (\cref{sec:critdeg}); at held degrees $2$, $3$ and $4$ chains of banks restore weak signals (\cref{sec:chains}). On the algorithmic side, supported propagation in the spirit of~\cite[Lemma~F.8 and Theorem~F.9]{KM} is refined by states whose support persists (\cref{sec:propagation}). \Cref{sec:phase} combines the two, and \cref{sec:methodlimits} records where each method stops.

The results concern the exact-solvency notion (D) on $\mathcal R(d)$; the hardness results hold also for (D$^\circ$) and (S). They do not move the windows for (I) and (W): on $\mathcal R(d)$ every writer lies in $F$, so (I) and (W) are solvable in polynomial time for every rational $E>3-2\sqrt2$ at every held degree, by $\BOX(E)$ for $E<1/2$ and trivially for $E\ge1/2$ (\cref{rem:degWI}).

Throughout this section $E$ is a fixed rational number with $0<E<1$ (for $E\ge1$ the all-ones vector is a (D)-$E$ vector by \cref{lem:relations}(c)), and
\[
\rho=\frac{1+E}{1-E},\qquad m=\lceil\rho\rceil,\qquad A=1+E .
\]
Then $\rho>1$, $m\ge2$, $\rho(1-E)=A$ and $E=(\rho-1)/(\rho+1)$. The map $E\mapsto\rho$ is an increasing bijection from $(0,1)$ onto $(1,\infty)$; the errors $1/7,1/5,1/4,1/3,1/2,3/5$ correspond to $\rho=4/3,3/2,5/3,2,3,4$. Held degree always means the number of held unit contracts, counted with multiplicity (\cref{def:Rd}); it is neither the number of distinct references nor a total network degree.

\subsection{The classification}\label{sec:degclass}

Put $d_H(E)=\bigl\lceil(m+1)\rho\bigr\rceil$ and, for every integer $d\ge1$,
\[
p_d=\inf\{E\in(0,1]\cap\mathbb Q:\ \text{(D)-}E\text{ on }\mathcal R(d)\text{ is solvable in polynomial time}\}.
\]
By \cref{lem:Rmono} the set in the definition of $p_d$ is closed upwards, so $p_d$ is the quantity $p(X)$ of \cref{sec:coincidence} for the family $X$ of problems (D)-$E$ on $\mathcal R(d)$. Always $d_H(E)\in[\rho^2+\rho,\ \rho^2+2\rho+1)$, because $\rho\le m<\rho+1$. For $d\ge1$ let $n=n(d)$ be the integer with $n(n+1)\le d<(n+1)(n+2)$, and for $d\in\{6,7\}$ and every $d\ge12$ put
\[
\bar p_d=\max\Bigl\{\frac{n-1}{n+1},\ \frac{d-n-2}{d+n+2}\Bigr\};
\]
thus $\bar p_6=\bar p_7=1/3$ and $\bar p_{12}=1/2$.

\begin{theorem}\label{thm:degmain}
Let $E$ be a fixed rational with $0<E<1$.
\begin{enumerate}[label=(\alph*)]
\item If $1/2\le E<1$, then (D)-$E$ on $\mathcal R(d)$ is solvable in polynomial time for every integer $d<(m+1)\rho$ and PPAD-complete for every integer $d\ge(m+1)\rho$. For $E=1/3$ it is solvable in polynomial time for $d\le5$ and PPAD-complete for $d\ge6$.
\item (D)-$E$ on $\mathcal R(d)$ is PPAD-complete for every $d\ge4$ if $E\le3/13$, for every $d\ge3$ if $E<1/7$, and for every $d\ge2$ if $E<1/9$. On $\mathcal R(1)$ an exact clearing vector that satisfies the solvency clause of (D) with $E=0$ is computable in linear time.
\item Let $d\in\{6,7\}$ or $d\ge12$. Then (D)-$\bar p_d$ on $\mathcal R(d)$ is PPAD-hard and (D)-$E$ on $\mathcal R(d)$ is solvable in polynomial time for every rational $E>\bar p_d$; in particular $p_d\le\bar p_d$. If PPAD is not contained in FP, then $p_d=\bar p_d$. Moreover $\lim_{d\to\infty}\sqrt d\,(1-\bar p_d)=2$.
\end{enumerate}
The hardness statements in (a) and (b) hold also for (D$^\circ$) and (S), and the polynomial-time statements hold also for (D$^\circ$). The polynomial algorithms run in time linear in the number of banks and held contracts, counted in incidence and word operations, and have polynomial bit complexity in the binary encoding of the input.
\end{theorem}

The FP, PPAD-hardness and PPAD-completeness statements in (a), (b) and (c) are unconditional; reading them as an exclusive alternative between polynomial and non-polynomial problems, and the identification $p_d=\bar p_d$ in (c), require that PPAD is not contained in FP. The proof is given in \cref{sec:phase}, after \cref{cor:asym}; the full phase diagram, including the degrees and errors that remain open, is \cref{tab:phase}. Part (a) for $E\ge1/2$ is \cref{thm:degE12}, the case $E=1/3$ is \cref{cor:deg13}, part (b) is \cref{thm:lowdeg}, and part (c) is \cref{cor:pd,cor:asym}.

\paragraph{What is inherited and what is new.} Pure-Circuit and its gate semantics (\cref{thm:pc}) are known. The sparse NAND receivers and the two-arm PURIFY topology appear with weighted contracts in the earlier construction of~\cite{DHK}, which attributes the NAND topology to Schuldenzucker, Seuken and Battiston and the PURIFY topology to Ioannidis, de Keijzer and Ventre (\cref{sec:literature}), and in \cref{thm:D}; from \cref{thm:D} we also take the decoding of bit $1$ through the solvency clause. Supported propagation with permanent states is the mechanism of~\cite[Lemma~F.8 and Theorem~F.9]{KM}, which gives (D)-$1/5$ at held degree $2$. The new content is the parameter design of \cref{thm:dH}, the restoring chains of \cref{thm:lowdeg}, the refined rule of \cref{thm:Pk}, which relaxes the rule of~\cite{KM}, the propagation rules of \cref{thm:A2,thm:A3}, and the combination of these into the classification of \cref{thm:degmain}. The direct expansion of the notionals of \cref{thm:D} into unit contracts, with an integer $C\ge\max\{3,2\rho\}$ in place of $2\rho$, has held degree $C^2\ge9$; \cref{thm:dH} lowers the degree to $d_H(E)<(\rho+1)^2\le C^2$, for instance from $16$ to $6$ at $E=1/3$, and \cref{thm:lowdeg} lowers it to $2$ for $E<1/9$. In the primary sources we checked we found no theorem that subsumes these degree-sensitive bounds (\cref{sec:literature}); this is not a claim of exhaustive priority.

\subsection{Hardness at the critical held degree}\label{sec:critdeg}

\begin{theorem}\label{thm:dH}
Let $E\in(0,1)$ be rational and $D=d_H(E)=\lceil(m+1)\rho\rceil$. Then $D\ge2m$, and for every integer $d\ge D$ the problems (D)-$E$, (D$^\circ$)-$E$ and (S)-$E$ on $\mathcal R(d)$ are PPAD-complete. Hardness holds already for networks in which every unit CDS has its own writer with external assets $3$ and the creditor has external assets $1$. On these networks every (D$^\circ$)-$E$ vector, hence every (D)-$E$ and every (S)-$E$ vector, decodes to a Pure-Circuit solution.
\end{theorem}

To see $D\ge2m$: for $m=2$, $3\rho>3$ gives $D\ge4$; for $m\ge3$, $\rho>m-1$ gives $(m+1)\rho>m^2-1\ge2m$.

\paragraph{Construction.} Reduce from \PCirc\ with gates NAND and PURIFY (\cref{thm:pc}). For every node $y$ create a core bank $b_y$, and for every PURIFY gate $g$ two further core banks $a_g,c_g$.
\begin{itemize}
\item $\mathrm{NAND}(u,v;w)$: $b_w$ holds $m$ unit CDSs on $b_u$ and $m$ on $b_v$.
\item $\mathrm{PURIFY}(u;v,w)$: $a_g$ holds $D$ unit CDSs on $b_u$; $b_v$ holds $m$ on $a_g$; $c_g$ holds $m$ on $b_u$; $b_w$ holds $D$ on $c_g$.
\end{itemize}
Every unit CDS gets its own writer with external assets $3$, whose largest possible liability is $1$. All core banks have $e=0$ and owe $1$ to one creditor $z$ with $e_z=1$. Every node is the output of exactly one gate, so each $b_y$ holds exactly the contracts listed for its gate. The nodes of a gate are pairwise distinct and $a_g,c_g$ are fresh, so no core bank references itself and every CDS has three distinct parties. The held degrees are $2m$ at NAND outputs, $D$ at $a_g$ and at the second PURIFY output, and $m$ at $c_g$ and at the first PURIFY output. Since $D\ge2m>m$, the network lies in $\mathcal R(D)$, and $e_w=3>(1+E)L^{\max}_w$.

\paragraph{Decoding.} For a node $y$ let $X_y=1-r_{b_y}$. Decode $y$ as $1$ if $X_y=0$, as $0$ if $X_y\ge1-E$ (that is, $r_{b_y}\le E$), and as $\bot$ otherwise. The two pure sets are disjoint because $E<1$.

\begin{lemma}[gates]\label{lem:dHgates}
Let $r$ be a (D$^\circ$)-$E$ vector of the network. Then \eqref{eq:core} holds at every core bank, and:
\begin{enumerate}[label=(\alph*)]
\item for $\mathrm{NAND}(u,v;w)$: if $X_u\ge1-E$ or $X_v\ge1-E$ then $X_w=0$; if $X_u=X_v=0$ then $X_w\ge1-E$;
\item for $g=\mathrm{PURIFY}(u;v,w)$: if $X_u=0$ then $X_v=X_w=0$; if $X_u\ge1-E$ then $X_v,X_w\ge1-E$; if $X_u\ge A/D$ then $X_v\ge1-E$; if $X_u<A/D$ then $X_w=0$.
\end{enumerate}
\end{lemma}
\begin{proof}
\eqref{eq:core} holds by \cref{lem:core}(c).

(a) If $X_u\ge1-E$, then $S_{b_w}\ge m(1-E)\ge\rho(1-E)=A$, so $X_w=0$ by (Sol); likewise for $v$. If $X_u=X_v=0$, then $S_{b_w}=0$ and (Res) gives $X_w\ge1-E$.

(b) \emph{Input $X_u=0$.} Then $S_{a_g}=S_{c_g}=0$, so $x_{a_g},x_{c_g}\ge1-E$ by (Res). Hence $S_{b_v}\ge m(1-E)\ge A$ and $S_{b_w}\ge D(1-E)\ge A$, and (Sol) gives $X_v=X_w=0$.
\emph{Input $X_u\ge1-E$.} Then $S_{a_g}\ge D(1-E)\ge A$ and $S_{c_g}\ge m(1-E)\ge A$, so $x_{a_g}=x_{c_g}=0$. Hence $S_{b_v}=S_{b_w}=0$ and $X_v,X_w\ge1-E$ by (Res).
\emph{Input $X_u\ge A/D$.} Then $S_{a_g}=DX_u\ge A$, so $x_{a_g}=0$, $S_{b_v}=0$ and $X_v\ge1-E$.
\emph{Input $X_u<A/D$.} Then $S_{c_g}=mX_u<mA/D$. The inequality $D\ge(m+1)\rho$ is equivalent to $(m+1)A\le D(1-E)$, that is, to
\begin{equation}\label{eq:dHsplit}
1-E-\frac{mA}D\ \ge\ \frac AD .
\end{equation}
By (Res), $x_{c_g}\ge1-E-S_{c_g}>1-E-mA/D\ge A/D$. So $S_{b_w}=Dx_{c_g}>A$, and $X_w=0$ by (Sol).
\end{proof}

\begin{proof}[Proof of \cref{thm:dH}]
Let $r$ be a (D$^\circ$)-$E$ vector and decode it. For NAND, \cref{lem:dHgates}(a) says that a decoded $0$ at either input forces a decoded $1$ at the output, and two decoded inputs $1$ force a decoded $0$; this is the NAND constraint. For PURIFY, the first two cases of \cref{lem:dHgates}(b) copy a pure input to both outputs. If the input is $\bot$, either $X_u\ge A/D$ and $v$ decodes to $0$, or $X_u<A/D$ and $w$ decodes to $1$; so at least one output is pure. Hence the decoded assignment solves the \PCirc\ instance, which is PPAD-hard for the gate set $\{\mathrm{NAND},\mathrm{PURIFY}\}$. For fixed $E$ the numbers $m$ and $D$ are constants, so the network has size linear in the instance and the construction and decoding take polynomial time. Every (D)-$E$ vector and every (S)-$E$ vector is a (D$^\circ$)-$E$ vector (\cref{lem:relations}(a)), so the same decoding applies to them. Every network of $\mathcal R(d)$ is non-degenerate (\cref{sec:Rclass}), so (D)-$E$ on $\mathcal R(d)$ lies in PPAD by~\cite[Appendix~B, Proposition~3]{DHK} (the membership cited in \cref{thm:D}); (D$^\circ$)-$E$ and (S)-$E$ lie in PPAD on all networks by \cref{thm:gap}. So all three problems are PPAD-complete on $\mathcal R(D)$, and \cref{lem:Rmono} extends hardness to every $d\ge D$.
\end{proof}

The decoder needs every (D$^\circ$)-$E$ vector to decode correctly, not only one exact solution; this holds because each implication in \cref{lem:dHgates} is pointwise. The equality case $X_u=A/D$ belongs to the first PURIFY branch, and equality in $D\ge(m+1)\rho$ loses no input value, because the solvency clause of (D) is inclusive. At $E=1/3$ we have $m=2$, $D=6$, and the split is at $X_u=2/9$.

\subsection{Restoring chains: held degrees 2, 3 and 4}\label{sec:chains}

\Cref{thm:dH} needs held degree at least $2m\ge4$, because its NAND output must reach assets $A$ from a single defaulting input. Below that, a bank that sees one defaulting input (assets at least $1-E$) is only known to have default at most $2E$. A chain of banks can amplify such a weak bound back to an exact $0$.

\begin{definition}[chain arms]\label{def:chain}
A \emph{chain arm} with multiplicities $(k_1,\dots,k_L)$, $k_i\ge1$, hanging from a core bank $u$ consists of pairwise distinct core banks $s_1,\dots,s_L$, all distinct from $u$, such that $s_1$ holds exactly $k_1$ unit CDSs, all on $u$, and each $s_i$ ($2\le i\le L$) holds exactly $k_i$ unit CDSs, all on $s_{i-1}$. Its \emph{output} is $s_L$, and $s_0:=u$. We write $(k_1,\dots,k_r)^{\times n}$ for $n$ consecutive copies of a sequence, so $(3,2)^{\times2}=(3,2,3,2)$. A \emph{chain PURIFY gadget} consists of two chain arms hanging from the same input bank $u$ with disjoint sets of banks; their outputs are the two outputs of the gate.
\end{definition}

The mechanism is the following. If a bank $Q$ holds $2$ contracts on $P$ and a bank $R$ holds $k$ contracts on $Q$, then $x_P\le h$ implies $x_R\le G_{2,k}(h):=\mathrm{up}(k\,\mathrm{lo}(2h))$ (\cref{lem:restore}(a)). For $k\in\{2,3\}$ and small $E$, the map $G_{2,k}$ is affine with slope $2k$ near its fixed point
\[
h^*_k=\frac{k(1-E)-A}{2k-1}\qquad\Bigl(h^*_2=\frac{1-3E}3,\quad h^*_3=\frac{2-4E}5\Bigr),
\]
and every start below $h^*_k$ reaches $0$ after finitely many pairs of stages (\cref{lem:restore}(b)). The weak bound $2E$ lies below $h^*_2$ exactly when $E<1/9$ and below $h^*_3$ exactly when $E<1/7$; at $E=1/9$ and $E=1/7$ the weak bound is itself a positive fixed point. In these ranges let $N_k(E)$ be the least $j\ge1$ with $G_{2,k}^{\,j}(2E)=0$; it exists by \cref{lem:restore}(b), and for fixed rational $E$ it is a constant, computed by iterating $G_{2,k}$ in exact arithmetic. For example $N_2(E)=1$ if and only if $E\le1/11$, and $N_3(E)=1$ if and only if $E\le1/8$.

\begin{theorem}\label{thm:lowdeg}
Let $E$ be rational.
\begin{enumerate}[label=(\alph*)]
\item If $0<E\le3/13$, then (D)-$E$, (D$^\circ$)-$E$ and (S)-$E$ on $\mathcal R(d)$ are PPAD-complete for every $d\ge4$.
\item If $0<E<1/7$, the same holds for every $d\ge3$.
\item If $0<E<1/9$, the same holds for every $d\ge2$. Since an exact rational clearing vector of a network in $\mathcal R(1)$ can be computed in linear time (see the proof), for $0<E<1/9$ each of the three problems on $\mathcal R(d)$ is in FP for $d\le1$ and PPAD-complete for every $d\ge2$.
\end{enumerate}
\end{theorem}

The endpoint $3/13$ is included in (a); the endpoints $1/7$ and $1/9$ are excluded in (b) and (c), and hardness at them is not asserted.

\paragraph{Construction.} As in \cref{thm:dH}: \PCirc\ with gates NAND and PURIFY, a core bank $b_y$ for every node, fresh core banks for each gate, every unit CDS with its own writer of external assets $3$, all core banks with $e=0$ owing $1$ to the creditor $z$ with $e_z=1$. The gates are realized as follows, with $N=N_3(E)$ in case (b) and $M=N_2(E)$ in case (c).
\begin{center}\small
\begin{tabular}{@{}lR{5.6cm}lll@{}}
\toprule
case & $\mathrm{NAND}(u,v;w)$ & arm from $b_u$ to $b_v$ & arm from $b_u$ to $b_w$ & $\theta$\\
\midrule
(a) & $b_w$ holds $2$ on $b_u$ and $2$ on $b_v$ & $(4,2,4,2)$ & $(2,4,2,4)$ & $(1-E)/3$\\
(b) & fresh $g,p$: $g$ holds $1$ on $b_u$ and $1$ on $b_v$; $p$ holds $2$ on $g$; $b_w$ holds $1$ on $b_u$, $1$ on $b_v$, $1$ on $p$ & $(3,2)^{\times(N+1)}$ & $(2,3)^{\times(N+1)}$ & $(1-E)/3$\\
(c) & fresh $g$ holding $1$ on $b_u$ and $1$ on $b_v$, and a chain arm $(2)^{\times 2M}$ hanging from $g$ with output $b_w$ & $(2)^{\times(2M+2)}$ & $(1)(2)^{\times(2M+1)}$ & $(1-E)/2$\\
\bottomrule
\end{tabular}
\end{center}
All arms have even length. Held degrees are at most $4$, $3$, $2$ respectively. Every node is the output of exactly one gate, the nodes of a gate are distinct and all other gate banks are fresh, so every CDS has three distinct parties, no core bank references itself, and $e_w=3>(1+E)L^{\max}_w$. The parameter $\theta$ is the split point of the PURIFY gadget: an input with default at least $\theta$ forces the output $b_v$ to decode to $0$, and an input with default at most $\theta$ forces $b_w$ to decode to $1$ (\cref{lem:lowgates}).

In case (c) the first stage of the arm to $b_w$ has multiplicity $1$. It is not an exact inverter: for a defaulting pure input it only gives default at most $2E$, and the proof restores this weak bound by the following pairs of stages, as it does for the weak NAND output. \Cref{app:chains} gives the complete proof of \cref{thm:lowdeg}, including the exact chain envelopes (\cref{lem:envelopes}), restoration (\cref{lem:restore}), the gate claims for arbitrary inputs (\cref{lem:lowgates}) and the linear-time exact algorithm at held degree $1$.

\subsection{Supported propagation}\label{sec:propagation}

All counts below are over held contracts, with multiplicity. For a core bank $b$ in a current labelling let $n_{\neg\mathsf L}(b)$ be the number of its held contracts whose reference is not labelled $\mathsf L$, $n_{\mathsf H}(b)$ the number whose reference is labelled $\mathsf H$, and, where defined, $n_{\mathsf S}(b)$ the number whose reference is a \emph{strong} $\mathsf H$ bank.

\paragraph{Engine $\mathcal P(k)$} ($k\ge0$). Labels $\mathsf U$ (unresolved), $\mathsf L$, $\mathsf H$; initially all banks are $\mathsf U$. Repeatedly pick a $\mathsf U$ bank $b$ and apply:
\begin{itemize}
\item[(P-L)] if $n_{\mathsf H}(b)\ge1$, label $b$ with $\mathsf L$;
\item[(P-H)] otherwise, if $n_{\neg\mathsf L}(b)\le k$, label $b$ with $\mathsf H$.
\end{itemize}

\paragraph{Engine $\mathcal Q(\nu)$} ($\nu\in\{1,2\}$). Labels $\mathsf U,\mathsf L,\mathsf H$; an $\mathsf H$ bank $j$ is \emph{strong} when $n_{\neg\mathsf L}(j)\le\nu-1$ and \emph{weak} otherwise. Repeatedly pick a $\mathsf U$ bank $b$ and apply:
\begin{itemize}
\item[(Q-L)] if $n_{\mathsf S}(b)\ge1$ or $n_{\mathsf H}(b)\ge2$, label $b$ with $\mathsf L$;
\item[(Q-H)] otherwise, if $n_{\neg\mathsf L}(b)\le\nu$, label $b$ with $\mathsf H$.
\end{itemize}
Both engines stop when no rule applies to any $\mathsf U$ bank.

\begin{lemma}[engine invariants]\label{lem:engines}
In both engines the labels $\mathsf L$ and $\mathsf H$ are permanent. For each bank, $n_{\neg\mathsf L}$ never increases, while $n_{\mathsf H}$ and $n_{\mathsf S}$ never decrease; in $\mathcal Q(\nu)$ a strong bank stays strong. Each engine stops after at most $|V|$ label assignments, whatever the order of rule applications, and can be implemented with $O(|V|+M)$ operations, $M$ the total number of held contracts. At termination:
\begin{itemize}
\item in $\mathcal P(k)$: every $\mathsf L$ bank has $n_{\mathsf H}\ge1$; every $\mathsf H$ bank has $n_{\neg\mathsf L}\le k$; every $\mathsf U$ bank has $n_{\mathsf H}=0$ and $n_{\neg\mathsf L}\ge k+1$, so all its non-$\mathsf L$ references are $\mathsf U$ banks;
\item in $\mathcal Q(\nu)$: every $\mathsf L$ bank has $n_{\mathsf S}\ge1$ or $n_{\mathsf H}\ge2$; every $\mathsf H$ bank has $n_{\neg\mathsf L}\le\nu$, and it is strong if $n_{\neg\mathsf L}\le\nu-1$ and weak if $n_{\neg\mathsf L}=\nu$; every $\mathsf U$ bank has $n_{\mathsf S}=0$, $n_{\mathsf H}\le1$ and $n_{\neg\mathsf L}\ge\nu+1$, so its non-$\mathsf L$ references are $\mathsf U$ banks except for at most one held contract referencing a weak bank.
\end{itemize}
\end{lemma}

Its proof, which also gives the implementation with $O(|V|+M)$ operations, is in \cref{app:degree}.

In every algorithm below, all writers and the creditor receive recovery $1$, and the value of a core bank depends only on its terminal label (and, in $\mathcal Q$, on strong versus weak). By \cref{lem:core}(b), correctness reduces to~\eqref{eq:core} at every core bank. Each value is one of at most four fixed rationals, so the output has polynomial bit length. The linear bounds count incidence processing and word operations on the input numbers; in the bit model the running time is polynomial in the length of the binary encoding of the input, including the cost of reading it. The hardness reductions fix $E$ in advance and make no uniform size claim in the binary length of $1/(1-E)$. In each proof, $S$ denotes the assets of the bank under consideration. References labelled $\mathsf L$ contribute $0$ to $S$, so only non-$\mathsf L$ references count.

\begin{theorem}[refined rule]\label{thm:Pk}
Let $k\ge1$ be an integer with $k<\rho$, and assume $E\ge1/2$ or $k+1\le\rho$. Run $\mathcal P(k)$ and give $\mathsf L$ banks $x=0$, $\mathsf H$ banks $x=h:=1-E$, and $\mathsf U$ banks $x=t:=(1-E)/(k+2)$. If every core bank has held degree $d<(k+2)\rho$, the result is a (D)-$E$ vector. In particular:
\begin{itemize}
\item for $E\ge1/2$, with $k=m-1$: every $d<(m+1)\rho$;
\item for $E=1/3$, with $k=1$: every $d\le5$;
\item for $1/3<E<1/2$, with $k=1$: every $d<3\rho$.
\end{itemize}
\end{theorem}

The hypotheses are $E\ge\min\{1/2,k/(k+2)\}$ and $E>(k-1)/(k+1)$, rewritten in terms of $\rho$: $E\ge k/(k+2)$ is $k+1\le\rho$, and $E>(k-1)/(k+1)$ is $k<\rho$. The rule relaxes the supported propagation of~\cite[Lemma~F.8]{KM}, which labels a bank $\mathsf H$ only when all its references are $\mathsf L$: here up to $k$ references may still be unresolved when the bank is labelled $\mathsf H$.

\begin{proof}
All values lie in $[0,h]$ because $t\le h$.

\emph{$\mathsf L$ bank.} It references an $\mathsf H$ bank, so $S\ge h=1-E$, $T\le E$, and $x=0$ satisfies~\eqref{eq:core}.

\emph{$\mathsf H$ bank.} At most $k$ non-$\mathsf L$ references, each of value at most $h$, so $0\le S\le kh$. Since $T\le1$, $T-h\le1-h=E$. For $h-T$: if $S\le1$, $h-T=h+S-1\le(k+1)h-1$; if $S>1$, $h-T=h$. When $E\ge1/2$ both are at most $E$, as $h+S-1\le h\le E$. When $E<1/2$ the hypothesis $k+1\le\rho$ gives $(k+1)(1-E)\le1+E$, i.e.\ $(k+1)h-1\le E$, and also $kh\le1+E-h=2E<1$, so the case $S>1$ does not occur. The (D) clause: $S\le kh<1+E$ because $k<\rho$.

\emph{$\mathsf U$ bank.} Its non-$\mathsf L$ references are $q$ $\mathsf U$ banks with $k+1\le q\le d$, so $S=qt$. Then $T-t\le1-(k+1)t-t=1-(k+2)t=E$ (and $T-t<0$ when $T=0$), and $t-T\le t\le E$: if $E\ge1/2$ then $t\le1-E\le E$; if $E<1/2$ then $k+1\le\rho$ gives $E\ge k/(k+2)\ge1/3>1/(k+3)$, which is $t\le E$. The (D) clause: $S\le dt<1+E$ is equivalent to $d<(k+2)\rho$.

For the special cases: if $E\ge1/2$ then $\rho\ge3$ and $k=m-1$ satisfies $1\le k<\rho$; if $1/3\le E<1/2$ then $2\le\rho<3$ and $k=1$ satisfies $k+1\le\rho$.
\end{proof}

\begin{theorem}[rule A2]\label{thm:A2}
Let $E>1/3$ and $t=(1-E)/4$. Run $\mathcal Q(2)$ and give $\mathsf L$ banks $x=0$, strong banks $x=4t=1-E$, weak banks $x=2t$, and $\mathsf U$ banks $x=t$. If every core bank has held degree $d<4\rho-1$, the result is a (D)-$E$ vector.
\end{theorem}

\begin{theorem}[rule A3]\label{thm:A3}
Let $E\ge1/4$ and $t=(1-E)/3$. Run $\mathcal Q(1)$ and give $\mathsf L$ banks $x=0$, strong banks $x=3t=1-E$, weak banks $x=2t$, and $\mathsf U$ banks $x=t$. If every core bank has held degree $d<3\rho-1$, the result is a (D)-$E$ vector.
\end{theorem}

The proofs of \cref{thm:A2,thm:A3} check~\eqref{eq:core} label by label, as in the proof of \cref{thm:Pk}; they are given in \cref{app:degree}.

Rule A2 needs $E>1/3$ strictly, while rule A3 includes $E=1/4$; at $E=1/4$ its degree bound $d<3\rho-1=4$ excludes held degree $4$. The degree bounds of \cref{thm:Pk,thm:A2,thm:A3} are exactly where these decoders stop working (\cref{prop:decodertight}); the networks that witness this have easily found (D)-$E$ vectors, so they are not hard instances.

\subsection{Thresholds and the phase diagram}\label{sec:phase}

Write $\mathrm{FP}(E,d)$ if (D)-$E$ on $\mathcal R(d)$ is solvable in polynomial time, and $\mathrm{HARD}(E,d)$ if it is PPAD-hard. \Cref{thm:dH,thm:Pk,thm:A2,thm:A3}, with \cref{lem:Rmono}, give the following; \cref{thm:lowdeg} enters only \cref{tab:phase} and the low-degree ranges after it.

\begin{theorem}[degree classification for $1/2\le E<1$]\label{thm:degE12}
Let $1/2\le E<1$ be rational. Then $\mathrm{FP}(E,d)$ for every integer $d<(m+1)\rho$, and $\mathrm{HARD}(E,d)$ for every integer $d\ge(m+1)\rho$.
\end{theorem}
\begin{proof}
The first part is \cref{thm:Pk} with $k=m-1$. For the second, an integer $d\ge(m+1)\rho$ satisfies $d\ge\lceil(m+1)\rho\rceil=d_H(E)$; apply \cref{thm:dH}.
\end{proof}

\begin{corollary}[$E=1/3$]\label{cor:deg13}
(D)-$1/3$ on $\mathcal R(d)$ is solvable in linear time for $d\le5$ and PPAD-complete for every $d\ge6$. More generally, for $E=(n-1)/(n+1)$ with integer $n\ge2$, (D)-$E$ on $\mathcal R(d)$ is solvable in linear time for $d\le n(n+1)-1$ and PPAD-complete for every $d\ge n(n+1)$.
\end{corollary}
\begin{proof}
At $E=1/3$, $\rho=2=m$ and $d_H=6$, so \cref{thm:dH} gives PPAD-completeness for every $d\ge6$, and \cref{thm:Pk} with $k=1$ gives a linear-time algorithm for $d\le5$. For $n\ge3$ we have $E\ge1/2$, $\rho=m=n$ and $d_H=n(n+1)$; apply \cref{thm:dH} and \cref{thm:Pk} with $k=m-1$.
\end{proof}

In particular, at $E=1/3$ the held degrees $6$ to $15$, which the direct unit expansion of \cref{thm:D} does not reach (\cref{rem:dHnew}), are PPAD-hard.

\begin{table}[t]
\centering\small
\begin{tabular}{@{}lllll@{}}
\toprule
error $E$ & $\rho$ & FP for held degree & PPAD-hard for held degree & unresolved\\
\midrule
$1/2\le E<1$ & $\rho\ge3$ & $d<(m+1)\rho$ & $d\ge(m+1)\rho$ & none\\
$1/3<E<1/2$ & $2<\rho<3$ & $d<4\rho-1$ & $d\ge4\rho$ & $d=\lceil4\rho\rceil-1$\\
$E=1/3$ & $2$ & $d\le5$ & $d\ge6$ & none\\
$1/4\le E<1/3$ & $5/3\le\rho<2$ & $d<3\rho-1$ & $d\ge3\rho$ & $d=\lceil3\rho\rceil-1$\\
$3/13<E<1/4$ & $8/5<\rho<5/3$ & $d\le2$ & $d\ge5$ & $3,4$\\
$1/5\le E\le3/13$ & $3/2\le\rho\le8/5$ & $d\le2$ & $d\ge4$ & $3$\\
$1/7\le E<1/5$ & $4/3\le\rho<3/2$ & $d\le1$ & $d\ge4$ & $2,3$\\
$1/9\le E<1/7$ & $5/4\le\rho<4/3$ & $d\le1$ & $d\ge3$ & $2$\\
$0<E<1/9$ & $1<\rho<5/4$ & $d\le1$ & $d\ge2$ & none\\
\bottomrule
\end{tabular}
\caption{Phase diagram of (D)-$E$ on $\mathcal R(d)$ for fixed rational $E\in(0,1)$, with $m=\lceil\rho\rceil$. ``FP'' lists the held degrees with a polynomial algorithm and ``PPAD-hard'' those with a hardness proof; both columns are unconditional. ``Unresolved'' means not settled by the results of this paper and the sources checked. The FP entries with $d\le2$ below $E=1/4$ are the exact algorithm at held degree $1$ (\cref{thm:lowdeg}(c)) and (D)-$E$ at held degree $2$ for $E\ge1/5$~\cite[Theorem~F.9]{KM}, reproved as the case $a=1$, $Q=2$ of \cref{thm:weightedD}.}\label{tab:phase}
\end{table}

\begin{corollary}[the threshold error $p_d$]\label{cor:pd}
Let $d\in\{6,7\}$ or $d\ge12$. Then (D)-$\bar p_d$ on $\mathcal R(d)$ is PPAD-hard, and (D)-$E$ on $\mathcal R(d)$ is in FP for every rational $E>\bar p_d$; hence $p_d\le\bar p_d$. If PPAD is not contained in FP, then $p_d=\bar p_d$.
\end{corollary}

The proof is in \cref{app:degree}.

If PPAD is not contained in FP, the infimum $p_d$ is not attained by a polynomial algorithm in these cases.

\begin{corollary}[asymptotics]\label{cor:asym}
For $d\ge12$,
\[
\frac{2}{n+2}\le1-\bar p_d\le\frac2{n+1},\qquad \frac2{\sqrt d+2}\le1-\bar p_d<\frac2{\sqrt d-1/2},
\]
so $\lim_{d\to\infty}\sqrt d\,(1-\bar p_d)=2$. If PPAD is not contained in FP, the same holds for $p_d$ by \cref{cor:pd}.
\end{corollary}
\begin{proof}
$1-\bar p_d=\min\{2/(n+1),\,2(n+2)/(d+n+2)\}$, and $d<(n+1)(n+2)$ gives $2(n+2)/(d+n+2)>2(n+2)/(n+2)^2=2/(n+2)$. From $n(n+1)\le d<(n+1)(n+2)$: $n\le\sqrt d$, so $n+2\le\sqrt d+2$; and $(n+\frac32)^2>(n+1)(n+2)>d$, so $n+1>\sqrt d-\frac12$.
\end{proof}

\begin{proof}[Proof of \cref{thm:degmain}]
(a) is \cref{thm:degE12} together with the PPAD membership of (D)-$E$ on $\mathcal R(d)$ shown in the proof of \cref{thm:dH}, and \cref{cor:deg13}. (b) is \cref{thm:lowdeg}. (c) is \cref{cor:pd,cor:asym}. The hardness statements for (D$^\circ$) and (S) are part of \cref{thm:dH,thm:lowdeg}. Every (D)-$E$ vector is a (D$^\circ$)-$E$ vector (\cref{lem:relations}(a)), so every algorithm for (D)-$E$ on $\mathcal R(d)$ also solves (D$^\circ$)-$E$ on $\mathcal R(d)$. The running times are those of \cref{lem:engines} and of the proof of \cref{thm:lowdeg}(c).
\end{proof}

The derivation of the entries of \cref{tab:phase} and of the known ranges of $p_d$ for the held degrees $2$ to $5$ and $8$ to $11$ is given in \cref{app:degree}, together with a comparison of $d_H(E)$ with the held degree of the direct unit expansion of \cref{thm:D} (\cref{rem:dHnew}).

\begin{remark}[(I) and (W) are different]\label{rem:degWI}
On $\mathcal R(d)$ every CDS writer lies in $F$, so for every rational $E$ with $3-2\sqrt2<E<1/2$, $\BOX(E)$ gives (I)-$E$ and (W)-$E$ vectors regardless of the held degree (\cref{thm:algws}, stated for every network whose CDS writers lie in $F$); for $E\ge1/2$ the vector of \cref{lem:relations}(c) is one. The degree phenomena of this section are specific to the exact-solvency notion (D). No statement here moves the windows of \cref{tab:landscape}.
\end{remark}

\begin{remark}[residual solutions need not satisfy exact solvency]\label{rem:residualD}
Let $0<E<1$. Take two core banks $b_1,b_2$ that hold no CDS and a third core bank $b_3$ that holds one unit CDS on each, every CDS with its own funded writer, and the creditor as in \cref{def:Rd}; this network lies in $\mathcal R(2)$. With recovery $1$ at writers and creditor, the defaults $x_{b_1}=x_{b_2}=1$ and $x_{b_3}=E$ give targets $T=(1,1,0)$, so every residual is at most $E$; but $S_{b_3}=2\ge1+E$ while $x_{b_3}=E>0$, so the vector is not a (D)-$E$ vector. Thus an algorithm for the residual condition alone need not return a (D)-$E$ vector, and the hardness results of this section do not by themselves give hardness of the residual problem on $\mathcal R(d)$ (\cref{sec:literature}).
\end{remark}

\subsection{Method limits}\label{sec:methodlimits}

\Cref{app:limits} shows that each method is tight at its own boundary, for its own decoder. The two-arm PURIFY with two stages per arm and the decoding of \cref{thm:dH} works exactly when its multiplicities satisfy an explicit inequality, and then needs held degree at least $(m+1)\rho$ (\cref{prop:twoarm}). For chain PURIFY gadgets with that decoding, the ranges of \cref{thm:lowdeg} are exactly the ranges in which gadgets of held degree $2$, $3$ and $4$ exist, and for $E\ge1/4$ no chain gadget has held degree below $d_H(E)$ (\cref{prop:chainlimits,cor:chainlimits}). The decoders of \cref{thm:Pk,thm:A2,thm:A3} fail exactly at their degree bounds, on networks that have easily found (D)-$E$ vectors (\cref{prop:decodertight}). These are method boundaries for the stated families, decodings and decoders. They do not exclude PURIFY gadgets that are not chains (for example with a bank holding contracts on several different banks, or with output paths that share banks), other decodings, other reductions, or other algorithms, and they are not complexity lower bounds. The degrees left open by \cref{tab:phase} are listed in \cref{sec:open}.

\section{Membership without non-degeneracy}\label{sec:membership}

PPAD membership of approximate clearing is usually proved for non-degenerate networks~\cite[Lemma~4.4]{SSB17}, \cite[Appendix~B]{DHK}: then every liability, or every asset, is bounded away from $0$, and the clearing map is Lipschitz. Without non-degeneracy the map is discontinuous where $A_i=\ell_i=0$. This section shows that (W) and (I) are nevertheless in PPAD on all networks (\cref{thm:member}), that the exempted notion (D$^\circ$) and the clipped-ratio notion (S) are in PPAD on all networks and PPAD-complete for every fixed error below~$1$ (\cref{thm:gap,cor:Dcirc}), and that (D) itself is not a total problem on all networks (\cref{prop:Dfail}). Throughout, $D_i=\sum_jc_{ij}$ and $N_{ik}=\sum_jc^k_{ij}$, so that $\ell_i(r)=D_i+\sum_kN_{ik}(1-r_k)$.

\subsection{(W) and (I)}

\begin{theorem}\label{thm:member}
Given a network with rational data and a rational $E>0$ in binary, finding a (W)-$E$ vector is in PPAD, and so is finding an (I)-$E$ vector. Such vectors always exist, with polynomially many bits. For every fixed rational $E$ at which (W) or (I) is PPAD-hard on non-degenerate networks, the problem on all networks is PPAD-complete.
\end{theorem}

The proof is in \cref{app:membership}. It applies the weak-approximation theorem of~\cite[Proposition~2.2]{EY10} to $\Phi\circ\mathrm{cap}$, where $\mathrm{cap}(x)_i=\min(x_i,1-\eta)$ with $\eta=E/2$, and $\mu$ denotes the least positive value of $D_i+\sum_kN_{ik}$.

Capping keeps every liability that is not identically zero at least $\eta\mu$, so the map is never evaluated near the discontinuity at $A_i=\ell_i=0$; the price is an additive $\eta$ in the error. The cap forbids $r_i=1$, which the (D) clause may require, so it does not apply to (D). This is why (D) needs a different argument.

\subsection{The exempted notion (D\texorpdfstring{$^\circ$}{-degree}) and the clipped-ratio notion (S)}

\begin{theorem}[empty-gap rounding]\label{thm:gap}
On arbitrary networks with rational data, finding a (D$^\circ$)-$E$ vector is in PPAD for every rational $E>0$ given in binary, and so is finding an (S)-$E$ vector. Polynomial-length rational solutions always exist.
\end{theorem}

The proof, in \cref{app:membership}, adds a tiny amount to every external asset, makes one query for a (D)-$(E/4)$ vector of the resulting non-degenerate network, and snaps to $1$ every recovery whose deficit lies below an empty interval among $n+1$ geometric scales. The empty interval gives at the same time a small perturbation and a lower bound on every remaining positive liability.

\begin{corollary}\label{cor:Dcirc}
For every fixed rational $0<E<1$, finding a (D$^\circ$)-$E$ vector and finding an (S)-$E$ vector are PPAD-complete on arbitrary networks.
\end{corollary}
\begin{proof}
Membership is \cref{thm:gap}. For hardness use the construction of \cref{thm:D}: there every bank owes a positive ordinary debt, so every liability is positive and every (D$^\circ$)-$E$ vector, in particular every (S)-$E$ vector (\cref{lem:relations}(a)), is decoded correctly by \cref{lem:Dgates}.
\end{proof}

\subsection{(D) is not total without non-degeneracy}

The displayed construction in the next proposition uses a rational error $E$, so that its data satisfy the convention of \cref{sec:model}. The nonexistence conclusion extends to every real target error in $(0,1)$ by the rational enlargement given after the proof.

\begin{proposition}\label{prop:Dfail}
Let $0<E<1$ and $C=(1+E)/(1-E)$. The network with banks $a,s,b$, external assets $(e_a,e_s,e_b)=(0,2C,0)$, an ordinary debt of $1$ from $b$ to $s$, and CDSs $(a,s,b;1)$ and $(s,b,a;C)$ has no (D)-$E$ vector. The conclusion does not depend on the convention used for $\Phi_a$ when $\ell_a=0$. The vector $(0,1,1-E)$ is a (W)-$E$ and an (I)-$E$ vector, and $(0,1,1)$ is a (D$^\circ$)-$E$ vector.
\end{proposition}
\begin{proof}
All CDS roles are distinct. The balances are $\ell_a=1-r_b$, $A_a=0$; $\ell_s=C(1-r_a)$, $A_s\ge2C$; $\ell_b=1$, $A_b=r_sC(1-r_a)$. Let $r$ be a (D)-$E$ vector. Since $A_s\ge2C\ge(1+E)\ell_s$, the clause gives $r_s=1$. If $r_b=1$, then $A_a=\ell_a=0$, so $A_a\ge(1+E)\ell_a$ and the clause gives $r_a=1$; then $A_b=0$, $\Phi_b=0$ and the residual of $b$ is $1>E$. If $r_b<1$, then $\ell_a>0$ and $\Phi_a=0$, so $r_a\le E$; then $A_b\ge C(1-E)=1+E=(1+E)\ell_b$, and the clause forces $r_b=1$, a contradiction. The value of $\Phi_a$ at $\ell_a=0$ was never used. For $(0,1,1-E)$: $\ell_a=E>0$ gives $\Phi_a=0$, $\Phi_s=1$, $\Phi_b=\min(1,C)=1$, so the residuals are $0,0,E$; and $T=\{s\}$. For $(0,1,1)$, bank $a$ has $\ell_a=0$ and is exempt, and $s$ and $b$ have residual $0$ with the clause satisfied.
\end{proof}

For an arbitrary real target error $E\in(0,1)$, choose a rational $\widehat E$ with $E\le\widehat E<1$ and use the displayed construction at $\widehat E$. Its data are rational and it has no (D)-$\widehat E$ vector; since every (D)-$E$ vector would be a (D)-$\widehat E$ vector by \cref{lem:relations}(b), it has no (D)-$E$ vector either. This supplies the real-error nonexistence assertion of \cref{cor:Wtotal} without using irrational contract notionals.

The same network has no exact clearing vector: with $r_s=1$, $r_b=1$ forces $\Phi_a=1$ and then $\Phi_b=0$, while $r_b<1$ forces $r_a=0$ and then $\Phi_b=1$.

\begin{remark}[the question of~\cite{DHK}]\label{rem:dhkquestion}
The last paragraph of~\cite[Appendix~A]{DHK} suggests leaving the recovery of a bank without liabilities unconstrained, notes that clearing vectors then always exist, and states that it is not known whether computing an $\eps$-approximate clearing recovery rate vector, in the sense of their Definition~3, is in PPAD without non-degeneracy. \Cref{prop:Dfail} shows that Definition~3 as written is not a total problem on arbitrary networks, so it is not in PPAD, which contains only total problems. With zero-liability banks exempted from both clauses, the problem becomes total: the correspondence of~\cite{SSB17}, which leaves the recovery of a zero-liability bank unconstrained, has nonempty convex compact values and a closed graph, so Kakutani's theorem gives an exact fixed point, and an exact fixed point is a (D$^\circ$)-$E$ vector for every $E\ge0$. \Cref{thm:gap,cor:Dcirc} settle the complexity of this exempted problem. The convention matters: under $\Phi_i=1$ at $\ell_i=0$, the network of \cref{prop:Dfail} has no exact clearing vector, while under the unconstrained convention it has $(0,1,1)$.
\end{remark}

\section{Exact existence, totality, and stalled boxes}\label{sec:existence}

\subsection{Existence of exact and approximate clearing vectors}

\begin{proposition}\label{prop:wsexact}
Every network in which every CDS writer lies in $F$, in particular every writer-solvent network, has an exact clearing vector.
\end{proposition}
\begin{proof}
Banks of $F$ have $\Phi_i\equiv1$. A bank outside $F$ writes no CDS, so its liability is the constant $D_i$. If $D_i=0$ then $\Phi_i\equiv1$; otherwise $\Phi_i=\min(1,A_i/D_i)$ with $A_i$ continuous. So $\Phi$ is a continuous self-map of $[0,1]^n$, and Brouwer's theorem gives a fixed point.
\end{proof}

\begin{proposition}[\cite{DHK,IKV22}]\label{prop:ndexact}\KN\
Every non-degenerate network has an exact clearing vector. The hypothesis cannot be dropped: there are degenerate networks with three banks and no exact clearing vector.
\end{proposition}
\begin{proof}
Under \cref{def:classes}, $\max(A_i(r),\ell_i(r))\ge e_i>0$ or $\ell_i(r)\ge c_{ij}>0$, so $f_i=A_i/\max(A_i,\ell_i)$ is continuous on the whole cube and coincides with $\Phi_i$; Brouwer's theorem applies. This argument is given in~\cite[Section~2.2]{DHK}, which attributes the observation to~\cite{IKV22}. A degenerate network without clearing vector is given in~\cite[Appendix~A]{DHK}; the network of \cref{prop:Dfail} is another one.
\end{proof}

\begin{corollary}\label{cor:Wtotal}
For every rational $E>0$, every network with rational data has a (W)-$E$ vector, an (I)-$E$ vector, a (D$^\circ$)-$E$ vector and an (S)-$E$ vector, each with a number of bits polynomial in the size of the network and of $E$. For every real $E>0$ such vectors exist. For every $E\in(0,1)$ there are networks without (D)-$E$ vectors.
\end{corollary}
\begin{proof}
The first sentence restates \cref{thm:member,thm:gap}. For real $E>0$ apply them to a rational $E'\in(0,E]$ and use \cref{lem:relations}(b). The last sentence is \cref{prop:Dfail}.
\end{proof}

So the notion (D) of~\cite[Definition~3]{DHK}, as written, is not total on arbitrary networks, while its exempted version (D$^\circ$) is total and lies in PPAD.

\subsection{Stalled boxes on inverter configurations}\label{sec:stalls}

Below the thresholds of \cref{sec:box} the algorithm $\BOX(E)$ can stall (\cref{rem:boxtests}). We say that $\BOX(E)$ \emph{stalls} in a round if some bank is unsettled, no bank settles in the round, and every unsettled box is unchanged by the round; all later rounds are then identical, and the stall can be detected. A natural idea for a global rule at a stall is to solve the affine system that the stall determines. This subsection, which is about networks whose CDS writers all lie in $F$, shows that this rule is exact on the unsettled banks when its solution lies in the stalled box, proves (in \cref{app:membership}) that the solution lies there in two cases, and shows by an example with three banks that it does not in general.

Consider a stall of $\BOX(E)$ with unsettled set $U$, stalled box $P=\prod_{i\in U}[lo_i,hi_i]$, widths $w=hi-lo$, settled points $p_k$ outside $U\cup F$, and the value $1$ on $F$. A bank $i\in U$ lies outside $F$, so it writes no CDS, its liability is the constant $D_i$, and $D_i=L_i^{\max}>e_i\ge0$. We call the stall an \emph{inverter stall} if
\begin{itemize}
\item[(H)] no bank of $U$ is owed an ordinary debt by another bank of $U$.
\end{itemize}
Then, with the settled banks at their points, every $i\in U$ has
\[
\Phi_i(r)=\min\bigl(1,\Psi_i(r_U)\bigr),\qquad \Psi(r_U)=\tilde a-Nr_U,
\]
where $\tilde a_i$ is $1/D_i$ times the assets of $i$ at $r_U=0$, and $N_{ij}\ge0$ is $1/D_i$ times the total notional of the CDSs held by $i$ with reference $j\in U$; $N_{ii}=0$ by distinctness of roles. Every unsettled bank's assets are nonincreasing in the recovery of every other unsettled bank, which is what the word ``inverter'' refers to.

\begin{proposition}[structure of an inverter stall]\label{prop:stall}
At an inverter stall:
\begin{enumerate}[label=(\alph*)]
\item the enclosures of the round are exact: $\lambda_i=\min(1,\Psi_i(hi))$ and $\mu_i=\min(1,\Psi_i(lo))$ for $i\in U$;
\item $\Psi(hi)=lo-E\mathbf 1$ and $\Psi(lo)\ge hi+E\mathbf 1$;
\item $Nw\ge w+2E\mathbf 1$, so the spectral radius of $N$ exceeds $1$;
\item $\tilde a<(I+N)\mathbf 1$ componentwise;
\item if $I+N$ is nonsingular, put $r^\#=(I+N)^{-1}\tilde a$, the unique solution of $\Psi(r)=r$. Then $r^\#\in P$ if and only if $0\le(I+N)^{-1}(w+E\mathbf 1)\le w$, and this holds if and only if $P$ contains a point $r_U$ at which the vector $(r_U,p,\mathbf 1_F)$ satisfies $\Phi_i=r_i$ for every $i\in U$.
\end{enumerate}
\end{proposition}

The proof is in \cref{app:membership}.

Part (c) shows that at a stall the affine part of the map is expansive: $\rho(N)>1$. Part (e) shows that the affine rule below succeeds exactly when the stalled box contains an exact fixed point of the unsettled banks.

\begin{proposition}[the affine rule]\label{prop:affinerule}
At an inverter stall with $I+N$ nonsingular and $r^\#\in P$, the vector $r^\ast$ with $r^\ast_U=r^\#$, $r^\ast_k=p_k$ on the settled banks and $r^\ast=1$ on $F$ is an (I)-$E$ vector, and $\Phi_i(r^\ast)=r^\ast_i$ for every $i\in U$. Given the stalled configuration, it is computed by Gaussian elimination in time polynomial in the size of the network and of the box endpoints. If no bank outside $F$ settled before the stall, $r^\ast$ is an exact clearing vector.
\end{proposition}
\begin{proof}
For $i\in U$, $\Psi_i(r^\#)=r^\#_i\in[0,1]$, so $\Phi_i(r^\ast)=\min(1,r^\#_i)=r^\#_i$. The vector $r^\ast$ lies in the boxes of all rounds, because $P$ lies in all earlier boxes and settled points are never changed. So, exactly as in the proof of \cref{lem:sound}, every settled bank has residual at most $E$, and banks of $F$ have $\Phi_i\equiv1=r^\ast_i$.
\end{proof}

The number of rounds before a stall is not bounded by these results, and the endpoints can grow with the number of rounds (\cref{lem:bits} bounds their size only in terms of it); so \cref{prop:affinerule} is not a running-time statement about $\BOX(E)$ below the thresholds.

\begin{proposition}[two cases in which the solution lies in the box]\label{prop:stallcases}
\begin{enumerate}[label=(\alph*)]
\item If $|U|=2$, then at every inverter stall $I+N$ is nonsingular and $r^\#\in P$.
\item \emph{Symmetric cycles.} Let $m\ge2$, let $E$ be rational with $0<E<3-2\sqrt2$, and put $d=(1-E)/(2E)$. Let $b_1,\dots,b_m$ be banks with no external assets, each owing $1$ to a sink and holding a CDS of notional $d$, written by an ex-ante solvent writer, on $b_{i-1}$ (indices modulo $m$). Then $\BOX(E)$ stalls in round $2$ with $P=[(1+E)/2,1-E]^m$, no bank outside $F$ settles before the stall (the writer and the sink lie in $F$ and are fixed at $1$ from the start), and $r^\#=\frac{1-E}{1+E}\mathbf 1$ lies in $P$; so $r^\#$, extended by $1$ on $F$, is an exact clearing vector.
\end{enumerate}
\end{proposition}

The proof is in \cref{app:membership}, which also reports a computer search for stalls and an exact grid of two-bank networks (\cref{obs:stalls}).

\begin{example}[the affine rule fails with three banks]\label{ex:stallfail}
Let $E=4/25$, inside the window $(\alpha,3-2\sqrt2)$. Let $b_1,b_2,b_3$ have no external assets, each owe $1$ to a sink $z$ with $e_z=1$, and hold CDSs written by one writer $s$ with $e_s=2L_s^{\max}+1$:
\begin{center}\small
\begin{tabular}{@{}lll@{}}
\toprule
holder & reference, notional & reference, notional\\
\midrule
$b_1$ & $b_2$, $2$ & $b_3$, $1/2$\\
$b_2$ & $b_1$, $3/2$ & $b_3$, $1$\\
$b_3$ & $b_1$, $3/2$ & $b_2$, $3/2$\\
\bottomrule
\end{tabular}
\end{center}
The network is non-degenerate and one-writer, and (H) holds. In round $1$ of $\BOX(4/25)$ the enclosures are $[2/5,1]$, $[2/5,1]$, $[12/25,1]$, so the boxes become $[14/25,21/25]$, $[14/25,21/25]$, $[16/25,21/25]$; in round $2$ the enclosures and boxes are the same, with every $\mu_i=1$ (for instance $\Psi_1(lo)=2\cdot\frac{11}{25}+\frac12\cdot\frac9{25}=\frac{53}{50}$). So $\BOX(4/25)$ stalls in round $2$, and no bank outside $F=\{s,z\}$ settles before the stall. Here $\det(I+N)=-1/8$ and $r^\#=(-1,1,3)\notin[0,1]^3$. By \cref{prop:stall}(e), the stalled box contains no exact clearing vector. The network does have exact clearing vectors, for example $(1/2,3/4,1)$ on $(b_1,b_2,b_3)$, at which $b_3$ is clipped: $\Psi_3=\frac32\cdot\frac12+\frac32\cdot\frac14=\frac98>1$.
\end{example}

\begin{proof}
All values are exact rational computations; they are written out above and are recomputed by the program \texttt{stall\_examples.py} (\cref{app:verification}). The affine system $r_1=\frac52-2r_2-\frac12r_3$, $r_2=\frac52-\frac32r_1-r_3$, $r_3=3-\frac32r_1-\frac32r_2$ has the unique solution $(-1,1,3)$. At $(1/2,3/4,1)$: $\Psi_1=2\cdot\frac14+0=\frac12$, $\Psi_2=\frac32\cdot\frac12+0=\frac34$, and $\Psi_3=\frac98$, so $\Phi=(1/2,3/4,1)$.
\end{proof}

So the affine rule is not a global rule for stalled configurations: it ignores the clipping $\min(1,\cdot)$, and an exact clearing vector may have to be clipped at a bank that is unsettled, or may lie outside the stalled box altogether.

\begin{question}\label{q:stall}
Which inverter stalls admit an exact clearing vector inside the stalled box, and is there a rule, computable in polynomial time, that resolves every stall of $\BOX(E)$ for $E$ in the writer-solvent window? A positive answer would also have to handle stalls that violate (H), and boxes that shrink forever without stalling.
\end{question}

\section{Robust selection at near-stalls on structured networks}\label{sec:selection}

By \cref{thm:transfer}, a polynomial algorithm for writer-solvent (I)-$E$ inside the window $(\alpha,3-2\sqrt2]$ would follow from a polynomial near-stall selector at target $E$. This section and \cref{app:selection} construct such selectors on two explicit classes of near-stalls, defined by the structure of the signed residual matrix of the unsettled banks, and composes them along strongly connected components. The optimization mechanisms are standard: coordinate descent on a quadratic potential, a perturbation bound in the spirit of near-potential games~\cite{COP13}, and the extragradient method~\cite{Kor76} for monotone variational inequalities; the variational-inequality view of truncated affine best responses in network games is developed in~\cite{PO19}. What is specific here is the financial normalization of a rounded near-stall, the transfer from projected to actual clearing residuals, and the composition with \cref{thm:roundedbox}. The results give (I)- and (W)-vectors only; they give no (D), (D$^\circ$) or (S) guarantee, and they do not move either window of \cref{tab:landscape}.

Let every CDS writer lie in $F$, and let $\BOX(E_0,M)$ stop with a nonempty unsettled set $U$. \Cref{app:selection} shows that every unsettled bank has a constant liability and an affine asset ratio $f(r)=a+Pr+N(\mathbf1-r)$ with nonnegative coefficients (\cref{lem:nearnormal}). With the signed residual matrix $A=I-P+N$ and the projection $G$ of $f$ onto the reached product of boxes, a point $r$ of that product with $\|r-G(r)\|_\infty\le E-E_0$ gives an (I)-$E$ vector whenever $E_0+\delta/2\le E$ (\cref{lem:projtransfer}). Two selectors find such points: coordinate descent on a quadratic potential, whose guarantee carries a skew allowance computed on the reached product, equal to $0$ when $\Omega A$ is symmetric for a supplied positive diagonal $\Omega$ (\cref{thm:potential}); and the extragradient method when the symmetric part of $A$ is positive semidefinite (\cref{thm:monotone}). They are combined along the strongly connected components of $U$ (\cref{thm:composition}).

The network-level promise behind \cref{thm:composition} is the following. Let $A^0=I-P^0+N^0$ be the signed residual matrix of the original network on the banks outside $F$ (every such bank writes no CDS, has positive constant liability, and has an asset ratio of the form above with all other banks outside $F$ as variables). The terminal matrix $A$ is the principal submatrix of $A^0$ on $U$, and each strongly connected component of the terminal graph lies inside a strongly connected component of the graph of $A^0$. Symmetry and positive semidefiniteness of the symmetric part pass to principal submatrices. So if every diagonal block $A^0_{B'B'}$ of a strongly connected component $B'$ is symmetric or has positive semidefinite symmetric part, then every terminal component satisfies the hypothesis of \cref{thm:composition}, with $\Omega_B=I$ ($\chi_B=0$, $R_B=|B|$) or with \cref{thm:monotone}.

\begin{corollary}[structured writer-solvent networks]\label{cor:structured}
Let $\mathcal S$ be the class of networks in which every CDS writer lies in $F$ and every diagonal block of $A^0$ on a strongly connected component is symmetric or has positive semidefinite symmetric part. For every rational $0<E<1/2$, an (I)-$E$ vector, and hence a (W)-$E$ vector, of a network of $\mathcal S$ can be computed in time polynomial in the encoding size and $1/E$, by $\BOX(E/2,M)$ with $M=\lceil(1-E)/E\rceil$ followed by \cref{thm:composition} with $\tau_B=E/2$.
\end{corollary}
\begin{proof}
With $E_0=E/2$ we have $\delta=(1-E)/M\le E$, so $E_0+\delta/2\le E$, and $C=2/E-1$. By the discussion above every terminal component satisfies the hypothesis of \cref{thm:composition} with $\tau_B=E/2=E-E_0$, and the bounds of \cref{thm:potential,thm:monotone} are polynomial in $n$ and $1/E$.
\end{proof}

This is an approximation scheme on a promise class, polynomial in $1/E$ and not in $\log(1/E)$; for $E\ge1/2$ the trivial vector of \cref{lem:relations}(c) applies. At $E=1/6$, inside the writer-solvent window, one can take $E_0=1/7$, $M=15$ and $\delta=1/21$, so that settled banks keep error $E_0+\delta/2=1/6$; a potential component with a supplied $\Omega_B$ of polynomially bounded ratio $R_B$ can use $\tau_B=1/84$ when its skew allowance on the reached product is $\chi_B\le1/84$ (a symmetric block with $\Omega_B=I$ has $\chi_B=0$ and can use $\tau_B=1/42$), and a monotone component can use $\tau_B=1/42$; here $C=6$. This gives a polynomial (I)-$1/6$ algorithm on the class $\mathcal S$, and on networks whose terminal components satisfy the a-posteriori conditions of \cref{thm:composition}. It does not improve the unrestricted bound $3-2\sqrt2$ of \cref{thm:algws}, and it says nothing about the hardness threshold $\alpha$ for unrestricted writer-solvent networks. Components with neither a usable potential certificate nor a monotone residual operator remain open (\cref{q:stall}); \cref{app:stalls} records what is known about exact stalls and two obstructions to simpler rules.

\section{Relation to the literature}\label{sec:literature}

\paragraph{Clearing with debts.} Eisenberg and Noe~\cite{EN01} introduced the clearing model for networks of ordinary debts: clearing vectors exist, a greatest clearing vector exists, and it can be computed in polynomial time. Rogers and Veraart~\cite{RV13} added default costs. According to their abstract, default costs change the multiplicity of clearing vectors and create incentives for solvent banks to rescue failing ones, which they do not have without default costs; a greatest clearing vector still exists, and it is reached by letting banks fail in succession until only solvent banks remain. We use no computational property of their model, and our model has no default costs.

\paragraph{Clearing with CDSs.} Schuldenzucker, Seuken and Battiston~\cite{SSB16,SSB20} showed that CDSs change the picture: clearing vectors need not be unique, and with default costs they need not exist~\cite[Theorem~1]{SSB20}. Without default costs they proved existence~\cite[Theorem~2]{SSB20}, in a model in which every reference entity owes an ordinary debt. In~\cite{SSB17} (extended in~\cite{SSB19}) they introduced the approximate problem with the clipped-ratio correspondence, leaving the recovery of a zero-liability bank unconstrained, and proved that it is PPAD-complete on networks that are non-degenerate in their sense; they asked whether polynomial-length approximate solutions exist without that assumption~\cite[footnote~4]{SSB17}.

Ioannidis, de Keijzer and Ventre showed that clearing vectors can be irrational, that exact clearing is FIXP-complete, and that strong approximation of clearing vectors is $\mathrm{FIXP}_a$-complete~\cite[Theorem~5]{IKV22} (Theorem~3 of the full version~\cite{IKV21}; a journal version is~\cite{IKV26a}). Both the conference version and the full version define the problem only on networks that are non-degenerate in their sense (\cite[Definition~3]{IKV21} and~\cite[Definition~3]{IKV22}): every CDS writer has positive external assets or owes a positive debt, and every reference entity owes a positive debt. They also proved PPAD-hardness of approximate clearing with an explicit constant for the notion of~\cite[Definition~4]{IKV23}, whose first clause fixes the strictly ex-ante solvent banks at recovery $1$, that is, for our notion (I): the arXiv version gives $\eps\le(3-\sqrt5)/16\approx0.048$~\cite[Theorem~2]{IKV23}, and the journal version gives $(2-\sqrt3)/4\approx0.067$~\cite[Definition~4, Theorem~2 and Corollary~2]{IKV26b}. For the journal version these three statements were checked in the publisher's indexed text, not in the typeset article, and their proofs were not audited (\cref{app:verification}); its abstract also reports hardness for networks with a central clearing authority~\cite[record~L5]{Supp}. Section~5 of the journal version, also checked in the indexed text, asks whether the constant $(2-\sqrt3)/4$ can be improved and notes that the notion of~\cite{DHK} is different. The thresholds $\EI$ and $\alpha$ of \cref{thm:general,thm:wsI}, which are constants for (I), answer the first question affirmatively, subject to our proofs; they do not determine the optimal thresholds. The arXiv version already covers networks with a \emph{central CDS debtor}~\cite[Definition~8, Corollary~3]{IKV23}: one bank writes all CDSs, owes no debt, and has external assets at least its total CDS notional, so that it lies in $F$; the constant there is $(3-\sqrt5)/16$. This is the published result closest to our one-writer setting, in which the writer's external assets strictly exceed its liabilities. Dohn, Hansen and Klinkby~\cite{DHK} introduced the exact-solvency notion (D), proved PPAD-hardness of (D) for $\eps\le5-2\sqrt6\approx0.101$, proved membership under non-degeneracy, and asked about membership without it. Their construction is sparse and weighted: in the arXiv version its NAND receivers hold two references, each with coefficient $1/(1-\gamma)$ for $\gamma=3-\sqrt6$, and its two-arm PURIFY also uses non-unit coefficients~\cite[Section~3.2]{DHK}; there the NAND topology is attributed to Schuldenzucker, Seuken and Battiston and the PURIFY topology to Ioannidis, de Keijzer and Ventre. With unit liabilities fixed, a coefficient $c$ can be written with contracts of notional $1$ only if $c$ is an integer, and then the multiplicity counts toward the held degree; so a construction with two weighted references per bank does not by itself give hardness on $\mathcal R(2)$. The conference version~\cite{DHK} was compared only through its public publisher page, that is, its publication data, its abstract and the one footnote shown there; its full text and proofs were not available to us, and our statements about its formulas refer to the arXiv version. That abstract states hardness for $\eps\approx0.101$. The same footnote is footnote~1 of Section~3 of the arXiv version, where it is attached to the discussion of the $\eps$-almost-fixed-point notion $\|f(r)-r\|_\infty\le\eps$, which is (W) on non-degenerate networks: it states that with that notion a reduction could not guarantee that a source bank is solvent however large its external assets are, and that their result and that of Ioannidis, de Keijzer and Ventre could be adapted to it but would give PPAD-hardness for smaller values of $\eps$. No constant for that notion is stated there.

\paragraph{Restricted classes.} Kanellopoulos and Melissourgos~\cite[Appendix~F.1]{KM} define the residual and exact-solvency approximation problems on a restricted financial class (Definitions~F.5 and~F.6), with repeated references counted with multiplicity; $\mathcal R(d)$ is a normal form within that class (\cref{sec:Rclass}). Their Theorem~F.9 gives (D)-$1/5$ at held degree two by supported propagation (Lemma~F.8); its error $1/3$ for arbitrary degree concerns their game models, not (D) clearing. Their Theorem~F.7 reduces the residual problem on the restricted class to generalized circuits, which yields algorithms; it cannot be reversed to give hardness for the financial class, since that would require financial gadgets, correct for every approximate solution and within a unit held-degree budget, for the arithmetic and comparison gates used in their circuit hardness theorem~\cite[Theorem~3.4]{KM}. They also relate the restricted class to public goods games. Hardness of (D) does not by itself give hardness of the residual condition (\cref{rem:residualD}), so \cref{sec:degree} gives no hardness for divisible public goods games with additive best-response error; for that model the hardness result of~\cite[Theorem~5.1]{PP21} is for inverse-polynomial error, and the results of~\cite{DDH24} concern indivisible public goods and a different equilibrium notion. These statements of~\cite{KM} were checked in its arXiv version~1 (Theorem~3.4 and Appendix~F.1), and those of~\cite{PP21} in its arXiv version~2; for~\cite{DDH24} only the published abstract and highlights were consulted.

\paragraph{Tools.} PPAD was introduced by Papadimitriou~\cite{Pap94}. Our membership proofs use the weak-approximation theorem of Etessami and Yannakakis~\cite[Proposition~2.2]{EY10}, as do~\cite[Appendix~B]{DHK}; \cite[Lemma~C.1]{SSB17} use the Lipschitz Brouwer problem instead. Our hardness proofs reduce from \PCirc~\cite{DFHM22,DFHM24}; the statement used in \cref{thm:pc} (three distinct nodes per gate, exactly one producing gate per node, the robust NAND and PURIFY constraints, and PPAD-completeness for $\{\mathrm{PURIFY},\mathrm{NAND}\}$; Definition~1, Theorem~2.1 and Corollary~2.2) was checked in the complete accepted manuscript of the journal version~\cite{DFHM24}, and the same numbering was confirmed in the full version of~\cite{DFHM22} (arXiv:2209.15149v1). The selectors of \cref{sec:selection} use coordinate descent on a quadratic potential, a perturbation bound in the spirit of near-potential games~\cite{COP13}, and the extragradient method~\cite{Kor76}, in the variational-inequality setting of network games~\cite{PO19}; for monotone linear complementarity problems, interior-point methods give a dependence on the accuracy that is polynomial in its logarithm~\cite{KMY89}.

\paragraph{Institutional setting.} One-writer networks model a market in which all protection is sold by a single counterparty that is solvent ex ante. McLaughlin studies the resources and capital that central counterparties need against losses other than member defaults~\cite{McL21}, and the trade-off between shorter settlement times and the benefits of multilateral netting in deferred net settlement~\cite{McL23}. They concern clearing infrastructures, not the recovery-rate fixed points studied here.

\paragraph{What is new, and how we checked.} We consulted~\cite{DHK} in its arXiv version (the section, appendix and definition numbers we cite refer to it) and the public publisher page of its conference version, \cite{SSB17,SSB19,SSB20}, \cite{IKV21,IKV23}, the published version of~\cite{IKV22}, the indexed statements of~\cite{IKV26b} named above, the abstract of~\cite{RV13}, the publication data of~\cite{IKV26a}, Theorem~3.4 and Appendix~F.1 of~\cite{KM}, the accepted manuscript of~\cite{DFHM24} and the numbering of the full version of~\cite{DFHM22}, the cited parts of~\cite{PP21} and the abstract of~\cite{DDH24}, together with web searches. ``Not found'' below means not found in these sources; it is not a proof of absence.
\begin{itemize}
\item \emph{Hardness constants.} The thresholds $\EI$ (\cref{thm:general}) and $\alpha$ (\cref{thm:wsI,cor:oneW}) improve the published constant $(2-\sqrt3)/4\approx0.067$, which is a constant for (I)~\cite{IKV26b}; our thresholds hold for (I) and for (W). For (W), \cite[footnote~1]{DHK} remarks that their reduction and that of Ioannidis, de Keijzer and Ventre could be adapted, with smaller constants that are not stated. Hardness of (D) for every error below $1$ (\cref{thm:D}) improves the published constant $5-2\sqrt6\approx0.101$ for (D)~\cite{DHK}. We did not find the calibration reduction (\cref{thm:calib}) or the coincidence of thresholds (\cref{thm:equal,cor:equalgen}) in the checked sources.
\item \emph{Algorithms and barriers.} In the checked sources we found no polynomial-time algorithm for (W) or (I) on general CDS networks at a constant error below $1/2$, the error achieved by $r_i=1$ for $i\in T$ and $r_i=1/2$ elsewhere (\cref{lem:relations}(c)); algorithms there concern restricted classes, such as covered networks and the class of~\cite[Appendix~F]{KM}. The one-pass algorithm with error $(3-\sqrt5)/2$ comes from our research record~\cite[record~L1; Section~2]{Supp}. The box algorithm and its thresholds (\cref{thm:algws,thm:alggen}), the local lower bounds of \cref{thm:locallower}, and the barriers of \cref{sec:coincidence,sec:barriers} are, as far as we found, new.
\item \emph{Exact solvency at bounded held degree.} In the primary sources we checked we found no theorem that subsumes the degree-sensitive bounds of \cref{sec:degree}. Some ingredients are known: \PCirc, the sparse NAND and two-arm PURIFY topologies (used with weighted contracts in~\cite{DHK} and in \cref{thm:D}), the decoding of bit $1$ through the solvency clause, and supported propagation~\cite{KM}, which already gives (D)-$1/5$ at held degree two. The new content is the unit-contract parameter design of \cref{thm:dH} and its optimality among two-stage two-arm PURIFY gadgets with the same decoding (\cref{prop:twoarm}), the restoring chains of \cref{thm:lowdeg} and their exact ranges among chain PURIFY gadgets (\cref{prop:chainlimits}), the refined rule of \cref{thm:Pk}, the rules A2 and A3, and the classification that results (with the exact values of $p_d$ under the assumption that PPAD is not contained in FP); the optimality statements are method boundaries, not lower bounds. Relative to our own earlier construction: the direct expansion of \cref{thm:D} into unit contracts needs held degree at least $9$ (\cref{sec:degclass}). The weighted algorithms of \cref{app:weighted} extend the mechanism of~\cite{KM} with coefficient-sensitive conditions.
\item \emph{Interfaces and finite precision.} The normalization behind \cref{thm:interfaces} was not found in the checked sources; the least-bound argument for separate saturated bounds on the two rails of a wire already appears in~\cite[record~L19]{Supp}. We did not find the rounded algorithm, the reduction of \cref{thm:transfer} or the financial normalization of \cref{sec:selection} in the checked sources; the optimization methods used there are standard~\cite{COP13,Kor76,KMY89,PO19}.
\item \emph{Exact existence.} Existence under non-degeneracy (\cref{prop:ndexact}) is known: \cite[Section~2.2]{DHK} gives the Brouwer argument and attributes it to~\cite{IKV22}, and~\cite[Appendix~A]{DHK} shows that the assumption cannot simply be dropped. We did not find \cref{prop:wsexact} stated. It is not literally covered by~\cite[Theorem~2]{SSB20}, whose model requires reference entities to owe a debt, nor by the non-degenerate statement, since a writer-solvent network may contain a reference entity with neither external assets nor debts; it follows from the same Brouwer argument.
\item \emph{Totality and membership without non-degeneracy.} We did not find \cref{thm:member} (in particular the existence of (W)-$E$ vectors on all networks), \cref{thm:gap}, \cref{cor:Dcirc} or \cref{prop:Dfail} stated. The nearest statements are the open question~\cite[footnote~4]{SSB17}, which concerns their $\eps$-solutions, of type (S) in our notation, and the open question of~\cite[Appendix~A]{DHK}, which concerns Definition~3, that is, (D). \Cref{prop:Dfail} shows that the latter is not total as written; with zero-liability banks exempted it is total and, by \cref{thm:gap}, in PPAD. The network of~\cite[Appendix~A]{DHK} has no exact clearing vector but does have Definition-3 vectors: by its equation~(19) the vector $(1-\eps,1/2,0)$ qualifies for every $\eps\in(0,1)$.
\end{itemize}

\paragraph{Conventions and versions.} The treatment of zero-liability banks differs between papers: \cite{SSB17} leave their recovery unconstrained; \cite{SSB20}, \cite{IKV23} and this paper set it to $1$; \cite{DHK} use $f_i=a_i/\max(a_i,\ell_i)$, which needs non-degeneracy to be defined. Statements about existence must be read with the convention and the version. The arXiv version of~\cite{IKV21} states, as its Theorem~1 citing earlier work, that every financial system has a clearing recovery rate vector, with recovery $1$ at zero liability; read literally for all networks, this conflicts with~\cite[Appendix~A]{DHK} and with the network of \cref{prop:Dfail}, which has no exact clearing vector (see the remark after that proposition). Both versions define their problem only on networks that are non-degenerate in the sense of their Definition~3, and the conference version states existence only for these networks, attributing it to~\cite{SSB16} \cite[Section~2.1]{IKV22}. On that class existence holds: a bank that owes a positive debt has liabilities bounded away from $0$, a bank that writes a CDS but owes no debt has $A_i\ge e_i>0$, so that $\Phi_i=\min(1,A_i/\ell_i)\to1$ as $\ell_i\to0$, and a bank with no liabilities has $\Phi_i\equiv1$; so $\Phi$ is continuous and Brouwer's theorem applies, as in \cref{prop:ndexact}. Existence also holds under \cref{def:classes} (\cref{prop:ndexact}) and, by~\cite[Theorem~2]{SSB20}, under the model assumption of~\cite{SSB20}. We have not consulted the text of the journal version~\cite{IKV26a} and make no statement about its formulation.

\section{Open problems}\label{sec:open}

\begin{enumerate}[label=\arabic*.,leftmargin=1.6em]
\item \emph{The two windows.} For general networks, (I) and (W) are PPAD-hard below $\EI$ and polynomial above $\elad$; the window is
\[
[\EI,\elad]\subseteq[0.25299586,\,0.28336554].
\]
For writer-solvent networks, (I), (W) and one-writer (W) are PPAD-hard below $\alpha$ and polynomial above $3-2\sqrt2$; the window is
\[
[\alpha,3-2\sqrt2]\subseteq[0.15767078,\,0.17157288].
\]
(Lower ends are rounded down and upper ends up.) By \cref{thm:equal,cor:equalgen}, within each class (on non-degenerate networks) all these problems have the same supremum of hard errors and the same infimum of polynomial errors, so determining the thresholds in either class reduces to (I). We do not conjecture where the thresholds lie.

\item \emph{Is $\alpha$ the true ceiling of hardness on writer-solvent networks?} Unless PPAD is contained in FP, \cref{thm:barrier,thm:nonuniform} exclude reductions for writer-solvent (I) from \PCirc\ whose gadgets are certified by interval propagation with the saturated wire interface, in the gadget model of \cref{def:gadget,def:unary}: disjoint input wires, output rails that are receivers of the gadget, and other banks entering as fixed banks whose boxes do not depend on the condition. By \cref{thm:interfaces} this extends to arbitrary fixed pairs of disjoint low and high intervals on each rail, including non-saturated interfaces and separate interfaces for the two rails of a wire. Open routes: gadgets whose correctness is proved on exact internal solution sets or by fixed-point arguments (beyond the two-receiver template of \cref{prop:tworeceiver}, from~\cite[record~L18; Theorem~9.1]{Supp}); certificates that keep correlations between receivers; outputs that are aliases of input rails; and source problems other than \PCirc. Unless PPAD is contained in FP, any successful route stays below $3-2\sqrt2$ (\cref{cor:ceiling}).

\item \emph{Beyond $\EI$ on general networks.} In the ideal limit and on $[1/5,27/100]$, the design family of \cref{sec:general} has a valid PURIFY design exactly for $E<\EI$, and allowing the single-bank two-rail restorers and single-bank seed splits of~\cite[record~L16; Theorem~B]{Supp} inside this restricted PURIFY architecture, with its rectangular certificates, does not move that endpoint (\cref{rem:EIendpoint}). The obstruction is the PURIFY window, not the cleaning of signals: a single restorer still lowers bounds above $\EI$ (\cref{rem:EIendpoint} gives an example at $E=13/50$). Within this architecture, a construction past $\EI$ needs a PURIFY gadget outside these families, for instance seeds or arms made of several banks that read each other, certificates that keep correlations, or encodings with more than two levels.

\item \emph{Algorithms inside the windows.} By \cref{cor:local}, a polynomial algorithm for (I) inside either window cannot have bounded radius. Inside the writer-solvent window, $\BOX(E)$ stalls on inverter cycles (\cref{rem:boxtests,prop:stallcases}); inside the general window, it did not settle ladder rings within $400$ rounds in the tests of \cref{rem:boxtests}, which rounded box endpoints outward. A global rule for stalled configurations is the natural next step. The affine rule of \cref{prop:affinerule} is exact on the unsettled banks whenever its solution lies in the stalled box, which holds for two unsettled banks and for symmetric cycles but not in general (\cref{ex:stallfail}), and \cref{prop:affinerule} gives an exact clearing vector of the whole network if, in addition, no bank with $e_i<L_i^{\max}$ settled before the stall; \cref{q:stall} asks for a rule that handles every stall. By \cref{thm:transfer} it suffices to select a point in every near-stalled box produced by the rounded algorithm, in polynomial time; on networks whose CDS writers lie in $F$, \cref{thm:composition} does this when every terminal component is monotone (positive semidefinite symmetric part) or comes with a supplied diagonal weight of polynomially bounded ratio whose skew allowance on the reached product is small enough, which is an a-posteriori certificate; symmetric or monotone diagonal blocks of the original residual matrix on its strongly connected components suffice (\cref{cor:structured}). Components with neither structure remain open. At exact inverter stalls the procedure \Select\ of \cref{app:stalls} is sound but not known to be complete, and the family of feasible clip sets need not be closed under intersection (\cref{prop:nointersection}). For (W) on writer-solvent networks the bounded-radius threshold lies in $[0.1609,3-2\sqrt2]$~\cite[record~L11; Proposition~5.1]{Supp}, and on general networks the $R$-local thresholds for $2\le R<\infty$ lie between $\elad$ and the values of \cref{rem:rounds}.

\item \emph{Exact solvency at bounded held degree.} On $\mathcal R(d)$ (\cref{tab:phase}): for every rational $E\in[1/4,1/2)$ other than $1/3$, exactly one held degree is unresolved, namely $\lceil(\lceil\rho\rceil+1)\rho\rceil-1$; at smaller errors, held degree $2$ is open for $1/9\le E<1/5$, held degree $3$ for $1/7\le E<1/4$, and held degree $4$ for $3/13<E\le1/4$. Consequently, assuming that PPAD is not contained in FP, $p_d$ is known only within intervals for $d\in\{2,3,4,5,8,9,10,11\}$; if the unresolved degree is in FP for every such $E$, then $p_5=1/4$, $p_8=1/3$, $p_9=5/13$, $p_{10}=3/7$ and $p_{11}=7/15$, and if it is PPAD-hard for every such $E$, these become $1/3$, $5/13$, $3/7$, $7/15$ and $1/2$. The limits of our methods are method boundaries, not lower bounds. With the decoding of \cref{thm:dH}, no chain PURIFY gadget of these held degrees is correct (\cref{cor:chainlimits}); this bounds that gadget family only, not other gadgets, decodings or reductions. For $1/4\le E<1/2$, $E\ne1/3$, the decoder that gives the polynomial column of \cref{tab:phase} (rule A3 for $1/4\le E<1/3$, rule A2 for $1/3<E<1/2$) fails exactly at the unresolved degree (\cref{prop:decodertight}), at an unresolved bank with one held contract on a weak bank, so a rule that handles that configuration is one possible route to an algorithm; in the open cases with $E\le1/4$ no polynomial algorithm is known, and at held degree $2$ the common decode of \cref{prop:weightedtight} stops at $1/5$. For the weighted classes of \cref{app:weighted} we have algorithms but no classification. We also do not know the complexity of (I) and (W) on $\mathcal R(d)$ below $3-2\sqrt2$: the reductions of \cref{sec:general,sec:onewriter} use notionals other than $1$.
\item \emph{Default costs.} All our results assume no default costs. With default costs clearing vectors need not exist~\cite[Theorem~1]{SSB20}. The complexity, and even the totality, of the exempted and clipped-ratio notions (D$^\circ$) and (S) in models with default costs are open.
\end{enumerate}

\section*{Contributions and use of AI}
AI tools assisted with mathematical exploration, proof development, computational checks and manuscript preparation. The author takes responsibility for the final content.

\bibliographystyle{plain}
{\small\raggedright
\bibliography{refs}
}

\clearpage
\appendix
% Preserve appendix-subsection reference types across cleveref/LaTeX versions.
\crefname{subappendix}{Appendix}{Appendices}
\Crefname{subappendix}{Appendix}{Appendices}
\crefalias{subsection}{subappendix}
\addtocontents{toc}{\protect\setcounter{tocdepth}{1}}
\section{The ladder construction for (I) and (W)}\label[appendix]{app:ladder}

This appendix gives the construction behind \cref{thm:general}, its proof, a direct construction for (W), and the scope of the design family; \cref{sec:ladderoverview} outlines them.

\subsection{Gadget architecture for (I) and (W)}\label{sec:architecture}

Fix $0<E<1/2$, write $H=1-E$, and let $t_0=1$ under (I) and $t_0=H$ under (W).

\paragraph{Banks and scales.} The networks have one \emph{source} $s$, which writes a number of CDSs and has $e_s=2L_s^{\max}+1$, and one \emph{sink} $z$ with $e_z=1$ and no liabilities, which is never a writer or a reference. Both lie in $T$, so $r_s\in[t_0,1]$ in every (I)-$E$ and every (W)-$E$ vector. All other banks are \emph{gadget banks}. A gadget bank $b$ has $e_b=0$ and a \emph{scale} $S_b$, a power of $2$. It owes $d_bS_b$ to $z$ with $d_b>0$, holds CDSs of notionals $\kappa\,S_b$, writes \emph{main} CDSs of notionals $\kappa' S_b$ to $z$, and may write CDSs held by later gadget banks, its \emph{consumers}. Dividing by $S_b$, its map takes the normalized form
\[
\Phi_b=\min\Bigl(1,\frac{a}{l+x}\Bigr),\qquad a=\sum\kappa\,r_w(1-r_k),\qquad l=d_b+\sum\kappa'(1-r_{k'}),\qquad x\in[0,\xi],
\]
where the sums run over the held and the main CDSs, and $x$ is the normalized liability towards consumers. The bound $x\le\xi$ is arranged by the scales: once the network is complete, fix the scales in reverse order of creation, giving each bank the least power of $2$ that is at least $1$ and at least $\xi^{-1}\sum_c\kappa_cS_c$, the sum over its consumers $c$ with their coefficients $\kappa_c$. Every consumer is created after its writer, so its scale is already fixed. In our gadgets all consumers of a gadget bank belong to the same gadget, so the scales can be fixed gadget by gadget, and they are bounded by a constant that depends only on the gadget.

\paragraph{Pairs and wires.} A \emph{pair} $(p,n)$ of gadget banks with \emph{bounds} $(g_p,g_n)$ represents bit $0$ by $r_p\le g_p$ and $r_n\ge H$, and bit $1$ by $r_p\ge H$ and $r_n\le g_n$. A statement ``$(p,n)$ is a pair with bounds $(g_p,g_n)$'' means: in every (I)-$E$ vector (respectively every (W)-$E$ vector) the pair represents bit $\beta$ whenever a specified condition $\mathcal C_\beta$ holds, for $\beta=0,1$. A \emph{wire} is an ordered pair of banks $[p,n]$; with a threshold $\gamma\in(0,H)$ it \emph{decodes} to $0$ if $r_p\le\gamma$ and $r_n\ge H$, to $1$ if $r_p\ge H$ and $r_n\le\gamma$, and to $\bot$ otherwise.

\begin{lemma}[one-coordinate refinement]\label{lem:step}
Let $(p,n)$ be a pair with bounds $g_p,g_n<1$, and let $D,\xi>0$. Let a new bank $P$ hold a CDS written by $p$ on $n$ with coefficient $k_P=(D+E+\xi)/(H(1-g_n))$, owe $D$, and write a main CDS of coefficient $1$ on $p$. Then $(P,n)$ is a pair, under the same conditions, with bounds
\[
\Bigl(E+\frac{k_Pg_pE}{D+1-g_p},\ g_n\Bigr).
\]
Symmetrically, a bank $N$ that holds a CDS written by $n$ on $p$ with coefficient $k_N=(D+E+\xi)/(H(1-g_p))$, owes $D$ and writes a main CDS of coefficient $1$ on $n$ makes $(p,N)$ a pair with bounds $(g_p,\ E+k_Ng_nE/(D+1-g_n))$.
\end{lemma}
\begin{proof}
We have $a_P=k_Pr_p(1-r_n)$ and $l_P=D+(1-r_p)+x$ with $x\in[0,\xi]$. At bit $0$, $a_P\le k_Pg_pE$ and $l_P\ge D+1-g_p$, so $r_P\le\Phi_P+E$ is at most the stated bound. At bit $1$, $a_P\ge k_PH(1-g_n)=D+E+\xi\ge l_P$, so $\Phi_P=1$ and $r_P\ge H$. The bank $n$ gains no contract, and $p$ gains one consumer, which the scales account for; the bounds on $p$ and $n$ were derived for arbitrary consumer liabilities in $[0,\xi]$. The case of $N$ is symmetric.
\end{proof}

\begin{lemma}[pair reader]\label{lem:reader}
Let $[p_{\rm in},n_{\rm in}]$ be a wire with threshold $\gamma\in(0,H)$, and let $D,\xi>0$. Let $P_1$ hold a CDS written by the source on $n_{\rm in}$ with coefficient $c=(D+E+\xi)/(t_0(1-\gamma))$, owe $D$, and write a main CDS of coefficient $1$ on $p_{\rm in}$; let $N_1$ be the same with $p_{\rm in}$ and $n_{\rm in}$ exchanged. Then $(P_1,N_1)$ is a pair with bounds $(g_1,g_1)$, $g_1=E+cE/(D+1-\gamma)$, where bit $\beta$ is the condition that the wire decodes to $\beta$. The banks $p_{\rm in},n_{\rm in}$ gain no contract.
\end{lemma}
\begin{proof}
Let $t=r_s\in[t_0,1]$. Then $a_{P_1}=ct(1-r_{n_{\rm in}})$ and $l_{P_1}=D+(1-r_{p_{\rm in}})+x$. If the wire decodes to $0$, then $a_{P_1}\le cE$ and $l_{P_1}\ge D+1-\gamma$, so $r_{P_1}\le g_1$; and $a_{N_1}=ct(1-r_{p_{\rm in}})\ge ct_0(1-\gamma)=D+E+\xi\ge l_{N_1}$, so $r_{N_1}\ge H$. Decoding to $1$ is symmetric. The input banks occur only as references.
\end{proof}

\begin{lemma}[source-funded inverse]\label{lem:inverse}
Let $X$ be a bank, $E<\tau<1$ and $\xi>0$. Let $n_X$ hold a CDS written by the source on $X$ with coefficient $c_\tau=(1+\xi)/(t_0(1-\tau))$ and owe $1$. Then $r_X\le\tau$ implies $r_{n_X}\ge H$, and $r_X\ge H$ implies $r_{n_X}\le t:=E+c_\tau E$.
\end{lemma}
\begin{proof}
If $r_X\le\tau$, the assets are at least $t_0c_\tau(1-\tau)=1+\xi\ge l_{n_X}$. If $r_X\ge H$, they are at most $c_\tau E$ while $l_{n_X}\ge1$.
\end{proof}

\paragraph{The gadgets.} Fix rationals $D,\xi>0$, a wire threshold $\gamma\in(0,H)$ and a PURIFY threshold $\tau\in(E,1)$. A \emph{word} is a finite sequence of steps $\stP$ and $\stN$; the step $\stP$ applies \cref{lem:step} to the first bank of the current pair and $\stN$ to the second. (In \cref{sec:directW} words may also contain the steps $\stp$ and $\stn$ of \cref{lem:conststep}; everything below holds for them verbatim.)

$\mathrm{PURIFY}(u;v,w)$:
\begin{enumerate}[label=(\arabic*)]
\item A pair reader on the input wire $u$ gives $(P_1,N_1)$. An input word $w_{\rm in}$ maps it to a pair $(p,n)$ with bounds $(\sigma_p,\sigma_n)$.
\item Two \emph{seeds} $V$ and $W$ hold CDSs written by $p$ on $n$ with coefficients
\[
c_V=\frac{D+E+\xi}{H(1-\sigma_n)},\qquad c_W=\frac{(1+\xi/D)\,c_V}{\tau-E},
\]
and each owes $D$ and writes a main CDS of coefficient $1$ on $p$.
\item Source-funded inverses $n_V$ of $V$ and $n_W$ of $W$, with threshold $\tau$ (\cref{lem:inverse}).
\item An arm word $w_{\rm arm}$ is applied to $(V,n_V)$ and, separately, to $(W,n_W)$; the final pairs are the output wires $v$ and $w$.
\end{enumerate}

$\mathrm{NAND}(u,v;w)$: a pair reader and the input word $w_{\rm in}$ on each input wire give pairs $(p_u,n_u)$ and $(p_v,n_v)$ with bounds $(\sigma_p,\sigma_n)$. The seed has two banks.
\begin{itemize}
\item $P_0$ holds CDSs written by $n_u$ on $p_u$ and by $n_v$ on $p_v$, each with coefficient $k_0=(D+E+1+\xi)/(H(1-\sigma_p))$; it owes $D$ and writes main CDSs of coefficient $1$ on $n_u$ and on $n_v$.
\item $N_0$ holds a CDS written by $p_u$ on $n_v$ with coefficient $k_A=(D+E+\xi)/(H(1-\sigma_n))$; it owes $D$ and writes main CDSs of coefficients $l_u$ on $p_u$ and $l_v$ on $p_v$, where $l_u,l_v>0$ are rationals with $l_u+l_v=1$.
\end{itemize}
An output word $w_{\rm out}$ is applied to $(P_0,N_0)$, and the final pair is the output wire $w$.

\paragraph{Design bounds.} The \emph{design bounds} of a gadget are the numbers obtained from the start values
\begin{gather*}
g_1,\qquad (\tau,t)\ \text{with}\ t=E+c_\tau E,\\
g_{P0}=E+\frac{2k_0\sigma_nE}{D+2(1-\sigma_n)},\qquad g_{N0}=E+\max\Bigl(\frac{k_A\sigma_p}{D+l_u(1-\sigma_p)},\frac{k_AE}{D+l_v(1-\sigma_p)}\Bigr)
\end{gather*}
by applying the bound formulas of \cref{lem:step} along the words, each result possibly rounded up to a rational. Rounding up is harmless because the lemmas use the bounds only as upper bounds. A network \emph{contains} a gadget if it contains its banks with exactly the contracts listed (debts and main CDSs to the sink included), and no bank outside the gadget holds a CDS written by a gadget bank or is owed a debt by one, except the sink. The input wire banks may be arbitrary.

\begin{lemma}[gate claims]\label{lem:gates}
Let $E<\tau<1$ and suppose the design bounds satisfy:
\begin{enumerate}[label=(D\arabic*)]
\item $w_{\rm in}$ maps $(g_1,g_1)$ to $(\sigma_p,\sigma_n)$, and the \emph{raw seed} $s_0=E+c_W\sigma_pE/(D+1-\sigma_p)$ satisfies $s_0\le\tau$;
\item $w_{\rm arm}$ maps $(\tau,t)$ to bounds that are both at most $\gamma$;
\item $w_{\rm out}$ maps $(g_{P0},g_{N0})$ to bounds that are both at most $\gamma$;
\item every design bound, including the start values, is below $1$.
\end{enumerate}
Then in every (I)-$E$ vector (respectively every (W)-$E$ vector) of every network that contains the gadget, with scales as above:
\begin{itemize}
\item[(P1)] $u=0$ gives $v=w=0$; (P2) $u=1$ gives $v=w=1$;
\item[(P3)] for every input, $r_V\le\tau$ gives $v=0$; (P4) $r_W\ge H$ gives $w=1$; (P5) $r_V>\tau$ gives $r_W\ge H$;
\item[(N1)] $u=0$ or $v=0$ gives $w=1$, whatever the other input; (N2) $u=v=1$ gives $w=0$.
\end{itemize}
In particular, for every input at least one PURIFY output decodes to $0$ or $1$.
\end{lemma}
\begin{proof}
\emph{Pairs.} By \cref{lem:reader} and repeated use of \cref{lem:step} (and of \cref{lem:conststep} for the steps of \cref{sec:directW}), $(p,n)$ is a pair with bounds $(\sigma_p,\sigma_n)$ whose bits are the decoded input values. Consumers stay within $[0,\xi]$ by the choice of scales.

\emph{Seeds.} At bit $1$ of $(p,n)$, $a_V\ge c_VH(1-\sigma_n)=D+E+\xi\ge l_V$, so $r_V\ge H$, and likewise $r_W\ge H$ because $c_W\ge c_V$. At bit $0$, $a_W\le c_W\sigma_pE$ and $l_W\ge D+1-\sigma_p$, so $r_W\le s_0\le\tau$, and likewise $r_V\le\tau$.

\emph{Arms.} By \cref{lem:inverse}, $(V,n_V)$ is a pair with bounds $(\tau,t)$ for the conditions ``$r_V\le\tau$'' (bit $0$) and ``$r_V\ge H$'' (bit $1$), and so is $(W,n_W)$. By (D2) the final pair $v$ decodes to $0$ whenever $r_V\le\tau$ and to $1$ whenever $r_V\ge H$, and similarly for $w$. This gives (P3) and (P4), and together with the seeds (P1) and (P2).

\emph{Split.} Suppose $r_V>\tau$. Then $\Phi_V>\tau-E$, and since $l_V\ge D+1-r_p$ we get $a_V>(\tau-E)(D+1-r_p)$. The seeds have the same writer and reference, so $a_W=(c_W/c_V)a_V>(1+\xi/D)(D+1-r_p)\ge D+1-r_p+\xi\ge l_W$. Hence $r_W\ge H$, which is (P5).

\emph{NAND.} If $u=0$, then $r_{p_u}\le\sigma_p$ and $r_{n_u}\ge H$, so $a_{P_0}\ge k_0H(1-\sigma_p)=D+E+1+\xi\ge D+(1-r_{n_u})+(1-r_{n_v})+x=l_{P_0}$ and $r_{P_0}\ge H$; moreover $a_{N_0}\le k_A\sigma_p$ and $l_{N_0}\ge D+l_u(1-\sigma_p)$. If $v=0$, the second CDS of $P_0$ saturates it in the same way, and $a_{N_0}\le k_AE$, $l_{N_0}\ge D+l_v(1-\sigma_p)$. If $u=v=1$, then $a_{P_0}\le2k_0\sigma_nE$ and $l_{P_0}\ge D+2(1-\sigma_n)$, while $a_{N_0}\ge k_AH(1-\sigma_n)=D+E+\xi\ge l_{N_0}$. So $(P_0,N_0)$ is a pair with bounds $(g_{P0},g_{N0})$, with bit $1$ for ``$u=0$ or $v=0$'' and bit $0$ for ``$u=v=1$'', and (D3) gives (N1) and (N2).
\end{proof}

\subsection{The ideal maps and the basin}\label{sec:basin}

In the limit $D=\xi=0$ the steps act on bounds by
\[
\stP:(p,n)\mapsto(f(p,n),n),\quad \stN:(p,n)\mapsto(p,f(n,p)),\quad f(x,o)=\min\Bigl(1,\ E+\frac{E^2x}{H(1-x)(1-o)}\Bigr),
\]
with $f=1$ if $x\ge1$ or $o\ge1$. A point $(p,n)\in[0,1]^2$ is \emph{cleanable} if for every $\delta>0$ some word maps it into $[0,y+\delta]^2$, where $y$ is defined below. For $o\in[0,1)$ put $a(o)=E^2/(H(1-o))$ and $Q_o(x)=x^2-(1+E-a(o))x+E$. If $x<1$ and $f(x,o)<1$, then
\begin{equation}\label{eq:fQ}
f(x,o)-x=\frac{Q_o(x)}{1-x}.
\end{equation}
Let $\omax=1-E^2/(H(1-\sqrt E)^2)$, the value with $a(\omax)=(1-\sqrt E)^2$, and let $g_E(x)=H(x-E)(1-x)^2-E^2x$.

\begin{lemma}\label{lem:facts}
Let $0<E\le27/100$.
\begin{enumerate}[label=(\roman*)]
\item For $0\le o<\omax$, $Q_o$ has roots $\xm(o)<\sqrt E<\xp(o)<1$ with $\xm(o)\xp(o)=E$; on $[0,\omax)$, $\xm$ is continuous and increasing and $\xp$ is continuous and decreasing.
\item For $o<\omax$ and $x\in[0,1]$, $f(x,o)<x$ if and only if $\xm(o)<x<\xp(o)$. For $o\ge\omax$, $f(x,o)\ge x$ for all $x$.
\item $f$ is nondecreasing in both arguments, so $\stP$ and $\stN$ are monotone in the componentwise order.
\item $g_E$ has exactly one root $y\in(E,\sqrt E)$, one root $z\in(\sqrt E,1)$ and one root above $1$; $F(x):=f(x,x)$ satisfies $F(y)=y$, and $F(x)<x$ exactly for $x\in(y,z)$.
\item $\sqrt E<\omax$ and $\xp(\sqrt E)>\omax$.
\end{enumerate}
\end{lemma}
\begin{proof}
(i) Put $b=1+E-a(o)$, which decreases in $o$. For $o<\omax$, $b>1+E-(1-\sqrt E)^2=2\sqrt E$, so the roots are real and distinct with product $E$; the larger root $(b+\sqrt{b^2-4E})/2$ increases with $b$. Also $Q_o(1)=a(o)>0$ and $b/2<1$, so $\xp(o)<1$.
(ii) For $x<1$ the untruncated value of $f(x,o)$ is $x+Q_o(x)/(1-x)$ by~\eqref{eq:fQ}. If $Q_o(x)<0$, this value is below $x<1$, so no truncation occurs and $f(x,o)<x$; if $Q_o(x)\ge0$, then $f(x,o)\ge x$ whether or not it is truncated; and $f(1,o)=1$. For $o\ge\omax$, $b\le2\sqrt E$: either $b^2\le4E$, or $b<-2\sqrt E$ and both roots are negative; in both cases $Q_o\ge0$ on $[0,1]$.
(iii) $a(o)$ is increasing and $x/(1-x)$ is increasing.
(iv) For $x<1$ the untruncated value of $F(x)$ is $x-g_E(x)/(H(1-x)^2)$, so, as in (ii), $F(x)<x$ exactly when $g_E(x)>0$. We have $g_E(E)=-E^3$, $g_E(1)=-E^2$, and with $s=\sqrt E$, $g_E(s)=s\bigl((1+s)(1-s)^4-s^4\bigr)$, which is positive for $0<s\le13/25$ (the quintic in brackets is positive at $s=1/2$ and has no root in $(0,13/25]$, by a Sturm count); note $27/100<(13/25)^2$. The cubic $g_E$ has positive leading coefficient $H$, so it has the three stated roots.
(v) $(\omax-s)(1-s)^3(1+s)=(1+s)(1-s)^4-s^4>0$. The rational function $Q_s(\omax)$ of $s$ has numerator $-s^5(s^5-4s^4+2s^3+2s^2-3s+1)$ up to a positive constant, neither the quintic nor the denominator has a root in $(0,13/25]$, and its value at $s=1/2$ is $-1/36$. So $\omax\in(\xm(s),\xp(s))$.
\end{proof}

\begin{proposition}[the basin]\label{prop:basin}
Let $0<E\le27/100$. A point $(p,n)\in[0,1]^2$ is cleanable if and only if it lies in
\[
B_E=\{p<\omax,\ n<\xp(p)\}\cup\{n<\omax,\ p<\xp(n)\}.
\]
The complement $[0,1]^2\setminus B_E$ is an up-set that $\stP$ and $\stN$ map into itself without decreasing any coordinate.
\end{proposition}
\begin{proof}
\emph{The complement.} $(p,n)\notin B_E$ means that [$p\ge\omax$ or $n\ge\xp(p)$] and [$n\ge\omax$ or $p\ge\xp(n)$]. This is an up-set because $\xp$ is decreasing. By \cref{lem:facts}(ii) the second bracket gives $f(p,n)\ge p$ and the first gives $f(n,p)\ge n$. So $\stP$ and $\stN$ map the point to a point that is at least as large, hence again in the complement. Moreover the first bracket gives $p\ge\omax>\sqrt E$ or $n\ge\xp(p)>\sqrt E$. Hence no point reachable from the complement enters $[0,y+\delta]^2$ for $\delta<\sqrt E-y$, which proves necessity.

\emph{Sufficiency.} Let $p<\omax$ and $n<\xp(p)$; the other case is symmetric. (1) If $n>\xm(p)$, the $\stN$-iterates decrease and stay above the fixed point $\xm(p)$ of $f(\cdot,p)$, so they converge to it; since $\xm(p)<\sqrt E$, finitely many steps give $n'<\sqrt E$. (2) Then $\xp(n')>\xp(\sqrt E)>\omax>p$ by \cref{lem:facts}(v), and in the same way finitely many $\stP$-steps give $p'<\sqrt E$. (3) Let $b=\max(p',n')<\sqrt E<z$. If $b\le y$ we are done. Otherwise $F(b)<b$, and by monotonicity $\stP\stN(b,b)=(F(b),f(b,F(b)))\le(F(b),F(b))$. Hence $(\stP\stN)^k(p',n')\le(F^k(b),F^k(b))$, and $F^k(b)$ decreases to a fixed point of $F$ in $[y,b)$, which is $y$.
\end{proof}

Along the word of this proof every bound is at most the larger coordinate of the starting point. We also record a simple membership test: if $\sqrt E<p<n<1$, then $(p,n)\in B_E$ if and only if $Q_p(n)<0$ (the basin test in the proof of~\cite[record~L15; Proposition~2.1]{Supp}). We use only the direction ``if'', which is immediate: $Q_p(n)<0$ forces two real roots, both positive, with sum $1+E-a(p)>2\sqrt E$, so $p<\omax$ and $n\in(\xm(p),\xp(p))$.

\begin{lemma}\label{lem:nest}
For $0<E_1<E_2\le27/100$ we have $y(E_1)<y(E_2)$, $z(E_2)<z(E_1)$ and $B_{E_2}\subseteq B_{E_1}$. Each $B_E$ is a down-set.
\end{lemma}
\begin{proof}
$g_{E_1}(x)-g_{E_2}(x)=(E_2-E_1)\Gamma(x)$ with $\Gamma(x)=(1-x)^2(1+x-E_1-E_2)+(E_1+E_2)x>0$ on $(0,1)$. So $g_{E_1}>0$ at $y(E_2)$ and $z(E_2)$, and by \cref{lem:facts}(iv) both points lie in $(y(E_1),z(E_1))$. The set $B_E$ is a down-set since cleanability passes to smaller points by monotonicity. If a word cleans $q$ at $E_2$ to within $\delta$, then, because $f_{E_1}\le f_{E_2}$ pointwise, the same word maps $q$ into $[0,y(E_2)+\delta]^2$ at $E_1$; choosing $\delta$ with $y(E_2)+\delta<z(E_2)<z(E_1)$, the iteration $(\stP\stN)^k$ at $E_1$ then converges to $(y(E_1),y(E_1))$ as in step (3) above.
\end{proof}

\subsection{The PURIFY window}\label{sec:window}

Let
\[
\tmin(E)=E+\sqrt{y-E},\qquad T(E)=\frac{1-3E}{1-2E},\qquad n_0^I(\tau)=E+\frac E{1-\tau}.
\]
The value $n_0^I(\tau)$ is the ideal low bound of a source-funded inverse under (I) ($t_0=1$, $\xi=0$ in \cref{lem:inverse}). The PURIFY design needs a threshold $\tau$ that is large enough for the raw seed to fall below it, and small enough for the arm start $(\tau,n_0^I(\tau))$ to be cleanable.

\begin{lemma}[the window]\label{lem:window}
Let $1/5\le E\le27/100$.
\begin{enumerate}[label=(\alph*)]
\item If $E\le\tau<T(E)$, then $(\tau,n_0^I(\tau))\in B_E$.
\item $\tmin(E)<T(E)$ if and only if $E<\EI$, where $\EI$ is the unique root of $P_I$ in $[1/5,27/100]$.
\end{enumerate}
\end{lemma}
\begin{proof}
(a) The identity $Q_\tau(n_0^I(\tau))=\frac E{1-\tau}\cdot\frac{n_0^I(\tau)-H}H$ holds for all $\tau<1$ (expand), and $n_0^I(\tau)<H$ is equivalent to $\tau<T(E)$ (note $n_0^I(T(E))=H$ and $n_0^I$ increases). So $Q_\tau(n_0^I(\tau))<0$, which by the test above gives $\tau<\omax$ and $n_0^I(\tau)<\xp(\tau)$.

(b) Put $\delta_E=T(E)-E=(1-4E+2E^2)/(1-2E)$, which is positive on the interval, and $v=E+\delta_E^2$. Then $\tmin<T(E)$ if and only if $y<v$. One checks that $E<v<\sqrt E$ on $[1/5,27/100]$ (the rational function $E-v^2$ has no root there and is positive at $1/5$), so $v\in(E,z)$, and by \cref{lem:facts}(iv) $y<v$ if and only if $g_E(v)>0$. Expanding,
\[
g_E(E+\delta_E^2)=-\frac{2E^2P_I(E)}{(2E-1)^6}.
\]
By a Sturm count $P_I$ has exactly one root in $[1/5,27/100]$, and $P_I(1/5)<0<P_I(27/100)$; so $P_I<0$ on $[1/5,\EI)$ and $P_I>0$ on $(\EI,27/100]$.
\end{proof}

\subsection{Proof of the main theorem}\label{sec:mainthm}

We need four facts about specific numbers. Let $E_{hi}=2531/10000$ and $\bar y=617/2000$. All of the following are evaluations of rational functions at rational points or Sturm counts, done in exact arithmetic by \texttt{code/sec3\_facts.py}.
\begin{enumerate}[label=(F\arabic*)]
\item $\EI<E_{hi}$, since $P_I(E_{hi})>0$.
\item $\bar y\in(E_{hi},\sqrt{E_{hi}})$ and $g_{E_{hi}}(\bar y)=2.35\ldots\cdot10^{-5}>0$, so $y(E_{hi})<\bar y$. With $a_{hi}=E_{hi}+E_{hi}^2/((1-E_{hi})(1-\bar y)^2)=0.43246\ldots$ we have $g_{E_{hi}}(a_{hi})=0.0154\ldots>0$, so $a_{hi}<z(E_{hi})$.
\item The point $(p_S,n_S)=\bigl(E+\frac{E\bar y(1+E)}{H(1-\bar y)^2},\ E+\frac{E(\bar y+E)}{H(1-\bar y)^2}\bigr)$ at $E=E_{hi}$ equals $(0.52705\ldots,0.65109\ldots)$, satisfies $\sqrt{E_{hi}}<p_S<n_S$, and $Q_{p_S}(n_S)=-0.0207\ldots<0$; hence it lies in $B_{E_{hi}}$.
\item $P_I(1/5)<0$, and $P_I$ has exactly one root in $[1/5,27/100]$.
\end{enumerate}

\begin{proofof}{thm:general}
\begin{proof}
Membership in PPAD on all networks is \cref{thm:member}. We prove hardness of (I)-$E$; hardness of (W)-$E$ follows from it by calibration (\cref{cor:generalW}), and \cref{sec:directW} gives a direct construction.

\emph{Step 0: from a design to hardness.} Fix a rational $E$. Suppose there are rationals $\gamma\in(0,H)$, $\tau\in(E,1)$, $D,\xi>0$, weights $l_u,l_v$ and words such that (D1)--(D4) hold. Reduce from \PCirc\ with NAND and PURIFY gates (\cref{thm:pc}). Each gate becomes a copy of its gadget, whose input wires are the output wires of the gadgets that compute the input nodes. One source writes all source CDSs, and all debts and main CDSs go to one sink. Gadgets use banks of other gadgets only as references, so the network contains every gadget, the scales can be fixed gadget by gadget, and \cref{lem:gates} applies to every gadget. The roles of every CDS are distinct: within a gadget this is clear from the construction, and a gadget reads only wires of other nodes because the nodes of a gate are distinct. Every gadget bank owes a positive debt and $s,z$ have positive external assets, so the network is non-degenerate. For fixed $E$ every gadget has a fixed size and fixed rational data, so the network has polynomial size. Given an (I)-$E$ vector, decode every wire with threshold $\gamma$. By \cref{lem:gates} the decoded assignment satisfies every NAND and PURIFY constraint. Finally, hardness at $E$ gives hardness at every smaller error (\cref{lem:relations}(b)); in particular $E=1/5$ covers all $E<1/5$.

\emph{Step 1: ideal choices.} Fix a rational $E\in[1/5,\EI)$ and first make every choice for the ideal maps ($D=\xi=0$, no rounding), under which the steps act by $f$ and the start values take their limits.
\begin{enumerate}[label=(\alph*)]
\item By \cref{lem:window}(b) choose a rational $\tau\in(\tmin(E),T(E))$. Then $\tau>E$.
\item By (F1), (F2) and \cref{lem:nest}, $y(E)\le y(E_{hi})<\bar y$, and $\tau>\tmin(E)>y(E)$. Choose a rational $\gamma$ with $y(E)<\gamma<\min(\bar y,\tau)$.
\item The ideal pair-reader start is $a=E+E^2/(1-\gamma)^2$. Since $\gamma>y$ and $y<\sqrt E<H$, $a>E+E^2y/(H(1-y)^2)=y$, the last equality being $g_E(y)=0$. Also $a\le a_{hi}<z(E_{hi})\le z(E)$ by monotonicity in $E$ and $\gamma$, (F2) and \cref{lem:nest}. On $(y,z)$, $F(x)<x$ and $F(x)\ge F(y)=y$, and the word $\stP\stN$ maps $(x,x)$ to a point at most $(F(x),F(x))$. Hence for every $\delta>0$ some word $(\stP\stN)^k$ maps $(a,a)$ into $[0,y+\delta]^2$, and every bound along it is at most $a$.
\item Let $\kappa(\sigma_p,\sigma_n)=E^2\sigma_p/(H(1-\sigma_n)(1-\sigma_p))$, so that the ideal raw seed is $E+\kappa(\sigma)/(\tau-E)$. The function $\kappa$ is continuous and increasing, and $\kappa(y,y)=y-E<(\tau-E)^2$. Choose $\delta>0$ with $y+\delta<\gamma$ and $\kappa(y+\delta,y+\delta)<(\tau-E)^2$, and let $w_{\rm in}=(\stP\stN)^k$ be as in (c), with ideal end point $\sigma\le(y+\delta,y+\delta)\le(\bar y,\bar y)$. Then the ideal raw seed satisfies $E+\kappa(\sigma)/(\tau-E)<\tau$.
\item Take $l_u=\sigma_p/(\sigma_p+E)$. The ideal NAND seed is then
\[
(g_{P0},g_{N0})=\Bigl(E+\frac{E\sigma_n(1+E)}{H(1-\sigma_p)(1-\sigma_n)},\ E+\frac{E(\sigma_p+E)}{H(1-\sigma_p)(1-\sigma_n)}\Bigr),
\]
which increases in $E$, $\sigma_p$ and $\sigma_n$. With $\sigma\le(\bar y,\bar y)$ and $E\le E_{hi}$ it is at most the point of (F3), which lies in $B_{E_{hi}}\subseteq B_E$. Since $B_E$ is a down-set, the seed lies in $B_E$, and the word $w_{\rm out}$ of \cref{prop:basin} maps it into $[0,\gamma)^2$, with every bound at most $\max(g_{P0},g_{N0})$.
\item The ideal arm start is $(\tau,n_0^I(\tau))$, which lies in $B_E$ by \cref{lem:window}(a). The word $w_{\rm arm}$ of \cref{prop:basin} maps it into $[0,\gamma)^2$, with every bound at most $\max(\tau,n_0^I(\tau))<1$.
\item Every ideal bound is therefore below $1$.
\end{enumerate}

\emph{Step 2: perturbation.} Let $D>0$ be rational, put $\xi=D^2$, and round every design bound up to the grid $D\mathbb Z$. Each design bound is obtained from the start values by a fixed finite sequence of maps of pairs $(g_p,g_n)\mapsto(\lceil\psi(g_p,g_n;D,\xi)\rceil_D,\,g_n)$ and their mirror images, where $\psi$ is the bound formula of \cref{lem:step}; the raw seed $s_0$ is a further rational function of $(\sigma_p,\sigma_n,D,\xi)$. Before truncation at $1$, these are rational functions whose denominators are positive at the ideal points (all ideal bounds are below $1$), and at $D=\xi=0$ the step formula is exactly $f$ and $s_0$ is $E+\kappa(\sigma)/(\tau-E)$. The start values $g_1$, $t$, $g_{P0}$, $g_{N0}$ (with $l_u$ fixed) are continuous at $D=\xi=0$ as well, and $c_W=(1+D)c_V/(\tau-E)$. So every design bound tends to its ideal value as $D\to0$. The ideal values satisfy the strict inequalities of Step~1: raw seed below $\tau$, final bounds below $\gamma$, all bounds below $1$. Hence for some small rational $D$ the perturbed design satisfies (D1)--(D4), and Step~0 applies.
\end{proof}
\end{proofof}

\subsection{A direct construction for (W)}\label{sec:directW}

Under (W) the source only satisfies $r_s\ge H$, and the inverse of \cref{lem:inverse} has the ideal low bound $n_0^1(\tau)=E+E/(H(1-\tau))$, which can exceed $H$. A second kind of refinement step handles such arm starts.

\begin{lemma}[constant-liability refinement]\label{lem:conststep}
Let $(p,n)$ be a pair with bounds $g_p,g_n<1$ and $\xi>0$. Let a new bank $P'$ hold a CDS written by $p$ on $n$ with coefficient $k=(1+\xi)/(H(1-g_n))$, owe $1$, and write no main CDS. Then $(P',n)$ is a pair, under the same conditions, with bounds $\bigl(E+(1+\xi)Eg_p/(H(1-g_n)),\ g_n\bigr)$. The step on the second rail is symmetric.
\end{lemma}
\begin{proof}
$a=kr_p(1-r_n)$ and $l=1+x$ with $x\in[0,\xi]$. At bit $1$, $a\ge kH(1-g_n)=1+\xi\ge l$. At bit $0$, $a\le kg_pE$ and $l\ge1$.
\end{proof}

In the ideal limit the new steps $\stp$ and $\stn$ act by $h(x,o)=\min(1,E+Ex/(H(1-o)))$ on the refined bound $x$, given the other bound $o$.

\begin{lemma}\label{lem:enlarged}
Let $1/5\le E\le27/100$. If $p<T(E)$ and $n<1$, a finite word of $\stn$-steps maps $(p,n)$ to a point of $B_E$, with every bound at most $\max(n,H)$.
\end{lemma}
\begin{proof}
With $p$ fixed, an $\stn$-step acts as $n\mapsto E+an$ with $a=E/(H(1-p))$. The identity $h(H,p)-H=E/(1-p)-(1-2E)$ shows $h(H,p)<H$ for $p<T(E)$, so $aH+E<H$; hence $a<1$ and the fixed point $E/(1-a)$ lies below $H$. Finitely many steps therefore give $n'<H$, and every intermediate bound is at most $\max(n,H)$. Now $Q_p(H)=E\bigl(2E-1+E/(1-p)\bigr)<0$ for $p<T(E)$, so $p<\omax$ and $H<\xp(p)$, whence $n'<\xp(p)$ and $(p,n')\in B_E$.
\end{proof}

\begin{proof}[Direct proof of \cref{thm:general} for (W)]
Use $t_0=H$ in \cref{lem:reader,lem:inverse} and run Steps 0--2 of the previous proof with the following changes. In (c) the ideal pair-reader start is $a=E+E^2/(H(1-\gamma)^2)$; it satisfies $a>y$ and $a\le a_{hi}<z(E)$ by the same argument, using (F2) directly. In (f) the ideal arm start is $(\tau,t)$ with $t=n_0^1(\tau)$. Since $n_0^1$ increases and $n_0^1(T(E))=1-E^2/(1-E)<1$, we have $t<1$, and $\tau<T(E)$. Let $w_{\rm arm}$ consist of the $\stn$-steps of \cref{lem:enlarged} followed by the word of \cref{prop:basin}; its ideal bounds are at most $\max(t,H)<1$. In Step~2 the constant-liability formula is also a rational function whose value at $\xi=0$ is $h$. Everything else is unchanged.
\end{proof}

\subsection{Scope of the construction}\label{sec:generalscope}

The historical auxiliary proofs cited in the following remark are reproduced in \cref{app:hardnessaux}, with their original family restrictions.

\begin{remark}[the endpoint of this design family]\label{rem:EIendpoint}
For the ideal recurrences ($D=\xi=0$) of this family (pair readers, words over the refinement steps of \cref{lem:step,lem:conststep}, the shared-input seed split of \cref{lem:gates}, and source-funded inverses), the following hold for every $E\in[\EI,27/100]$ under either notion (the family was not analysed above $27/100$): a valid PURIFY needs $\tau>\tmin(E)$ (the seed is bounded below through the fixed point $(y,y)$, which every reachable point dominates), and its arm start is cleanable only if $\tau<T(E)$ (the cleanable set for both kinds of steps is exactly $B_E\cup\{\min(p,n)<T(E)\}$). Proofs are in~\cite[record~L13; Theorem~5.4]{Supp}, \cite[record~L14; Theorem~7.2]{Supp} with the audit~\cite[record~L15; Theorem~2.3]{Supp}, and~\cite[record~L15; Proposition~4.1]{Supp}. So, in the ideal limit and on $[1/5,27/100]$, $\EI$ is the exact supremum of the errors at which the family has a valid design, and it is not attained. With finite $D,\xi$ the step is slightly weaker below $H$ and slightly stronger above $H$, and the transfer of the endpoint to finite designs uses the remark after~\cite[record~L15; Proposition~4.1]{Supp}, which needs the cleaned bounds to stay at most $H$. The endpoint does not move if, inside this architecture and with its rectangular certificates, the alphabet of steps is enlarged to all single-bank two-rail restorers (on $[1/5,27/100]$; the cleanable set does not change), or the seeds are taken from a broader class of single-bank seed splits (on $[1/4,27/100]$), for ideal steps under the dual-rail semantics~\cite[record~L16; Theorem~B; Lemma~9.1, Theorem~9.2 and Section~9.3]{Supp}.

These are statements about complete PURIFY designs in this restricted architecture, not about individual restorers, which can still lower bounds above $\EI$. For example, let $E=13/50>\EI$ and let $(p,n)$ be a pair with bounds $(7/20,7/20)$; under (I) the pair reader of \cref{lem:reader} with $\gamma=1/100$ and the parameters below already gives one, with $g_1=16147183/49054500<7/20$. \Cref{lem:step} with $D=1/1000$ and $\xi=10^{-6}$, so that $k_P=20077/37000$, makes $(P,n)$ a pair with bounds $(1155661/3441000,\ 7/20)$, and $7/20-1155661/3441000=48689/3441000>0$; in the ideal limit the new bound is $f(7/20,7/20)=621/1850$. The step also improves the bound $g_1$ of the reader pair itself: with $(g_p,g_n)=(g_1,g_1)$ the new bound is $0.32697\ldots<g_1=0.32916\ldots$ (\texttt{code/scope\_examples.py}). This is consistent with the above: \cref{prop:basin} holds for every $E\le27/100$, so the basin is not empty above $\EI$; what fails on $[\EI,27/100]$ is the PURIFY window of \cref{lem:window}(b). The example is not a gadget and gives no hardness. Within this architecture, passing $\EI$ needs a PURIFY outside these families.
\end{remark}

\begin{remark}[finite certificates]\label{rem:EIcert}
\CO\ The proof is existential in $D$. Explicit instances were built in exact rational arithmetic in~\cite[record~L15; Section~3.4]{Supp}: (W) at $E=2529958621/10^{10}$ with one source ($\EI-E=7.21\cdot10^{-11}$; closed test network $\mathrm{NAND}(x_0,x_1;x_2)$, $\mathrm{PURIFY}(x_2;x_0,x_1)$ with $333$ banks and $563$ CDSs; it is not one-writer, since $332$ banks write CDSs), and (I) at $E=25299586217212/10^{14}$ ($\EI-E=4.15\cdot10^{-15}$; $291$ banks and $579$ CDSs). The producing program checked every gate claim on the actual contracts, by interval propagation in creation order and by a contract-level fixed-point check; the further checks, by a separate contract checker and by a rebuild from the printed parameters, are listed in the check index of the support bundle~\cite{Supp}. The certificates are not needed for \cref{thm:general}.
\end{remark}

\section{Auxiliary details for the hardness-design comparisons}\label[appendix]{app:hardnessaux}

This appendix supplies the historical proof details used in
\cref{rem:EIendpoint,rem:leak}. These are limits and comparisons for the
specified gadget families, not upper bounds on the clearing problems.
The sources are \cite[record~L13; Theorem~5.4]{Supp}, \cite[record~L14; Theorem~7.2]{Supp},
\cite[record~L15; Theorem~2.3, Proposition~4.1 and the following remark]{Supp},
\cite[record~L16; Sections~9.1--9.3]{Supp}, and \cite[record~L3; Theorem~4.3]{Supp}.

\subsection{The enlarged ideal basin and the shared-expression endpoint}

Use $E\in[1/5,27/100]$, $H=1-E$, and the functions $f,Q_o,g_E$, the
roots $y,z$ and the set $B_E$ of \cref{sec:basin}. Put
\[
 h(x,o)=\min\left(1,E+\frac{Ex}{H(1-o)}\right),\qquad
 T=\frac{1-3E}{1-2E},\qquad
 C_E=B_E\cup\{\min(p,n)<T\}.
\]
At $o=1$ use the trivial bound $h=1$; this convention is never needed
in a successful cleaning step. Old steps apply $f$ to one coordinate;
constant-liability steps apply $h$.

\paragraph{Scalar facts.}
Before truncation,
\[
\begin{aligned}
 h(x,o)-f(x,o)&=\frac{Ex(H-x)}{H(1-o)(1-x)},\\
 h(H,p)-H&=\frac E{1-p}-(1-2E),\\
 Q_p(H)&=E\left(2E-1+\frac E{1-p}\right).
\end{aligned}
\]
In particular $h\ge f$ when $x\le H$. Also $T<\omax$: at $p=T$ the
roots of $Q_T$ are $H$ and $E/H$, which are distinct since
$H^2>E$ throughout the stated interval. Thus its root sum exceeds
$2\sqrt E$, the defining condition for $T<\omax$.
For $p\ge T$ one has $f(p,H)\ge p$. Here are exact sign details for
this last assertion, avoiding an appeal to a numerical plot. Write
\[
 Q_H(x)=x^2-(1+E-E/H)x+E.
\]
Then
\[
 Q_H(T)=\frac{2E^4}{H(1-2E)^2}>0,\qquad
 Q'_H(T)=\frac{1-5E+5E^2-2E^3}{H(1-2E)}.
\]
On $[1/5,11/50]$ the last numerator is positive: its derivative
$-5+10E-6E^2$ is negative, and its value at $11/50$ is positive.
So $Q_H(x)>0$ for $x\ge T$ there. On $[11/50,27/100]$ the
discriminant has numerator
$E^4-2E^3+7E^2-6E+1$, negative at $11/50$ and strictly decreasing
on this interval, since its derivative is at most
$4(27/100)^3+14(27/100)-6<0$.
Thus $Q_H>0$ everywhere on this second interval. The sign rule of
\cref{lem:facts}(ii) proves the claim.
Finally $T>y$: $T\ge19/46>9/25$, whereas
\[
 g_E(9/25)=\frac{775E^2-8704E+2304}{15625}>0
\]
on $[1/5,27/100]$ (the numerator decreases and is positive at
$27/100$), so $y<9/25$ by \cref{lem:facts}(iv).

\paragraph{Exact enlarged basin.}
If $p<T$, repeated constant-liability updates to $n$ follow
$n\mapsto E+an$, $a=E/(H(1-p))$. Its value at $H$ is below $H$,
so $a<1$ and its fixed point is below $H$; finitely many updates give
$n'<H$. The negative value $Q_p(H)<0$ gives
$p<\omax$ and $H<\xp(p)$, so $(p,n')\in B_E$ and
\cref{prop:basin} completes cleaning. The swapped case is identical.

Conversely, suppose $(p,n)\notin C_E$. If both coordinates are at
most $H$, take $b=(p,n)$. Nonmembership of $B_E$ makes both old
steps nondecreasing at $b$, and $h\ge f$ there. If $p\le H<n$,
take $b=(p,H)$. The scalar facts give
$f(p,H)\ge p$, $h(p,H)\ge p$ and
$f(H,p)=h(H,p)\ge H$. The case $n\le H<p$ is symmetric. If both
coordinates exceed $H$, take $b=(H,H)$, where the steps give $1$.
In every case $b\le(p,n)$ is a common post-fixed lower bound for
all four monotone maps and $b\ge(T,T)>(y,y)$. No reachable point
can approach $(y,y)$. This proves that $C_E$ is exactly the
cleanable set, including the strict boundary $\min(p,n)=T$.

\paragraph{The endpoint of the shared-expression design.}
Every bound in a successful design is at least $E$, so its decoding
threshold satisfies $\gamma\ge E$. The ideal pair-reader start is
\[
 a=E+\frac{E^2}{t_0(1-\gamma)^2}>y,
 \qquad t_0\in\{1,H\}.
\]
Indeed $t_0(1-\gamma)^2\le H$, while
$y<(1-y)^2$ follows from $y<9/25$; compare with
$y-E=E^2y/[H(1-y)^2]$. Each permitted step is monotone and
post-fixed at $(y,y)$, with $f(y,y)=y$ and $h(y,y)>y$.
A finite word from $(a,a)$ stays above $(y,y)$ and cannot become
exactly $(y,y)$: the last strict coordinate cannot be removed by an
old step, which is strictly increasing below truncation, and a
constant-liability step is strictly above $y$ at $(y,y)$.
Thus the cleaned bounds $\sigma$ satisfy
\[
 \kappa(\sigma):=
 \frac{E^2\sigma_p}{H(1-\sigma_p)(1-\sigma_n)}>y-E.
\]
The raw shared-expression seed bound is
$E+\kappa(\sigma)/(\tau-E)$. Its required inequality $s_0\le\tau$
therefore implies $\tau>E+\sqrt{y-E}=\tmin(E)$.

The ideal source-funded inverse has low bound at least
$n_0^I(\tau)=E+E/(1-\tau)$ under either notion, with equality for
(I). If $\tau\ge T$, this is at least $H$. The common lower barrier
$(\min(\tau,H),H)$ above then keeps the second arm bound at least
$H$, contradicting a final dual-rail bound $\gamma<H$.
Consequently $\tau<T$. Conversely, when
$\tmin(E)<\tau<T$, the finite ideal words constructed in the proof
of \cref{thm:general} work. By \cref{lem:window}, the window is
nonempty exactly when $E<\EI$. Its boundary is strict and is not
attained. For the one-source inverses used here,
$n_0^W(\tau)\ge n_0^I(\tau)$; for $\tau<T$ the constant-liability
steps supply the additional cleaning needed under (W).

\paragraph{Finite debts, loads and rounding.}
For the old finite-debt bound without consumer load,
\[
 f_D(x,o)=E+\frac{(D+E)xE}{H(1-o)(D+1-x)},\qquad
 \frac{\partial f_D}{\partial D}
 =\frac{xE(H-x)}{H(1-o)(D+1-x)^2}.
\]
Its minimum over $D\ge0$ is one of the ideal endpoints $f$ or $h$,
with truncation understood. Positive consumer load and upward
rounding only raise the bound. Reader and seed bounds have the
form $E+(D+E+\xi)u/(D+v)$ with $u\ge0$ and $v\ge E$, provided
the cleaned bound $\sigma_p\le H$ (and $\gamma<H$). Their
finite value is no smaller than the ideal value because
$D(v-E)+\xi v\ge0$; the second seed has the additional factor
$1+\xi/D\ge1$. Inverse bounds also only increase. Induction along
the words compares any such successful finite design with an
ideal design using the better of the two endpoint steps. Hence
the ideal impossibility applies to this stated finite-design
scope. It does not assert the comparison without the condition
on the cleaned bound.

\subsection{The broader single-bank comparisons of \texorpdfstring{\cite[record~L16]{Supp}}{research record L16}}

\paragraph{Primitive class and its exact restriction.}
A primitive has only the two old rails as informative
counterparties or references and no informative helper. At source
recovery $1$, its assets are a nonnegative linear combination of
\[
 1,p,n,1-p,1-n,p(1-n),n(1-p),
\]
and its liabilities a nonnegative linear combination of
$1,1-p,1-n$. Correctness is required for every point of the
rectangular input domains, using saturation to force the high
state. All claims below retain this local-certificate restriction.

For $E\le x,o\le H$, compare the high point $(H,o)$ and the low
point $(x,H)$. Each asset basis term at the low point is at least
$m=xE/[H(1-o)]$ times its high value, and each liability basis
term at the high point is at least $q=E/(1-x)$ times its low value.
Both facts follow term by term from $x/H\le1$ and
$E/(1-o)\le1$; the product asset and the liability $1-p$ attain
the respective ratios. Therefore
\[
 A_{\rm low}\ge mA_{\rm high}\ge m\ell_{\rm high}
              \ge mq\ell_{\rm low}.
\]
If the low denominator vanishes, the zero-liability convention
makes the map saturated, and the same lower bound on the
adversarial recovery holds. Otherwise its largest admissible recovery is at least
\[
 \min\left(1,E+\frac{E^2x}{H(1-x)(1-o)}\right)=f(x,o).
\]
For (W) the all-ones source assignment is allowed in the local
certificate, so the same lower bound applies.

Apply this inequality to the lower barriers in the enlarged-basin
proof. Outside $C_E$ their coordinates lie in $[T,H]\subset[E,H]$;
each new primitive must respect the old lower barrier on that
smaller rectangular domain. The same holds for simultaneous
independent refinements, since neither new output is an informative
input of the other. Thus this larger primitive set leaves $C_E$
unchanged. This excludes no correlated multi-bank certificate.

\paragraph{The broader independent-seed comparison.}
Here restrict further to $1/4\le E\le27/100$ and cleaned bounds
$\sigma_p,\sigma_n\in[y,H]$. Let $V,W$ be two banks in the preceding
primitive class, independent of each other's recoveries. Assume
$V$ saturates on bit~$1$ and the implication
$r_V>\tau\Rightarrow r_W\ge H$ holds for every admissible local
choice. Attach one-bank unary inverses and rectangular refinement
words, as in \cite[record~L16; Section~9.3]{Supp}.
A unary inverse at threshold $q<H$ cannot improve the lower bound
$E+E/(1-q)$: compare its asset terms $1,r,1-r$ at $r=q$ and $r=H$;
the smallest low-to-high ratio is $E/(1-q)$, and the high-to-low
liability ratio is at least $1$. Threshold $q\ge H$ already gives
overlapping high/low requirements. The same arm barrier therefore
requires $\tau<T$ and a raw low bound $s_W<T$.

Retain the split threshold $\tau>E$ of the architecture, so $E<\tau<T$. Put
\[
 u=\tau-E,\qquad b=\frac{E^2}{H(1-\sigma_n)},\qquad
 p_* =\frac{u}{b+u}.
\]
Since $u<T-E\le E/H$ on the stated interval, one has $p_*<H$:
indeed $\sigma_n\ge y>E$ gives $u<E/(1-\sigma_n)$, equivalent
to $p_*<H$. For the inequality used here,
\[
 \frac EH-(T-E)=
 \frac{2E^3-8E^2+6E-1}{(1-E)(1-2E)}>0;
\]
the numerator is $1/32$ at $E=1/4$ and increases up to
$27/100$, since its derivative $6E^2-16E+6$ is positive there.
Comparing $(H,\sigma_n)$ with $(p,H)$ as above gives
$A_V(p,H)/\ell_V(p,H)\ge bp/(1-p)$ for $0<p\le H$;
a zero denominator again means saturation.
Thus points $p>p_*$ sufficiently close to $p_*$ admit
$r_V>\tau$. Because $W$ is independent of $r_V$, the split
condition forces $W$ to saturate at every such point.
If $\sigma_p>p_*$, this forces saturation inside the bit-$0$
rectangle and contradicts $s_W<T$. Otherwise compare $W$ at
$(\sigma_p,H)$ with its saturated point $(p,H)$ and let
$p\downarrow p_*$; the termwise asset and liability ratios give
\[
 s_W\ge\min\left(1,E+
 \frac{\sigma_p(1-p_*)}{p_*(1-\sigma_p)}\right)
 =\min\left(1,E+
 \frac{E^2\sigma_p}{H(1-\sigma_n)(1-\sigma_p)(\tau-E)}\right).
\]
Truncation cannot help because $s_W<T<1$. As
$\sigma\ge(y,y)$, this implies
$E+(y-E)/(\tau-E)<T$. Together with $\tau<T$ it gives
$\tmin(E)<T$, hence $E<\EI$.
These are precisely the independent-seed and rectangular
assumptions behind the broader-family sentence of
\cref{rem:EIendpoint}; they are not conditions satisfied by all
possible PURIFY circuits.

\subsection{The historical direct-(W) inverter comparison}

For completeness, the direct-(W) sentence in \cref{rem:leak} uses
$w=H=1-E$ and the following replacements in the proof of
\cref{thm:wsI}:
\[
 h_w(g)=E+\frac{E}{w(1-g)},\qquad
 t=\frac{H+\sqrt{H^2-4E/w}}2,\qquad
 y_-=1-t,\quad y_+=E+t.
\]
Each source-funded notional is divided by $w$. Its payment lies
between $w$ and $1$ times that notional, so the inverter lower
and upper bounds are exactly the bounds defining $h_w$.
The sufficient strict parameter inequalities are
\[
 wt^2>2E,\qquad w^2t^3>E.
\]
Choose rational $\gamma_a\in(y_-,y_+)$, $\gamma_d$ and $\varphi$
so that
\[
 E+\frac{2E}{w(1-\gamma_a)}\le\gamma_d<y_+,\qquad
 \frac{E}{wt}<1-\varphi<wt(1-\gamma_a).
\]
They exist by taking $\gamma_a$ sufficiently close to $y_-$.
The seed upper bounds are
$g_A=E+E/[w(1-\varphi)]$ and
$g_B=E+(1-\varphi)/[w(1-\gamma_a)]$, both below $y_+$.
The same fixed-point argument as \cref{lem:hmap} supplies the
even input buffers and odd PURIFY tails, and $y_+<H$ follows by
squaring the inequality for the displayed discriminant. For
NAND, one low buffered input supplies assets at least
$w/[w(1-\gamma_a)]=1$, while two high inputs give recovery at
most $E+2E/[w(1-\gamma_a)]\le\gamma_d$.
For PURIFY, input $\rho\le\varphi$ saturates the first seed and
$\rho\ge H$ bounds it by $g_A$; input $\rho\le\gamma_a$
saturates the second and $\rho\ge\varphi$ bounds it by $g_B$.
The odd tails give exactly the four implications of the proof of
\cref{thm:wsI}. Their inclusive split covers $\rho=\varphi$.
Distinct parties, rational encoding, fixed gadget size and
non-degeneracy are unchanged.

For $w=H$, the discriminant is nonnegative when $H^3\ge4E$;
$t$ decreases with $E$. The first strict inequality is lost at
the root of $2H^3=9E$, in $(1409/10000,141/1000)$.
Eliminating $t$ from
\[
 Ht^2-H^2t+E=0,\qquad H^2t^3=E
\]
gives
\[
 E(E-1)^3\bigl(E^5-5E^4+12E^3-15E^2+9E-1\bigr)=0.
\]
Write this quintic as $P_W$. Its third derivative is
$60(E-1)^2+12>0$, so $P_W''$ increases on $[0,1/2]$ and is at
most $P_W''(1/2)=-13/2$. Thus $P_W'$ decreases and is at least
$P_W'(1/2)=13/16>0$. Together with $P_W(0)=-1$ and
$P_W(1/2)=31/32$, this proves uniqueness of its root.
Exact substitution gives
\[
 P_W(281/2000)=\frac{-6890437304599}{32000000000000000}<0,
 \qquad
 P_W(703/5000)=\frac{1027442210558743}{3125000000000000000}>0.
\]
For the first condition, $2(1-E)^3-9E$ is strictly decreasing;
its values at $1409/10000$ and $141/1000$ are respectively
$11169071/500000000000>0$ and $-660221/500000000<0$.
The discriminant numerator $H^3-4E$ decreases and at $703/5000$
equals $9040706073/125000000000>0$.
At the quintic root, the positive solution of $H^2t^3=E$
has $t>H/2$ because $8E>H^5$ (already $E>1/8$). The resultant
above guarantees a common root; any nonreal cube roots cannot
be roots of the real quadratic, whose discriminant is positive.
Hence the common root is precisely the larger quadratic root
used here. Both strict conditions hold near $0$ and decrease
to their respective zeros, so the second is binding and the
direct construction reaches every rational error below
$0.14053956\ldots$. This comparison is
superseded by calibration for hardness purposes, and is not
needed for \cref{cor:oneW}.

\section{Proofs for writer-solvent hardness, calibration and the barrier}\label[appendix]{app:onewriter}

This appendix contains two perturbation estimates, which are also used in \cref{app:membership}, and the proofs for \cref{sec:onewriter,sec:coincidence}.

\subsection{Two perturbation estimates}

\begin{lemma}[clipped ratios]\label{lem:clip}
For $A,B\ge0$ and $L,K>0$,
\[
\bigl|\min(1,A/L)-\min(1,B/K)\bigr|\le\frac{|A-B|+|L-K|}{\min(L,K)} .
\]
\end{lemma}
\begin{proof}
Changing the numerator from $A$ to $B$ at denominator $L$ changes the clipped ratio by at most $|A-B|/L$. Now fix the numerator $B$ and suppose $L\le K$. If $B\le L$ the change is $B(K-L)/(LK)\le(K-L)/L$; if $L<B\le K$ it is $1-B/K\le(K-L)/K$; if $B>K$ it is $0$. The case $K<L$ is symmetric. The triangle inequality completes the proof.
\end{proof}

\begin{lemma}[rescaling one income stream]\label{lem:rescale}
Let $A,B\ge0$, $L>0$ and $u>0$, and put $\varphi=\min(1,(A+B)/L)$ and $\psi=\min(1,(A+uB)/L)$. Then $|\psi-\varphi|\le|u-1|$.
\end{lemma}
\begin{proof}
If $u\le1$, then $u(A+B)\le A+uB\le A+B$; since $\min(1,uz)\ge u\min(1,z)$ for $z\ge0$, we get $u\varphi\le\psi\le\varphi$, and $\varphi-\psi\le(1-u)\varphi\le1-u$. If $u\ge1$, then $A+B\le A+uB\le u(A+B)$, so $\varphi\le\psi\le\min(1,u\varphi)$ and $\psi-\varphi\le(u-1)\varphi\le u-1$.
\end{proof}

The bound of \cref{lem:rescale} involves neither the size of the notionals nor a lower bound on liabilities, so the calibration copies of \cref{sec:onewriter} need no such bounds.

\subsection{Writer-solvent hardness}

\begin{proofof}{thm:wsI}
\begin{proof}
Membership is \cref{thm:member}. For hardness we first show that the two conditions
\begin{equation}\label{eq:Acond}
t^2>2E\qquad\text{and}\qquad t^3>E
\end{equation}
suffice, and then that they hold for $E<\alpha$.

\emph{Parameters.} Assume~\eqref{eq:Acond}. Choose rationals $\gamma_a\in(\lambda_-,\lambda_+)$, a decoding threshold $\gamma_d$ and $\varphi$ with
\[
\gamma_N:=E+\frac{2E}{1-\gamma_a}\le\gamma_d<\lambda_+,\qquad \frac Et<1-\varphi<t(1-\gamma_a).
\]
This is possible: as $\gamma_a\downarrow\lambda_-$, $1-\gamma_a\to t$, so $\gamma_N\to E+2E/t<E+t=\lambda_+$ by the first condition, and $t(1-\gamma_a)\to t^2>E/t$ by the second. Since $t\le1$ we get $1-\varphi<1-\gamma_a$ and $1-\varphi>E$, so $\gamma_a<\varphi<1-E$. Put
\[
g_A=E+\frac E{1-\varphi},\qquad g_B=E+\frac{1-\varphi}{1-\gamma_a}.
\]
Then $g_A<E+t=\lambda_+$ and $g_B<E+t=\lambda_+$. Also $\gamma_d\ge\gamma_N>h(\gamma_a)>h(\lambda_-)=\lambda_-$, so $\gamma_d\in(\lambda_-,\lambda_+)$. By \cref{lem:hmap} fix an even $m$ with $h^m(\gamma_d)\le\gamma_a$ and odd $m_A,m_B$ with $h^{m_A}(g_A)\le\gamma_d$ and $h^{m_B}(g_B)\le\gamma_d$.

\emph{Network.} Every node $x$ of a \PCirc\ instance with NAND and PURIFY gates gets a receiver $b_x$, decoded as $0$ if $r\le\gamma_d$, as $1$ if $r\ge1-E$, and as $\bot$ otherwise; this is well defined since $\gamma_d<\lambda_+<1-E$. For $\mathrm{NAND}(u,v;x)$ build chains of length $m$ from $b_u$ and from $b_v$ with start code $\gamma_d$, ending in $e_u$ and $e_v$, and let $b_x$ hold CDSs of notional $c=1/(1-\gamma_a)$ on $e_u$ and on $e_v$. For $\mathrm{PURIFY}(u;v,x)$ build a chain of length $m$ from $b_u$ with start code $\gamma_d$, ending in $x_1$; receivers $A$ and $B$ holding one CDS each on $x_1$, of notionals $1/(1-\varphi)$ and $1/(1-\gamma_a)$; a chain of length $m_A$ from $A$ with start code $g_A$, ending in $b_v$; and a chain of length $m_B$ from $B$ with start code $g_B$, ending in $b_x$. Finally set $e_s=2L_s^{\max}+1$. Every CDS is held by a newly created receiver and references a different bank, so all roles are distinct. The network is non-degenerate and one-writer, and it has polynomial size for fixed $E$.

\emph{Correctness.} Let $r$ be an (I)-$E$ vector. Consider $\mathrm{NAND}(u,v;x)$. If $b_u$ decodes to $0$, then $r_{b_u}\le\gamma_d$, so $r_{e_u}\le h^m(\gamma_d)\le\gamma_a$ (even chain), hence $A_{b_x}\ge c(1-\gamma_a)=1$ and $r_{b_x}\ge1-E$; likewise for $v$. If both inputs decode to $1$, then $r_{e_u},r_{e_v}\ge1-E$, so $A_{b_x}\le2cE$ and $r_{b_x}\le E+2cE=\gamma_N\le\gamma_d$. Consider $\mathrm{PURIFY}(u;v,x)$ and let $\rho=r_{x_1}$.
\begin{itemize}
\item If $\rho\le\varphi$, then $\Phi_A\ge1$, so $r_A\ge1-E$, and the odd chain gives $r_{b_v}\le\gamma_d$.
\item If $\rho\ge1-E$, then $\Phi_A\le E/(1-\varphi)$, so $r_A\le g_A$ and $r_{b_v}\ge1-E$.
\item If $\rho\le\gamma_a$, then $\Phi_B\ge1$, $r_B\ge1-E$ and $r_{b_x}\le\gamma_d$.
\item If $\rho\ge\varphi$, then $\Phi_B\le(1-\varphi)/(1-\gamma_a)$, so $r_B\le g_B$ and $r_{b_x}\ge1-E$.
\end{itemize}
If $b_u$ decodes to $0$ then $\rho\le\gamma_a<\varphi$ and both outputs decode to $0$; if it decodes to $1$ then $\rho\ge1-E>\varphi$ and both decode to $1$; and for every $\rho$ either $\rho\le\varphi$ and $v$ decodes to $0$, or $\rho\ge\varphi$ and $x$ decodes to $1$. So the decoded assignment solves the instance.

\emph{The threshold.} On $(0,3-2\sqrt2]$ the number $t$ is real and decreasing in $E$, so $t^2-2E$ and $t^3-E$ are decreasing. At a zero of $t^2-2E$ the relation $t^2-(1-E)t+E=0$ gives $t=3E/(1-E)$, and then $9E^2/(1-E)^2=2E$, that is, $2E^2-13E+2=0$, whose root in $(0,1)$ is $\alpha$ (there $t=3\alpha/(1-\alpha)=0.56155\ldots$ exceeds $(1-\alpha)/2$, so it is indeed the larger root). Eliminating $t$ from $t^2-(1-E)t+E=0$ and $t^3=E$ gives $E^3-5E^2+7E-1=0$. This cubic has a single real root, $0.16071\ldots$ (its local maximum and minimum, at $E=1$ and $E=7/3$, are $2$ and $22/27$), and it exceeds $\alpha$. Hence~\eqref{eq:Acond} holds for every $E<\alpha$.
\end{proof}
\end{proofof}

\begin{remark}[the leak form of $\alpha$]\label{rem:leak}
Since $E/(1-\lambda_-)=\lambda_--E$, the first condition in~\eqref{eq:Acond} reads $E+2E/(1-\lambda_-)<\lambda_+$. When both inputs of the NAND receiver are high, its low bound must stay below the larger fixed point $\lambda_+$ of the inverter map $h$; at the best input code $\lambda_-$ this low bound is $E+2E/(1-\lambda_-)$. At $E=\alpha$ the two sides are equal. The PURIFY is not binding: its condition $t^3>E$ is equivalent to $b_-<Hb_+-1$ with $b_\pm=(\lambda_\pm-E)/E$, because $Hb_+-1-b_-=((2-E)t-1)/E$, both $t^3-E$ and $(2-E)t-1$ are strictly decreasing on $(0,3-2\sqrt2]$ (as $t$ is), positive near $0$ and negative at $3-2\sqrt2$, and eliminating $t$ from either $t^3=E$ or $(2-E)t=1$ gives the same cubic $E^3-5E^2+7E-1$, whose only real root is $0.16071\ldots$; so both conditions hold exactly below that root, and at $E=\alpha$ the slack is $Hb_+-1-b_-=(15-\sqrt{153})/12>0$~\cite[record~L16; Section~7.1]{Supp}. Applied directly to (W), the same construction with notionals divided by $1-E$ reaches only $0.14053956\ldots$, the root in $(0,1/2)$ of $E^5-5E^4+12E^3-15E^2+9E-1$~\cite[record~L3; Theorem~4.3]{Supp}; the calibration below removes this loss.
\end{remark}

The historical direct-(W) comparison in \cref{rem:leak} is proved in \cref{app:hardnessaux}.

\subsection{Calibration copies}

\paragraph{Normalization.} Let $\mathcal N$ be a non-degenerate network with ex-ante solvent set $T$ and $U=N\setminus T$. Form $\mathcal N'$ as follows.
\begin{enumerate}[label=(\arabic*)]
\item Delete every CDS whose reference lies in $T$.
\item For every ordinary debt owed by a bank of $T$ to a bank $i\in U$, add its notional to $e_i$ and delete the debt.
\item Redirect every debt and every CDS written by a bank of $U$ and held by a bank of $T$ to a new sink $z$.
\item Replace every CDS written by a bank of $T$ and held by a bank of $U$ by a CDS with the same holder, reference and notional, written by a new source $s$.
\item Delete the banks of $T$ with their remaining contracts. Give $z$ external assets $1$ and no liabilities, and give $s$ external assets $2L_s^{\max}+1$.
\end{enumerate}

\begin{lemma}\label{lem:normal}
For every $r_U\in[0,1]^U$, every $i\in U$ has the same liabilities and the same assets in $\mathcal N'$ at $(r_U,r_s=1)$ as in $\mathcal N$ at $(r_U,\mathbf 1_T)$, hence the same map value. The network $\mathcal N'$ is non-degenerate and has pairwise distinct CDS roles. If $\mathcal N$ is writer-solvent, no bank of $\mathcal N'$ other than $s$ writes a CDS.
\end{lemma}
\begin{proof}
At $(r_U,\mathbf 1_T)$ the liabilities of $i$ are its debts plus $c(1-r_k)$ for each CDS it writes; step (3) only changes creditors and holders, and step (1) deletes terms $c(1-1)=0$. Its assets are $e_i$, the debts $r_jc_{ji}$ and CDS income $r_jc(1-r_k)$ from writers $j\in U$ (unchanged, except for deleted terms that are $0$), the debts $1\cdot c_{ti}$ from $t\in T$ (moved into $e_i$ by step (2)), and the CDS income $1\cdot c(1-r_k)$ from writers $t\in T$ (paid by $s$ at recovery $1$ after step (4)). So both sums agree term by term, including when $\ell_i=0$. A CDS of step (4) has writer $s$, holder $i$ and reference $k\in U\setminus\{i\}$; a redirected CDS has holder $z$; nobody references $s$ or $z$. Every $i\in U$ keeps its debts (they are redirected, not deleted) and its external assets only grow, and $s,z$ have positive external assets. If every writer of $\mathcal N$ lies in $T$, then $U$ writes no CDS.
\end{proof}

\begin{proofof}{thm:calib}
\begin{proof}
Put $\delta=\eta-\eps$, $H_\eps=1-\eps$,
\[
K=\Bigl\lceil\frac\eps{2\delta H_\eps}\Bigr\rceil,\qquad t_j=H_\eps+\frac{j\eps}K\quad(j=0,\dots,K).
\]
Normalize the input to $\mathcal N'$. The target network consists of $K+1$ disjoint copies of $\mathcal N'\setminus\{s,z\}$ that share one source $s$ and one sink $z$; in copy $j$ every notional of a CDS written by $s$ is divided by $t_j$. If $s$ writes no CDS in the copies (for instance if $\mathcal N$ has no CDS, or if $U=\emptyset$ and the copies are empty), add one \emph{padding} CDS of notional $1$, written by $s$ and held by $z$, whose reference is a fresh bank $f$ with $e_f=1$ and no other contract; it touches no retained bank, $f$ and $z$ lie in $T$, and $s$ becomes a writer. Only then, with all copies and the padding in place, compute $L_s^{\max}$ in the target and set $e_s=2L_s^{\max}+1$; so $e_s>L_s^{\max}$ and $s\in T$. The target is non-degenerate, its CDS roles are distinct (no CDS references $s$ or $z$, the padding CDS has three distinct parties, and no CDS connects two copies), and it is one-writer if $\mathcal N$ is writer-solvent. If $U=\emptyset$, the decoder below returns $\mathbf 1_T$, an exact clearing vector of $\mathcal N$. The program \texttt{code/calibration\_padding.py} builds $\mathcal N'$ and the target for six inputs, among them one with no CDS, one with $U=\emptyset$ and one that is not writer-solvent, and checks these properties and \cref{lem:normal} on all six, and the decoding below on the five writer-solvent ones, in exact arithmetic.

Let $r$ be a (W)-$\eps$ vector of the target. Since $s\in T$, $\Phi_s\equiv1$ and $t:=r_s\in[H_\eps,1]=[t_0,t_K]$. For adjacent centres $a=t_j<b=t_{j+1}$ and $t\in[a,b]$, the functions $t/a-1$ and $1-t/b$ meet at $t=2ab/(a+b)$ with common value $(b-a)/(a+b)\le(\eps/K)/(2H_\eps)\le\delta$. So the index $j$ that minimizes $|t/t_j-1|$, found by rational comparisons, satisfies $|t/t_j-1|\le\delta$. Let $r^{(j)}$ be the restriction of $r$ to copy $j$.

A bank $i$ of copy $j$ has the liabilities of $\mathcal N'$, which involve no notional written by $s$, and assets $A+uB$ with $u=t/t_j$, where $A+B$ is its asset value in $\mathcal N'$ at $(r^{(j)},r_s=1)$ and $B\ge0$ is its income from $s$ there. By \cref{lem:rescale} (or trivially if its liability is $0$), its map value in the target differs from its map value in $\mathcal N'$ by at most $|u-1|\le\delta$. Hence $|r^{(j)}_i-\Phi'_i(r^{(j)},1)|\le\eps+\delta=\eta$, and by \cref{lem:normal} the vector $(r^{(j)},\mathbf 1_T)$ is an (I)-$\eta$ vector of $\mathcal N$; banks of $T$ have $\Phi_i\equiv1$. The number of copies depends only on $(\eps,\eta)$, and the new notionals are old ones divided by rationals of constant size, so the reduction is polynomial for fixed $\eps<\eta$.
\end{proof}
\end{proofof}

\begin{remark}[an explicit instance]\label{rem:cert157}
\CO\ At $\eps=157/1000$ and $\eta=197/1250$ we have $\delta=3/5000$, $H_\eps=843/1000$ and $K=156$, so there are $157$ copies with centres $t_j=843/1000+157j/156000$. The worst relative calibration error is exactly $157/263173$, leaving the slack $\delta-157/263173=4519/1315865000>0$. The base circuit $\mathrm{NAND}(x_0,x_1;x_2)$, $\mathrm{PURIFY}(x_2;x_0,x_1)$ at error $\eta$ uses the design of~\cite[record~L16; Section~7.1]{Supp} rather than the construction of \cref{thm:wsI}: receivers with debt $1$, wire bound $591/1250$, input buffers of $28$ inverters, a NAND output tail of $12$ inverters, and a PURIFY whose two seeds hold CDSs on the buffered input with coefficients $S$ (the least admissible inverse coefficient for the buffered input, rounded up) and $B=7/2$, with threshold $\tau=\eta+S/B$; one arm is an inverse of the first seed with coefficient at least $1/(1-\tau)$ followed by a tail of $4$ inverters, and the other is a tail of $9$ inverters applied to the second seed. All coefficients are rounded up to the mesh $10^{-14}$. It has $115$ banks, $114$ CDSs and $113$ debts, and the calibrated target has $113\cdot157+2=17743$ banks and $114\cdot157=17898$ CDSs. The instance was built in~\cite[record~L16; Section~8]{Supp}; its checks, by separate contract checkers and by a rebuild from the printed parameters, are listed in the check index of the support bundle~\cite{Supp}. The numbers $K=156$, $157/263173$ and $4519/1315865000$ are recomputed by \texttt{code/constants.py}.
\end{remark}

\subsection{The comparison lemma and the transfer bound}

\begin{lemma}[comparison]\label{lem:comparison}
Let $A$ be separable affine on $[0,1]^K$. Let $x^1,\dots,x^m\in[0,1]^K$ with weights $w_j\ge0$, $\sum_jw_j=1$, and $z^1,\dots,z^p\in[0,1]^K$ with weights $c_i\ge0$, $C=\sum_ic_i\le1$. Put $\bar x=\sum_jw_jx^j$ and $\bar z=\sum_ic_iz^i$. If $\bar z_k\le\bar x_k\le1-C+\bar z_k$ for every $k$, then $\sum_jw_jA(x^j)\ge\sum_ic_iA(z^i)$.
\end{lemma}
\begin{proof}
$A$ is affine, so $\sum_jw_jA(x^j)=a_0+\sum_k(a_k\bar x_k+a'_k(1-\bar x_k))$ and $\sum_ic_iA(z^i)=Ca_0+\sum_k(a_k\bar z_k+a'_k(C-\bar z_k))$. Compare term by term: $a_0\ge Ca_0$, $a_k\bar x_k\ge a_k\bar z_k$, and $a'_k(1-\bar x_k)\ge a'_k(C-\bar z_k)$.
\end{proof}

\begin{lemma}\label{lem:psi}
Let $0<E<1/2$.
\begin{enumerate}[label=(\roman*)]
\item If $E\le3-2\sqrt2$, then $\lambda_+<H$ and $\kappa(\lambda_\pm)=E/(1-\lambda_\pm)=\lambda_\pm-E$, so $\psi(\lambda_\pm)=\lambda_\pm$.
\item If $E\le3-2\sqrt2$, then $\psi(x)\ge x$ for $x\in[0,\lambda_-]\cup[\lambda_+,H)$. If $E>3-2\sqrt2$, then $\psi(x)>x$ for all $x\in[0,H)$.
\item For every $x\in[0,H)$ with $\psi(x)\ge x$, also $E+\min(1,2\kappa(x))>x$.
\item If $\alpha\le E\le3-2\sqrt2$, then $E+2\kappa(\lambda_-)\ge\lambda_+$.
\item $E+\Lambda(x)>x$ for all $x\in[0,H)$, and $E+\Lambda(x)\ge\lambda_+$ for all $x$ if $\alpha\le E\le3-2\sqrt2$.
\end{enumerate}
\end{lemma}
\begin{proof}
Let $D=\sqrt{H^2-4E}$ when $E\le3-2\sqrt2$, so that $\lambda_\pm=E+(H\pm D)/2$.

(i) $\lambda_+<H$ if and only if $D<1-3E$, which holds since $1-3E>0$ and $(1-3E)^2-D^2=8E^2$. For a root $\lambda<1$, $\lambda(1-\lambda)=\lambda-((1+E)\lambda-2E)=E(2-\lambda)\ge EH$, so $E/(1-\lambda)\le\lambda/H$, and $\kappa(\lambda)=E/(1-\lambda)=\lambda-E<1$.

(ii) If $\kappa(x)=x/H$, then $\psi(x)=E+\min(1,x/H)>x$. If $\kappa(x)=E/(1-x)\ge1$, then $\psi(x)=E+1>x$. Otherwise $(\psi(x)-x)(1-x)=x^2-(1+E)x+2E$, which is nonnegative outside $(\lambda_-,\lambda_+)$ and positive everywhere if there are no real roots.

(iii) If $\kappa(x)\ge1$ the left side is $E+1>x$; if $\kappa(x)=0$ then $x=0<E$; otherwise $\min(1,2\kappa(x))>\kappa(x)$ and $E+\min(1,2\kappa(x))>\psi(x)\ge x$.

(iv) By (i), $E+2\kappa(\lambda_-)=2\lambda_--E$, and $2\lambda_--E-\lambda_+=(H-3D)/2$. Now $H\ge3D$ is equivalent to $9E\ge2H^2$, that is, to $2E^2-13E+2\le0$, which holds for $E\ge\alpha$.

(v) For the first claim: $E+(1+x)/(2H)>x$ because $1+2HE-(2H-1)x\ge2-2H^2>0$; $E+(1+E)/(2(1-x))>x$ because $2(1-x)(x-E)\le(1-E)^2/2<1+E$; and $E+1>x$. For the second, $\Lambda(x)\ge\min(1,1/(2H),(1+E)/2)$, and each of the three is at least $(H+D)/2=\lambda_+-E$: for $1$ this is clear; for $1/(2H)$ it means $H(H+D)\le1$, which holds at $E=\alpha$ (there $D=H/3$ and $2H^2=9\alpha$, so $H(H+D)=6\alpha<1$) and $H(H+D)$ decreases in $E$; for $(1+E)/2$ it means $2E\ge D$, that is, $3E^2+6E-1\ge0$, which holds for $E\ge2/\sqrt3-1=0.1547\ldots$.
\end{proof}

\subsection{The family \texorpdfstring{$\mathfrak F$}{F} and three-status certificates}

\paragraph{The family $\mathfrak F$ and three-status certificates.} The gadgets of \cref{thm:wsI} have a special form. A NAND gadget belongs to the family $\mathfrak F$ if its receivers are arranged in pairwise disjoint stages: input chains $U_1,\dots,U_a$ and $V_1,\dots,V_b$, where each receiver of $U_i$ reads only banks of $U_{i-1}$ ($U_0$ being the rails of $u$), and similarly for $V$; one seed receiver, reading only banks of $U_a$ and $V_b$; and an output chain $O_1,\dots,O_c$ ($c\ge0$), where each receiver of $O_i$ reads only banks of $O_{i-1}$ ($O_0$ being the seed), the output wire consisting of banks of $O_c$. In a \emph{three-status certificate} each bank gets, in each condition, a status $\mathrm H$ (box $[H,1]$), $\mathrm L(\lambda)$ (box $[0,\lambda]$, $\lambda<H$) or $\mathrm A$ (box $[0,1]$): input rails get their condition boxes, and a receiver $b$ with input box $\Pi_b(\sigma)$ is $\mathrm H$ if $A_b\ge1$ on $\Pi_b(\sigma)$, and otherwise $\mathrm L(\lambda_b)$ with $\lambda_b=E+\min(1,\max_{\Pi_b(\sigma)}A_b)$ if $\lambda_b<H$, and $\mathrm A$ if not. Statuses are computed stage by stage. Such a gadget is \emph{correct} if the output wire carries $1$ in $\sigma_1,\sigma_2$ and $0$ in $\sigma_3$, with low bound at most $\gamma_{\rm out}$. The NAND gadget of \cref{thm:wsI} (two input chains of single inverters and a two-CDS seed, with $c=0$) is a correct member of $\mathfrak F$ with the uniform interface $\gamma_u=\gamma_v=\gamma_{\rm out}=\gamma_d$, the decoding threshold of that proof. Indeed, with $g_0=\gamma_d$ and $g_j=h(g_{j-1})$, the $j$-th inverter of the $u$-chain is $\mathrm H$ for odd $j$ and $\mathrm L(g_j)$ for even $j$ in $\sigma_1$, the reverse in $\sigma_3$, and $\mathrm A$ in $\sigma_2$, since its assets range over $[0,1/(1-g_{j-1})]$ there; so the end of the even chain is $\mathrm L(g_m)$ with $g_m\le\gamma_a$ in $\sigma_1$ and $\mathrm H$ in $\sigma_3$. The seed holds CDSs of notional $1/(1-\gamma_a)$ on the two chain ends, so its assets are at least $1$ on the box of $\sigma_1$ (and of $\sigma_2$), making it $\mathrm H$, and at most $2E/(1-\gamma_a)$ on the box of $\sigma_3$, making it $\mathrm L(\gamma_N)$ with $\gamma_N\le\gamma_d<H$.

\begin{lemma}\label{lem:threestatus}
Every NAND gadget of $\mathfrak F$ that is correct under three-status certificates is certified correct by an interval certificate.
\end{lemma}
\begin{proof}
Update the receivers once each, stage by stage. We show by induction that each new interval box lies in the three-status box of the same bank and condition. Suppose this holds for the inputs of $b$ (for input rails the boxes coincide). The extremes $\alpha_b^j,\beta_b^j$ are monotone in the input boxes, so $\alpha_b^j$ is at least, and $\beta_b^j$ at most, the corresponding extremes over $\Pi_b(\sigma_j)$. If $b$ is $\mathrm H$, then $\alpha_b^j\ge1$ and $P_b^j\subseteq[1-E,1]=[H,1]$. If $b$ is $\mathrm L(\lambda_b)$, then $\min(1,\beta_b^j+E)\le E+\min(1,\max_{\Pi_b(\sigma_j)}A_b)=\lambda_b$, using $\min(1,y+E)\le E+\min(1,y)$. If $b$ is $\mathrm A$ there is nothing to show.
\end{proof}

\subsection{Comparison regions}\label{app:compregions}

Fix parameters $c_1,s,c_\ast\in[0,1]$. For a bank with boxes $[b_j,t_j]=[\Bot^j,\Top^j]$ in the three conditions consider
\begin{align*}
&\text{(P1)}\quad c_1b_j\le t_3\ \text{ and }\ c_1(1-t_j)\le1-b_3\qquad(j=1,2);\\
&\text{(P2)}\quad \tfrac s2(b_1+b_2)\le t_3\ \text{ and }\ \tfrac s2\bigl((1-t_1)+(1-t_2)\bigr)\le1-b_3;\\
&\text{(P3)}\quad c_1b_3\le t_j\ \text{ and }\ c_1(1-t_3)\le1-b_j\qquad(j=1,2);\\
&\text{(P4)}\quad c_\ast b_3\le\tfrac12(t_1+t_2)\ \text{ and }\ c_\ast(1-t_3)\le\tfrac12\bigl((1-b_1)+(1-b_2)\bigr).
\end{align*}
They say that certain weighted comparisons between the conditions are possible at that coordinate.

\begin{lemma}\label{lem:K}
Let $b\in\mathcal R$, suppose every bank that $A_b$ depends on satisfies (P1)--(P4), and write $\alpha_j=\alpha_b^j$, $\beta_j=\beta_b^j$. Then
\begin{align*}
&\text{(K1)}\quad \beta_3\ge c_1\alpha_j\ \text{ and }\ \beta_j\ge c_1\alpha_3\qquad(j=1,2);\\
&\text{(K2)}\quad \beta_3\ge\tfrac s2(\alpha_1+\alpha_2);\\
&\text{(K3)}\quad \tfrac12(\beta_1+\beta_2)\ge c_\ast\alpha_3.
\end{align*}
\end{lemma}
\begin{proof}
We apply \cref{lem:comparison} to points chosen in the boxes. (K2): take one point $x\in\Pi_b^3$ of weight $1$ and points $z^1\in\Pi_b^1$, $z^2\in\Pi_b^2$ of weight $s/2$ each. At a coordinate $k$ we need $\bar z_k=\tfrac s2(z^1_k+z^2_k)$ and $x_k$ with $0\le x_k-\bar z_k\le1-s$. As the points range over the boxes, $\bar z_k$ ranges over $[\tfrac s2(b_1+b_2),\tfrac s2(t_1+t_2)]$ and $x_k$ over $[b_3,t_3]$; suitable values exist if and only if $\tfrac s2(b_1+b_2)\le t_3$ and $b_3-\tfrac s2(t_1+t_2)\le1-s$, which is (P2). Then $\beta_3\ge A_b(x)\ge\tfrac s2(A_b(z^1)+A_b(z^2))\ge\tfrac s2(\alpha_1+\alpha_2)$. (K3): take $x^1\in\Pi_b^1$, $x^2\in\Pi_b^2$ of weight $1/2$ each and $z\in\Pi_b^3$ of weight $c_\ast$; the condition is (P4). (K1): one point on each side with weight $c_1$; the condition is (P1) for $\beta_3\ge c_1\alpha_j$ and (P3) for $\beta_j\ge c_1\alpha_3$. Constants and banks of $T$ enter through $a_0$ and need no condition.
\end{proof}

\subsection{The invariants survive every update}\label{app:invariants}

Consider the conditions
\begin{align*}
&\text{(C1)}\quad c_1^2-(1-E)c_1+E\ge0;\\
&\text{(C2)}\quad s(1-E-c_\ast)\le E\ \text{ and }\ \tfrac s2(1-E-c_1)\le E;\\
&\text{(C3)}\quad c_\ast(1-E-s)\le E,\quad c_\ast\le\frac{1+E+c_1}{2H},\quad c_\ast\bigl(1-E-\max(c_1,\tfrac s2)\bigr)\le\frac{1+E}2.
\end{align*}
For $E\le3-2\sqrt2$, (C1) says that $c_1\le\lambda_--E$ or $c_1\ge\lambda_+-E$, since $\lambda_\pm-E$ are the roots of $x^2-(1-E)x+E$; for $E>3-2\sqrt2$ it always holds.

\begin{lemma}[closure]\label{lem:closure}
Assume (C1)--(C3). If every bank that $A_b$ depends on satisfies (P1)--(P4), then the boxes $P_b^1,P_b^2,P_b^3$ satisfy (P1)--(P4).
\end{lemma}
\begin{proof}
Write $a_j=\min(1,\alpha_j)$. Then $\Bot(P_b^j)=\max(0,a_j-E)$ and $\Top(P_b^j)=\min(1,\beta_j+E)$, so
\[
1-\Top(P_b^j)=\max(0,1-E-\beta_j),\qquad 1-\Bot(P_b^j)=\min(1,1+E-a_j).
\]
We use (K1)--(K3) of \cref{lem:K}, and twice the fact that an affine function of $a\in[0,1]$ that is nonnegative at $a=0$ and at $a=1$ is nonnegative on $[0,1]$.

\emph{(P1), first part.} If $\Bot(P_b^j)=0$ there is nothing to show. Otherwise $c_1\Bot(P_b^j)\le c_1a_j\le\min(1,c_1\alpha_j)\le\min(1,\beta_3)\le\Top(P_b^3)$ by (K1).

\emph{(P1), second part.} By (K1), $1-\Top(P_b^j)\le\max(0,1-E-c_1\alpha_3)\le\max(0,1-E-c_1a_3)$. If the maximum is $0$ we are done, since $1-\Bot(P_b^3)=\min(1,1+E-a_3)$ is positive. Otherwise we need $c_1(1-E-c_1a)\le\min(1,1+E-a)$ for $a=a_3\in[0,1]$. The bound by $1$ holds since $c_1(1-E)\le1$. The bound by $1+E-a$ is affine in $a$; at $a=0$ it reads $c_1(1-E)\le1+E$, and at $a=1$ it reads $c_1(1-E-c_1)\le E$, which is (C1).

\emph{(P3).} The same two arguments with the roles of $j\in\{1,2\}$ and $3$ exchanged, using the other half of (K1).

\emph{(P2), first part.} $\tfrac s2(\Bot(P_b^1)+\Bot(P_b^2))\le\tfrac s2(a_1+a_2)\le\min(1,\tfrac s2(\alpha_1+\alpha_2))\le\min(1,\beta_3)\le\Top(P_b^3)$ by (K2), using $s\le1$.

\emph{(P2), second part.} We need $\tfrac s2\bigl(\max(0,1-E-\beta_1)+\max(0,1-E-\beta_2)\bigr)\le\min(1,1+E-a_3)$; the bound by $1$ is clear since the left side is at most $s(1-E)\le1$. If $\beta_1,\beta_2<1-E$, the left side is $s(1-E-\tfrac12(\beta_1+\beta_2))\le s(1-E-c_\ast a_3)$ by (K3), and $s(1-E-c_\ast a)\le1+E-a$ holds at $a=0$ and, by (C2), at $a=1$. If $\beta_1\ge1-E>\beta_2$, the left side is $\tfrac s2(1-E-\beta_2)\le\tfrac s2(1-E-c_1a_3)$ by (K1), and $\tfrac s2(1-E-c_1a)\le1+E-a$ holds at $a=0$ and, by (C2), at $a=1$. The case $\beta_2\ge1-E>\beta_1$ is symmetric, and if both are at least $1-E$ the left side is $0$.

\emph{(P4), first part.} If $\Bot(P_b^3)=0$ we are done; otherwise it equals $a_3-E$. If $\beta_1,\beta_2<1-E$, then $\tfrac12(\Top(P_b^1)+\Top(P_b^2))=E+\tfrac12(\beta_1+\beta_2)\ge E+c_\ast a_3>c_\ast(a_3-E)$ by (K3). If $\beta_1\ge1-E>\beta_2$, then $\tfrac12(\Top(P_b^1)+\Top(P_b^2))=\tfrac12(1+\beta_2+E)\ge\tfrac12(1+E+c_1a_3)$ by (K1), and $\tfrac12(1+E+c_1a)-c_\ast(a-E)$ is nonnegative at $a=0$ and, by the second inequality of (C3), at $a=1$. The case $\beta_2\ge1-E>\beta_1$ is symmetric. If both are at least $1-E$, the right side is $1\ge c_\ast\Bot(P_b^3)$.

\emph{(P4), second part.} We need $c_\ast\max(0,1-E-\beta_3)\le\tfrac12\bigl(\min(1,1+E-a_1)+\min(1,1+E-a_2)\bigr)$. Note $\min(1,1+E-a_j)=1+E-\max(a_j,E)$. If $a_1,a_2\ge E$, put $\sigma=\tfrac12(a_1+a_2)\in[E,1]$; by (K2), $\beta_3\ge s\sigma$, and $c_\ast(1-E-s\sigma)\le1+E-\sigma$ holds at $\sigma=E$ (as $c_\ast\le1$) and, by the first inequality of (C3), at $\sigma=1$. If $a_1<E\le a_2$, the right side is $1+\tfrac E2-\tfrac{a_2}2$, and by (K1) and (K2) $\beta_3\ge\max(c_1,\tfrac s2)a_2$; the inequality $c_\ast(1-E-\max(c_1,\tfrac s2)a)\le1+\tfrac E2-\tfrac a2$ holds at $a=E$, where the left side is at most $c_\ast\le1$ and the right side is $1$, and, by the third inequality of (C3), at $a=1$. The case $a_2<E\le a_1$ is symmetric, and if $a_1,a_2<E$ the right side is $1\ge c_\ast(1-E)$.
\end{proof}

\begin{lemma}[base]\label{lem:base}
(a) Every bank whose box is the same in the three conditions, in particular every receiver before its first update and every fixed bank, satisfies (P1)--(P4).
(b) An input rail with bound $\gamma$ satisfies (P1)--(P4) if $c_1\le\kappa(\gamma)$, $s/2\le\kappa(\gamma)$ and $c_\ast\le\Lambda(\gamma)$.
\end{lemma}
\begin{proof}
(a) With $b_j=b$, $t_j=t$ and $b\le t$, every inequality reads $xb\le t$ or $x(1-t)\le1-b$ with $x\le1$.
(b) A $+$rail of $u$ has boxes $[0,\gamma]$, $[0,1]$, $[H,1]$ in $\sigma_1,\sigma_2,\sigma_3$. Then (P1) reads $0\le1$ and $c_1(1-\gamma)\le E$ for $j=1$, and $0\le1$, $0\le E$ for $j=2$; (P3) reads $c_1H\le\gamma$, $0\le1$ for $j=1$ and $c_1H\le1$, $0\le1$ for $j=2$; (P2) reads $0\le1$ and $\tfrac s2(1-\gamma)\le E$; (P4) reads $c_\ast H\le\tfrac12(\gamma+1)$ and $0\le1$. So it suffices that $c_1,\tfrac s2\le\min(E/(1-\gamma),\gamma/H)=\kappa(\gamma)$ and $c_\ast\le(1+\gamma)/(2H)$. For a $-$rail of $u$ (boxes $[H,1]$, $[0,1]$, $[0,\gamma]$) the same computation gives $c_1,\tfrac s2\le\kappa(\gamma)$ and $c_\ast\le(1+E)/(2(1-\gamma))$. Rails of $v$ are the same with $\sigma_1$ and $\sigma_2$ exchanged, which leaves (P1)--(P4) invariant. Both bounds on $c_\ast$ follow from $c_\ast\le\Lambda(\gamma)$.
\end{proof}

\subsection{Proof of the barrier theorem and two consequences}\label{app:barrierproof}

\begin{proofof}{thm:barrier}
\begin{proof}
\emph{Reduction to parameters.} Suppose $c_1,s,c_\ast\in[0,1]$ satisfy (C1)--(C3) and the base conditions of \cref{lem:base}(b) for $\gamma=\gamma_u$ and $\gamma=\gamma_v$; since $\kappa$ and $\Lambda$ are nondecreasing it suffices to check them at $\gamma_{\rm in}$. By \cref{lem:base} all banks satisfy (P1)--(P4) before any update, and by \cref{lem:closure} every update preserves this. So in the certificate every receiver satisfies (P1)--(P4), and \cref{lem:K} applies to every receiver.

Take an output rail $o$. If it is a $+$rail, its boxes in $\sigma_1,\sigma_2$ lie in $[H,1]$, so $\min(1,\alpha_j)-E\ge H$, that is, $\alpha_j\ge1$ for $j=1,2$; by (K2), $\beta_3\ge s$, so its top in $\sigma_3$ is at least $\min(1,E+s)$. If it is a $-$rail, its box in $\sigma_3$ lies in $[H,1]$, so $\alpha_3\ge1$; by (K3), $\max(\beta_1,\beta_2)\ge c_\ast$, so its top in $\sigma_1$ or $\sigma_2$ is at least $\min(1,E+c_\ast)$. These tops are at most $\gamma_{\rm out}<H<1$, so in all cases
\[
\gamma_{\rm out}\ge E+\min(s,c_\ast).
\]
It remains to choose the parameters.

\emph{Case (a): $E\le3-2\sqrt2$ and $\lambda_-\le\gamma_{\rm in}$.} Take $c_1=\lambda_--E$ and $s=c_\ast=\lambda_+-E$. Since $(\lambda-E)(1-\lambda)=E$ for $\lambda=\lambda_\pm$, (C1), the first part of (C2) and the first part of (C3) hold with equality. The second part of (C2) is $(\lambda_+-E)(1-\lambda_-)\le2E$; as $E/(1-\lambda_-)=\lambda_--E$, it is equivalent to $\lambda_++E\le2\lambda_-$, which is \cref{lem:psi}(iv) (equivalently $3\sqrt{1-6E+E^2}\le1-E$, by the identity $(1-E)^2-9(1-6E+E^2)=-4(2E^2-13E+2)$). The second part of (C3) is $\lambda_+-E\le(1+\lambda_-)/(2H)$, which follows from $E+\Lambda(\lambda_-)\ge\lambda_+$, \cref{lem:psi}(v). For the third part of (C3), $\max(c_1,s/2)\ge c_1$ gives a left side at most $(\lambda_+-E)(1-\lambda_-)\le2E\le(1+E)/2$. Base conditions: $c_1=\kappa(\lambda_-)\le\kappa(\gamma_{\rm in})$ by \cref{lem:psi}(i); $s/2\le\lambda_--E$ is again $\lambda_++E\le2\lambda_-$; and $c_\ast=\lambda_+-E\le\Lambda(\lambda_-)\le\Lambda(\gamma_{\rm in})$ by \cref{lem:psi}(v). Hence $\gamma_{\rm out}\ge\lambda_+$. If $\gamma_{\rm in}<\lambda_+$ this exceeds $\gamma_{\rm in}$; the case $\gamma_{\rm in}\ge\lambda_+$ is covered by (c).

\emph{Case (b): $E\le3-2\sqrt2$ and $\gamma_{\rm in}<\lambda_-$.} Take $c_1=\kappa(\gamma_{\rm in})$ and $s=c_\ast=x$ with
\[
x=\min\Bigl(2c_1,\ \lambda_--E,\ \Lambda(\gamma_{\rm in}),\ \frac{1+E+c_1}{2H}\Bigr).
\]
Since $c_1\le\kappa(\lambda_-)=\lambda_--E$, (C1) holds. As $x\le\lambda_--E$, the smaller root of $y^2-(1-E)y+E$, we have $x(1-E-x)\le E$, which gives the first parts of (C2) and (C3). The second part of (C2): $\tfrac x2(1-E-c_1)\le c_1(1-E-c_1)\le E$ by (C1). The second part of (C3) holds by the choice of $x$. For the third, $x(1-E)\le(\lambda_--E)(1-E)\le\tfrac12(1-E)^2\le\tfrac12(1+E)$, using $\lambda_-\le(1+E)/2$. The base conditions hold by the choice of $c_1$ and $x$. Finally $E+x>\gamma_{\rm in}$: $E+\min(1,2\kappa(\gamma_{\rm in}))>\gamma_{\rm in}$ by \cref{lem:psi}(ii) and (iii); $E+(\lambda_--E)=\lambda_->\gamma_{\rm in}$; $E+\Lambda(\gamma_{\rm in})>\gamma_{\rm in}$ by \cref{lem:psi}(v); and $E+(1+E)/(2H)>H>\gamma_{\rm in}$, because $2EH+1+E-2H^2=-1+7E-4E^2>0$ for $E>(7-\sqrt{33})/8=0.1569\ldots$, which holds as $E\ge\alpha$.

\emph{Case (c): $E\le3-2\sqrt2$ and $\gamma_{\rm in}\ge\lambda_+$, or $E>3-2\sqrt2$.} Take $c_1=\min(1,\kappa(\gamma_{\rm in}))$ and $s=c_\ast=x$ with
\[
x=\min\Bigl(1,\ 2c_1,\ \Lambda(\gamma_{\rm in}),\ \frac{1+E+c_1}{2H}\Bigr).
\]
If $E\le3-2\sqrt2$, then $c_1\ge\kappa(\lambda_+)=\lambda_+-E$, and each of the four terms defining $x$ is at least $\lambda_+-E$ (for $\Lambda(\gamma_{\rm in})$ by \cref{lem:psi}(v), and $(1+E+c_1)/(2H)\ge1/(2H)\ge\lambda_+-E$ as in its proof); so $c_1$ and $x$ are at least the larger root $\lambda_+-E$, which gives (C1) and $x(1-E-x)\le E$. If $E>3-2\sqrt2$ these hold because $y^2-(1-E)y+E>0$ for all $y$. The second part of (C2): if $1-E-c_1\le0$ it is clear, and otherwise $\tfrac x2(1-E-c_1)\le c_1(1-E-c_1)\le E$ by (C1). The second part of (C3) holds by the choice of $x$; for the third, the left side is at most $x(1-E-\tfrac x2)\le\tfrac12(1-E)^2\le\tfrac12(1+E)$. The base conditions hold by the choice of $c_1$ and $x$. Finally $E+x>\gamma_{\rm in}$: $E+1>\gamma_{\rm in}$; $E+\min(1,2\kappa(\gamma_{\rm in}))>\gamma_{\rm in}$ by \cref{lem:psi}(ii) and (iii), since $\psi(\gamma_{\rm in})\ge\gamma_{\rm in}$ in this case; $E+\Lambda(\gamma_{\rm in})>\gamma_{\rm in}$ by \cref{lem:psi}(v); and $E+(1+E+c_1)/(2H)>H>\gamma_{\rm in}$ as in case (b).
\end{proof}
\end{proofof}

\begin{corollary}\label{cor:uniform}
Let $\alpha\le\eta<1$. No NAND gadget for writer-solvent (I)-$\eta$ that is certified by an interval certificate has $\gamma_{\rm out}\le\gamma_{\rm in}$; in particular none has a uniform interface $\gamma_u=\gamma_v=\gamma_{\rm out}$.
\end{corollary}
\begin{proof}
For $\alpha\le\eta<1/2$ this is \cref{thm:barrier}. For $1/2\le\eta<1$ no NAND gadget is certified at all: every receiver box $P_b^j$ has top at least $\min(1,\eta)\ge1/2\ge H>\gamma_{\rm out}$, while every output rail needs a box inside $[0,\gamma_{\rm out}]$ in some condition.
\end{proof}

\begin{corollary}[$\alpha$ is the exact endpoint of $\mathfrak F$]\label{cor:Fendpoint}
For every rational $0<E<\alpha$ there is a NAND gadget in $\mathfrak F$, correct under three-status certificates, with a uniform interface $\gamma_u=\gamma_v=\gamma_{\rm out}$. For $\alpha\le E<1$ there is none, and none certified by any interval certificate. Consequently, Pure-Circuit reductions for writer-solvent (I) built from uniform interval-certified NAND gadgets reach exactly the errors below $\alpha$, and calibration (\cref{thm:calib}) turns them into one-writer (W) hardness for exactly the same errors.
\end{corollary}
\begin{proof}
For $E<\alpha$ take the NAND gadget of \cref{thm:wsI} (with rational parameters chosen as in its proof), which lies in $\mathfrak F$ by the paragraph before \cref{lem:threestatus}; alternatively, the gadget of~\cite[record~L16; Section~7.1]{Supp}, which was shown to be a correct member of $\mathfrak F$ with a uniform interface for every $E<\alpha$ in~\cite[record~L17; Proposition~7.8]{Supp}. For $E\ge\alpha$ apply \cref{lem:threestatus} and \cref{cor:uniform}. The last sentence follows because \cref{thm:calib} needs $\eps<\eta$.
\end{proof}

\begin{remark}[sanity checks]\label{rem:sanity5}\CO\
The program \texttt{dagbarrier.py} of~\cite[record~L19]{Supp} checks (C1)--(C3) and the base conditions on $4001$ high-precision points of $[\alpha,3-2\sqrt2]$ for case (a) (the smallest slacks are the equalities of the proof) and on $29618$ points for cases (b) and (c). In floating point it builds $1600$ random gadgets at $E\in\{0.158,0.16,0.165,0.17\}$ (acyclic and cyclic, $3$ to $9$ receivers, single- and dual-rail inputs, rereads allowed), runs interval propagation, and finds (P1)--(P4) at every bank; over these gadgets the smallest optimum of the linear program for the best NAND-type final reader is $0.5656$ at $E=0.158$, against the bound $\lambda_+-E=0.5597$. The checks of these computations by separate implementations are listed in the check index of the support bundle~\cite{Supp}.
\end{remark}

\section{Proofs for the box algorithm}\label[appendix]{app:box}

This appendix contains the proofs for \cref{sec:box}: soundness, termination and running time of $\BOX(E)$, the local lower bounds with the related remarks, and the finite-precision algorithm.

\subsection{Soundness}

\begin{proofof}{lem:sound}
\begin{proof}
Throughout, unsettled boxes and the points of settled banks outside $F$ lie in $[E,1-E]$: this holds initially, a new box $[\lambda_i+E,\mu_i-E]$ lies in $[E,1-E]$ because $0\le\lambda_i\le\mu_i\le1$, and a settled point lies in the old box.

(a) If $\ell_i^+>0$, some liability term has a positive upper value. A debt gives $\ell_i^-\ge c_{ij}>0$. A CDS term $c^k_{ij}(1-r_k)$ with positive upper value has $k\notin F$ (for $k\in F$ the factor is $0$), so its lower value is $c^k_{ij}(1-hi_k)\ge c^k_{ij}E>0$.

(b) Every term of $A_i$ and of $\ell_i$ is a nonnegative constant times a product of factors $r_j$ or $1-r_k$, nondecreasing in each factor. Replacing each factor by its least (largest) value over the box gives $A_i^-\le A_i(r)\le A_i^+$ and $\ell_i^-\le\ell_i(r)\le\ell_i^+$, whether or not different terms share variables. If $\ell_i^+=0$, then $\ell_i\equiv0$ on the box and $\Phi_i\equiv1$. Otherwise $\ell_i(r)\ge\ell_i^->0$ by (a), so $A_i^-/\ell_i^+\le A_i(r)/\ell_i(r)\le A_i^+/\ell_i^-$, and applying $y\mapsto\min(1,y)$ gives (b). No monotonicity of $\Phi$ itself is used.

(c) We show by induction that every unsettled box at the start of a round has the form $[\lambda'+E,\mu'-E]$ with $\lambda'\le\lambda_i$ and $\mu'\ge\mu_i$ for the values $\lambda_i,\mu_i$ of the current round. In round $1$ take $\lambda'=0$, $\mu'=1$. In a later round, $\lambda',\mu'$ are the values of the previous round, computed from boxes that contain the current ones; $A_i^\pm$ and $\ell_i^\pm$ are monotone in the endpoints, so $\lambda'\le\lambda_i$ and $\mu'\ge\mu_i$. Hence $[\lambda_i+E,\mu_i-E]\subseteq[\lambda'+E,\mu'-E]$. For a bank settled in the round, $m=(\lambda_i+\mu_i)/2$ lies in $[\mu_i-E,\lambda_i+E]$ because $\mu_i-\lambda_i\le2E$. If $m$ lies below the box, then $p_i=\lambda'+E\le\lambda_i+E$ and $p_i>m\ge\mu_i-E$; if it lies above, then $p_i=\mu'-E\ge\mu_i-E$ and $p_i<m\le\lambda_i+E$; otherwise $p_i=m$.

When the algorithm stops, the output lies in all boxes of all rounds, since they are nested. For $i\in F$, $r_i=1=\Phi_i(r)$. A bank $i\notin F$ settled in some round has $\Phi_i(r)\in[\lambda_i,\mu_i]$ by (b) applied to the boxes of that round, and $p_i\in[\mu_i-E,\lambda_i+E]$ by (c); so $|r_i-\Phi_i(r)|\le E$. As $T\subseteq F$, the output is an (I)-$E$ vector.
\end{proof}
\end{proofof}

\subsection{Termination and running time}

At the start of a round let $U$ be the set of unsettled banks and, if $U\ne\emptyset$,
\[
\rho_r=\max_{k\in U}\frac{hi_k}{lo_k},\qquad \rho_n=\max_{k\in U}\frac{1-lo_k}{1-hi_k}\qquad(\text{both at least }1).
\]

\begin{lemma}[ratio bound]\label{lem:ratio}
In a round with $U\ne\emptyset$, every unsettled bank $i$ satisfies $\mu_i\le\min(1,K\lambda_i)$, where $K=\rho_r\rho_n^2$. If no bank outside $F$ writes a CDS, the same holds with $K=\max(\rho_r,\rho_n)$.
\end{lemma}
\begin{proof}
For $k\in U$ the factor $r_k$ ranges over $[lo_k,hi_k]$ with $hi_k\le\rho_rlo_k$, and $1-r_k$ over $[1-hi_k,1-lo_k]$ with $1-lo_k\le\rho_n(1-hi_k)$; factors of settled banks are constants. The asset terms are constants, $c\,r_j$ and $c\,r_j(1-r_k)$, so each upper value is at most $\rho_r\rho_n$ times the lower value, and $A_i^+\le\rho_r\rho_nA_i^-$. The liability terms are constants and $c(1-r_k)$, so $\ell_i^+\le\rho_n\ell_i^-$. If $\ell_i^+=0$ then $\lambda_i=\mu_i=1$. Otherwise $A_i^+/\ell_i^-\le\rho_r\rho_n^2A_i^-/\ell_i^+$; if $A_i^-/\ell_i^+\ge1$ then $\lambda_i=\mu_i=1$, and otherwise $\mu_i\le\min(1,K\lambda_i)$. If no bank outside $F$ writes a CDS, an unsettled bank has constant liabilities, and every CDS it holds is written by a bank of $F$, whose point is $1$; its asset terms are then constants, $c\,r_j$ and $c(1-r_k)$, so $A_i^+\le\max(\rho_r,\rho_n)A_i^-$ and $\ell_i^+=\ell_i^-$.
\end{proof}

For $K>0$ put
\[
R_1(K)=\frac{1-E-1/K}{E},\qquad R_2(K)=\frac{1-E}{1/K+E}.
\]
Both are increasing in $K$.

\begin{lemma}[new boxes]\label{lem:newbox}
Let $K\ge1$ and $0\le\lambda\le\mu\le\min(1,K\lambda)$ with $\mu-\lambda>2E$. Then $K>1/(1-2E)$, and the box $[\ell',h']=[\lambda+E,\mu-E]$ satisfies
\[
\frac{1-\ell'}{1-h'}\le R_1(K),\qquad \frac{h'}{\ell'}\le R_2(K)\le R_1(K).
\]
\end{lemma}
\begin{proof}
The function $\lambda\mapsto\min(1,K\lambda)-\lambda$ is largest at $\lambda=1/K$, so $2E<\mu-\lambda\le1-1/K$, which gives $K>1/(1-2E)$. Note that $1/(1-2E)>(1+E)/(1-E)$. Both ratios increase with $\mu$, so we may take $\mu=\min(1,K\lambda)$. If $K\lambda\ge1$, then $\lambda\ge1/K$, and $(1-\lambda-E)/E\le R_1(K)$, $(1-E)/(\lambda+E)\le R_2(K)$. If $K\lambda<1$, the function $(1-\lambda-E)/(1+E-K\lambda)$ has derivative of the sign of $K(1-E)-(1+E)>0$, so it is below its value $R_1(K)$ at $\lambda=1/K$; likewise $(K\lambda-E)/(\lambda+E)$ has derivative of the sign of $E(K+1)>0$, so it is below $R_2(K)$. Finally, with $u=1/K<1-2E$, $R_1(K)\ge R_2(K)$ is equivalent to $(1-E-u)(u+E)\ge(1-E)E$, that is, to $u(1-2E-u)\ge0$.
\end{proof}

\begin{proofof}{thm:algws}
\begin{proof}
Put $K_0=t$ and $K_{s+1}=R_1(K_s)$ as long as $K_s>1/(1-2E)$. We show by induction on $s$: (i) if some bank is unsettled at the start of round $s+1$, then $\max(\rho_r,\rho_n)\le K_s$ in that round; (ii) if moreover some bank is still unsettled at the start of round $s+2$, then $K_s>1/(1-2E)$. In round $1$ every unsettled box is $[E,1-E]$, whose two ratios equal $t=K_0$. Suppose (i) holds in round $s+1$, and let $i$ stay unsettled in that round. By \cref{lem:ratio} (second case), $\mu_i\le\min(1,K_s\lambda_i)$, and $\mu_i-\lambda_i>2E$. \Cref{lem:newbox} with $K=K_s\ge1$ gives $K_s>1/(1-2E)$, which is (ii), and shows that the new box of $i$ has both ratios at most $R_1(K_s)=K_{s+1}$; so (i) holds in round $s+2$. Consequently, if $K_s\le1/(1-2E)$ for some $s$, the algorithm stops after round $s+1$ at the latest.

Now $R_1(K)-K=-(EK^2-(1-E)K+1)/(EK)$. The discriminant of the quadratic is $(1-E)^2-4E=E^2-6E+1$, which is negative on $(3-2\sqrt2,1/2)$; so the quadratic is at least its minimum $\Delta$, and $R_1(K)<K$ for $K>0$. If $K_0,\dots,K_s$ all exceed $1/(1-2E)$, then $K_s\le K_0=t$, so $EK_s\le1-E$ and $K_{s+1}\le K_s-\Delta/(1-E)$. Therefore $K_s\le1/(1-2E)$ for some $s\le\max(0,\lceil(1-E)(t-1/(1-2E))/\Delta\rceil)$, and the algorithm stops after at most $R_{\rm ws}(E)$ rounds.

For the finer bound let $s^\ast$ be the least $s$ with $K_s\le1/(1-2E)$, and put $g_s=h_E^s(E)$. We claim $K_s=(1-g_s)/E$ and $g_s<1$ for $0\le s\le s^\ast$. For $s=0$ this holds. If it holds for some $s<s^\ast$, then $R_1((1-g_s)/E)=(1-E-E/(1-g_s))/E=(1-g_{s+1})/E$, and $K_{s+1}=(1-E-1/K_s)/E>(1-E-(1-2E))/E=1$ because $1/K_s<1-2E$, so $g_{s+1}<1$. For $s\le s^\ast$, since $1-g_s>0$, the inequality $K_s\le1/(1-2E)$ is equivalent to $E/(1-g_s)\ge1-2E$, that is, to $g_{s+1}\ge1-E$. Hence $g_{s+1}<1-E$ for $s<s^\ast$ and $g_{s^\ast+1}\ge1-E$. So $R_E=s^\ast+1$, and the algorithm stops after at most $R_E$ rounds. Also $g_s\ge E$, since $h_E(g)\ge E$ for $g<1$.

Writer-solvent and one-writer networks have all writers in $T\subseteq F$, and an (I)-$E$ vector is a (W)-$E$ vector. The running time is polynomial for fixed $E$ by \cref{lem:bits}.
\end{proof}
\end{proofof}

\begin{proofof}{thm:alggen}
\begin{proof}
Put $\kappa_E(K)=R_2(K)R_1(K)^2$, $K_0=t^3$, and $K_{s+1}=\kappa_E(K_s)$ as long as $K_s>1/(1-2E)$. Exactly as in the previous proof, now with the first case of \cref{lem:ratio} and both bounds of \cref{lem:newbox}, we get by induction on $s$: (i) if some bank is unsettled at the start of round $s+1$, then $\rho_r\rho_n^2\le K_s$ in that round; (ii) if moreover some bank is still unsettled at the start of round $s+2$, then $K_s>1/(1-2E)$. In round $1$ both ratios equal $t$, so $\rho_r\rho_n^2=t^3=K_0$; and a bank that stays unsettled gets a new box with $(1-\ell')/(1-h')\le R_1(K_s)$ and $h'/\ell'\le R_2(K_s)$, so $\rho_r\rho_n^2\le\kappa_E(K_s)=K_{s+1}$ in the next round.

With $u=1/K$ we have $\kappa_E(K)/K=(1-E)u(1-E-u)^2/(E^2(u+E))$, and
\[
\frac{\kappa_E(K)}K-1=\frac{p_E(u+E)}{E^2(u+E)},\qquad p_E(\gamma)=(1-E)(\gamma-E)(1-\gamma)^2-E^2\gamma .
\]
We claim $p_E(\gamma)<0$ for all $\gamma\le1$. Expanding, $p_E(\gamma)=a_3\gamma^3+a_2\gamma^2+a_1\gamma+a_0$ with $a_3=1-E$, $a_2=-(1-E)(2+E)$, $a_1=1+E-3E^2$, $a_0=-E(1-E)$. Its discriminant $a_2^2a_1^2-4a_3a_1^3-4a_2^3a_0-27a_3^2a_0^2+18a_3a_2a_1a_0$ equals $E^2(E-1)Q_5(E)$, an identity between polynomials of degree $8$. On $[0,1/2]$, $Q_5'(E)=25E^4-180E^3+156E^2-64E+20\ge q(E):=20-64E+156E^2-180E^3$, and $q'(E)=-64+312E-540E^2<0$ for all $E$ (its discriminant is $-40896$), so $q\ge q(1/2)=9/2>0$. So $Q_5$ increases on $[0,1/2]$ from $Q_5(0)=-4$ to $Q_5(1/2)=59/32$, and $\elad$ is its only root there. For $E\in(\elad,1/2)$, $Q_5(E)>0$, the discriminant is negative, and $p_E$ has a single real root. Since $p_E(1)=-E^2<0$ and $p_E(\gamma)\to+\infty$ as $\gamma\to\infty$, that root exceeds $1$, and $p_E<0$ on $(-\infty,1]$. Hence $\kappa_E(K)/K<1$ for every $u\in[0,1-2E]$, and by compactness $c(E)<1$.

If $K_s>1/(1-2E)$, then $u_s=1/K_s<1-2E$ and $K_{s+1}\le c(E)K_s$. So $K_s\le1/(1-2E)$ for some $s\le\max(0,\lceil\log(t^3(1-2E))/\log(1/c(E))\rceil)$, and the algorithm stops in the next round. \Cref{lem:sound} gives correctness and \cref{lem:bits} the running time.
\end{proof}
\end{proofof}

\begin{lemma}[running time]\label{lem:bits}
For a rational $x=p/q$ in lowest terms let its \emph{size} be the larger of the bit lengths of $|p|$ and $q$. Let $B\ge1$ bound the sizes of the input numbers and of $E$, and let $m\ge1$ be the number of contracts (if $m=0$ every bank lies in $F$). After $s$ rounds every box endpoint and every settled point has size at most $(18(m+1))^s(B+2)$, and each round uses $O(n+m)$ arithmetic operations. Hence, for fixed $E$, $\BOX(E)$ runs in polynomial time on the networks of \cref{thm:algws,thm:alggen}.
\end{lemma}
\begin{proof}
A product of numbers has size at most the sum of their sizes; a sum of $k$ numbers has size at most the sum of their sizes plus $\lceil\log_2k\rceil$; and for $x\in[0,1]$, $1-x$ has size at most that of $x$. Let $\beta\ge B$ bound the sizes of the current endpoints and points. Each term of $A_i^\pm$ or $\ell_i^\pm$ is a product of at most three numbers of size at most $\beta+1$, so a sum of at most $m+1$ such terms has size at most $4(m+1)(\beta+1)$. Hence $\lambda_i,\mu_i$ have size at most $8(m+1)(\beta+1)$, and $\lambda_i+E$, $\mu_i-E$ and $(\lambda_i+\mu_i)/2$ at most $17(m+1)(\beta+1)$; a clamped point is one of these or an old endpoint. So the new bound $\beta'$ satisfies $\beta'+1\le18(m+1)(\beta+1)$. The initial endpoints have size at most $B$. By \cref{thm:algws,thm:alggen} the number of rounds is bounded by a constant that depends only on $E$.
\end{proof}

The bounds are polynomial for each fixed rational $E$ above the threshold. They are not uniform as $E$ approaches the threshold: the number of rounds, and with it the exponent of the size bound, grows as the contraction factor approaches $1$. For example, the finer bounds of the proofs (the integer $R_E$ of \cref{thm:algws}, and the first $s$ with $\kappa_E^s(t^3)\le1/(1-2E)$, plus one, in \cref{thm:alggen}) are $208$ rounds for writer-solvent networks at $E=429/2500=0.1716$ and $1670$ rounds for general networks at $E=141683/500000=0.283366$ (\cite[record~L21]{Supp}; recomputed by \texttt{code/constants.py}); the closed formulas $R_{\rm ws}$ and $R_{\rm gen}$ are much larger there.

\subsection{Local lower bounds}

\begin{proofof}{thm:locallower}
\begin{proof}
An algorithm that achieves $E$ also achieves every larger error, and both hypotheses are open conditions in $E$, so we may assume $E$ rational.

(a) Put $L=R+1$, $g_0=g_1=E$ and $g_{i+1}=h_E(g_i)$ for $i\ge1$, so that $g_i=h_E^{i-1}(E)<1-E$ for $1\le i\le L$ by hypothesis. Take banks $b_0,\dots,b_L$, writers $s_1,\dots,s_L$, and a private sink for each $b_i$ (external assets $1$, no liabilities). Every $b_i$ owes $1$ to its sink; for $i\ge1$, $b_i$ has $e=0$ and holds a CDS written by $s_i$ on $b_{i-1}$ of notional $c_i=1/(1-g_{i-1})$, and $e_{s_i}=2c_i+1$, so $s_i\in T$. Two networks differ only in $b_0$: in $X_{\rm hi}$, $e_{b_0}=2$, so $b_0\in T$; in $X_{\rm lo}$, $e_{b_0}=0$, so $\Phi_{b_0}\equiv0$. Both are non-degenerate and writer-solvent. In the dependency graph, giving $b_i$, $s_i$ and the sink of $b_i$ the layer $i$, every edge joins layers that differ by at most one; so $b_0$ is at distance $L>R$ from $b_L$, and $V_R(b_L)$ is the same in both networks. Let $r$ be an (I)-$E$ vector; then $r_{s_i}=1$, $\ell_{b_i}=1$ and $A_{b_i}=c_i(1-r_{b_{i-1}})$. In $X_{\rm hi}$, $r_{b_0}=1$, so $r_{b_1}\le E=g_1$; in $X_{\rm lo}$, $r_{b_0}\le E=g_0$, so $A_{b_1}\ge1$ and $r_{b_1}\ge1-E$. Inductively, if $r_{b_i}\le g_i$ in one network and $r_{b_i}\ge1-E$ in the other, then in the first $A_{b_{i+1}}\ge c_{i+1}(1-g_i)=1$ and $r_{b_{i+1}}\ge1-E$, while in the second $A_{b_{i+1}}\le c_{i+1}E$ and $r_{b_{i+1}}\le E+E/(1-g_i)=g_{i+1}$. So $r_{b_L}\ge1-E$ in one network and $r_{b_L}\le g_L<1-E$ in the other, for every (I)-$E$ vector, while an $R$-local algorithm outputs the same value at $b_L$ in both.

(b) Let $p_E(\gamma)=(1-E)(\gamma-E)(1-\gamma)^2-E^2\gamma$ as in \cref{thm:alggen}. We first find a rational $\gamma_1\in(E,1-E)$ with $p_E(\gamma_1)>0$. For $\gamma\le E$ we have $p_E(\gamma)<0$ (for $0\le\gamma\le E$ both terms are nonpositive and not both zero, and for $\gamma<0$, $p_E(\gamma)\le(1-E)(\gamma-E)-E^2\gamma=\gamma(1-E-E^2)-E(1-E)<0$). For $\gamma\in[1-E,1]$, $(1-E)(\gamma-E)(1-\gamma)^2\le(1-E)^2E^2<E^2\gamma$, so $p_E(\gamma)<0$. Since $Q_5$ increases on $[0,1/2]$ and vanishes only at $\elad$ (proof of \cref{thm:alggen}), $E<\elad$ gives $Q_5(E)<0$, and the discriminant $E^2(E-1)Q_5(E)$ is positive, so $p_E$ has three distinct real roots, none in $(-\infty,E]\cup[1-E,1]$. As $p_E(1-E)<0<p_E(+\infty)$, an odd number of them exceeds $1$, and as $p_E(E)<0$ and $p_E(1-E)<0$, an even number lies in $(E,1-E)$. They cannot all exceed $1$, because their sum is $-a_2/a_3=2+E<3$. So two roots lie in $(E,1-E)$ and $p_E>0$ between them.

Next choose rationals $D>0$, $0<\xi<1$ and $\kappa\ge(D+E+\xi)/((1-E)(1-\gamma_1))$ with
\begin{equation}\label{eq:ladcond}
E+\frac{\kappa\gamma_1E}{D+1-\gamma_1}\le\gamma_1;
\end{equation}
this is possible because the left side tends to $E+E^2\gamma_1/((1-E)(1-\gamma_1)^2)=\gamma_1-p_E(\gamma_1)/((1-E)(1-\gamma_1)^2)<\gamma_1$ as $D,\xi\to0$ with $\kappa$ at its lower bound.

For $\beta\in\{0,1\}$ and $L\ge1$ the \emph{ladder} $\mathcal N^\beta_L$ has banks $P_k,N_k$ ($0\le k\le L$), each with a private sink (external assets $1$, no liabilities). Put $S_k=(\xi/\kappa)^k$. The banks $P_0,N_0$ owe $1$ to their sinks; in $\mathcal N^1$, $e_{P_0}=2$ and $e_{N_0}=0$, and in $\mathcal N^0$ these are exchanged. For $1\le k\le L$, $P_k$ owes $DS_k$ to its sink, writes a CDS on $P_{k-1}$ of notional $S_k$ to its sink, and holds a CDS written by $P_{k-1}$ on $N_{k-1}$ of notional $\kappa S_k=\xi S_{k-1}$; symmetrically, $N_k$ owes $DS_k$, writes a CDS on $N_{k-1}$ of notional $S_k$ to its sink, and holds a CDS written by $N_{k-1}$ on $P_{k-1}$ of notional $\kappa S_k$. So each $P_k,N_k$ with $k<L$ writes one further CDS, of notional $\xi S_k$. The ladders are non-degenerate, and $P_0\in T$ in $\mathcal N^1$ since $e_{P_0}=2>1+\xi$.

\emph{Claim.} In every (W)-$E$ vector of $\mathcal N^1$, $r_{P_k}\ge1-E$ and $r_{N_k}\le\gamma_1$ for all $k$; in $\mathcal N^0$ the same holds with $P$ and $N$ exchanged. For $k=0$, $\Phi_{P_0}=1$ and $\Phi_{N_0}=0$. For $k\ge1$, using the claim for $k-1$,
\[
A_{P_k}\ge\kappa S_k(1-E)(1-\gamma_1),\quad \ell_{P_k}\le S_k(D+E+\xi),\qquad
A_{N_k}\le\kappa S_k\gamma_1E,\quad \ell_{N_k}\ge S_k(D+1-\gamma_1),
\]
where $\xi S_k$ bounds the CDS written to the next layer. So $\Phi_{P_k}\ge1$ by the choice of $\kappa$, and $\Phi_{N_k}\le\gamma_1-E$ by~\eqref{eq:ladcond}.

Every edge of the dependency graph joins banks of the same or adjacent layers (a sink belongs to the layer of its bank), so $P_L$ is at distance at least $L$ from layer $0$. Take $L=R+1$. The two ladders differ only in $e_{P_0},e_{N_0}$, which are not part of $V_R(P_L)$, so an $R$-local algorithm outputs the same value at $P_L$ in both, while the claim gives $r_{P_L}\ge1-E$ in $\mathcal N^1$ and $r_{P_L}\le\gamma_1<1-E$ in $\mathcal N^0$. Every (I)-$E$ vector is a (W)-$E$ vector, which gives the statement for (I).
\end{proof}
\end{proofof}

\begin{proofof}{cor:local}
\begin{proof}
$R$ rounds of $\BOX(E)$ form an $R$-local algorithm: the round-$1$ values of $i$ depend on $e_i$, on the contracts of $i$, and on whether its debtors, the writers of the CDSs it holds and the references of the CDSs it holds or writes lie in $F$, all of which is part of $V_1(i)$; round $s$ uses the round-$(s-1)$ boxes of the same neighbours, which depend on their views of radius $s-1$, contained in $V_s(i)$. If $\BOX(E)$ settles every bank within $R$ rounds on every network of the class, its output at $i$ depends on $V_R(i)$ only.

(a) If $h_E^s(E)\ge1-E$ for some $s\le R$, then $E>3-2\sqrt2$. Indeed, for $E\le3-2\sqrt2$ we have $E<\lambda_-$ (as $2(\lambda_--E)=1-E-\sqrt{(1-E)^2-4E}>0$), and $h_E$ is increasing with $h_E(\lambda_-)=\lambda_-$, so it maps $[0,\lambda_-]$ into itself and all iterates from $E$ stay below $\lambda_-<1-E$ (\cref{lem:hmap}); and \cref{thm:algws} shows that every bank is settled within $R_E\le s\le R$ rounds. Otherwise \cref{thm:locallower}(a) applies. For the second sentence, if $E\le3-2\sqrt2$ the iterates stay below $1-E$ for all $s$, and if $E>3-2\sqrt2$ \cref{thm:algws} gives the finite radius $R_E$.
(b) For $E<\elad$ use \cref{thm:locallower}(b); for $E>\elad$, \cref{thm:alggen}. The quintic $Q_5$ has no rational root (none of $\pm1,\pm2,\pm4,\pm1/5,\pm2/5,\pm4/5$ is a root), so $E\ne\elad$.
\end{proof}
\end{proofof}

\begin{remark}[the hypothesis of the writer-solvent local lower bound]\label{rem:L7}
The local lower bound~\cite[record~L7; Theorem~6.1]{Supp} was stated under the hypothesis $h_E^R(E)<1-E$ alone. Its proof uses that the iterates $g_i$, $i\le R+1$, stay below $1-E$ (the notionals $1/(1-g_{i-1})$ must be positive), which holds exactly when $h_E^s(E)<1-E$ for all $1\le s\le R$. With the literal hypothesis the statement is false: for $E=3/10$ the iterates of $h_E$ from $E$ are $51/70$, $267/190$ and $-339/770$, so $h_E^3(E)<1-E$, yet the first iterate already exceeds $1-E=7/10$, $R_E=1$, and $\BOX(3/10)$ settles every writer-solvent network in one round. The other conclusions of~\cite[record~L7; Theorem~6.1]{Supp} are unaffected: $1$-local algorithms for (I) on writer-solvent networks exist exactly for rational $E>1-1/\sqrt2$, radius $2$ fails below $0.237286\ldots$, and bounded radius fails for all $E\le3-2\sqrt2$.
\end{remark}

\begin{remark}[how many rounds]\label{rem:rounds}
\CO\ For each $R$ the table gives the infimum of the errors $E$ at which the bound sequences of the two proofs guarantee that $R$ rounds suffice, rounded to $20$ decimal places. For writer-solvent networks this is the infimum of the $E$ with $R_E\le R$; at this value $h_E^R(E)=1-E$, and by \cref{cor:local}(a) it is also the infimum of the rational $E$ for which an $R$-local algorithm for (I) exists. For general networks it is the infimum of the $E$ with $\kappa_E^s(t^3)\le1/(1-2E)$ for some $s\le R-1$, an upper bound on the $R$-local threshold. Both conditions are monotone in $E$, which justifies bisection. The entries for $R=2$ are certified roots: $2E^3-8E^2+6E-1$ for writer-solvent networks, and
\[
-2E^9+22E^8-90E^7+185E^6-235E^5+199E^4-115E^3+44E^2-10E+1
\]
for general networks, the expansion of $(1-2E)\bigl((1-E)^4-E^3\bigr)^2-E^3(1-E)^2\bigl(E^2+(1-E)^3\bigr)$; the other entries are high-precision numerical values.
\begin{center}\footnotesize
\begin{tabular}{@{}rll@{}}
\toprule
rounds $R$ & writer-solvent: root of $h_E^R(E)=1-E$ & general: $\kappa_E^{R-1}(t^3)=1/(1-2E)$\\
\midrule
1 & $0.29289321881345247560=1-1/\sqrt2$ & $0.38196601125010515180=(3-\sqrt5)/2$\\
2 & $0.23728621957824145633$ & $0.34158115468590270132$\\
3 & $0.21342239329511376033$ & $0.32222411198999461037$\\
5 & $0.19317531940858405206$ & $0.30445151483662148931$\\
10 & $0.17917491162275650449$ & $0.29115447650580809028$\\
100 & $0.17168642690504855151$ & $0.28348866574560776001$\\
$\infty$ & $0.17157287525380990240=3-2\sqrt2$ & $0.28336553249330853976=\elad$\\
\bottomrule
\end{tabular}
\end{center}
For general networks and (I), the $1$-local threshold is exactly $(3-\sqrt5)/2$~\cite[record~L3; Theorem~7.1(a)]{Supp}. Two rounds already beat it, which answers the question of~\cite[record~L3]{Supp} whether two refinement rounds can do so.
\end{remark}

\begin{remark}[tests and stalls]\label{rem:boxtests}
\CO\ The exact implementation of~\cite[record~L19]{Supp} was run on $400$ random general networks ($191$ of them degenerate) at $E\in\{0.2834,\allowbreak 0.29,0.3,0.35,0.4\}$ and on $400$ random writer-solvent networks ($181$ of them degenerate) at $E\in\{0.1716,0.175,0.18,0.2,0.25\}$, with exact box endpoints: all were solved, within $3$ rounds, and every output was checked exactly against the definition of (I). A separate implementation of $\BOX(E)$ in~\cite[record~L21]{Supp} ran $107$ networks ($7$ exported by~\cite[record~L19]{Supp}, $4$ synthetic and $96$ new random ones) with $1922$ exact checks of the enclosures and of the ratio bound of \cref{lem:ratio}, and checked $131080$ rational instances of the contraction inequalities behind \cref{thm:algws,thm:alggen} on dense grids. Below the thresholds $\BOX$ can stall. On a writer-solvent cycle of $m\ge2$ receivers, in which every receiver has liability $1$ and holds a CDS of notional $d$ on its predecessor, with $1/(1-E)\le d<(1-2E)/E$, round $1$ gives every receiver the box $[E(1+d),1-E]$, and this box is a fixed point of the update whenever $Ed^2-(1-E)d+1\le0$; for $E\le3-2\sqrt2$ this holds at $d=(1-E)/(2E)$. Degenerate rings with the ladder structure of \cref{thm:locallower}(b), eight banks plus their sinks and liabilities consisting of CDSs only, with $\gamma=451/1000$ at $E=0.26$ and $\gamma=227/500$ at $E=0.28$, so that $p_E(\gamma)>0$, did not settle within $400$ rounds; rings of the same structure with three other choices of the two CDS notionals were solved at $E\in\{0.2834,0.29,0.3\}$. These ring runs rounded box endpoints outward to the grid $2^{-60}\mathbb Z$, which preserves the properties used in \cref{lem:sound}, and the solved outputs were checked exactly~\cite[record~L19]{Supp}. \Cref{sec:stalls} analyses the stalls of the first kind.
\end{remark}

The historical $1$-local lower bound and writer-solvent (W) lower bound $0.1609$ cited in \cref{sec:local} and \cref{rem:rounds} are proved in \cref{app:localaux}.

\subsection{Finite precision}

\begin{proofof}{thm:roundedbox}
\begin{proof}
The soundness argument of \cref{lem:sound}(a),(b) applies to the enclosures without change: every coordinate outside $F$ stays in $[E_0,1-E_0]$, so a liability with positive upper bound has a positive lower bound (a positive debt, or a CDS whose reference lies outside $F$ and contributes at least $E_0$ times its notional), while references in $F$ contribute identically zero. So no ratio has a zero denominator.

For every unsettled bank and the enclosure computed on the current product we maintain
\begin{equation}\label{eq:roundinv}
l_i\le\lambda_i+E_0,\qquad h_i\ge\mu_i-E_0 .
\end{equation}
Initially this follows from $0\le\lambda_i\le\mu_i\le1$. Enclosures tighten when the product shrinks, because the formulas use nonnegative terms and monotone interval arithmetic. After a continuing update the new lower endpoint is at most $\lambda_i+E_0$ and the new upper endpoint at least $\mu_i-E_0$, by~\eqref{eq:roundinv} and the directions of rounding; hence~\eqref{eq:roundinv} holds again for the next, tighter enclosure. When a box continues, $\lambda_i+E_0<\mu_i-E_0$ and both numbers lie in the old box by~\eqref{eq:roundinv}, so the new box contains $[\lambda_i+E_0,\mu_i-E_0]$ and is a nonempty interval of positive width with endpoints in $G$. A settlement takes a point of the old box. This proves (1).

For a settlement let $p^\ast$ be the clamped midpoint. Exactly as in \cref{lem:sound}(c), \eqref{eq:roundinv} gives $\mu_i-E_0\le p^\ast\le\lambda_i+E_0$. The old box has endpoints in $G$, so it contains a point $p$ of $G$ with $|p-p^\ast|\le\delta/2$. Hence $|p-y|\le E_0+\delta/2$ for every $y\in[\lambda_i,\mu_i]$. Every later product lies in the current one, on which $\Phi_i$ takes values in $[\lambda_i,\mu_i]$; this proves (2).

Let $V=\sum_{i\notin F}(h_i-l_i)/\delta$, with settled banks contributing $0$. It is a nonnegative integer, at most $nM$ initially. Every round that changes some box or settles some bank decreases $V$ by at least $1$: boxes are nested with endpoints in $G$, a strict change of an endpoint lowers a width by a multiple of $\delta$, and a settled bank had positive width before. So at most $nM$ rounds change something, and the next round stops the algorithm. A point of $G$ is $E_0+j(1-2E_0)/M$, whose numerator and denominator have bit length $O(\mathrm{size}(E_0)+\log M)$. In a round the data of the network and these grid points enter sums, products, quotients and comparisons of polynomial bit length; rounding uses exact integer division and returns the endpoints to $G$, so no denominators accumulate from round to round. This proves (3).

At a terminal round with $U\ne\emptyset$ no bank settles, so $\mu_i-\lambda_i>2E_0$, and no box changes. If $\lambda_i+E_0\ge l_i+\delta$, then $\lfloor\lambda_i+E_0\rfloor_G\ge l_i+\delta$, and the lower endpoint would change; together with~\eqref{eq:roundinv} this gives the middle part of~\eqref{eq:nearstall}, and the last part is symmetric. This proves (4).

For (5), let $\mathcal B$ be a product of boxes reached by the algorithm, with the settled banks fixed at their points and the banks of $F$ at $1$. On the set of all vectors with $r_F=\mathbf1$ and $r_i\in[E_0,1-E_0]$ for $i\notin F$, every liability $\ell_i$ with $i\notin F$ is either identically zero or bounded below by a positive constant (as in the first paragraph), so $\Phi$ is continuous there. For an unsettled bank, \eqref{eq:roundinv} and (1) give $\Phi_i(\mathcal B)\subseteq[\lambda_i,\mu_i]\subseteq[l_i-E_0,h_i+E_0]$. Let $\Gamma$ send $r\in\mathcal B$ to the vector whose unsettled coordinates are the projections of $\Phi_i(r)$ onto $[l_i,h_i]$ and whose other coordinates are those of $r$. This is a continuous self-map of the nonempty compact convex set $\mathcal B$, so by Brouwer's theorem it has a fixed point $r$. Every unsettled bank has $|r_i-\Phi_i(r)|\le E_0$, because $r_i$ is the projection of a number of $[l_i-E_0,h_i+E_0]$ onto $[l_i,h_i]$; settled banks have residual at most $E_0+\delta/2$ by (2); and $r_F=\mathbf1$ with $T\subseteq F$.
\end{proof}
\end{proofof}

\begin{remark}[exact boxes]\label{rem:exactbox}
The same argument with $\delta=0$ applies to the exact algorithm $\BOX(E)$ of \cref{sec:box}: by the proof of \cref{lem:sound}(c), every unsettled box at the start of a round has the form $[\lambda'+E,\mu'-E]$ with $\lambda'\le\lambda_i$ and $\mu'\ge\mu_i$, so $\Phi_i$ maps the current product into $[lo_i-E,hi_i+E]$, and every product reached by $\BOX(E)$ contains an (I)-$E$ vector. The boxes need not contain an exact clearing vector (\cref{ex:stallfail}); what they always contain is an approximate one, and the problem at a stall is to compute it.
\end{remark}

\section{Proofs of two historical local bounds}\label[appendix]{app:localaux}

The following arguments supply the two historical local claims cited in
\cref{sec:local}. They reproduce the mathematical content of
\cite[record~L3; Theorem~7.1(a)]{Supp} and \cite[record~L11; Proposition~5.1]{Supp}; they are not new
results. The locality model is exactly the one defined in \cref{sec:local}.

\paragraph{The historical general-network $1$-local bound.}
For every $0<E<(3-\sqrt5)/2$, no deterministic $1$-local algorithm outputs an
(I)-$E$ vector for every non-degenerate network. By monotonicity in the error,
it suffices to consider rational $E$ in this range. Put $H=1-E$ and $t=H/E$.
Then
\[
  1-2E>t^{-3},
\]
because $t^3(t-1)>t+1$ is equivalent to
$(t^2+1)(t^2-t-1)>0$. Choose a rational $d\in(0,H^2)$ sufficiently small that
\[
  1-2E>Q(d):=\frac{E^3}{H^3-d(1-2E)},
  \qquad c'=\frac{H^2-d}{E}>0.
\]
This is possible by continuity at $d=0$; the denominator is positive there
and remains positive for a sufficiently small choice.

Use banks $s,h,p,y,q,p',y',q'$ and a private sink for every bank other than
$s$, each sink having external assets $1$ and no liabilities. Bank $h$ owes
$d$ to its sink and writes to that sink a CDS on $q$ of notional $c'$.
Bank $p$ owes $1$ to its sink, holds a CDS written by $s$ on $p'$ of notional
$2/H$, and writes to $h$ a CDS on $y$ of notional $1$. Banks $y$ and $q$ each
owe $1$ to their sinks and hold CDSs written by $s$ on $y'$ and $q'$,
respectively, of notional $1/H$. Banks $p',y',q'$ each owe $1$ to their own
sinks. Set $e_s=2L_s^{\max}+1$; unspecified external assets are zero.

The networks $X_{\rm hi}$ and $X_{\rm lo}$ differ only in the external
assets of $(p',y',q')$, which are $(0,2,0)$ and $(2,0,2)$, respectively.
All CDS roles are distinct and every non-sink bank either owes an ordinary
debt or has positive external assets, so both networks are non-degenerate.
The neighbours of $h$ in the dependency graph are $p,y,q$ and its sink:
$\mathcal S_h=\{h,p,y,q\}$, and the only other $\mathcal S$-set containing
$h$ is that of its sink. Thus $s,p',y',q'$ are outside $V_1(h)$, and the two
views $V_1(h)$ are identical, including all contracts incident to viewed
banks. A $1$-local algorithm must output the same value $x$ at $h$.

In an (I)-$E$ vector all banks of external assets $2$ and liability $1$,
and also $s$ and the sinks, have recovery $1$. In $X_{\rm hi}$ this gives
$r_p\ge H$, $r_y\le E$ and $r_q\ge H$: for example
$A_p\ge(2/H)H=2\ge\ell_p$, whereas $A_y=0$. Consequently
\[
 A_h=r_p(1-r_y)\ge H^2,
 \qquad \ell_h=d+c'(1-r_q)\le d+c'E=H^2,
\]
so $x\ge H$. In $X_{\rm lo}$, the same reasoning gives $r_p\le E$,
$r_y\ge H$ and $r_q\le E$, and hence
\[
 x\le E+\frac{E^2}{d+c'H}
   =E+Q(d)<H.
\]
This contradicts the common output at $h$. Together with the one-round
upper bound, this proves the asserted threshold. For completeness, the
upper bound follows directly from \cref{lem:ratio}: in the initial product
$K=((1-E)/E)^3$, and $\mu_i-\lambda_i\le1-1/K\le2E$ when
$E\ge(3-\sqrt5)/2$. Thus every bank settles in the first round. For the
binary-rational computational model it is enough to take any rational
$E>(3-\sqrt5)/2$; the displayed constant is the infimum.

\paragraph{The historical writer-solvent (W) bound at $0.1609$.}
We give the universal-radius argument from \cite[record~L11; Proposition~5.1]{Supp},
including the finite rational certificate used there. Set
\[
 E=\frac{1609}{10000},\qquad H=1-E=\frac{8391}{10000},
 \qquad \widehat g=\frac{69533}{125000}.
\]
We construct two networks that differ only in the external assets of a
bank $v_0$, set to $0$ in one and $2$ in the other. Bank $v_0$ owes $1$ to a
private sink and holds nothing; thus its recoveries are at most $E$ and at
least $H$, respectively. Every subsequent signal bank $v_j$ has zero
external assets, owes $1$ to its own private sink and holds the CDSs
described below. Every sink has external assets $1$ and no liabilities.
All writers have no ordinary debt or holdings and have external assets
$2L_w^{\max}+1$, so every writer recovery in a (W)-$E$ vector lies in
$[H,1]$. All newly specified banks are distinct except for the explicitly
shared writer. Therefore all CDS roles are distinct, the networks are
non-degenerate and writer-solvent, and all data are rational.

Suppose a signal has recovery at most $g$ in one network and at least $H$
in the other. A stage of type $\mathrm u$ introduces a private writer $w$
and a CDS on that signal of notional
\[
 c_{\mathrm u}(g)=\frac1{H(1-g)}.
\]
Writing $t=r_w\in[H,1]$, its output is at least $H$ on the low input,
because its assets are at least $c_{\mathrm u}H(1-g)=1$, and is at most
$E+Etc_{\mathrm u}$ on the high input.

A stage of type $\mathrm B$ uses a new private main writer $w$, a CDS on
the signal of notional
\[
 c_{\mathrm B}(g)=\frac1{H(1-g+E^2)},
\]
and a CDS on $w$ of notional $Ec_{\mathrm B}$ written by a second private
writer. If $1-g\ge HE$, its assets on a low input are at least
\[
 c_{\mathrm B}\{t(1-g)+HE(1-t)\}
 \ge c_{\mathrm B}H(1-g+E^2)=1.
\]
On a high input its assets are at most
$c_{\mathrm B}\{tE+E(1-t)\}=Ec_{\mathrm B}$. Thus it reverses the two
states and has constant low bound $E+Ec_{\mathrm B}$.

A stage of type $\mathrm C$ directly follows a $\mathrm u$ stage and
reuses its writer $w$, with recovery $t$. It holds a CDS on the preceding
signal of notional
\[
 c_{\mathrm C}(g)=\max\left\{
 \frac1{H-Ec_{\mathrm u}(g)},
 \frac1{H\bigl(H-EHc_{\mathrm u}(g)+E^2\bigr)}\right\},
\]
and a CDS on $w$ of notional $Ec_{\mathrm C}$ written by a fresh private
bias writer. Both denominators will be positive below. On the low output
of the preceding $\mathrm u$ stage its assets are at least
\[
 c_{\mathrm C}\{t(H-Etc_{\mathrm u})+HE(1-t)\}.
\]
The expression in braces is concave in $t$, and its endpoint values at
$t=1,H$ are precisely the two denominators in the definition of
$c_{\mathrm C}$. Thus these assets are at least $1$. On the high output
of the preceding stage the assets are at most $Ec_{\mathrm C}$.
A block $\mathrm{uC}$ therefore restores the original orientation with
constant low bound
\[
 M(g)=E+Ec_{\mathrm C}(g).
\]
This function is increasing wherever the displayed denominators are
positive, because $c_{\mathrm u}$ increases and those denominators
decrease with $g$.

Here are exact checks that certify arbitrarily many such blocks, not just
finitely many simulated lengths:
\[
\begin{aligned}
 c_{\mathrm u}(\widehat g)&=\frac{1250000000}{465423597},\\
 H-Ec_{\mathrm u}(\widehat g)
   &=\frac{1894119402427}{4654235970000}>0,\\
 H\bigl(H-EHc_{\mathrm u}(\widehat g)+E^2\bigr)
   &=\frac{7794073860498319}{18489000000000000}>0,\\
 \widehat g-M(\widehat g)
   &=\frac{13962037107}{473529850606750000}>0,\\
 (1-\widehat g)(1-2E)-E
   &=\frac{87526097}{625000000}>0.
\end{aligned}
\]
Also $E<\widehat g$, the initial $\mathrm B$ stage is permitted since
$1-E-HE=70408881/100000000>0$, and
\[
 \widehat g-\bigl(E+Ec_{\mathrm B}(E)\bigr)
 =\frac{31514994811064111}{181453027617750000}>0.
\]
For even $R\ge2$ use $R/2$ blocks $\mathrm{uC}$; for odd $R\ge1$ use an
initial $\mathrm B$ followed by $(R-1)/2$ such blocks. At every block
endpoint the low bound is at most $\widehat g$, by induction using the
monotonicity and invariant inequality for $M$. All denominators stay
positive. The intermediate $\mathrm u$ bound is below $1$, since
$H-Ec_{\mathrm u}>0$. The terminal signal $v_R$ consequently has recovery
at least $H$ in one network and at most some rational $g_R\le\widehat g$
in the other.

Use a fresh terminal writer $w_*$, common to finitely many probe banks.
Every probe owes $1$ to a private sink and holds a CDS on $v_R$ written by
$w_*$, with a positive rational notional $c$. Put $A_*=1-g_R>E$,
\[
 L=\frac{2E}{A_*-E},\qquad U=\frac{1-2E}{E}.
\]
The certified inequality $A_*(1-2E)>E$ says exactly that $L<U$.
For a common recovery $t=r_{w_*}\in[H,1]$, a probe with $L<tc<U$ has a
clearing-map gap greater than $2E$ between the two networks: its asset
ratios are bounded below by $tcA_*$ on the low input and above by $tcE$
on the high input, and
\[
 \min(1,tcA_*)-tcE
   =\min\{tc(A_*-E),1-tcE\}>2E.
\]
The open intervals $(L/c,U/c)$, with positive rational $c$, cover
$[H,1]$: for any $t>0$, density gives a rational $c\in(L/t,U/t)$.
Compactness gives a finite rational subcover, chosen as the probe
notionals. Its existence suffices for the local impossibility argument;
for fixed $E,R$ it is a finite fixed network construction.

To verify locality, put $v_j$ and its sink in layer $j$. A private
$\mathrm u$ or $\mathrm B$ main writer is put in its output layer. The
writer shared by a $\mathrm u$ stage $j$ and the following
$\mathrm C$ stage $j+1$ is put in layer $j$. Every private bias writer is
put in its output layer. Put the terminal probes, their sinks and $w_*$
in layer $R+1$. Each set $\mathcal S_i$ meets only one layer or two
adjacent layers. In particular the shared writer's own set contains
references $v_{j-1},v_j$, and the next signal's set contains that writer
and $v_j$, without creating an edge across two layers. Hence no dependency
edge changes the layer by more than one. The terminal probes and $w_*$
are at distance at least $R+1$ from $v_0$; all their radius-$R$ views,
including incident contracts, are identical in the two networks.

If an $R$-local algorithm succeeded on both networks, it would assign the
same recovery $t$ to $w_*$ and the same recovery to each corresponding
probe. Some probe in the finite cover would then have clearing-map
values separated by more than $2E$, so its common recovery could not be
within $E$ of both values. This is a contradiction. Monotonicity of (W)
extends the impossibility to every smaller error. Since $R$ was
arbitrary, the bound holds at every fixed radius (and rules out radius
zero as well, since a radius-zero algorithm is also radius one).

\section{Proofs for the escape routes}\label[appendix]{app:routes}

This appendix contains the proofs for \cref{sec:barriers}, computational evidence on two further routes, and a two-receiver template judged on exact solution sets (\cref{prop:tworeceiver}).

\subsection{Unary gadgets}

\begin{proofof}{thm:unary}
\begin{proof}
For a bank with boxes $[b_0,t_0]$ in $\tau_0$ and $[b_1,t_1]$ in $\tau_1$ consider
\[
\text{(U)}\qquad cb_0\le t_1,\qquad c(1-t_0)\le1-b_1,\qquad cb_1\le t_0,\qquad c(1-t_1)\le1-b_0 .
\]
These are (P1) and (P3) for the pair of conditions $(\tau_0,\tau_1)$ in place of $(\sigma_j,\sigma_3)$. Write $\alpha^i,\beta^i$ for the minimum and maximum of $A_b$ over the boxes of $\tau_i$. If every bank that $A_b$ depends on satisfies (U), the argument for (K1) in \cref{lem:K} gives $\beta^1\ge c\alpha^0$ and $\beta^0\ge c\alpha^1$, and the arguments for (P1) and (P3) in \cref{lem:closure}, which use only (K1), $c\le1$ and (C1), show that the updated boxes of $b$ satisfy (U). Before any update every bank satisfies (U): a receiver or a fixed bank because its two boxes are equal, and a rail of $u$ because $c\le\kappa(\gamma)$: for a $+$rail the boxes are $[0,\gamma]$ and $[H,1]$, and (U) reads $0\le1$, $c(1-\gamma)\le E$, $cH\le\gamma$, $0\le1$; for a $-$rail the boxes are exchanged and (U) reads $cH\le\gamma$, $0\le1$, $0\le1$, $c(1-\gamma)\le E$. So (U) holds at every bank of the certificate.

Let $o$ be an output rail of a wire with bound $\gamma_o$, with box inside $[H,1]$ in the condition $\tau_a$ and inside $[0,\gamma_o]$ in the other condition $\tau_b$. The first gives $\alpha^a\ge1$ for $A_o$, hence $\beta^b\ge c$, and the top of $o$ in $\tau_b$ is at least $\min(1,E+c)$. It is at most $\gamma_o<H<1$, so $\gamma_o\ge E+c$.

For (a) take $c=\min(1,\kappa(\gamma))$, so that $E+c=\psi(\gamma)$. If $E>3-2\sqrt2$, then $y^2-(1-E)y+E>0$ for all $y$, so $c(1-E-c)<E$. If $E\le3-2\sqrt2$ and $\gamma\le\lambda_-$, then $c\le\kappa(\lambda_-)=\lambda_--E$ by \cref{lem:psi}(i), the smaller root of $y^2-(1-E)y+E$; if $\gamma\ge\lambda_+$, then $c\ge\min(1,\kappa(\lambda_+))=\lambda_+-E$, the larger root. Either way $c(1-E-c)\le E$. For (b) take $c=\lambda_--E=\kappa(\lambda_-)\le\kappa(\gamma)$, for which $c(1-E-c)=E$. For the last sentence: $\psi(x)>x$ for $0\le x<\lambda_-$ (by the proof of \cref{lem:psi}(ii), $\psi(x)-x$ is positive when $\kappa(x)=x/H$ or $\kappa(x)\ge1$, and otherwise has the sign of $x^2-(1+E)x+2E>0$), $\psi(x)>x$ on $[0,H)$ if $E>3-2\sqrt2$, and $\psi(x)\ge x$ on $[\lambda_+,H)$ by \cref{lem:psi}(ii).
\end{proof}
\end{proofof}

We call $\psi(\gamma)=E+\min(1,\kappa(\gamma))$ an \emph{uncapped transfer bound}: it adds $E$ to what one contract can transfer, without the cap at $1$. The actual box top of the best single receiver is $\min\{1,E+\kappa(\gamma)\}$. A receiver that reads the input through a CDS on the $+$rail with notional $1/(1-\gamma)$ (best if $\kappa(\gamma)=E/(1-\gamma)$), or through a debt of notional $1/H$ owed by the $-$rail (best if $\kappa(\gamma)=\gamma/H$), saturates in one condition and has box top $\min\{1,E+\kappa(\gamma)\}$ in the other. In case (a) no unary gadget does better: the proof above, with $c=\min(1,\kappa(\gamma))$, shows that in that condition every output rail has box top at least $\min(1,E+c)=\min\{1,E+\kappa(\gamma)\}$. The two expressions agree whenever $\psi(\gamma)<H$, and they can differ otherwise: at $E=1/5$ and $\gamma=79/100$ we have $\kappa(\gamma)=20/21$ and $\psi(\gamma)=121/105>1$, while the tight endpoint is $\min\{1,E+\kappa(\gamma)\}=1$ (\texttt{code/scope\_examples.py}). So for $\gamma\notin(\lambda_-,\lambda_+)$, and for every $\gamma$ if $E>3-2\sqrt2$: if $\psi(\gamma)<H$, no unary gadget beats the best single receiver, whose output bound is $\psi(\gamma)$; if $\psi(\gamma)\ge H$, no unary gadget with input bound $\gamma$ and at least one output wire is certified correct, because (a) forces every output bound to be at least $\psi(\gamma)\ge H$, while a wire needs a bound below $H$. (The case $\psi(\gamma)\ge H$ does not occur for $\gamma\in(\lambda_-,\lambda_+)$: there $\gamma(1-\gamma)>E(2-\gamma)\ge EH$, so $\kappa(\gamma)=E/(1-\gamma)$, which is below $1$ as $\gamma<\lambda_+<H$, and $(\psi(\gamma)-\gamma)(1-\gamma)=\gamma^2-(1+E)\gamma+2E<0$, so $\psi(\gamma)<\gamma<H$.) Between the fixed points $\lambda_\pm$, chains of inverters lower a bound towards $\lambda_-$, but no unary gadget gets below $\lambda_-$.

\subsection{Asymmetric inputs}

For two inputs we use weighted versions of (P1)--(P4).

\begin{lemma}[weighted invariants]\label{lem:weighted}
Let $c_1,c_2,c_\ast\in[0,1]$ and $s_1,s_2>0$ with $S=s_1+s_2\le1$, and put $w_j=s_j/S$. For a bank with boxes $[b_j,t_j]$ in $\sigma_j$ consider
\begin{align*}
&\text{(Q1)}\quad c_jb_j\le t_3,\quad c_j(1-t_j)\le1-b_3,\quad c_jb_3\le t_j,\quad c_j(1-t_3)\le1-b_j\qquad(j=1,2);\\
&\text{(Q2)}\quad s_1b_1+s_2b_2\le t_3\ \text{ and }\ s_1(1-t_1)+s_2(1-t_2)\le1-b_3;\\
&\text{(Q3)}\quad c_\ast b_3\le w_1t_1+w_2t_2\ \text{ and }\ c_\ast(1-t_3)\le w_1(1-b_1)+w_2(1-b_2).
\end{align*}
Assume, for $j\in\{1,2\}$ and $j'=3-j$,
\begin{align*}
&\text{(W1)}\ c_j(1-E-c_j)\le E; &&\text{(W2)}\ S(1-E-c_\ast)\le E;\\
&\text{(W3)}\ s_j(1-E-c_j)\le E; &&\text{(W4)}\ c_\ast(1-E)\le w_{j'}+w_j(E+c_j);\\
&\text{(W5)}\ c_\ast(1-E-S)\le E; &&\text{(W6)}\ c_\ast\bigl(1-E-\max(c_j,s_j)\bigr)\le w_{j'}+w_jE.
\end{align*}
If every bank that $A_b$ depends on satisfies (Q1)--(Q3), then, with $\alpha_j=\alpha_b^j$ and $\beta_j=\beta_b^j$,
\[
\text{(L1)}\ \beta_3\ge c_j\alpha_j,\ \ \beta_j\ge c_j\alpha_3;\qquad \text{(L2)}\ \beta_3\ge s_1\alpha_1+s_2\alpha_2;\qquad \text{(L3)}\ w_1\beta_1+w_2\beta_2\ge c_\ast\alpha_3,
\]
and the boxes $P_b^1,P_b^2,P_b^3$ satisfy (Q1)--(Q3). A bank whose box is the same in the three conditions satisfies (Q1)--(Q3).
\end{lemma}
For $c_1=c_2$ and $s_1=s_2=s/2$ these are (P1)--(P4) and (C1)--(C3), and the lemma reduces to \cref{lem:K}, \cref{lem:closure} and \cref{lem:base}(a).
\begin{proof}
\emph{Comparison.} Let $w'_j\ge0$ with $\sum_jw'_j=1$ and $c'_j\ge0$ with $C'=\sum_jc'_j\le1$ ($j=1,2,3$), and suppose every bank $k$ that $A_b$ depends on satisfies $\sum_jc'_jb_j\le\sum_jw'_jt_j$ and $\sum_jc'_j(1-t_j)\le\sum_jw'_j(1-b_j)$. As $x^j_k$ and $z^j_k$ range over $[b_j,t_j]$, the averages $\bar x_k=\sum_jw'_jx^j_k$ and $\bar z_k=\sum_jc'_jz^j_k$ range independently over $[\sum_jw'_jb_j,\sum_jw'_jt_j]$ and $[\sum_jc'_jb_j,\sum_jc'_jt_j]$; the two inequalities say exactly that some choice has $0\le\bar x_k-\bar z_k\le1-C'$. By \cref{lem:comparison}, $\sum_jw'_j\beta_j\ge\sum_jw'_jA_b(x^j)\ge\sum_jc'_jA_b(z^j)\ge\sum_jc'_j\alpha_j$. Write $\chi_j$ for the $j$-th unit vector of $\mathbb R^3$. Then (Q1) is this property for $(w',c')=(\chi_3,c_j\chi_j)$ and $(\chi_j,c_j\chi_3)$, (Q2) for $(\chi_3,(s_1,s_2,0))$, and (Q3) for $((w_1,w_2,0),c_\ast\chi_3)$; they give (L1), (L2) and (L3).

\emph{Closure.} Write $a_j=\min(1,\alpha_j)$. Then $\Bot(P_b^j)=\max(0,a_j-E)$, $1-\Top(P_b^j)=\max(0,1-E-\beta_j)$ and $1-\Bot(P_b^j)=\min(1,1+E-a_j)=1+E-\max(a_j,E)$. We use that an affine function on an interval that is nonnegative at both ends is nonnegative on it.

(Q1): the proofs of (P1) and (P3) in \cref{lem:closure} apply word for word with $c_j$ in place of $c_1$, using (L1) and (W1).

(Q2), first part: $s_1\Bot(P_b^1)+s_2\Bot(P_b^2)\le s_1a_1+s_2a_2\le\min(S,s_1\alpha_1+s_2\alpha_2)\le\min(1,\beta_3)\le\Top(P_b^3)$ by (L2).

(Q2), second part: we need $\sum_js_j\max(0,1-E-\beta_j)\le\min(1,1+E-a_3)$. The left side is at most $S(1-E)\le1$. If $\beta_1,\beta_2<1-E$, the left side is $S(1-E)-S(w_1\beta_1+w_2\beta_2)\le S(1-E-c_\ast a_3)$ by (L3), and $S(1-E-c_\ast a)\le1+E-a$ holds at $a=0$ and, by (W2), at $a=1$. If $\beta_{j'}\ge1-E>\beta_j$, the left side is $s_j(1-E-\beta_j)\le s_j(1-E-c_ja_3)$ by (L1), and $s_j(1-E-c_ja)\le1+E-a$ holds at $a=0$ and, by (W3), at $a=1$. If $\beta_1,\beta_2\ge1-E$, the left side is $0$.

(Q3), first part: we need $c_\ast\Bot(P_b^3)\le w_1\Top(P_b^1)+w_2\Top(P_b^2)$; assume $\Bot(P_b^3)=a_3-E>0$. If $\beta_1,\beta_2<1-E$, the right side is $E+w_1\beta_1+w_2\beta_2\ge E+c_\ast a_3>c_\ast(a_3-E)$ by (L3). If $\beta_{j'}\ge1-E>\beta_j$, the right side is $w_{j'}+w_j(\beta_j+E)\ge w_{j'}+w_j(E+c_ja_3)$ by (L1), and $w_{j'}+w_j(E+c_ja)-c_\ast(a-E)$ is nonnegative at $a=E$ and, by (W4), at $a=1$. If $\beta_1,\beta_2\ge1-E$, the right side is $1\ge c_\ast$.

(Q3), second part: we need $c_\ast\max(0,1-E-\beta_3)\le w_1(1+E-\max(a_1,E))+w_2(1+E-\max(a_2,E))$; the right side is positive. If $a_1,a_2\ge E$, put $\sigma=w_1a_1+w_2a_2\in[E,1]$; by (L2), $\beta_3\ge S\sigma$, and $c_\ast(1-E-S\sigma)\le1+E-\sigma$ holds at $\sigma=E$ (the left side is at most $c_\ast\le1$) and, by (W5), at $\sigma=1$. If $a_{j'}<E\le a_j$, the right side is $w_{j'}+w_j(1+E-a_j)$, and $\beta_3\ge\max(c_j,s_j)a_j$ by (L1) and (L2); the inequality $c_\ast(1-E-\max(c_j,s_j)a)\le w_{j'}+w_j(1+E-a)$ holds at $a=E$ (the left side is at most $1=w_{j'}+w_j$) and, by (W6), at $a=1$. If $a_1,a_2<E$, the right side is $1\ge c_\ast(1-E)$.

\emph{Constant boxes.} With $b_j=b\le t_j=t$, every inequality reads $xb\le t$ or $x(1-t)\le1-b$ with $x\in\{c_1,c_2,S,c_\ast\}\subseteq[0,1]$.
\end{proof}

\begin{proofof}{thm:asym}
\begin{proof}
Take $c_1=s_1=\lambda_+-E$ and $c_2=s_2=\lambda_--E$, so that $S=\lambda_++\lambda_--2E=H$ and $w_j=s_j/H$, and $c_\ast=(1-2E)/H^2=1-E^2/H^2\in[0,1]$. We use $(\lambda-E)(1-\lambda)=E$ for $\lambda=\lambda_\pm$, and $D=\sqrt{1-6E+E^2}$, $\lambda_\pm=(1+E\pm D)/2$.
\begin{itemize}
\item (W1), (W3): $c_j(1-E-c_j)=s_j(1-E-c_j)=(\lambda-E)(1-\lambda)=E$.
\item (W2): $E-H(1-E-c_\ast)=E(2-4E+E^2)/H>0$. (W5): $1-E-S=0$.
\item (W4): $w_2+w_1(E+c_1)=(s_2+s_1\lambda_+)/H$, and $s_2+s_1\lambda_+=\lambda_--E+\lambda_+^2-E\lambda_+=\lambda_--E+\lambda_+-2E=1-2E$, using $\lambda_+^2=(1+E)\lambda_+-2E$. So $w_2+w_1(E+c_1)=(1-2E)/H=c_\ast(1-E)$, and in the same way $w_1+w_2(E+c_2)=(s_1+s_2\lambda_-)/H=(1-2E)/H$.
\item (W6): here $\max(c_j,s_j)=c_j$ and $c_\ast(1-E-c_j)=(1-2E)(1-\lambda)/H^2$ with $\lambda=\lambda_+$ for $j=1$ and $\lambda=\lambda_-$ for $j=2$. We have $H^2(w_2+w_1E)=H(s_2+Es_1)=H^2(1+E-D)/2$ and $H^2(w_1+w_2E)=H(s_1+Es_2)=H^2(1+E+D)/2$, while $(1-2E)(1-\lambda_\pm)=(1-2E)(H\mp D)/2$. With $H(1+E)-(1-2E)=E(2-E)$ and $H^2-(1-2E)=E^2$ this gives
\begin{align*}
H^2(w_2+w_1E)-(1-2E)(1-\lambda_+)&=\tfrac E2\bigl(H(2-E)-ED\bigr)>0,\\
H^2(w_1+w_2E)-(1-2E)(1-\lambda_-)&=\tfrac12\bigl(HE(2-E)+E^2D\bigr)>0,
\end{align*}
the first because $H(2-E)>1>ED$.
\end{itemize}
\emph{Base.} The computation of \cref{lem:base}(b) shows that a rail of $u$ satisfies (Q1)--(Q3) if $c_1,s_1\le\kappa(\gamma_u)$ and $c_\ast\le\min\bigl((w_1\gamma_u+w_2)/H,(w_1E+w_2)/(1-\gamma_u)\bigr)$ (the first bound for a $+$rail, the second for a $-$rail; the conditions that involve $c_2$ hold because the box of the rail in $\sigma_2$ is $[0,1]$), and a rail of $v$ if $c_2,s_2\le\kappa(\gamma_v)$ and $c_\ast\le\min\bigl((w_1+w_2\gamma_v)/H,(w_1+w_2E)/(1-\gamma_v)\bigr)$. All these bounds are nondecreasing in $\gamma_u,\gamma_v$, so it suffices to check them at $\gamma_u=\lambda_+$, $\gamma_v=\lambda_-$: $\kappa(\lambda_\pm)=\lambda_\pm-E$ by \cref{lem:psi}(i); $(w_1\lambda_++w_2)/H=(w_1+w_2\lambda_-)/H=(1-2E)/H^2=c_\ast$ by the computation for (W4); and the two remaining bounds are at least $c_\ast$ by the computation for (W6). Receivers before their first update and fixed banks satisfy (Q1)--(Q3) by the last sentence of \cref{lem:weighted}. By that lemma, (Q1)--(Q3) hold at every bank of every interval certificate, and (L1)--(L3) at every receiver.

\emph{Conclusion.} Let $o$ be an output rail. If it is a $+$rail, its boxes in $\sigma_1,\sigma_2$ lie in $[H,1]$, so $\alpha_1,\alpha_2\ge1$, and (L2) gives $\beta_3\ge S=H$; its top in $\sigma_3$ is $\min(1,\beta_3+E)=1>\gamma_{\rm out}$. If it is a $-$rail, then $\alpha_3\ge1$ and (L3) gives $w_1\beta_1+w_2\beta_2\ge c_\ast$, so $\beta_j\ge c_\ast$ for some $j\in\{1,2\}$, and its top in $\sigma_j$ is at least $\min(1,E+c_\ast)\ge H>\gamma_{\rm out}$, because $c_\ast-(1-2E)=(1-2E)(1-H^2)/H^2\ge0$. In both cases the output rail is not in its required box.
\end{proof}
\end{proofof}

\subsection{Nonuniform bounds and all standard gate types}

\begin{proofof}{lem:unarycircuits}
\begin{proof}
Every node has exactly one predecessor, the input of its generating gate, so every weak component of the graph of arcs (input $\to$ output) consists of one directed cycle with trees directed away from it. On a cycle, a PURIFY gate whose input lies on the cycle has at most one output on the cycle, because the cycle leaves each node along exactly one arc. If the cycle contains an even number of NOT gates, give one of its nodes the value $0$ and propagate around the cycle (NOT flips, COPY and PURIFY copy); the value returns consistently. If the number is odd, give every node of the cycle the value $\bot$; NOT and COPY gates with input $\bot$ impose nothing, and a PURIFY gate with input $\bot$ is satisfied once its output off the cycle is pure. Finally propagate outwards along the trees: a node whose generating gate has a Boolean input gets the forced value, and a node whose generating gate has input $\bot$ gets $0$. Every gate constraint then holds. Without NOT gates, $0$ is copied everywhere.
\end{proof}
\end{proofof}

\begin{proofof}{thm:nonuniform}
\begin{proof}
Draw an arc from every input node of a gate to every output node of that gate, and call the outputs of binary gates \emph{binary outputs}. \Cref{thm:barrier,thm:asym} apply to binary gates after relabeling, which changes no bound.

If $E>3-2\sqrt2$, consider the gate whose output wire has the least bound $\gamma_\ast$. A binary gate has $\gamma_{\rm out}>\gamma_{\rm in}\ge\gamma_\ast$ (\cref{thm:barrier}), and a unary gate with input bound $\gamma\ge\gamma_\ast$ has output bounds at least $\psi(\gamma)>\gamma$ (\cref{thm:unary}). Both contradict the choice of $\gamma_\ast$, so the instance has no gates and no nodes. Let now $\alpha\le E\le3-2\sqrt2$.

\emph{Step 1: every bound is at least $\lambda_-$.} Otherwise let $\gamma_\ast<\lambda_-$ be the least bound and consider the gate whose output wire has it. A binary gate has $\gamma_{\rm out}>\gamma_{\rm in}\ge\gamma_\ast$ by \cref{thm:barrier}. A unary gate with input bound $\gamma\ge\gamma_\ast$ has, by \cref{thm:unary}, output bounds at least $\psi(\gamma)>\gamma$ if $\gamma<\lambda_-$, at least $\lambda_-$ if $\lambda_-<\gamma<\lambda_+$, and at least $\psi(\gamma)\ge\gamma\ge\lambda_-$ if $\gamma=\lambda_-$ or $\gamma\ge\lambda_+$. In every case the output bound exceeds $\gamma_\ast$, a contradiction.

\emph{Step 2: every node reachable from a binary output (by a directed path, possibly of length $0$) has bound at least $\lambda_+$.} A binary gate has input bounds at least $\lambda_-$ by Step~1, so its output bound is at least $\lambda_+$ by \cref{thm:barrier}. A unary gate with input bound $\gamma\ge\lambda_+$ has output bounds at least $\psi(\gamma)\ge\gamma\ge\lambda_+$. Induction along a directed path gives the claim.

\emph{Step 3: no input of a binary gate is reachable from a binary output.} Such an input would have bound at least $\lambda_+$ by Step~2, the other input of its gate has bound at least $\lambda_-$ by Step~1, and the gadget of the gate would contradict \cref{thm:asym}.

\emph{Step 4: a solution.} Let $M$ be the set of nodes reachable from binary outputs, binary outputs included, and $U$ the set of the other nodes. Arcs into a node $y$ come only from the inputs of its generating gate. Hence: if $y\in M$ is not a binary output, its generating gate is unary and its input $x$ lies in $M$, and the depth of $y$ (the length of a shortest path from a binary output to $y$) exceeds that of $x$; if $y\in U$, its generating gate is unary and its input lies in $U$. By Step~3 the inputs of binary gates lie in $U$. So every gate is a unary gate with all its nodes in $U$, a binary gate with inputs in $U$ and output in $M$, or a unary gate with all its nodes in $M$. The gates of the first kind form a circuit over NOT, COPY and PURIFY in which every node of $U$ is the output of exactly one gate; \cref{lem:unarycircuits} solves it. Next, treat the nodes of $M$ in order of increasing depth. For a binary output $z$ of a gate with inputs $x,y\in U$, give $z$ the value forced by the values of $x,y$ if there is one, and $0$ otherwise. For a node of $M$ generated by a unary gate whose input already has a Boolean value, give it the value forced by the gate (for PURIFY, both outputs at once). By induction on the depth every node of $M$ gets a Boolean value, and every gate of the second and third kind is satisfied. Reachability, depths and the propagation take polynomial time.
\end{proof}
\end{proofof}

\begin{remark}[sanity checks]\label{rem:sanity7}\CO\
In~\cite[record~L19]{Supp} the identities in the proof of \cref{thm:asym} were checked symbolically; the conditions (W1)--(W6), the base conditions and both final contradictions on $6000$ points of $(0,3-2\sqrt2]$ at $50$ digits; the parameter choice of \cref{thm:unary} on a $300\times200$ grid of $(E,\gamma)$; in floating point, (Q1)--(Q3) at every bank of $1500$ random NAND gadgets with $\gamma_u=\lambda_+$, $\gamma_v=\lambda_-$ (the smallest final-reader optima found by linear programming agree with $H$ and $c_\ast$ to six digits) and (U) at every bank of $3000$ random unary gadgets; and the procedure of Step~4 on the $20000$ random circuits over the seven gate types, among $144585$ generated, that satisfy the conclusion of Step~3. The checks of these computations by separate implementations are listed in the check index of the support bundle~\cite{Supp}.
\end{remark}

\subsection{Arbitrary two-interval interfaces}

\paragraph{The domain of a produced rail.} Let $E<1$. Every updated box $P_b^j$ of \cref{def:cert} has upper endpoint $\min(1,\beta_b^j+E)\ge E$ and lower endpoint $\max(0,\min(1,\alpha_b^j)-E)\le1-E=H$. An output rail of a gadget is certified inside $L_q$ in one condition and inside $U_q$ in another, and its initial box $[0,1]$ lies in neither interval (as $l_q<h_q$, $l_q<1$ and $h_q>0$), so it has been updated. Hence every rail that is the output of a correctly certified gadget satisfies
\begin{equation}\label{eq:ifdomain}
E\le l_q<h_q\le H .
\end{equation}
In a realization of a \PCirc\ instance every node is the output of a gate, so~\eqref{eq:ifdomain} holds for every rail. For $E\ge1/2$ it is impossible, and the instance has no nodes and no gates; this proves (a) and (c) for $E\ge1/2$. From now on let $0<E<1/2$.

\paragraph{The effective bound of a rail.} For endpoints satisfying~\eqref{eq:ifdomain} put
\begin{equation}\label{eq:ifgamma}
z_q=\frac{l_q-E}{h_q+E},\qquad \gamma_q=E+z_q,
\end{equation}
and for a wire $x$ let $\Gamma_x=\min_q\gamma_q$ over its rails. Since $l_q<h_q\le H$ and $y\mapsto(y-E)/(y+E)$ is increasing,
\[
0\le z_q<\frac{h_q-E}{h_q+E}\le\frac{H-E}{H+E}=1-2E,\qquad E\le\gamma_q<H .
\]
For the standard interface ($l_q=\gamma$, $h_q=H$) we get $\gamma_q=\gamma$. With $\delta_q=H-h_q\ge0$, \eqref{eq:ifgamma} gives
\begin{equation}\label{eq:ifdelta}
h_q=H-\delta_q,\qquad l_q=\gamma_q-z_q\delta_q .
\end{equation}

\begin{lemma}[base-cost dominance]\label{lem:basecost}
Let $l,h$ satisfy $E\le l<h\le H$, and let $\gamma=E+z$ with $z=(l-E)/(h+E)$. For every $w\in[0,1]$,
\[
\min\Bigl(1,\frac{wl+1-w}{h},\frac{w(1-h)+1-w}{1-l}\Bigr)\ \ge\ \min\Bigl(1,\frac{w\gamma+1-w}{H},\frac{wE+1-w}{1-\gamma}\Bigr).
\]
In particular $\kappa_\ast(l,h):=\min(l/h,(1-h)/(1-l))\ge\kappa(\gamma)$ and $\Lambda_\ast(l,h):=\min\bigl(1,(1+l)/(2h),(2-h)/(2(1-l))\bigr)\ge\Lambda(\gamma)$, with $\kappa$ and $\Lambda$ as in \cref{sec:coincidence}.
\end{lemma}
\begin{proof}
All denominators are positive. Put $\delta=H-h$, so that $h=H-\delta$ and $l=\gamma-z\delta$, and put $d=1-\gamma=H-z>0$ and $n=1-wH$. For the first fractions,
\[
\frac{wl+1-w}{h}-\frac{w\gamma+1-w}{H}=\frac{w(lH-\gamma h)+(1-w)(H-h)}{hH}=\frac{\delta\,\bigl(1-w+wE(1+z)\bigr)}{hH}\ \ge0,
\]
using $lH-\gamma h=\delta(\gamma-zH)$ and $\gamma-zH=E+z-z(1-E)=E(1+z)$. The second fractions are $(1-wh)/(1-l)=(n+w\delta)/(d+z\delta)$ and $(1-wH)/(1-\gamma)=n/d$. If $n/d\ge1$, then $\gamma\ge wH$; the first comparison at $w=1$ gives $l/h\ge\gamma/H\ge w$, hence $1-wh\ge1-l$ and $(1-wh)/(1-l)\ge1$. If $n/d<1$, then $wH>\gamma=E+z$, and
\[
wd-zn=wH-z(1+wE)>E+z-z-zwE=E(1-wz)>0,
\]
since $z<1$ and $w\le1$; hence $(n+w\delta)/(d+z\delta)-n/d=\delta(wd-zn)/(d(d+z\delta))\ge0$. In both cases the second generalized fraction is at least $\min(1,n/d)$, and the first is at least its counterpart; taking the minimum with $1$ proves the inequality. For $w=1$ both first fractions are below $1$ ($l<h$ and $\gamma<H$), which gives $\kappa_\ast\ge\kappa(\gamma)$; $w=1/2$ gives $\Lambda_\ast\ge\Lambda(\gamma)$.
\end{proof}

The cap at $1$ is needed: the second generalized fraction alone need not dominate its counterpart (for $E=4/25$, $l=3/10$, $h=3/5$ and $w=0$ the two second fractions are $10/7$ and $950/623$).

\paragraph{Input rails.} Consider a rail $q$ of the input wire $u$ of a NAND gadget, with the conditions $\sigma_1,\sigma_2,\sigma_3$ of \cref{def:gadget}. For a $+$rail its boxes are $[b,l]$, $[0,1]$, $[h,t]$ (dropping the index $q$). Substituting into (P1)--(P4) of \cref{app:compregions}, with all parameters in $[0,1]$: (P1) for $j=1$ reads $c_1b\le t$, which holds as $b<t$, and $c_1(1-l)\le1-h$; for $j=2$ it reads $0\le t$ and $0\le1-h$. (P3) for $j=1$ reads $c_1h\le l$ and $c_1(1-t)\le1-b$, the latter true as $b<t$; for $j=2$ it reads $c_1h\le1$ and $c_1(1-t)\le1$. (P2) reads $\tfrac s2b\le t$ and $\tfrac s2(1-l)\le1-h$. (P4) reads $c_\ast h\le\tfrac12(l+1)$ and $c_\ast(1-t)\le\tfrac12(2-b)$, the latter true because $1-t\le1-h<1-b/2$. For a $-$rail the boxes are $[h,t]$, $[0,1]$, $[b,l]$, and the same substitution leaves $c_1h\le l$, $c_1(1-l)\le1-h$, $\tfrac s2h\le l$ and $c_\ast(1-l)\le\tfrac12(2-h)$, all other inequalities following from $b\le l<h\le t$. Rails of $v$ are the same with $\sigma_1$ and $\sigma_2$ exchanged, which leaves (P1)--(P4) invariant. So an input rail satisfies (P1)--(P4) whenever
\begin{equation}\label{eq:ifbase}
c_1\le\kappa_\ast(l_q,h_q),\qquad \tfrac s2\le\kappa_\ast(l_q,h_q),\qquad c_\ast\le\Lambda_\ast(l_q,h_q).
\end{equation}
For the weighted invariants (Q1)--(Q3) of \cref{lem:weighted}, the same substitution shows that a rail of input $j$ (with $j'=3-j$) satisfies them whenever $c_j,s_j\le\kappa_\ast(l_q,h_q)$ and
\[
c_\ast\le\min\Bigl(1,\frac{w_jl_q+w_{j'}}{h_q},\frac{w_j(1-h_q)+w_{j'}}{1-l_q}\Bigr),
\]
the first fraction coming from a $+$rail and the second from a $-$rail. (For a $+$rail of $u$, (Q1) leaves $c_1(1-l)\le1-h$ and $c_1h\le l$, (Q2) leaves $s_1(1-l)\le1-h$, and (Q3) leaves $c_\ast h\le w_1l+w_2$, while $c_\ast(1-t)\le1-w_1b$ holds since $b<t$; the conditions involving $c_2,s_2$ hold because the box of the rail in $\sigma_2$ is $[0,1]$. The other cases are symmetric.) For the unary invariant (U) in the proof of \cref{thm:unary}, a rail with boxes $[b,l]$ and $[h,t]$ in the two conditions, in either order, satisfies (U) whenever $c\le\kappa_\ast(l_q,h_q)$. By \cref{lem:basecost} and the monotonicity of $\kappa$, $\Lambda$ and of the saturated weighted costs in $\gamma$, every one of these conditions holds at every rail of a wire $x$ as soon as the corresponding saturated condition of \cref{lem:base}(b), of the proof of \cref{thm:asym}, or of \cref{thm:unary} holds with the bound $\Gamma_x$.

\paragraph{Output rails.} Let $o$ be an output rail, certified inside $U_o=[h_o,t_o]$ in a condition $\tau$ and inside $L_o=[b_o,l_o]$ in a condition $\tau'$, and let $\alpha^\tau,\beta^{\tau'}$ be the extremes of $A_o$ at its last update (\cref{def:cert}). The high certificate gives $\max(0,\min(1,\alpha^\tau)-E)\ge h_o>0$, hence $\alpha^\tau\ge h_o+E$; the low certificate gives $\min(1,\beta^{\tau'}+E)\le l_o<1$, hence $l_o\ge\beta^{\tau'}+E$. So a comparison $\beta^{\tau'}\ge c\,\alpha^\tau$ yields
\begin{equation}\label{eq:ifout}
l_o\ge E+c\,(h_o+E),\qquad\text{that is,}\qquad z_o\ge c\quad\text{and}\quad\gamma_o\ge E+c .
\end{equation}
For a $+$output of a NAND gadget, (K2) of \cref{lem:K} gives $\beta^3\ge\frac s2(\alpha^1+\alpha^2)\ge s(h_o+E)$; for a $-$output, (K3) gives $\max(\beta^1,\beta^2)\ge c_\ast(h_o+E)$, and the rail is low in both $\sigma_1$ and $\sigma_2$. In the weighted setting (L2) and (L3) of \cref{lem:weighted} give the same with $S=s_1+s_2$ in place of $s$. The comparisons are applied at the last update of $o$, when all banks satisfied the invariants; later updates of other banks do not change the certified box of $o$.

\begin{proposition}[gate inequalities]\label{prop:ifgates}
Let $0<E<1/2$ and consider gadgets whose input and output rails all satisfy~\eqref{eq:ifdomain}, certified correct with two-interval interfaces. When $E\le3-2\sqrt2$ let $\lambda_\pm$ be as in \cref{sec:onewriter}, and let $\psi$ be as in \cref{sec:coincidence}.
\begin{enumerate}[label=(\alph*)]
\item \emph{Binary gates.} If $\alpha\le E$, every NAND gadget satisfies $\Gamma_{\rm out}>\min(\Gamma_u,\Gamma_v)$, and if moreover $E\le3-2\sqrt2$ and $\min(\Gamma_u,\Gamma_v)\ge\lambda_-$, then $\Gamma_{\rm out}\ge\lambda_+$.
\item \emph{Unary gates.} If $c\in[0,1]$ satisfies $c\le\kappa(\Gamma_{\rm in})$ and $c(1-E-c)\le E$, then every output wire has $\Gamma_{\rm out}\ge E+c$. Consequently every output wire has $\Gamma_{\rm out}\ge\psi(\Gamma_{\rm in})$ if $E>3-2\sqrt2$, or if $E\le3-2\sqrt2$ and $\Gamma_{\rm in}\notin(\lambda_-,\lambda_+)$, and $\Gamma_{\rm out}\ge\lambda_-$ if $E\le3-2\sqrt2$ and $\lambda_-<\Gamma_{\rm in}<\lambda_+$.
\item \emph{Asymmetric inputs.} If $E\le3-2\sqrt2$, no NAND gadget has $\Gamma_u\ge\lambda_+$ and $\Gamma_v\ge\lambda_-$, or $\Gamma_v\ge\lambda_+$ and $\Gamma_u\ge\lambda_-$.
\end{enumerate}
\end{proposition}
\begin{proof}
(a) Put $g=\min(\Gamma_u,\Gamma_v)\in[E,H)$ and choose $c_1,s,c_\ast$ exactly as in the three cases of the proof of \cref{thm:barrier}, with $g$ in place of $\gamma_{\rm in}$. That proof verifies (C1)--(C3), which involve only $E$ and the parameters, and the saturated base conditions $c_1\le\kappa(g)$, $s/2\le\kappa(g)$, $c_\ast\le\Lambda(g)$. Every rail $q$ of $u$ or $v$ has $\gamma_q\ge g$, so by \cref{lem:basecost} and monotonicity it satisfies~\eqref{eq:ifbase}, hence (P1)--(P4); receivers before their first update and fixed banks satisfy them by \cref{lem:base}(a), and \cref{lem:closure} preserves them under every update. By~\eqref{eq:ifout} every output rail has $\gamma_o\ge E+\min(s,c_\ast)$, so $\Gamma_{\rm out}\ge E+\min(s,c_\ast)$. The proof of \cref{thm:barrier} shows $E+\min(s,c_\ast)>g$ in all three cases, and $E+\min(s,c_\ast)=\lambda_+$ in its case (a) ($\lambda_-\le g$, $E\le3-2\sqrt2$); for $g\ge\lambda_+$ its case (c) applies. That proof uses only $\gamma_{\rm in}\in[0,H)$ (through \cref{lem:psi} and the inequality $E+(1+E)/(2H)>H>\gamma_{\rm in}$), and $g\in[E,H)$.

(b) The invariant (U) holds at every input rail by the paragraph on input rails, at receivers and fixed banks before their first update because their two boxes coincide, and it is preserved by updates as in the proof of \cref{thm:unary}, which uses only (K1), $c\le1$ and $c(1-E-c)\le E$. At an output rail, high in one condition and low in the other, (U) gives $\beta^{\tau'}\ge c\,\alpha^\tau$, and~\eqref{eq:ifout} gives $\gamma_o\ge E+c$. The consequences follow with the parameter choices of the proof of \cref{thm:unary}, with $\Gamma_{\rm in}$ in place of $\gamma$.

(c) Take the parameters of the proof of \cref{thm:asym}: $c_1=s_1=\lambda_+-E$, $c_2=s_2=\lambda_--E$, $S=H$, $w_j=s_j/H$ and $c_\ast=(1-2E)/H^2$. That proof verifies (W1)--(W6) and the saturated weighted base conditions at $\gamma_u=\lambda_+$, $\gamma_v=\lambda_-$; since these conditions are nondecreasing in the bounds, they hold at $\Gamma_u,\Gamma_v$, and by \cref{lem:basecost} at every input rail. By \cref{lem:weighted}, (Q1)--(Q3) hold at every bank and (L1)--(L3) at every receiver. If an output rail $o$ is a $+$rail, it is high in $\sigma_1,\sigma_2$ and low in $\sigma_3$, and (L2) with~\eqref{eq:ifout} gives $z_o\ge S=H$, contradicting $z_o<1-2E<H$. If it is a $-$rail, it is high in $\sigma_3$ and low in $\sigma_1,\sigma_2$, and (L3) gives $\max(\beta^1,\beta^2)\ge c_\ast(h_o+E)$, so $z_o\ge c_\ast\ge1-2E$, again a contradiction. Relabeling the polarities of a wire exchanges the roles of its intervals but changes no interval and no $\gamma_q$.
\end{proof}

\begin{proof}[Proof of \cref{thm:interfaces}]
The case $E\ge1/2$ was settled after~\eqref{eq:ifdomain}. Let $0<E<1/2$. Every rail satisfies~\eqref{eq:ifdomain}, so every wire has $\Gamma\in[E,H)$, and the relabeling of \cref{thm:nonuniform} turns every binary gadget into a NAND gadget without changing any interval. If $E>3-2\sqrt2$, a wire with the least $\Gamma$ would be the output of a gate; a binary gate contradicts minimality by \cref{prop:ifgates}(a), and a unary gate by \cref{prop:ifgates}(b), since $\psi(\gamma)>\gamma$ on $[0,H)$ for $E>3-2\sqrt2$ (\cref{lem:psi}(ii)). So there are no nodes and no gates.

Let $\alpha\le E\le3-2\sqrt2$. First, every $\Gamma$ is at least $\lambda_-$: otherwise take the least $\Gamma<\lambda_-$ and its producing gate; a binary gate has output $\Gamma$ larger than the smaller input $\Gamma$, and a unary gate has output $\Gamma$ larger than its input $\Gamma$ if that is below $\lambda_-$ (as $\psi(x)>x$ there, by the last part of \cref{thm:unary}), at least $\lambda_-$ if it lies in $(\lambda_-,\lambda_+)$, and at least its input if it is $\lambda_-$ or at least $\lambda_+$ (\cref{lem:psi}(ii)); each case contradicts leastness. Second, every binary output has $\Gamma\ge\lambda_+$ by \cref{prop:ifgates}(a), and so has every node reachable from a binary output, since unary gates do not decrease $\Gamma$ at or above $\lambda_+$. Third, no input of a binary gate is reachable from a binary output: it would have $\Gamma\ge\lambda_+$ while the other input has $\Gamma\ge\lambda_-$, contradicting \cref{prop:ifgates}(c). This proves (b), and Step~4 of the proof of \cref{thm:nonuniform}, which uses only this structural property and \cref{lem:unarycircuits}, gives the solution in (c). The last sentence of the theorem follows from (c) as in the discussion after \cref{cor:ceiling}: every instance that such a reduction realizes is solvable in polynomial time, and for $E\ge1/2$ none is realized.
\end{proof}

\begin{remark}[exact checks]\label{rem:ifcheck}\CO\
The program \texttt{interface\_exact\_checks.py} of the supporting archive verifies in exact rational arithmetic the algebraic identities in the proof of \cref{lem:basecost} and the inequality at five designed interfaces and five weights, including $l=E$, $h=H$, nearly touching intervals, different interfaces on the two rails of a wire, and the interface above at which the uncapped second fraction fails to dominate. The program \texttt{integration\_checks.py} of this version recomputes that example and tests the inequality of \cref{lem:basecost} at $4000$ random rational interfaces and five weights. The proof does not depend on these checks.
\end{remark}

\subsection{Computational evidence on two further routes}

\paragraph{Non-saturated interfaces.} \CO\ Before \cref{thm:interfaces} was proved, a floating-point scan in~\cite[record~L19]{Supp} over a grid of $9\times80\times79$ uniform non-saturated interfaces $(E,l,h)$ (in the notation of \cref{def:interfaces}, with the same $l<h$ on every rail) with $E\in[\alpha,3-2\sqrt2]$ found, at every grid point, parameters satisfying (C1)--(C3) and the base conditions of \cref{app:invariants} for which the interface violates the condition $l\ge E+\min(s,c_\ast)(h+E)$ that a certified output would then need (compare~\eqref{eq:ifout}), so that no uniform gadget with that interface is interval-certified at these grid points; the smallest margin, $0.00129$, occurs at $E=\alpha$, $h=H$ and $l=0.43716$, just below $\lambda_-=0.43845$. \Cref{thm:interfaces} now proves the barrier for all two-interval interfaces, uniform or not, within the interval-certificate model, so this scan is only historical evidence.

\paragraph{Gadgets judged on exact solution sets.} \CO\ Searches over gadgets with two or three receivers, with arbitrary positive and negative reads of the two input rails and of each other (feedback allowed), judged on the \emph{exact} internal solution sets (for each condition, the extreme output value is the best of $2^m$ linear programs, one per clipping pattern), were run at $E=0.16$ with input bound $\gamma=\lambda_-+10^{-3}=0.452938$. The best single-receiver seed has output bound $E+2E/(1-\gamma)=0.744942$; a gadget with output bound below $\lambda_+=0.708062$ would beat what \cref{thm:barrier} allows for interval certificates. Nelder--Mead searches from $28$ starts and differential evolution with $21140$ and $32616$ evaluations found nothing below the seed: the best value found is $0.744942$, and the best from random starts $0.746601$. These are floating-point local and evolutionary searches over $14$ to $27$ parameters; a failed search is weak evidence, not a proof of impossibility.

\subsection{A two-receiver template judged on exact solution sets}\label{app:tworeceiver}

The barriers of \cref{sec:barriers} concern interval certificates. The following result, from~\cite[record~L18; Theorem~9.1]{Supp}, concerns one specific template whose correctness is judged on its exact set of internal approximate solutions, so that the correlation between its two receivers is retained. Write $H=1-E$ and
\[
P_E(p)=p^2-(1+E)p+2E ;
\]
for $E\le3-2\sqrt2$ its roots are $\lambda_\pm$ (\cref{sec:onewriter}).

\begin{proposition}[a two-receiver mutual-inhibition template~{\cite[record~L18; Theorem~9.1]{Supp}}]\label{prop:tworeceiver}
Let $\alpha\le E<1/2$ and $0\le\gamma<H$. Consider two receivers $m$ and $z$, neither of them in $T$, with liabilities normalized to $1$, whose maps depend on the recoveries $x,y$ of two input banks as
\[
\Phi_m=\min\{1,\ b+a_x(1-x)+a_y(1-y)+c(1-z)\},\qquad \Phi_z=\min\{1,\ e+d(1-m)\},
\]
with coefficients $a_x,a_y,b,c,d,e\ge0$. For fixed inputs $x,y\in[0,1]$, call $(m,z)\in[0,1]^2$ an \emph{internal $E$-solution} if $|m-\Phi_m|\le E$ and $|z-\Phi_z|\le E$. If every internal $E$-solution has $m\ge H$ whenever $x\le\gamma$ or $y\le\gamma$, then at $x=y=H$ there is an internal $E$-solution with $m>\gamma$.
\end{proposition}

Read as a NAND gadget with input rails $x,y$ and output rail $m$, all with the bound $\gamma$, the template cannot be correct on its exact internal solution sets: if it outputs $1$ whenever an input carries $0$, then it does not output $0$ when both inputs carry $1$. The intercepts $b$ and $e$ may come from external assets or constant payments. The helper $z$ reads only $m$, not the input rails. The proposition does not cover two or more helpers, a helper that also reads the input rails, gadgets of other shapes with correlated receivers, or certificates of other kinds.

\begin{proof}
If $a_x=a_y=0$, the maps do not depend on the inputs. The map $(m,z)\mapsto(\Phi_m,\Phi_z)$ is a continuous self-map of $[0,1]^2$, so by Brouwer's theorem it has a fixed point, which is an internal $E$-solution for all inputs. By the hypothesis, applied at $x=0$, it has $m\ge H>\gamma$, and it is the required solution at $x=y=H$. Let now $a_x+a_y>0$, put $a=\min(a_x,a_y)$ and $q=1-\gamma$, and suppose by symmetry that $a=a_x$.

\emph{A necessary condition.} At the inputs $x=\gamma$, $y=1$ the hypothesis applies, and $\Phi_m=\min\{1,b+aq+c(1-z)\}$. For $m\in[0,1]$ put
\begin{align*}
z^+(m)&=\min\{1,\ e+d(1-m)+E\},\\
R^-(m)&=\max\bigl\{0,\ \min\{1,\ b+aq+c(1-z^+(m))\}-E\bigr\}.
\end{align*}
Then $|z^+(m)-\Phi_z(m)|\le E$, and $|R^-(m)-\Phi_m|\le E$ at $z=z^+(m)$, so a fixed point $m=R^-(m)$ gives the internal $E$-solution $(m,z^+(m))$. The map $R^-$ is continuous and maps $[0,H]$ into itself. If $R^-(H)<H$, then $R^-(m)-m$ is nonnegative at $m=0$ and negative at $m=H$, so $R^-$ has a fixed point below $H$, contrary to the hypothesis. Hence $R^-(H)\ge H$, that is, $b+aq+c(1-z^+(H))\ge1$. Since $1-z^+(H)=\max\{0,H-e-dE\}$, this reads
\begin{equation}\label{eq:twonec}
b+aq+cu\ge1,\qquad u=\max\{0,\ H-e-dE\}.
\end{equation}

\emph{A comparison inequality.} Let $p\in[E,H)$ satisfy $P_E(p)\ge0$, and put $v=1-\max\{0,\min\{1,e+d(1-p)\}-E\}$. We show that
\begin{equation}\label{eq:twocomp}
v\ge(p-E)u .
\end{equation}
If $u=0$, this holds because $v\ge0$. Let $u>0$, and suppose first that $e=0$. As a function of $d$, $v$ equals $1$ for $d\le E/(1-p)$, equals $1+E-d(1-p)$ for $E/(1-p)\le d\le1/(1-p)$, and equals $E$ for $d\ge1/(1-p)$, while $u=\max\{0,H-dE\}$ does not increase with $d$. For $d\le E/(1-p)$ we have $v=1>(p-E)u$, since $0\le p-E<1$ and $u\le1$. On the part of the middle range on which $u>0$, the difference $v-(p-E)u$ is affine in $d$. It is nonnegative at $d=E/(1-p)$, by the first range, and at the other end of that part: if that end is $d=1/(1-p)$, then
\[
E-(p-E)\Bigl(H-\frac E{1-p}\Bigr)=\frac{H\,P_E(p)}{1-p}\ge0,
\]
and if $u$ vanishes before, the difference equals $v\ge0$ there. Where $u=0$ the inequality holds because $v\ge0$. For $d\ge1/(1-p)$ we have $v=E$ while $u$ does not increase, so the difference does not decrease and is nonnegative, as it is at $d=1/(1-p)$. This proves~\eqref{eq:twocomp} when $e=0$. If $e>0$, replace $d$ by $d'=d+e/E$ and $e$ by $0$. This does not change $u$, and since $p<H$ we have $d'(1-p)=d(1-p)+e(1-p)/E\ge e+d(1-p)$, so the replacement can only lower $v$; the case $e=0$ therefore gives~\eqref{eq:twocomp} in general.

\emph{A false-state solution.} Let $x=y=H$, so that $a_x(1-x)+a_y(1-y)=E(a_x+a_y)$. For $m\in[0,1]$ put
\begin{align*}
z^-(m)&=\max\bigl\{0,\ \min\{1,e+d(1-m)\}-E\bigr\},\\
R^+(m)&=\min\bigl\{1,\ b+E(a_x+a_y)+c(1-z^-(m))+E\bigr\}.
\end{align*}
Then $|z^-(m)-\Phi_z(m)|\le E$, and $|R^+(m)-\Phi_m|\le E$ at $z=z^-(m)$, so a fixed point $m=R^+(m)$ gives the internal $E$-solution $(m,z^-(m))$. The map $R^+$ is continuous with values in $[E,1]$, so it has a fixed point, and every fixed point is at least $E$; this proves the claim if $\gamma<E$. Moreover, if $R^+(p)>p$ for some $p\in[0,1)$, then $R^+(m)-m$ is positive at $m=p$ and nonpositive at $m=1$, so $R^+$ has a fixed point above $p$.

Let $\gamma\ge E$, and suppose that $p\in[E,H)$ satisfies $P_E(p)\ge0$ and $2E>(p-E)q$. With $v=1-z^-(p)$ as above,
\[
b+E(a_x+a_y)+c\,v\ >\ (p-E)(b+aq+cu)\ \ge\ p-E .
\]
Indeed, termwise, $b\ge(p-E)b$; $E(a_x+a_y)\ge2Ea>(p-E)qa$ if $a>0$, and $E(a_x+a_y)>0=(p-E)qa$ if $a=0$; and $cv\ge(p-E)cu$ by~\eqref{eq:twocomp}. The last inequality is~\eqref{eq:twonec}, multiplied by $p-E\ge0$. Hence $R^+(p)>\min\{1,p\}=p$, and $R^+$ has a fixed point above $p$. It remains to find such a $p$ with $p\ge\gamma$.

If $P_E(\gamma)\ge0$, take $p=\gamma$; then $2E-(\gamma-E)(1-\gamma)=P_E(\gamma)+E>0$. Otherwise $P_E$ has real roots, so $E\le3-2\sqrt2$, and $\lambda_-<\gamma<\lambda_+$. Take $p=\lambda_+$, which lies in $[E,H)$ by \cref{lem:psi}(i) and satisfies $P_E(p)=0$. Since $E/(1-\lambda_-)=\lambda_--E$ and, for $E\ge\alpha$, $2\lambda_-\ge\lambda_++E$ (\cref{lem:psi}(iv)),
\[
\frac{2E}{1-\gamma}>\frac{2E}{1-\lambda_-}=2(\lambda_--E)\ge\lambda_+-E=p-E ,
\]
so $2E>(p-E)q$. In both cases the fixed point lies above $p\ge\gamma$. For $E>3-2\sqrt2$, $P_E$ has no real root and only the first case occurs.
\end{proof}

\begin{remark}[an exact example]\label{rem:tworeceiverexample}\CO\
At $E=4/25$ and $\gamma=9/20$ take $a_x=a_y=17/10$, $c=1/5$, $d=3$ and $b=e=0$. Then $u=9/25$, and~\eqref{eq:twonec} holds with the margin $\frac{17}{10}\cdot\frac{11}{20}+\frac15\cdot\frac9{25}-1=\frac7{1000}$. An exact vertex enumeration in~\cite[record~L18; Section~9.4]{Supp} found that the least main recovery over the internal $E$-solutions at the inputs $x=\gamma$, $y=1$ is $H=21/25$, and the same guarantee holds for all inputs with $x\le\gamma$ or $y\le\gamma$, at which the main assets are only larger. At $x=y=H$ the pair $(m,z)=(21/25,8/25)$ has $\Phi_m=17/25$ and $\Phi_z=12/25$, so both residuals are exactly $E$ while $m>\gamma$. In~\cite[record~L18; Section~9.4]{Supp} these maps are realized in a writer-solvent network with one CDS writer, which has external assets $71/5$, one sink, two input banks with external assets $H$ and unit debts, and the two receivers; there the false-state pair, completed by the recovery $H$ of the input banks and by $1$ at the writer and the sink, is an (I)-$E$ vector. The program \texttt{tworeceiver\_checks.py} of the support bundle recomputes these numbers in exact arithmetic and checks the true-state guarantee of the example.
\end{remark}

\section{Proofs for the degree classification}\label[appendix]{app:degree}

This appendix contains the remaining proofs for \cref{sec:degree}, a comparison of the critical held degree with the direct unit expansion, and the derivation of the entries of \cref{tab:phase} and of the known ranges of $p_d$. The restoring chains are proved in \cref{app:chains} and the method boundaries in \cref{app:limits}.

\subsection{The critical held degree and the direct unit expansion}

\begin{remark}[the design]\label{rem:dHnew}
The gadget topology is the PURIFY of \cref{thm:D}: two auxiliary receivers, each feeding one output. \Cref{thm:D} uses notionals $C=2\rho$ and $C^2$ and reads bit $0$ as $r\le(1+E)/2$. Here bit $0$ is read as $r\le E$, which lets every short stage use only $m=\lceil\rho\rceil$ contracts, and the long multiplicity $D$ is placed on the \emph{first} stage of one arm and on the \emph{second} stage of the other. The following table compares $d_H(E)$ with the held degree $C^2$ of the direct unit expansion of \cref{thm:D}, $C=\max\{3,\lceil2\rho\rceil\}$.
\begin{center}\small
\begin{tabular}{@{}lccccccccc@{}}
\toprule
$E$ & $1/7$ & $1/5$ & $1/4$ & $1/3$ & $2/5$ & $1/2$ & $11/20$ & $3/5$ & $3/4$\\
\midrule
$C^2$ & $9$ & $9$ & $16$ & $16$ & $25$ & $36$ & $49$ & $64$ & $196$\\
$d_H(E)=\lceil(m+1)\rho\rceil$ & $4$ & $5$ & $5$ & $6$ & $10$ & $12$ & $18$ & $20$ & $56$\\
\bottomrule
\end{tabular}
\end{center}
Among two-arm PURIFY gadgets with two stages per arm and this decoding, $d_H(E)$ is the least possible held degree (\cref{prop:twoarm}); for $E\ge1/4$ the same holds for all chain PURIFY gadgets (\cref{cor:chainlimits}(ii)). These are statements about these gadget families with this decoding, not lower bounds for other reductions.
\end{remark}

\subsection{Engine invariants and the rules A2 and A3}

\begin{proofof}{lem:engines}
\begin{proof}
Rules only relabel $\mathsf U$ banks, so $\mathsf L$ and $\mathsf H$ are permanent and at most $|V|$ assignments occur. Since $\mathsf L$ is permanent, the set of non-$\mathsf L$ references of a bank shrinks over time, so $n_{\neg\mathsf L}$ never increases; since $\mathsf H$ is permanent, $n_{\mathsf H}$ never decreases. A strong bank has $n_{\neg\mathsf L}\le\nu-1$ and this persists; hence strong status and $n_{\mathsf S}$ are monotone. A bank labelled $\mathsf L$ met (P-L) or (Q-L) at that moment, and the quantities in these conditions never decrease, so they still hold at termination. A bank labelled $\mathsf H$ had $n_{\neg\mathsf L}\le k$ (resp.\ $\le\nu$) at that moment, and this persists. At termination no rule applies to a $\mathsf U$ bank, which gives the stated properties of $\mathsf U$ banks; in $\mathcal Q(\nu)$ the at most one held contract referencing an $\mathsf H$ bank cannot reference a strong bank, since $n_{\mathsf S}=0$.

Implementation: keep the counters $n_{\neg\mathsf L},n_{\mathsf H},n_{\mathsf S}$ for every bank and, for every bank $j$, the list of held contracts referencing $j$. When $j$ is labelled $\mathsf L$, decrement $n_{\neg\mathsf L}$ of each holder of a contract on $j$; when $j$ is labelled $\mathsf H$, increment $n_{\mathsf H}$ of each such holder; when $j$ becomes strong (at its $\mathsf H$ labelling, or when its own $n_{\neg\mathsf L}$ drops to $\nu-1$), increment $n_{\mathsf S}$ of each such holder. Each event touches each contract on $j$ once, and each bank becomes strong at most once. Start with every core bank on a queue, put a bank on the queue again whenever one of its counters changes, and test the rules when it is dequeued. The total work is $O(|V|+M)$.
\end{proof}
\end{proofof}

\begin{proofof}{thm:A2}
\begin{proof}
In terms of $t$: $E=1-4t$, $A=2-4t$, and $E>1/3$ is $t<1/6$. Every value is in $\{0,t,2t,4t\}$.

\emph{$\mathsf L$ bank.} Either a strong reference ($S\ge4t$) or two held contracts on $\mathsf H$ banks ($S\ge2\cdot2t$). So $S\ge4t=1-E$, $T\le E$, and $x=0$ is admissible.

\emph{Strong bank} ($n_{\neg\mathsf L}\le1$). $S\in\{0,t,2t,4t\}$, so $S\le4t<1$ and $T=1-S$. The residual is $|4t+S-1|$; as $S$ ranges over $[0,4t]$, $4t+S-1$ ranges over $[4t-1,8t-1]=[-E,\,8t-1]$, and $8t-1\le1-4t$ because $t\le1/6$. The (D) clause: $S\le4t<A$.

\emph{Weak bank} ($n_{\neg\mathsf L}=2$). $S$ is a sum of two values from $\{t,2t,4t\}$, so $2t\le S\le8t$. If $S\le1$, the residual is $|2t+S-1|$ with $2t+S-1\ge4t-1=-E$ and $2t+S-1\le\min\{10t,\,1+2t\}-1$. If $8t\le1$ then $10t-1\le1-4t$ since $t\le1/7$; if $8t>1$ then $2t+S-1\le2t\le1-4t$ since $t\le1/6$. If $S>1$, then $T=0$ and the residual is $2t\le E$. The (D) clause: $S\le8t<2-4t=A$ because $t<1/6$ (strictly; this is where $E>1/3$ is needed).

\emph{$\mathsf U$ bank} ($n_{\mathsf S}=0$, $n_{\mathsf H}\le1$, $n_{\neg\mathsf L}\ge3$). Its non-$\mathsf L$ references are $\mathsf U$ banks (value $t$), plus at most one held contract on a weak bank (value $2t$). Hence $3t\le S\le(d+1)t$. If $S\le1$, the residual is $|t+S-1|$ with $t+S-1\in[4t-1,t]$, and $t\le1-4t=E$ because $t\le1/5$. If $S>1$, the residual is $t\le E$. The (D) clause: $(d+1)t<A=2-4t$ is equivalent to $(d+5)t<2$, that is, $d<8/(1-E)-5=(3+5E)/(1-E)=4\rho-1$.
\end{proof}
\end{proofof}

In the weak case, $8t\le1$ gives $t\le1/8<1/7$, which is what the inequality $10t-1\le1-4t$ needs. At $E=1/3$ rule A2 fails, and not only through the degree bound: a weak bank with two strong references has assets $8t=A$ and positive default $2t$.

\begin{proofof}{thm:A3}
\begin{proof}
Now $E=1-3t$, $A=2-3t$, and $E\ge1/4$ is $t\le1/4$. Values lie in $\{0,t,2t,3t\}$.

\emph{$\mathsf L$ bank.} A strong reference gives $S\ge3t$, two held contracts on $\mathsf H$ banks give $S\ge4t$; in both cases $S\ge1-E$.

\emph{Strong bank} ($n_{\neg\mathsf L}=0$). $S=0$, $T=1$, residual $1-3t=E$; the (D) clause is vacuous.

\emph{Weak bank} ($n_{\neg\mathsf L}=1$). $S\in\{t,2t,3t\}$, so $S\le3t<1$. The residual is $|2t+S-1|$ with $2t+S-1\in[3t-1,5t-1]$; $3t-1=-E$ and $5t-1\le1-3t$ because $t\le1/4$. The (D) clause: $S\le3t<A$.

\emph{$\mathsf U$ bank} ($n_{\mathsf S}=0$, $n_{\mathsf H}\le1$, $n_{\neg\mathsf L}\ge2$). As in \cref{thm:A2}, $2t\le S\le(d+1)t$. If $S\le1$, $t+S-1\in[3t-1,t]$ and $t\le1-3t=E$. If $S>1$, the residual is $t\le E$. The (D) clause: $(d+1)t<2-3t$, i.e.\ $d<6/(1-E)-4=(2+4E)/(1-E)=3\rho-1$.
\end{proof}
\end{proofof}

\subsection{The threshold error and the phase diagram}

\begin{proofof}{cor:pd}
\begin{proof}
\emph{$d\in\{6,7\}$:} $n=2$ and $p:=\bar p_d=1/3$. $\mathrm{HARD}(1/3,d)$ by \cref{cor:deg13}. For rational $E>1/3$, $\rho>2$ and $4\rho-1>7\ge d$, so $\mathrm{FP}(E,d)$ by \cref{thm:A2}.

\emph{$d\ge12$:} then $n\ge3$ and $p:=\bar p_d\ge(n-1)/(n+1)\ge1/2$. Let $\rho^*=\rho(p)$.

(i) If $d<n(n+2)$, then $(d-n-2)/(d+n+2)<(n-1)/(n+1)$ (cross-multiplying, this is $d<n(n+2)$), so $p=(n-1)/(n+1)$ and $\rho^*=n$; then $d_H(p)=n(n+1)\le d$. If $d\ge n(n+2)$, then $p=(d-n-2)/(d+n+2)$ and $\rho^*=d/(n+2)\in[n,n+1)$; if $\rho^*=n$ then $d_H(p)=n(n+1)\le d$, and if $\rho^*>n$ then $m=n+1$ and $d_H(p)=\lceil(n+2)\rho^*\rceil=d$. In all cases $\mathrm{HARD}(p,d)$ by \cref{thm:dH}. Note that $p$ is rational and $p<1$.

(ii) Let $E>p$ be rational with $E<1$, so $E>1/2$ and $\rho>\rho^*$. If $\rho>n+1$, then $m\ge n+2$ and $(m+1)\rho>(n+3)(n+1)>d$. If $n<\rho\le n+1$, then $m=n+1$ and $(m+1)\rho=(n+2)\rho$; in case $\rho^*=n$ this exceeds $(n+2)n\ge d$ (because $\rho>n$ and $\rho^*=n$ forces $d\le n(n+2)$), and in case $\rho^*=d/(n+2)$ it exceeds $(n+2)\rho^*=d$. In all cases $d<(m+1)\rho$ and $\mathrm{FP}(E,d)$ by \cref{thm:degE12}. For rational $E\ge1$, $\mathrm{FP}(E,d)$ holds trivially by \cref{lem:relations}(c).

So in both cases every rational $E>p$ lies in the set defining $p_d$, and $p_d\le p$. If PPAD is not contained in FP, then $p$ and, by \cref{lem:Rmono}, every smaller error are PPAD-hard and hence not in that set; therefore $p_d=p$.
\end{proof}
\end{proofof}

For the hard column of \cref{tab:phase}: \cref{thm:dH} gives $d_H=\lceil4\rho\rceil$ for $2<\rho<3$ and $d_H=\lceil3\rho\rceil$ for $1<\rho\le2$, which is $4$ for $\rho\le4/3$, $5$ for $4/3<\rho\le5/3$ and $6$ for $5/3<\rho\le2$; \cref{thm:lowdeg} lowers this to $4$ for $E\le3/13$ ($\rho\le8/5$), to $3$ for $E<1/7$ ($\rho<4/3$) and to $2$ for $E<1/9$ ($\rho<5/4$). For the FP column, the rows with $E\ge1/4$ use \cref{thm:Pk,thm:A2,thm:A3}. For the unresolved column in the second and fourth rows: the largest FP degree is $\lceil4\rho-1\rceil-1=\lceil4\rho\rceil-2$ (resp.\ $\lceil3\rho\rceil-2$) and the smallest hard degree is $\lceil4\rho\rceil$ (resp.\ $\lceil3\rho\rceil$), so exactly one degree remains, also when $4\rho$ (resp.\ $3\rho$) is an integer.

Per degree, and assuming that PPAD is not contained in FP for the lower ends, the known ranges of $p_d$ are: $p_1=0$; $p_2\in[1/9,1/5]$; $p_3\in[1/7,1/4]$; $p_4\in[3/13,1/4]$; $p_5\in[1/4,1/3]$; $p_6=p_7=1/3$; $p_8\in[1/3,5/13]$; $p_9\in[5/13,3/7]$; $p_{10}\in[3/7,7/15]$; $p_{11}\in[7/15,1/2]$; and \cref{cor:pd} for $d\ge12$. (For $d=2$: hard for every $E<1/9$ by \cref{thm:lowdeg}(c); FP for $E\ge1/5$ by~\cite[Theorem~F.9]{KM} and \cref{thm:weightedD}. For $d=3$: hard for $E<1/7$ by \cref{thm:lowdeg}(b); FP for $E\ge1/4$ by \cref{thm:A3}. For $d=4$: hard for $E\le3/13$ by \cref{thm:lowdeg}(a); FP for $E>1/4$ by \cref{thm:A3}. For $d=5$: hard for $E\le1/4$ by \cref{thm:dH}; FP for $E\ge1/3$ by \cref{thm:Pk}. For $d=8,\dots,11$: hard for $E\le(d-4)/(d+4)$ or $E\le1/3$, whichever is larger, that is, $4\rho\le d$ or $\rho\le2$; FP for $E>(d-3)/(d+5)$, that is, $d<4\rho-1$, and at $E=1/2$ for $d\le11$ by \cref{thm:Pk}.) For $d\in\{4,5,8,9,10,11\}$ the lower end is attained by a PPAD-hard error; for $d\in\{2,3\}$ it is the supremum of an open interval of PPAD-hard errors, and $1/9$, $1/7$ themselves are not known to be hard. Among $d\in\{2,3,4,5,8,9,10,11\}$ the upper end is attained by an algorithm except for $d=4$ and $d=8,9,10$, where the algorithm needs a strict inequality. For $d\in\{6,7\}$ and $d\ge12$, $p_d=\bar p_d$, and $\bar p_d$ is a PPAD-hard error (\cref{cor:pd}).

\section{Restoring chains: complete proofs}\label[appendix]{app:chains}

This appendix proves \cref{thm:lowdeg}. We use the notation of \cref{sec:Rclass}: $A=1+E$, the admissible interval $\mathrm{Adm}(S)=[\mathrm{lo}(S),\mathrm{up}(S)]$ of a core bank with assets $S$, and the consequences (Sol) and (Res) of~\eqref{eq:core}. For $0<E\le1/2$ we use
\begin{equation}\label{eq:upbounds}
S\ge1-E\ \Longrightarrow\ \mathrm{up}(S)\le2E,\qquad S\ge1\ \Longrightarrow\ \mathrm{up}(S)\le E .
\end{equation}
Indeed, if $S\ge A$ then $\mathrm{up}(S)=0$; if $1\le S<A$ then $\mathrm{up}(S)=E$; and if $1-E\le S<1$ then $S\ge E$ (as $E\le1/2$), so $\mathrm{up}(S)=1+E-S\le2E$.

\subsection{Chain envelopes}

\begin{lemma}[chain envelopes]\label{lem:envelopes}
Let values $x$ satisfy~\eqref{eq:core} at the banks of a chain arm (\cref{def:chain}) with multiplicities $(k_1,\dots,k_L)$ hanging from $u$, and let $0\le l_0\le x_u\le h_0\le1$. For $1\le i\le L$ put
\[
l_i=\mathrm{lo}(k_ih_{i-1}),\qquad h_i=\mathrm{up}(k_il_{i-1}).
\]
\begin{enumerate}[label=(\alph*)]
\item Then $l_i\le x_{s_i}\le h_i$ for all $i$.
\item \emph{(Exactness.)} Fix $x_u=X$ and take $l_0=h_0=X$. For every $i$ and every $y\in[l_i,h_i]$ there are values at $s_1,\dots,s_L$ satisfying~\eqref{eq:core} with $x_{s_i}=y$.
\end{enumerate}
\end{lemma}
\begin{proof}
(a) Inductively, $S_{s_i}=k_ix_{s_{i-1}}\in[k_il_{i-1},k_ih_{i-1}]$ and $x_{s_i}\in\mathrm{Adm}(S_{s_i})$; use that $\mathrm{lo}$ and $\mathrm{up}$ are nonincreasing.

(b) Let $R_i$ be the set of values of $x_{s_i}$ over all assignments to $s_1,\dots,s_i$ satisfying~\eqref{eq:core} there. The banks of the arm are pairwise distinct and $s_i$ references only $s_{i-1}$, so every such partial assignment extends to $s_{i+1},\dots,s_L$, because each $\mathrm{Adm}(S)$ is nonempty; it therefore suffices to show $R_i=[l_i,h_i]$. This holds for $i=0$. If $R_{i-1}=[l,h]$, then $R_i=\bigcup_{S\in[k_il,k_ih]}\mathrm{Adm}(S)$, which lies in $[\mathrm{lo}(k_ih),\mathrm{up}(k_il)]$ by monotonicity. Conversely let $v$ lie in this interval. If $v\ge\mathrm{lo}(k_il)$, then $v\in\mathrm{Adm}(k_il)$. Otherwise $\mathrm{lo}(k_ih)\le v<\mathrm{lo}(k_il)$, and continuity of $\mathrm{lo}$ gives $S\in[k_il,k_ih]$ with $\mathrm{lo}(S)=v$, so $v\in\mathrm{Adm}(S)$.
\end{proof}

So for a fixed input $X$ the output is forced to decode to $1$ if and only if $h_L=0$, forced to decode to $0$ if and only if $l_L\ge1-E$, and can decode to $\bot$ otherwise: the interval $[l_L,h_L]$ meets $(0,1-E)$ if and only if $h_L>0$ and $l_L<1-E$. The two arms of a chain PURIFY gadget share only $u$, so both outputs can be $\bot$ for the same $X$ if and only if each can. The hardness proofs below use only part (a); part (b) is used in \cref{prop:chainlimits}(e). In (b) the upper endpoint $\mathrm{up}$ jumps down at $S=A$; the argument shows that this creates no gap in $R_i$, because values below $\mathrm{lo}(k_il)$ are reached through the continuous lower endpoint.

\subsection{Restoration}

\begin{lemma}[restoration]\label{lem:restore}
For integers $a,b\ge1$ let $G_{a,b}(h)=\mathrm{up}\bigl(b\,\mathrm{lo}(ah)\bigr)$ for $h\in[0,1]$.
\begin{enumerate}[label=(\alph*)]
\item $G_{a,b}$ is nondecreasing. If $Q$ holds exactly $a$ unit CDSs, all on $P$, and $R$ holds exactly $b$ unit CDSs, all on $Q$, then $x_P\le h$ implies $x_R\le G_{a,b}(h)$ for all values satisfying~\eqref{eq:core} at $Q$ and $R$.
\item Let $k\in\{2,3\}$, assume $E\le1/7$ if $k=2$ and $E\le1/5$ if $k=3$, and put
\[
h^*_k=\frac{k(1-E)-A}{2k-1},\qquad z_k=\frac{1-E-1/k}2 .
\]
If $0\le h<h^*_k$, then $G_{2,k}^{\,j}(h)=0$ for every $j\ge J+2$, where $J=0$ if $h\le z_k$ and $J=\bigl\lceil\log_{2k}\bigl((h^*_k-z_k)/(h^*_k-h)\bigr)\bigr\rceil$ otherwise.
\end{enumerate}
\end{lemma}
\begin{proof}
(a) Monotonicity is inherited from $\mathrm{lo}$ and $\mathrm{up}$. By \cref{lem:envelopes}(a) applied to the arm $(a,b)$ hanging from $P$, $x_Q\ge\mathrm{lo}(ax_P)\ge\mathrm{lo}(ah)$ and $x_R\le\mathrm{up}(bx_Q)\le\mathrm{up}(b\,\mathrm{lo}(ah))$.

(b) Write $G=G_{2,k}$, $h^*=h^*_k$, $z=z_k$ and $z'=(1-E-A/k)/2$. Evaluating $\mathrm{up}$ at $k(1-E-2h)$ gives $G(h)=0$ for $h\le z'$, $G(h)=E$ for $z'<h\le z$, and
\[
G(h)=1+E-k(1-E-2h)=h^*+2k(h-h^*)\qquad\text{for } z<h\le\tfrac{1}{2}\bigl(1-E-E/k\bigr).
\]
The hypothesis on $E$ is exactly $E\le z'$, so $G(E)=0$; also $G(0)=0$ since $k(1-E)\ge A$. Moreover $z<h^*\le(1-E-E/k)/2$: the left inequality is $E<1/6$ for $k=2$ and $E<2/9$ for $k=3$, and the right one holds for all $E\ge0$. Let $h_0=h$ and $h_{j+1}=G(h_j)$. If some $h_j\le z$, then $h_{j+1}\le E$, $h_{j+2}\le G(E)=0$ by monotonicity, and all later terms are $0$. While $h_0,\dots,h_{j-1}>z$, induction shows that each of them lies in $(z,h^*)$, inside the affine piece, and that $h^*-h_j=(2k)^j(h^*-h_0)>0$. As $h_j>z$ means $h^*-h_j<h^*-z$, some $h_j$ with $j\le J$ satisfies $h_j\le z$.
\end{proof}

For $k=2$ the condition $2E<h^*_2$ is $E<1/9$, and for $k=3$ the condition $2E<h^*_3$ is $E<1/7$; in these ranges the hypotheses of \cref{lem:restore}(b) hold, which justifies the definition of $N_k(E)$ in \cref{sec:chains}. At $E=1/9$ the weak bound $2/9$ is a fixed point of $G_{2,2}$, and at $E=1/7$ the weak bound $2/7$ is a fixed point of $G_{2,3}$; so the strict inequalities cannot be made inclusive by this argument. For example $N_2(111/1000)=5$, $N_2(1111/10000)=7$, $N_3(71/500)=3$ and $N_3(1/7-10^{-5})=6$.

\subsection{The gates}

\begin{lemma}[gates]\label{lem:lowgates}
In each case of \cref{thm:lowdeg}, with $E$ in the stated range and $\theta$ as in the table of \cref{sec:chains}, let $r$ be a (D$^\circ$)-$E$ vector. Then~\eqref{eq:core} holds at every core bank, the NAND claims of \cref{lem:dHgates}(a) hold, and for $\mathrm{PURIFY}(u;v,w)$: if $X_u=0$ then $X_v=X_w=0$; if $X_u\ge1-E$ then $X_v,X_w\ge1-E$; if $X_u\ge\theta$ then $X_v\ge1-E$; if $X_u\le\theta$ then $X_w=0$.
\end{lemma}
\begin{proof}
\eqref{eq:core} holds by \cref{lem:core}(c). In all cases $E<1/4$, so $2(1-E)\ge A$, and we use:
\begin{itemize}
\item[(Sol$_2$)] a bank holding at least two contracts on a bank with default at least $1-E$ has default $0$, by (Sol);
\item[(Res$_0$)] a bank all of whose held contracts reference banks with default $0$ has default at least $1-E$, by (Res);
\end{itemize}
and~\eqref{eq:upbounds}. Write the arm to $b_v$ as $(s_1,\dots,s_L)$ and the arm to $b_w$ as $(t_1,\dots,t_L)$, so $s_L=b_v$, $t_L=b_w$.

\emph{Pure inputs.} Every stage after the first has multiplicity at least $2$. If $X_u=0$, then the first bank of each arm has default at least $1-E$ by (Res$_0$), and by (Sol$_2$) and (Res$_0$) alternately every bank at an even position has default $0$; so $X_v=X_w=0$. If $X_u\ge1-E$ and the first multiplicity is at least $2$, then the first bank has default $0$ by (Sol$_2$), and alternately every bank at an even position has default at least $1-E$. The only first multiplicity $1$ is that of the $w$-arm in case (c): there $S_{t_1}=X_u\ge1-E$, so $x_{t_1}\le2E$ by~\eqref{eq:upbounds}; \cref{lem:restore}(a) applied to the pairs $(t_{2j},t_{2j+1})$, of multiplicities $(2,2)$, gives $x_{t_{2j+1}}\le G_{2,2}^{\,j}(2E)$; so $x_{t_{2M+1}}=0$ and $X_w=x_{t_{2M+2}}\ge1-E$ by (Res$_0$).

\emph{Thresholds, case (a).} Let $X_u\ge\theta$. Then $S_{s_1}=4X_u\ge4(1-E)/3\ge1$ (as $E\le1/4$), so $x_{s_1}\le E$ by~\eqref{eq:upbounds}; $x_{s_2}\ge1-3E$ by (Res); $S_{s_3}\ge4(1-3E)\ge A$, which is equivalent to $E\le3/13$, so $x_{s_3}=0$; and $X_v\ge1-E$ by (Res$_0$). Let $X_u\le\theta$. Then $x_{t_1}\ge1-E-2X_u\ge(1-E)/3$ by (Res), so $S_{t_2}\ge4(1-E)/3\ge1$ and $x_{t_2}\le E$; then $x_{t_3}\ge1-3E$, $S_{t_4}\ge A$ and $X_w=0$.

\emph{Case (b).} Let $X_u\ge\theta$. Then $S_{s_1}=3X_u\ge1-E$, so $x_{s_1}\le2E$; \cref{lem:restore}(a) on the pairs $(s_{2j},s_{2j+1})$, of multiplicities $(2,3)$, gives $x_{s_{2j+1}}\le G_{2,3}^{\,j}(2E)$, so $x_{s_{2N+1}}=0$ and $X_v=x_{s_{2N+2}}\ge1-E$ by (Res$_0$). Let $X_u\le\theta$. Then $x_{t_1}\ge1-E-2\theta=(1-E)/3$, $S_{t_2}=3x_{t_1}\ge1-E$ and $x_{t_2}\le2E$; the pairs $(t_{2j+1},t_{2j+2})$ give $x_{t_{2j+2}}\le G_{2,3}^{\,j}(2E)$, so $X_w=x_{t_{2N+2}}=0$.

\emph{Case (c).} The same argument with multiplicities $(2,2)$ and $M$ in place of $N$: $X_u\ge\theta$ gives $S_{s_1}=2X_u\ge1-E$, and $X_u\le\theta$ gives $x_{t_1}\ge1-E-X_u\ge(1-E)/2$ and $S_{t_2}=2x_{t_1}\ge1-E$.

\emph{NAND.} Case (a) is \cref{lem:dHgates}(a) with $m=2$. In cases (b) and (c), if $X_u=X_v=0$ then $S_g=0$ and $x_g\ge1-E$. In case (b), $x_p=0$ by (Sol$_2$), so $S_{b_w}=0$ and $X_w\ge1-E$; in case (c), the chain $(2)^{\times2M}$ from $g$ alternates as above and its output, at an even position, has default at least $1-E$. If instead $X_u\ge1-E$ (the case $X_v\ge1-E$ is symmetric), then $S_g\ge1-E$ and $x_g\le2E$ by~\eqref{eq:upbounds}. In case (b), $x_p\ge1-5E$ by (Res), so $S_{b_w}\ge X_u+x_p\ge2-6E\ge A$ because $E\le1/7$, and $X_w=0$. In case (c), \cref{lem:restore}(a) on consecutive pairs of the chain shows that its bank at position $2j$ has default at most $G_{2,2}^{\,j}(2E)$, so $X_w=0$ at position $2M$.
\end{proof}

In every case the claims hold for every input value $X_u\in[0,1]$, including inputs that decode to $\bot$, and for every (D$^\circ$)-$E$ vector, because each step is a pointwise implication of~\eqref{eq:core}. At $X_u=\theta$ both threshold claims apply, in each of the three cases, and they give two pure outputs, which a PURIFY gate with input $\bot$ permits. At $E=3/13$ the inequality $4(1-3E)\ge A$ holds with equality, and the inclusive solvency clause gives the required zero.

\subsection{Proof of the low-degree theorem}

\begin{proof}[Proof of \cref{thm:lowdeg}]
Decode as in \cref{thm:dH}. By \cref{lem:lowgates} the NAND constraint holds, pure PURIFY inputs are copied to both outputs, and a $\bot$ input has $X_u\ge\theta$ (then $v$ decodes to $0$) or $X_u\le\theta$ (then $w$ decodes to $1$). For fixed $E$ the arm lengths are constants, so the network has size linear in the instance. The rest of the proof of \cref{thm:dH} (membership, the extension to (D$^\circ$) and (S), and \cref{lem:Rmono} for larger $d$) applies verbatim.

\emph{Held degree $1$.} In $\mathcal R(1)$ a core bank $b$ holds no contract or one contract, on a core bank $k(b)$. Put $x_b=1$ if $b$ holds nothing, $x_b=1/2$ if $b$ lies on a cycle of the map $b\mapsto k(b)$, and otherwise $x_b=1-x_{k(b)}$, evaluated from the end of the path $b,k(b),k(k(b)),\dots$, which reaches a bank holding nothing or a cycle. This takes linear time, and every core bank satisfies $x_b=T_b(x)$: for a bank holding nothing $S_b=0$, and otherwise $S_b=x_{k(b)}\in[0,1]$ and $x_b=1-S_b$. Moreover $S_b\ge1$ forces $x_b=0$. So, with recovery $1$ at writers and creditor, $r=1-x$ satisfies~\eqref{eq:core} with $E=0$, and \cref{lem:core}(b), which holds for $E=0$, shows that $r$ is a (D)-$0$ vector, that is, an exact clearing vector satisfying the solvency clause; hence it is a (D)-$E$ vector for every $E\ge0$ (\cref{lem:relations}(b)), and therefore a (D$^\circ$)-$E$ vector (\cref{lem:relations}(a)). It is also an (S)-$E$ vector: if $\ell_i(r)>0$ then $r_i=\min(1,A_i/\ell_i)$ lies between $\cl{A_i/\ell_i-E}$ and $\cl{A_i/\ell_i+E}$. The values lie in $\{0,1/2,1\}$. This is also the case $a=Q=1$, $E=0$ of \cref{thm:weightedD}.
\end{proof}

\section{Method boundaries for the restricted class}\label[appendix]{app:limits}

The statements of this appendix are boundaries of specific methods: a gadget family with a fixed decoding, or a propagation rule with a fixed decoder. They are not complexity lower bounds and do not exclude other gadgets, decodings, reductions or algorithms. Notation is that of \cref{sec:Rclass,sec:degree,app:chains}; in particular \emph{the decoding of \cref{thm:dH}} reads a default $0$ as bit $1$, a default at least $1-E$ as bit $0$, and every other default as $\bot$.

\subsection{Two-arm PURIFY gadgets}

\begin{proposition}[method boundary: two-arm gadgets]\label{prop:twoarm}
Consider a PURIFY gadget with input bank $u$ and two disjoint \emph{arms}: in arm $j\in\{1,2\}$ a bank $y_j$ holds $f_j$ unit CDSs on $u$ and the output $o_j$ holds $s_j$ unit CDSs on $y_j$ ($f_j,s_j$ positive integers). Use the decoding of \cref{thm:dH}. The PURIFY constraint holds for every input value $X_u\in[0,1]$ and every choice of values satisfying~\eqref{eq:core} at $y_1,o_1,y_2,o_2$ if and only if
\begin{equation}\label{eq:twoarm}
f_1,s_1,f_2,s_2\ge\rho\quad\text{and}\quad \frac{f_i}{f_j}+\frac1{s_i}\le\frac1\rho\ \text{ for some ordered pair }(i,j)\text{ with }\{i,j\}=\{1,2\}.
\end{equation}
Every such gadget has held degree $\max\{f_1,s_1,f_2,s_2\}\ge(m+1)\rho$, and the design of \cref{thm:dH}, $(f_1,s_1,f_2,s_2)=(D,m,m,D)$, attains $\lceil(m+1)\rho\rceil$.
\end{proposition}
\begin{proof}
Values satisfying~\eqref{eq:core} can be chosen bank by bank in the order $y_j$ then $o_j$, since each bank's assets depend only on earlier choices. We use the admissible intervals $\mathrm{Adm}(S)$, which equal $[\mathrm{lo}(S),\mathrm{up}(S)]$ with $\mathrm{up}(S)\ge E$ for $S<A$, and $\{0\}$ for $S\ge A$.

\emph{Necessity of $f_j,s_j\ge\rho$.} Suppose $f_j<\rho$. Take $X_u=1-E$ (decoded $0$). Then $S_{y_j}=f_j(1-E)<A$, and we may choose $x_{y_j}=\mathrm{up}(S_{y_j})\ge E>0$. Then $S_{o_j}>0$. If $S_{o_j}\ge A$, then $X_{o_j}=0$, decoded $1$. Otherwise we may choose $X_{o_j}=\mathrm{lo}(S_{o_j})<1-E$, decoded $1$ or $\bot$. Either way $o_j$ is not decoded $0$, violating PURIFY. Suppose $s_j<\rho$. Take $X_u=0$ (decoded $1$). Then $S_{y_j}=0$ and we may choose $x_{y_j}=1-E$. Then $S_{o_j}=s_j(1-E)\in(0,A)$, and we may choose $X_{o_j}=\mathrm{up}(S_{o_j})\ge E>0$, so $o_j$ is not decoded $1$.

\emph{The input $\bot$.} For arm $j$ put $H_j=A/f_j$ and $L_j=(1-E-A/s_j)/f_j$; note $L_j<H_j$ because $1-E<A$. If $X_u\ge H_j$, then $y_j$ has assets at least $A$, so $x_{y_j}=0$ and $X_{o_j}\ge1-E$: $o_j$ is pure. If $X_u\le L_j$, then $x_{y_j}\ge1-E-f_jX_u\ge A/s_j$ by (Res), so $o_j$ has assets at least $A$ and $X_{o_j}=0$: pure. Conversely let $X_u\in(L_j,H_j)$ with $X_u>0$. Then $S_{y_j}=f_jX_u\in(0,A)$ and $\mathrm{lo}(S_{y_j})<A/s_j$. Choose $x_{y_j}=\mathrm{lo}(S_{y_j})$ if this is positive, and otherwise $x_{y_j}=\tfrac12\min\{\mathrm{up}(S_{y_j}),A/s_j\}$; in both cases $x_{y_j}$ is admissible and $0<s_jx_{y_j}<A$. Then $S_{o_j}\in(0,A)$, so $\mathrm{lo}(S_{o_j})<1-E$ and $\mathrm{up}(S_{o_j})>0$. Choose $X_{o_j}=\mathrm{lo}(S_{o_j})$ if this is positive, and otherwise $X_{o_j}=\tfrac12\min\{\mathrm{up}(S_{o_j}),1-E\}$; then $0<X_{o_j}<1-E$, so $o_j$ decodes to $\bot$. The two arms share only $u$, so these choices combine. Hence the PURIFY constraint holds for all $\bot$ inputs if and only if every $X\in(0,1-E)$ satisfies $X\le\max(L_1,L_2)$ or $X\ge\min(H_1,H_2)$. If $\max(L_1,L_2)<\min(H_1,H_2)$, the interval $(\max(L_1,L_2,0),\min(H_1,H_2,1-E))$ consists of such bad inputs, and it is nonempty because $L_j<1-E$ and $H_j>0$ for both $j$. So the condition is $\max_iL_i\ge\min_jH_j$. As $L_i<H_i$, this holds if and only if $L_i\ge H_j$ for some $i\ne j$, that is, $1-E-A/s_i\ge Af_i/f_j$, which is the second part of~\eqref{eq:twoarm} after division by $A$.

\emph{Sufficiency.} With the first part of~\eqref{eq:twoarm}, the necessity arguments reverse: $X_u=0$ gives $x_{y_j}\ge1-E$ and $S_{o_j}\ge s_j(1-E)\ge A$; $X_u\ge1-E$ gives $S_{y_j}\ge A$, $x_{y_j}=0$ and $X_{o_j}\ge1-E$. With the second part, $L_i\ge H_j$ covers every $\bot$ input as shown.

\emph{Degree.} Since $f_i\ge\rho$ is an integer, $f_i\ge m$, so~\eqref{eq:twoarm} gives $(m+1)/\max\{f_j,s_i\}\le m/f_j+1/s_i\le1/\rho$. For $(D,m,m,D)$ with $(i,j)=(2,1)$, \eqref{eq:twoarm} is~\eqref{eq:dHsplit}.
\end{proof}

\begin{remark}[exact checks]\label{rem:twoarmcheck}\CO\
The exact checker \texttt{gadget\_checks.py} of the supporting archive (\cref{app:verification}) confirms the gadget claims of \cref{thm:dH} at eleven rational errors from $1/10$ to $3/4$, by an exhaustive case split of~\eqref{eq:core} and exact Fourier--Motzkin elimination, and finds explicit bad assignments for multiplicities $(D-1,m,m,D-1)$, as \cref{prop:twoarm} predicts. At $E=1/3$, replacing $D=6$ by $5$ fails at the input $X_u=7/30$: the $(5,2)$ arm admits the defaults $(1/6,1/3)$ and the $(2,5)$ arm admits $(1/5,1/6)$, so both outputs are $\bot$.
\end{remark}

\subsection{Chain PURIFY gadgets}

Call a chain PURIFY gadget (\cref{def:chain}) \emph{correct at $E$} if every assignment of $u$ and of the arm banks that satisfies~\eqref{eq:core} at the arm banks obeys the PURIFY constraint under the decoding of \cref{thm:dH}.

\begin{proposition}[method boundary: chain gadgets]\label{prop:chainlimits}
\begin{enumerate}[label=(\alph*)]
\item For $1/7\le E<1/3$, no chain PURIFY gadget with all multiplicities in $\{2,3\}$ is correct.
\item For $3/13<E\le1/4$, none with all multiplicities in $\{2,3,4\}$ is correct.
\item For $E\ge1/4$ and $m=\lceil\rho\rceil$, none with all multiplicities in $\{m,m+1,\dots,\lceil(m+1)\rho\rceil-1\}$ is correct. For $1/4<E<1/3$ this set is $\{2,3,4,5\}$.
\item If $X_u=1/3$, an arm all of whose multiplicities equal $2$ admits default $1/3$ at every bank. Hence no chain PURIFY gadget with all multiplicities equal to $2$ is correct at any $E\in(0,2/3)$.
\item Let a chain arm contain a multiplicity $k<\rho$, and assume one of: (i) $E\ge1/5$; (ii) $1/9\le E<1/5$ and all multiplicities lie in $\{1,2\}$; (iii) $1/7\le E<1/5$ and all multiplicities lie in $\{1,2,3\}$. (For $E<1/3$, $k<\rho$ means $k=1$.) Then for $X_u=0$ or for $X_u=1-E$ some assignment of the arm banks satisfying~\eqref{eq:core} makes the output decode to $\bot$. So the arm belongs to no correct chain PURIFY gadget.
\end{enumerate}
\end{proposition}
\begin{proof}
For (a)--(c) we give a set $C\subset(0,1-E)$ and $X\in C$ such that $\mathrm{Adm}(ky)\cap C\neq\emptyset$ for every $y\in C$ and every allowed multiplicity $k$. Setting $X_u=X$ and choosing values in $C$ bank by bank along each arm gives an assignment satisfying~\eqref{eq:core} at every arm bank in which $u$, $v$ and $w$ all decode to $\bot$.

(a) $C=\{c\}$ with $c\in[(1-E)/3,(1+E)/4]$, nonempty if and only if $E\ge1/7$. Here $2c<1$ and $3c<1$ (as $E<1/3$), and for $kc<1$ one has $c\in\mathrm{Adm}(kc)$ if and only if $|(k+1)c-1|\le E$, which holds for $k=2,3$.

(b) $C=\{E,1-3E\}$ and $X=E$. From $y=E$: $\mathrm{Adm}(2E)=[1-3E,1-E]\ni1-3E$; $\mathrm{Adm}(3E)=[\max(0,1-4E),1-2E]\ni E$ since $E\ge1/5$; $\mathrm{Adm}(4E)\ni E$ (it is $[\max(0,1-5E),1-3E]$ if $4E<1$, and $[0,E]$ if $4E=1$). From $y=1-3E$: $\mathrm{Adm}(2-6E)=[\max(0,5E-1),7E-1]\ni E$ since $1/6\le E\le1/4$; $\mathrm{Adm}(3-9E)=[\max(0,8E-2),10E-2]\ni E$ since $2/9\le E\le2/7$; and $4-12E\in[1,A)$, which is exactly $3/13<E\le1/4$, so $\mathrm{Adm}(4-12E)=[0,E]$. Both elements of $C$ lie in $(0,1-E)$.

(c) $C=\{c\}$ with $c=(1-E)/(m+1)\in(0,1-E)$. First, $c\le E$: for $m=2$ (that is, $E\le1/3$) this is $E\ge1/4$, and for $m\ge3$ we have $E>1/3$ and $c\le(1-E)/4<E$. For $k=m$: $mc<1$ and $(m+1)c-1=-E$, so $c\in\mathrm{Adm}(mc)$. Let $m<k\le\lceil(m+1)\rho\rceil-1$, so $k<(m+1)\rho$ and $kc<(m+1)\rho c=\rho(1-E)=A$. If $kc\ge1$, then $\mathrm{Adm}(kc)=[0,E]\ni c$. If $kc<1$, then $-E=(m+1)c-1<(k+1)c-1<c\le E$, so $c\in\mathrm{Adm}(kc)$.

(d) The value $1/3$ has assets $2/3$ and residual $|1/3-(1-2/3)|=0$; for $E<2/3$ it decodes to $\bot$, and so do $u$ and both outputs.

(e) By \cref{lem:envelopes}(b), for a fixed input the set of possible defaults of $s_i$ is the interval obtained from that of $s_{i-1}$ by the map $[l,h]\mapsto[\mathrm{lo}(k_ih),\mathrm{up}(k_il)]$. Since $\mathrm{lo}(k)=0$ for $k\ge1$ and $\mathrm{up}(0)=1$, this map sends $[v,1]$ to $[0,\mathrm{up}(k_iv)]$ and $[0,w]$ to $[\mathrm{lo}(k_iw),1]$. The input $X_u=0$ gives $[0,0]$ at position $0$. The input $X_u=1-E$ gives $\{1-E\}$, whose image under any stage equals that of $[1-E,1]$ (as $\mathrm{lo}(k(1-E))=0$), so we may use $[1-E,1]$ instead. Hence for each input the intervals alternate between the forms $[v,1]$ with $v\le1-E$ and $[0,w]$, and at every position the two inputs have opposite forms (the interval $[0,1]$ has both forms). Each input has the form $[0,w]$ at position $L$ or $L-1$; call the parameter $w$ at the last such position its \emph{final parameter}. If the final parameter of an input is positive, its output can decode to $\bot$: the output interval is either $[0,w]$ with $w>0$, or $[\mathrm{lo}(k_Lw),1]$ with $\mathrm{lo}(k_Lw)<1-E$, and both meet $(0,1-E)$. Finally, every stage maps $[0,1]$ to itself.

\emph{Case (i).} Let $j$ be the last position with $k_j<\rho$, and let $R$ be the input whose interval at position $j-1$ has the form $[v,1]$. At position $j$, $R$ has $[0,w]$ with $w=\mathrm{up}(k_jv)\ge\mathrm{up}(k_j(1-E))$, and every later multiplicity $k$ satisfies $k\ge\rho$. If $E<1/3$, then $\rho<2$, so $k_j=1$, $w\ge\mathrm{up}(1-E)=2E$ and every later $k$ is at least $2$; hence $kw\ge4E\ge1-E$. If $E\ge1/3$, then $k_j(1-E)<\rho(1-E)=A$, so $w\ge E$, and every later $k$ satisfies $kw\ge\rho E\ge1-E$; the last inequality is $E(1+E)\ge(1-E)^2$, that is, $E\ge1/3$. In both cases the next stage, if any, maps $[0,w]$ to $[\mathrm{lo}(kw),1]=[0,1]$. So $R$ ends with $[0,w]$, $w>0$, if $j=L$, and with $[0,1]$ otherwise; both meet $(0,1-E)$.

\emph{Cases (ii) and (iii).} Here every multiplicity $1$ is below $\rho$ and every larger multiplicity is at least $2>\rho$. A stage of multiplicity $1$ maps $[v,1]$ to $[0,\mathrm{up}(v)]$ with $\mathrm{up}(v)\ge\mathrm{up}(1-E)=2E$. Let $j$ be the last position of multiplicity $1$, let $\delta=2$ in (ii) and $\delta=3$ in (iii), and let $R$ be the input whose interval at position $j-1$ has the form $[v,1]$. At position $j$, $R$ has $[0,w_j]$ with $w_j\ge2E$. All later multiplicities lie in $\{2,\dots,\delta\}$. If $R$ has $[0,w]$ at some position and the next two multiplicities are $a,b$, it has $[0,\mathrm{up}(b\,\mathrm{lo}(aw))]$ two positions later, and $\mathrm{up}(b\,\mathrm{lo}(aw))\ge\mathrm{up}(\delta\,\mathrm{lo}(2w))=G_{2,\delta}(w)$ by monotonicity. The piecewise form of $G_{2,\delta}$ derived in the proof of \cref{lem:restore}(b) uses no hypothesis on $E$. If $\delta=3$, or $\delta=2$ and $E<1/6$, then $z_\delta<h^*_\delta$ (this is $E<2/9$, resp.\ $E<1/6$), so $h^*_\delta$ lies in the affine piece and $G_{2,\delta}(h^*_\delta)=h^*_\delta$; moreover $2E\ge h^*_\delta$, which is $E\ge1/7$, resp.\ $E\ge1/9$. As $G_{2,\delta}$ is nondecreasing, the parameters of $R$ at positions $j,j+2,\dots$ are all at least $h^*_\delta>0$. If $\delta=2$ and $1/6\le E<1/5$, then $z'=(1-3E)/4<E$ and $G_{2,2}(h)\ge E$ for every $h>z'$, so these parameters are all at least $E>0$. Either way the final parameter of $R$ is positive.
\end{proof}

\begin{corollary}[method boundary: exact limits of chain gadgets]\label{cor:chainlimits}
With the decoding of \cref{thm:dH}:
\begin{enumerate}[label=(\roman*)]
\item no chain PURIFY gadget is correct if all its multiplicities are at most $2$ and $1/9\le E\le1/3$; at most $3$ and $1/7\le E<1/3$; at most $4$ and $3/13<E\le1/4$; or at most $5$ and $1/4<E<1/3$;
\item for every $E\ge1/4$, no chain PURIFY gadget of held degree less than $d_H(E)$ is correct, so the degree of \cref{thm:dH} cannot be lowered by chain gadgets;
\item for every rational $E\in(0,1)$, a correct chain PURIFY gadget of held degree at most $2$, $3$, $4$ exists exactly when $E<1/9$, $E<1/7$, $E\le3/13$ respectively; \cref{thm:lowdeg} supplies one in each of these ranges.
\end{enumerate}
\end{corollary}
\begin{proof}
(i) An arm containing a multiplicity $1$ is excluded by \cref{prop:chainlimits}(e): part (ii) or (iii) when $E<1/5$, part (i) when $E\ge1/5$ (this covers the last two ranges, as $3/13>1/5$). If no arm contains a $1$, apply \cref{prop:chainlimits}(d), (a), (b) or (c) respectively.

(ii) An arm with a multiplicity below $\rho$ is excluded by \cref{prop:chainlimits}(e)(i). If all multiplicities are at least $\rho$, hence at least $m$, \cref{prop:chainlimits}(c) applies.

(iii) Existence in the stated ranges is \cref{lem:lowgates}, whose PURIFY claims use only~\eqref{eq:core} at the arm banks. Nonexistence for $E<1/3$ is (i); multiplicities at most $4$ are covered for $3/13<E\le1/4$ by its third range and for $1/4<E<1/3$ by its fourth. For $E\ge1/3$ use (ii): $d_H$ is nondecreasing in $E$, so $d_H(E)\ge d_H(1/3)=6>4$.
\end{proof}

These limits concern chain PURIFY gadgets with the decoding of \cref{thm:dH} only. PURIFY gadgets that are not chains (for example with a bank holding contracts on several different banks, as the NAND gadgets of \cref{thm:lowdeg} do, or with output paths that share banks), other decodings, and other reductions are not covered. Part (e) of \cref{prop:chainlimits} needs the exactness of \cref{lem:envelopes}(b), which uses that the banks of an arm are pairwise distinct.

\begin{remark}[the held-degree-$3$ NAND]\label{rem:nand3}
The NAND gadget of \cref{thm:lowdeg}(b) works exactly for $E\le1/7$. For $E>1/7$ take $X_u=1-E$, $X_v=0$, $x_g=\mathrm{up}(1-E)\in\mathrm{Adm}(1-E)$ (this is $2E$ for $E\le1/2$ and $1$ for $E>1/2$) and $x_p=\mathrm{lo}(2x_g)\in\mathrm{Adm}(2x_g)$. Then $S_{b_w}=1-E+x_p$, which equals $1-E+\max(0,1-5E)$ for $E\le1/2$ and $1-E$ for $E>1/2$; in both cases $S_{b_w}<A$ (for $E\le1/5$ this is $2-6E<1+E$), and since $S_{b_w}\ge1-E$, $\mathrm{lo}(S_{b_w})\le E\le\mathrm{up}(S_{b_w})$. So $X_w=E$ is admissible: the output does not decode to $1$ although $u$ decodes to $0$.
\end{remark}

\subsection{Propagation decoders}

\begin{proposition}[method boundary: each decoder is tight]\label{prop:decodertight}
The degree bounds of \cref{thm:Pk,thm:A2,thm:A3} are exactly where these decoders stop working.
\begin{enumerate}[label=(\alph*)]
\item For $k\ge1$ and every integer $d\ge(k+2)\rho$, let two core banks $p,q$ each hold $d$ unit CDSs on the other. Then $\mathcal P(k)$ labels both $\mathsf U$, the decoded vector of \cref{thm:Pk} violates the (D) clause at both, and $x_p=x_q=1/(d+1)$ is an exact clearing vector.
\item For $E>1/3$ and every integer $d\ge4\rho-1$, let core banks $c_0,\dots,c_{n-1}$ ($n\ge4$) each hold one unit CDS on each of the next three ($c_i$ on $c_{i+1},c_{i+2},c_{i+3}$, indices modulo $n$); let $y$ hold one CDS on $c_0$ and one on $c_1$; and let $\hat y$ hold one CDS on $y$ and $d-1$ on $c_0$. Then, in every order of rule applications, $\mathcal Q(2)$ leaves all $c_i$ and $\hat y$ unresolved and labels $y$ weak; rule A2 decodes $\hat y$ to $t$ although $S_{\hat y}=(d+1)t\ge A$; and changing only $x_{\hat y}$ to $0$ gives a (D)-$E$ vector.
\item For $E\ge1/4$ and every integer $d\ge3\rho-1$, the same holds for $\mathcal Q(1)$ and rule A3 with each $c_i$ holding one CDS on each of the next two banks ($n\ge3$), $y$ holding one CDS on $c_0$, and $\hat y$ holding one CDS on $y$ and $d-1$ on $c_0$.
\end{enumerate}
\end{proposition}
\begin{proof}
(a) Initially $n_{\mathsf H}=0$ and $n_{\neg\mathsf L}=d>k$ at both banks (as $(k+2)\rho>k+1$), so no rule applies and the engine stops at once. Both get $x=t$ and $S=dt\ge A$ because $d\ge(k+2)\rho$. With $x=1/(d+1)$ on both, $S=d/(d+1)<1$ and $T=1/(d+1)$.

(b) A bank $c_i$ references only banks $c_j$. Its counters change only after one of them is labelled, and initially no rule applies to it ($n_{\neg\mathsf L}=3>2$, $n_{\mathsf H}=n_{\mathsf S}=0$). So the first $c_i$ to be labelled would be labelled with unchanged counters, which is impossible; hence no $c_i$ is ever labelled. The bank $y$ has $n_{\neg\mathsf L}=2$ throughout and never an $\mathsf H$ reference, so (Q-L) never applies to it and, at termination, (Q-H) has been applied: $y$ is $\mathsf H$ with $n_{\neg\mathsf L}=2$, that is, weak. The bank $\hat y$ has $n_{\neg\mathsf L}=d\ge3$, $n_{\mathsf H}\le1$ and $n_{\mathsf S}=0$ throughout, so it stays $\mathsf U$. Its decoded assets are $2t+(d-1)t=(d+1)t$, and $(d+1)t\ge A=2-4t$ is equivalent to $d\ge4\rho-1$ (proof of \cref{thm:A2}). All other banks have held degree at most $3<4\rho-1$, so by the case analysis of \cref{thm:A2} they satisfy~\eqref{eq:core}. No bank references $\hat y$, so setting $x_{\hat y}=0$ changes only the constraint at $\hat y$, where $S_{\hat y}\ge A>1-E$ makes $x_{\hat y}=0$ admissible.

(c) Identical, with $n_{\neg\mathsf L}(c_i)=2>1$, $n_{\neg\mathsf L}(y)=1$, $n_{\neg\mathsf L}(\hat y)=d\ge2$, and $(d+1)t\ge2-3t$ equivalent to $d\ge3\rho-1$.
\end{proof}

So the failures of the decoders at their boundaries are not complexity lower bounds: every network in (a)--(c) has an easily found (D)-$E$ vector.

\begin{remark}[exact checks]\label{rem:decodercheck}\CO\
The program \texttt{algorithm\_tables.py} of the supporting archive enumerates every terminal local configuration of $\mathcal P(k)$ and of rules A2 and A3 up to the claimed degree at several rational errors, and the first failing degree of each decoder; the fixture \texttt{A2-boundary-d9} of \texttt{fixtures.py} realizes (b) at $E=2/5$, $d=9$ on the original contracts (decoded assets $3/2\ge7/5$; the repaired vector is a (D)-$2/5$ vector at every bank).
\end{remark}

\section{Weighted propagation for (D)}\label[appendix]{app:weighted}

The algorithms of \cref{sec:propagation} were analysed for unit notionals. This appendix records two weight-sensitive algorithms for (D) on a weighted extension of $\mathcal R(d)$. Their unit specializations are superseded by \cref{thm:Pk,thm:A2,thm:A3} except at held degrees $1$ and $2$, where \cref{thm:weightedD} gives the exact algorithm and (D)-$1/5$; the weighted statements themselves are not implied by the unit results. The supported propagation mechanism is that of~\cite[Lemma~F.8 and Theorem~F.9]{KM}.

\paragraph{The weighted core class.} There is a core $V$. A core bank $i$ has zero external assets, owes a fixed positive rational debt $L_i$ to an outside creditor, writes no CDS, and holds only CDSs written by outside writers; every reference is a core bank different from the holder. All outside writers satisfy $e_s\ge L^{\max}_s$, the creditor has positive external assets and no liabilities, and there are no other payments into the core. After dividing by $L_i$, a held CDS has a \emph{normalized notional} $\omega>0$. For given rationals $0<a\le Q$ let $\mathcal W(a,Q)$ be the class of such networks in which every normalized notional is at least $a$ and every core bank's total normalized held notional is at most $Q$; a promise of at most $d$ held contracts with normalized notionals in $[a,b]$ allows $Q=db$. With $x_i=1-r_i$ put
\[
S_i(x)=\sum_{(j,\omega)\in N_i}\omega\,x_j,\qquad T_i(x)=\max\{0,1-S_i(x)\},
\]
where $N_i$ is the list of held contracts of $i$, with their references $j$ and normalized notionals $\omega$, repetitions counted. The proof of \cref{lem:core}(a),(b) applies verbatim: a vector with recovery $1$ at all writers and at the creditor is a (D)-$E$ vector if and only if every core bank satisfies $|x_i-T_i(x)|\le E$ and $S_i(x)\ge1+E\Rightarrow x_i=0$. The engine $\mathcal P(k)$ of \cref{sec:propagation}, which counts held contracts, and \cref{lem:engines} apply verbatim to these networks; for $k=0$ rule (P-H) labels a bank $\mathsf H$ exactly when all its references are labelled $\mathsf L$, which includes banks that hold nothing.

\subsection{Coefficient-sensitive propagation}

For $0<a\le Q$ put
\[
R(a,Q)=\min\Bigl\{\frac1{a+2},\ \frac{Q-a}{Q+a+2}\Bigr\},\qquad \Delta(a,Q)=\frac{Q-a-1}{Q+a+1}.
\]

\begin{theorem}[weighted propagation]\label{thm:weightedD}
Let $0\le E<1$ and $0<a\le Q$ be rational. If
\[
E\ge1-a,\qquad E\ge R(a,Q)\qquad\text{and}\qquad E>\Delta(a,Q),
\]
then the following algorithm produces a (D)-$E$ vector of every network of $\mathcal W(a,Q)$: run $\mathcal P(0)$, give $\mathsf L$ banks $x=0$, $\mathsf H$ banks $x=1$ and $\mathsf U$ banks $x=t:=(1-E)/(1+a)$, and give all writers and the creditor recovery $1$. The propagation takes $O(|V|+M)$ operations on the $M$ listed contracts, and the rational arithmetic has polynomial bit complexity even when $a$, $Q$ and $E$ are part of the input.
\end{theorem}
\begin{proof}
By \cref{lem:engines} with $k=0$, at termination every $\mathsf L$ bank has a reference labelled $\mathsf H$; every $\mathsf H$ bank has only references labelled $\mathsf L$, possibly none; and every $\mathsf U$ bank has no reference labelled $\mathsf H$ and at least one reference labelled $\mathsf U$. Because $0\le E<1$ and $a>0$, $0<t<1$.

At an $\mathsf H$ bank every referenced default is $0$, so $T_i=1=x_i$. Its assets are $0$, so (D) does not demand $x_i=0$.

At an $\mathsf L$ bank some referenced default is $1$ with normalized notional at least $a$. Thus $0\le T_i\le\max(0,1-a)$, and the residual is at most $\max\{0,1-a\}\le E$, because $E\ge0$ and $E\ge1-a$. The (D) implication holds because $x_i=0$.

At a $\mathsf U$ bank all positive contributions come from references labelled $\mathsf U$. If $s$ is the sum of their normalized notionals, then $a\le s\le Q$ and $T_i=\max(0,1-st)$. The upper end of this target is at most $\max(0,1-at)$, and $1-at-t=1-(1+a)t=E$, so $T_i-x_i\le E$. For the other direction it suffices to show
\begin{equation}\label{eq:weightedC}
t-\max(0,1-Qt)\le E .
\end{equation}
If $E\ge1/(a+2)$, then $t\le E$ and~\eqref{eq:weightedC} follows. If $E\ge(Q-a)/(Q+a+2)$, then $(1+Q)t\le1+E$; if $Qt\le1$ this is precisely~\eqref{eq:weightedC}, and if $Qt>1$ it implies $t\le1+E-Qt<E$, which again gives~\eqref{eq:weightedC}. The hypothesis $E\ge R(a,Q)$ guarantees that one of the two alternatives holds. For (D), the strict hypothesis $E>\Delta(a,Q)$ is equivalent to $Q(1-E)/(1+a)<1+E$, so every $\mathsf U$ bank has assets $st\le Qt<1+E$ and the premise of (D) never holds there. Writers and creditor have residual $0$ at recovery $1$ and satisfy (D). This proves the (D)-$E$ property at every bank.

The algorithm uses exact rational comparisons, sums and the single rational number $t$; the number of terms in every sum and the sizes of the input numbers bound all intermediate bit lengths polynomially, and a small positive slack in $E>\Delta(a,Q)$ is tested exactly.
\end{proof}

\begin{proposition}[method boundary: common decode]\label{prop:weightedtight}
Consider algorithms that run $\mathcal P(0)$, use $x=0$ on $\mathsf L$ banks and $x=1$ on $\mathsf H$ banks, and choose one common real $t\in[0,1]$ for all $\mathsf U$ banks. For $0\le E<1$ there is a common $t$ for which such an algorithm produces a (D)-$E$ vector on every network of $\mathcal W(a,Q)$ if and only if the three conditions of \cref{thm:weightedD} hold, and then $t=(1-E)/(1+a)$ works. The same holds for the subclass with at most $d$ held contracts per core bank and normalized notionals in $[a,b]$, with $Q=db$ (for integers $d\ge1$ and rationals $0<a\le b$).
\end{proposition}
\begin{proof}
Sufficiency is \cref{thm:weightedD}. For necessity, the following three disjoint components can occur in one network of the class, with one funded writer for all CDSs and unit debts to the creditor.
\begin{enumerate}[label=(\arabic*)]
\item A bank $h$ that holds nothing and a bank $l$ holding one CDS of normalized notional $a$ on $h$. Propagation gives $x_h=1$, $x_l=0$, and the residual of $l$ is $\max\{0,1-a\}$.
\item Two banks referencing each other, each holding one CDS of normalized notional $a$ on the other. Both remain $\mathsf U$. The residual condition $t+E\ge T=\max(0,1-at)$ gives $(1+a)t\ge1-E$, also when $T=0$; so $t\ge t_-:=(1-E)/(1+a)$.
\item Two banks referencing each other, each holding one CDS of normalized notional $Q$ on the other (for the subclass, $d$ parallel CDSs of normalized notional $b$). Both remain $\mathsf U$. Since $E<1$, component (2) forces $t>0$, so (D) requires $Qt<1+E$; the residual requires $t-\max(0,1-Qt)\le E$.
\end{enumerate}
For $t\ge0$ the last inequality is equivalent to $t\le t_+:=\max\{E,(1+E)/(1+Q)\}$: if $Qt\le1$ it reads $(1+Q)t\le1+E$, if $Qt\ge1$ it reads $t\le E$, and the two ranges join at $t=1/Q$. Consequently a common $t$ forces $t_-\le t_+$ and $Qt_-<1+E$. The first says $E\ge1/(a+2)$ or $E\ge(Q-a)/(Q+a+2)$, which is $E\ge R(a,Q)$; the second is $E>\Delta(a,Q)$; and component (1) forces $E\ge1-a$.
\end{proof}

The endpoint $E=\Delta(a,Q)$ fails whenever it is the binding condition, because the component (3) then has assets exactly $1+E$ at the smallest feasible $t$, while its default $t$ is positive.

\paragraph{Consequences.} For unit notionals ($a=1$, $Q=d$): at $d=1$, $E=0$ is permitted, which gives an exact rational clearing vector satisfying the solvency clause ($t=1/2$); at $d=2$, $E=1/5$, which is~\cite[Theorem~F.9]{KM}; at $d=3$, $E=1/3$; and for $d\ge4$ every rational $E>(d-2)/(d+2)$. For $d\ge3$ these are superseded by \cref{thm:Pk,thm:A3}. For at most two held contracts with normalized notionals in $[1,B]$ ($a=1$, $Q=2B$) the theorem gives $E=(2B-1)/(2B+3)$ for $1\le B\le3/2$, $E=1/3$ for $3/2\le B<2$, and every rational $E>(B-1)/(B+1)$ for $B\ge2$. For every rational $a\in[4/5,1]$ and at most two held contracts with normalized notionals in $[a,(1+3a)/4]$, it gives (D)-$1/5$: here $Q=(1+3a)/2$, so $R(a,Q)\le(Q-a)/(Q+a+2)=1/5$, $1-a\le1/5$, and $t=4/(5(1+a))$ gives $Qt=2(1+3a)/(5(1+a))<6/5$. For $a=4/5$ these are the normalized notionals in $[4/5,17/20]$.

\subsection{Relaxed propagation with weights}

\begin{theorem}[weighted relaxed rule]\label{thm:weightedPk}
Let $k\ge1$ be an integer, and let $0<E<1$, $0<a\le b$ and $Q\ge0$ be rationals. Consider the networks of the weighted core class with all normalized notionals in $[a,b]$ and every core bank's total normalized held notional at most $Q$. Put $m_0=\min(1,a)$, $s=(k+1)a$,
\[
h=\frac{1-E}{m_0},\qquad t=\frac{1-E}{1+s},\qquad R(s,Q)=\min\Bigl\{\frac1{s+2},\frac{Q-s}{Q+s+2}\Bigr\}.
\]
Run $\mathcal P(k)$ and give $\mathsf L$ banks $x=0$, $\mathsf H$ banks $x=h$ and $\mathsf U$ banks $x=t$, with recovery $1$ at writers and creditor. The result is a (D)-$E$ vector if
\[
E\ge1-a,\qquad E\ge\frac{1+kb-m_0}{1+kb+m_0},\qquad E\ge R(s,Q),\qquad E>\frac{Q-s-1}{Q+s+1}.
\]
If $Q<s$, no $\mathsf U$ bank remains and the last two conditions can be omitted. The procedure has the combinatorial bound of \cref{lem:engines} and polynomial rational bit complexity.
\end{theorem}
\begin{proof}
The first condition gives $h\le1$, and $h\ge1-E$. Since $m_0\le a$, $ah\ge1-E$. Also $0<t\le h$, because $m_0\le1+s$.

At an $\mathsf L$ bank, one reference labelled $\mathsf H$ contributes at least $ah\ge1-E$, so $T\le E$, the residual of $x=0$ is at most $E$, and (D) holds.

At an $\mathsf H$ bank at most $k$ held contracts reference banks not labelled $\mathsf L$, each with normalized notional at most $b$ and default at most $h$, so the assets satisfy $0\le S\le kbh$. The inequality $(1+kb)h\le1+E$ is exactly the second condition. Since $T\le1$, $T-h\le1-h\le E$. If $S\le1$, then $h-T=h+S-1\le(1+kb)h-1\le E$; if $S>1$, then $kbh>1$ and $(1+kb)h\le1+E$ give $h<E$, so $h-T=h<E$. Moreover $kbh\le1+E-h<1+E$, so the premise of (D) does not hold.

At a $\mathsf U$ bank all positive contributions come from references labelled $\mathsf U$, at least $k+1$ held contracts, so their normalized notionals sum to some $q\in[s,Q]$ and $T=\max(0,1-qt)$. Since $(1+s)t=1-E$, $T-t\le E$. To show $t-T\le E$ it suffices that $t-\max(0,1-Qt)\le E$. This holds if $t\le E$, which is equivalent to $E\ge1/(s+2)$, or if $(1+Q)t\le1+E$, which is equivalent to $E\ge(Q-s)/(Q+s+2)$; in the latter case the clipped subcase $Qt>1$ implies $t<E$. This is the third condition. Finally $Qt<1+E$ is exactly the strict last condition, so the premise of (D) never holds at a $\mathsf U$ bank. If $Q<s$, a $\mathsf U$ bank would have $q\ge s>Q$, so none exists.
\end{proof}

For unit notionals ($a=b=1$, $Q=d$) the conditions become $E\ge k/(k+2)$ and $E>(d-k-2)/(d+k+2)$, the third being implied by the second; \cref{thm:Pk} refines this for $E\ge1/2$. A heterogeneous consequence: with $k=1$, at most five held contracts, normalized notionals in $[99/100,101/100]$ and $Q=101/20$, the theorem gives (D)-$17/50$, with $h=2/3$ and $t=33/149$; the second condition holds with equality and the strict last condition holds with $Qt=1.118\ldots<1.34$. The two weighted theorems are incomparable, and the better of their conditions may be used; for instance at held degree two with unit notionals \cref{thm:weightedD} gives (D)-$1/5$, while \cref{thm:weightedPk} needs $E\ge1/3$.

\section{Proofs for membership and stalled boxes}\label[appendix]{app:membership}

This appendix contains the proofs of the membership theorems of \cref{sec:membership} and of the structure results for stalled boxes of \cref{sec:stalls}, and the computational evidence on stalls.

\subsection{Membership of (W) and (I)}

\begin{proofof}{thm:member}
\begin{proof}
If $E\ge1$, use \cref{lem:relations}(c). Let $0<E<1$, $\eta=E/2$, $\mathrm{cap}(x)_i=\min(x_i,1-\eta)$ and $G=\Phi\circ\mathrm{cap}:[0,1]^n\to[0,1]^n$. Let $M$ be the sum of all external assets, debts and CDS notionals. If $D_i+\sum_kN_{ik}=0$ for every $i$, all liabilities vanish identically and the all-ones vector is exact. Otherwise let $\mu$ be the least positive value of $D_i+\sum_kN_{ik}$. Both $M$ and $\mu$ have polynomially many bits.

\emph{$G$ is Lipschitz.} Let $y,y'$ be capped points, $d=\|y-y'\|_\infty$. If $D_i+\sum_kN_{ik}=0$, then $\Phi_i\equiv1$. Otherwise $\ell_i(y)\ge D_i+\eta\sum_kN_{ik}\ge\eta\mu$, and likewise at $y'$. Further $|\ell_i(y)-\ell_i(y')|\le Md$, and
\[
A_i(y)-A_i(y')=\sum_j(y_j-y'_j)\ell_{ji}(y)+\sum_jy'_j\bigl(\ell_{ji}(y)-\ell_{ji}(y')\bigr)
\]
gives $|A_i(y)-A_i(y')|\le2Md$, since the coefficients in the first sum add up to at most $M$, and the second sum is at most $d$ times the total CDS notional held by $i$. By \cref{lem:clip}, $|\Phi_i(y)-\Phi_i(y')|\le3Md/(\eta\mu)$. Since $\mathrm{cap}$ is $1$-Lipschitz, $G$ is Lipschitz in the sup norm with constant $3M/(\eta\mu)$, a number with polynomially many bits, and $G(x)$ is computed exactly by rational arithmetic in polynomial time.

\emph{Path argument.} Weak approximation of the fixed points of a polynomially computable, polynomially continuous map of a polytope into itself is in PPAD~\cite[Proposition~2.2]{EY10} (numbered as in~\cite[Appendix~B]{DHK}; it is Proposition~2, part~2, in the preprint version of~\cite{EY10}); a map with a Lipschitz constant of polynomial bit length is polynomially continuous. So in PPAD we find a rational $x\in[0,1]^n$ with $\|G(x)-x\|_\infty<E/2$, and the path construction gives such an $x$ with polynomially many bits. Let $s=\mathrm{cap}(x)$. Then $\Phi(s)=G(x)$ and $\|x-s\|_\infty\le\eta$, so $\|\Phi(s)-s\|_\infty<E/2+\eta=E$: $s$ is a (W)-$E$ vector. Existence alone also follows from Brouwer's theorem applied to the continuous map $G$.

\emph{(I).} The set $T$ is computable. Fix $r_T=\mathbf 1$ and use only the coordinates $r_U$, $U=N\setminus T$. For $i\in U$, $\ell_i(r_U,\mathbf 1_T)=D_i+\sum_{k\in U}N_{ik}(1-r_k)$, since CDSs on banks of $T$ pay nothing. If $U$ is empty or every reduced row total $D_i+\sum_{k\in U}N_{ik}$ vanishes, the all-ones vector is an (I)-$E$ vector. Otherwise repeat the argument on $[0,1]^U$ with $\mu$ the least positive reduced row total; $M$ remains an upper bound. Output $(\mathrm{cap}(x_U),\mathbf 1_T)$; banks of $T$ have $\Phi_i\equiv1$.

\emph{Completeness.} Our hardness reductions produce non-degenerate networks, which form a subclass.
\end{proof}
\end{proofof}

\subsection{Empty-gap rounding}

\begin{proofof}{thm:gap}
\begin{proof}
For $E\ge1$, or if all contract notionals vanish, the all-ones vector works. So let $0<E<1$ and suppose some notional is positive. Let $n$ be the number of banks, $M$ the sum of all external assets and contract notionals, and $\mu$ the least positive \emph{individual} debt or CDS notional. Put
\[
q=\frac{E\mu}{32M},\qquad t_j=q^j\quad(0\le j\le n+1),\qquad \delta=\frac{E\mu q^{n+1}}{32}.
\]
Let $\mathcal N^\delta$ be the network with $\delta$ added to every external asset. It is non-degenerate, and its data have polynomial bit length. Make one query for a (D)-$(E/4)$ vector $r$ of $\mathcal N^\delta$; this problem is in PPAD by~\cite[Appendix~B]{DHK} (see the last paragraph of the proof).

\emph{Empty interval and snapping.} There are $n$ deficits $1-r_i$ but $n+1$ pairwise disjoint intervals $(t_{j+1},t_j]$, $j=0,\dots,n$, so one of them contains no deficit. Fix it, put $b=t_j$ and $a=t_{j+1}=qb$, and define $s_i=1$ if $1-r_i\le a$ and $s_i=r_i$ otherwise. Every unsnapped deficit exceeds $a$ and avoids $(a,b]$, so it exceeds $b$. Hence
\[
s\ge r,\qquad\|s-r\|_\infty\le a\le q\le E/32 .
\]
If $\ell_i(s)>0$, then either $i$ owes a positive debt, or some CDS written by $i$ with notional at least $\mu$ has an unsnapped reference $k$, with $1-s_k>b$; in both cases
\begin{equation}\label{eq:floor}
\ell_i(s)\ge\mu b .
\end{equation}
The empty interval gives at the same time a small perturbation and a lower bound on every remaining positive liability. With a single tolerance $a$ this fails: a surviving deficit could lie just above $a$, so that a liability of order $\mu a$ would have to absorb a perturbation of order $Ma$, and $M/\mu$ need not be small.

\emph{Transfer of the residual.} Write $A_i,L_i$ for the balances of the original network at $s$ and $B_i,K_i$ for the balances of $\mathcal N^\delta$ at $r$. Liabilities decrease in the recoveries, so $0\le K_i-L_i\le Ma$; and $|A_i(s)-A_i(r)|\le2Ma$ as in the proof of \cref{thm:member}, so $|A_i-B_i|\le2Ma+\delta$. For a bank with $L_i>0$ we have $K_i\ge L_i\ge\mu b$, and \cref{lem:clip} gives
\[
|\Phi_i(s)-\Phi^\delta_i(r)|\le\frac{\delta+3Ma}{\mu b}\le\frac E{32}+\frac{3E}{32}=\frac E8,
\]
because $Ma/(\mu b)=Mq/\mu=E/32$ and $\delta/(\mu b)=Eq^{n+1-j}/32\le E/32$ (as $j\le n$). Hence
\begin{equation}\label{eq:gapres}
|s_i-\Phi_i(s)|\le|s_i-r_i|+|r_i-\Phi^\delta_i(r)|+|\Phi^\delta_i(r)-\Phi_i(s)|\le\frac E{32}+\frac E4+\frac E8=\frac{13E}{32}<E .
\end{equation}

\emph{Transfer of the clause.} Let $L_i>0$ and $A_i\ge(1+E)L_i$. Since $B_i=A_i(r)+\delta\ge A_i-2Ma$ and $K_i\le L_i+Ma$, and $Ma\le\frac E{32}\mu b\le\frac E{32}L_i$,
\[
B_i-\bigl(1+\tfrac E4\bigr)K_i\ge\tfrac{3E}4L_i-\bigl(3+\tfrac E4\bigr)Ma\ge EL_i\Bigl(\frac34-\frac{3+E/4}{32}\Bigr)=EL_i\,\frac{84-E}{128}>0 .
\]
The (D) clause of the query forces $r_i=1$, so $s_i=1$. With~\eqref{eq:gapres} this proves that $s$ is a (D$^\circ$)-$E$ vector, for every valid answer of the query.

\emph{The (S) strengthening.} Let $L_i>0$. If $s_i<1$, then $s_i=r_i<1$, so the query clause gives $B_i/K_i<1+E/4$, and the query residual then gives $|r_i-B_i/K_i|\le E/2$. Writing $A_i/L_i-B_i/K_i=(A_i-B_i)/L_i+(B_i/K_i)(K_i-L_i)/L_i$,
\[
\Bigl|\frac{A_i}{L_i}-\frac{B_i}{K_i}\Bigr|\le\frac{\delta+(3+E/4)Ma}{\mu b}\le\frac{E(16+E)}{128}\le\frac{17E}{128},
\]
so $|s_i-A_i/L_i|\le81E/128<E$, and $s_i$ lies in the clipped interval $[\cl{A_i/L_i-E},\cl{A_i/L_i+E}]$. If $s_i=1$, then~\eqref{eq:gapres} gives $A_i/L_i\ge1-E$, which is what is needed for $1$ to lie in that interval.

\emph{Complexity.} If the input has length $B$, then $M$, $\mu$ and $q$ have bit length polynomial in $B$ and in the length of $E$, and $q^{n+1}$ multiplies that length by at most $n+1$. So $\delta$ may be extremely small, but it has polynomially many bits, and the query has polynomial length. The decoder performs $O(n^2)$ rational comparisons and changes some coordinates to $1$. This is a polynomial-time reduction with one query.

\emph{The query is in PPAD.} Put $e=E/4$ and, on $[0,1+e]^n$, $g_i(x)=A^\delta_i(\cl x)/\max\{A^\delta_i(\cl x)/(1+e),\ell_i(\cl x)\}$, where $\cl x$ is applied coordinatewise. The denominator is at least $\delta/(1+e)>0$, the map is polynomially computable and Lipschitz with a constant of polynomial bit length, so~\cite[Proposition~2.2]{EY10} gives a rational $x$ with $\|g(x)-x\|_\infty\le e$. Put $r=\cl x$. If $A^\delta_i(r)\ge(1+e)\ell_i(r)$, then $g_i(x)=1+e$, so $x_i\ge1$ and $r_i=1$. Otherwise $g_i(x)=A^\delta_i(r)/\ell_i(r)<1+e$ (here $\ell_i(r)>0$), and comparing the cases $x_i\le1$ and $x_i>1$ gives $|r_i-\Phi^\delta_i(r)|\le e$. This is the argument of~\cite[Appendix~B]{DHK} for the non-degenerate network $\mathcal N^\delta$.
\end{proof}
\end{proofof}

\subsection{Stalled boxes}

\begin{proofof}{prop:stall}
\begin{proof}
(a) By (H) and since every CDS held by $i$ is written by a bank of $F$, each $r_j$ with $j\in U$ enters the assets of $i$ only through terms $c(1-r_j)$. The interval bounds of $\BOX$ are therefore attained, $A_i^-$ at $r_U=hi$ and $A_i^+$ at $r_U=lo$, and the liabilities are constant.

(b) If $\Psi_i(hi)>1$, then $\lambda_i=\mu_i=1$ and $i$ would settle. So $\lambda_i=\Psi_i(hi)$, and since the new box $[\lambda_i+E,\mu_i-E]$ equals the old one, $\Psi_i(hi)=lo_i-E$ and $\min(1,\Psi_i(lo))=hi_i+E$; the latter gives $\Psi_i(lo)\ge hi_i+E$.

(c) $\Psi(lo)-\Psi(hi)=N(hi-lo)=Nw$, and by (b) this is at least $w+2E\mathbf 1$. No bank settles, so $w>0$, and the Collatz--Wielandt bound $\rho(N)\ge\min_i(Nw)_i/w_i>1$ applies to the nonnegative matrix $N$.

(d) By (b), $\tilde a=\Psi(hi)+N\,hi=lo-E\mathbf 1+N\,hi$, so $\tilde a-(I+N)\mathbf 1=(lo-(1+E)\mathbf 1)+N(hi-\mathbf 1)$. Here $lo\le(1-E)\mathbf 1$ and $hi\le\mathbf 1$, so the first term is at most $-2E\mathbf 1$ and the second is at most $0$.

(e) From $\tilde a=lo-E\mathbf 1+N\,hi$ we get $r^\#-lo=(I+N)^{-1}(Nw-E\mathbf 1)=w-(I+N)^{-1}(w+E\mathbf 1)$, which gives the first equivalence. Every point of $P$ has coordinates at most $1-E<1$, so at a point of $P$ the equations $\Phi_i=r_i$ say $\Psi_i(r_U)=r_i$; their only solution is $r^\#$.
\end{proof}
\end{proofof}

\begin{proofof}{prop:stallcases}
\begin{proof}
(a) Write $N=\begin{pmatrix}0&a\\b&0\end{pmatrix}$. By \cref{prop:stall}(c), $aw_2\ge w_1+2E$ and $bw_1\ge w_2+2E$, so $ab\,w_1w_2>w_1w_2$ and $ab>1$; hence $\det(I+N)=1-ab\ne0$. With $v=(I+N)^{-1}(w+E\mathbf 1)$,
\[
v_1=\frac{a(w_2+E)-(w_1+E)}{ab-1},\qquad w_1-v_1=\frac{a(bw_1-w_2-E)+E}{ab-1}.
\]
Both numerators are positive: $a(w_2+E)\ge w_1+2E+aE>w_1+E$, and $bw_1-w_2-E\ge E$. So $0<v_1<w_1$, and symmetrically $0<v_2<w_2$; \cref{prop:stall}(e) gives $r^\#\in P$.

(b) The roles of every CDS are distinct since $m\ge2$. Each $b_i$ has $\ell=1$ and $A=d(1-r_{i-1})$, and (H) holds. In round $1$ all boxes are $[E,1-E]$, so $\lambda_i=dE=(1-E)/2$ and $\mu_i=\min(1,d(1-E))=1$, because $(1-E)^2\ge4E>2E$ for $E\le3-2\sqrt2$; the width $(1+E)/2$ exceeds $2E$ since $E<1/3$, so no bank settles, and the new box is $[(1+E)/2,1-E]$, which has positive width since $E<1/3$. In round $2$, $\lambda_i=d(1-(1-E))=(1-E)/2$ again and $\mu_i=\min(1,d(1-E)/2)=\min(1,(1-E)^2/(4E))=1$, so the boxes are unchanged and $\BOX(E)$ stalls. Here $N=dS$ with $S$ the cyclic shift; the eigenvalues of $I+dS$ are $1+d\omega$ with $\omega^m=1$, and they do not vanish because $d>1$. The vector $\frac d{1+d}\mathbf 1=\frac{1-E}{1+E}\mathbf 1$ solves $r=d(\mathbf 1-Sr)$, so it is $r^\#$. It is at most $1-E$, and it is at least $(1+E)/2$ because $2(1-E)-(1+E)^2=1-4E-E^2>0$ for $E<\sqrt5-2=0.2360\ldots$, which holds since $E\le3-2\sqrt2$. No bank outside $F$ settled, so \cref{prop:affinerule} applies; directly, $\Phi_i=\min(1,d\cdot2E/(1+E))=(1-E)/(1+E)$.
\end{proof}
\end{proofof}

\begin{observation}[computational evidence]\label{obs:stalls}\CO\
The note~\cite[record~Lsolo]{Supp} reports a search over $456$ test networks in which $\BOX(E)$ stalled $28$ times, always at an inverter stall: on symmetric cycles with $m=2,\dots,8$, on asymmetric networks with several references, on figure-eight graphs and on critical two-bank networks. According to the note, $I+N$ was nonsingular at all $28$ stalls, $r^\#$ lay in the stalled box and had residual $0$, the identities of \cref{prop:stall}(b) held exactly in rational arithmetic, and a floating-point linear-programming search found a point $r\in P$ with $\Psi(r)=r$ in every case; in separate tests, two-bank networks in which one bank holds a debt of the other never stalled, and adding such a debt holding, even of notional $0.05$, to a stalling cycle removed the stall. The programs of this search are named in the note but were not archived, so these numbers could not be re-checked (\cref{app:verification}). On a grid of $37\times37$ two-bank inverter networks at $E=4/25$, $35$ stalled, and $r^\#$ lay in $P$ at each of them, as \cref{prop:stallcases}(a) requires (\texttt{code/stall\_examples.py}).
\end{observation}

\section{Exact stalls, selectors and two obstructions}\label[appendix]{app:stalls}

This appendix collects results on the stalls of $\BOX$ that complement \cref{sec:stalls,sec:rounded,sec:selection}. None of them is used in the proofs of the main theorems. They are corrections and examples: they rule out some simple global rules, and they do not give hardness results.

\subsection{Selectors at exact inverter stalls}\label{sec:selectorcore}

Consider an inverter stall of $\BOX(E)$ as in \cref{sec:stalls}: unsettled set $U$, stalled box $P=\prod_{i\in U}[lo_i,hi_i]$, widths $w=hi-lo>0$, and, with the settled banks at their points and the banks of $F$ at $1$, the affine map $\Psi(r_U)=\tilde a-Nr_U$ with $N\ge0$, $N_{ii}=0$. Put $A=I+N$. A \emph{selector} is a point $r\in P$ at which every bank of $U$ has residual at most $E$; by \cref{lem:sound} and \cref{rem:exactbox} selectors exist, and together with the settled points and $1$ on $F$ every selector is an (I)-$E$ vector. Let $\mathrm{Sat}=\{i\in U:hi_i=1-E\}$.

\begin{proposition}[clip characterization]\label{prop:clipchar}
For $r\in P$ put $v=hi-r$. Then $r$ is a selector if and only if
\[
0\le v\le w,\qquad Av\ge w,\qquad v_i>0\ \Longrightarrow\ (Av)_i\le w_i+2E .
\]
\end{proposition}
\begin{proof}
By \cref{prop:stall}(b), $\Psi(hi)=lo-E\mathbf1$, so $\tilde a=lo-E\mathbf1+N\,hi$ and $r-\Psi(r)=w+E\mathbf1-Av$. Since $r_i\le1-E$, the inequality $r_i-\min(1,\Psi_i(r))\le E$ is equivalent to $r_i-\Psi_i(r)\le E$, that is, to $(Av)_i\ge w_i$. The other inequality, $\min(1,\Psi_i(r))-r_i\le E$, holds automatically if $r_i=1-E$; if $r_i<1-E$, then $r_i+E<1$ and it is equivalent to $\Psi_i(r)-r_i\le E$, that is, to $(Av)_i\le w_i+2E$. A coordinate with $v_i>0$ has $r_i<hi_i\le1-E$, so it is in the latter case. If $v_i=0$ and $hi_i<1-E$, then \cref{prop:stall}(a) and the stall condition $hi_i=\mu_i-E$ give $\min(1,\Psi_i(lo))=\mu_i=hi_i+E<1$, so $\Psi_i(lo)=hi_i+E$, so $(Nw)_i=\Psi_i(lo)-\Psi_i(hi)=w_i+2E$ and $(Av)_i=(Nv)_i\le(Nw)_i=w_i+2E$; if $v_i=0$ and $hi_i=1-E$, the inequality is automatic. This proves both directions.
\end{proof}

For $C\subseteq\mathrm{Sat}$ let
\[
\mathrm{LP}(C)=\bigl\{r\in P:\ r_i=1-E\ (i\in C),\ \ (Ar)_i\le\tilde a_i+E\ (i\in U),\ \ (Ar)_i\ge\tilde a_i-E\ (i\notin C)\bigr\},
\]
a bounded polyhedron with rational data, and let $\mathcal F=\{C\subseteq\mathrm{Sat}:\mathrm{LP}(C)\ne\emptyset\}$. Note that $(Ar)_i-\tilde a_i=r_i-\Psi_i(r)$.

\begin{proposition}[selectors and clip sets]\label{prop:clipsets}
\begin{enumerate}[label=(\alph*)]
\item The set of selectors is the union of the sets $\mathrm{LP}(C)$, $C\subseteq\mathrm{Sat}$. If $r$ is a selector, then $r\in\mathrm{LP}(C_r)$ for $C_r=\{i:\Psi_i(r)-r_i>E\}\subseteq\mathrm{Sat}$; a point of $\mathrm{LP}(C)$ need not have $C_r=C$.
\item Let $G(r)$ be the projection of $\Psi(r)$ onto $P$. Every fixed point of $G$ is a selector, and $G$ has a fixed point. The converse fails: not every selector is a fixed point of $G$.
\item For every selector $r$, $\mathrm{LP}(C_r)$ has a vertex, which is a selector whose bit length is polynomial in the encoding length of the configuration $(\tilde a,N,lo,hi,E)$.
\end{enumerate}
\end{proposition}
\begin{proof}
(a) Let $r\in\mathrm{LP}(C)$. For every $i$, $r_i-\min(1,\Psi_i)=\max(r_i-1,r_i-\Psi_i)\le E$. For $i\notin C$, $\min(1,\Psi_i)-r_i\le\Psi_i-r_i\le E$; for $i\in C$, $\min(1,\Psi_i)-r_i\le1-(1-E)=E$. So $r$ is a selector. Conversely let $r$ be a selector. If $\Psi_i(r)-r_i>E$, then $\min(1,\Psi_i)-r_i\le E$ forces $1-r_i\le E$, so $r_i=hi_i=1-E$ and $i\in\mathrm{Sat}$. If $\Psi_i(r)\le1$ then $r_i-\Psi_i\le E$ is the residual bound, and if $\Psi_i(r)>1$ then $r_i-\Psi_i<0$; so $(Ar)_i\le\tilde a_i+E$ for all $i$, and $(Ar)_i\ge\tilde a_i-E$ for $i\notin C_r$. A coordinate at its saturated top need not have $\Psi_i-r_i>E$, so $C_r$ can be a proper subset of $C$.

(b) $\Psi$ is antitone on $P$, so $\Psi_i(r)\ge\Psi_i(hi)=lo_i-E\ge0$, and as $hi_i<1$ the projections of $\Psi_i(r)$ and of $\Phi_i(r)=\min(1,\Psi_i(r))$ onto $[lo_i,hi_i]$ coincide. By \cref{rem:exactbox}, $\Phi_i(P)\subseteq[lo_i-E,hi_i+E]$, so at a fixed point $|r_i-\Phi_i(r)|\le E$. $G$ is a continuous self-map of $P$, so it has a fixed point. For the converse see \cref{ex:notfixed}.

(c) $\mathrm{LP}(C_r)$ contains $r$ and is bounded, so it has a vertex, which lies in $\mathrm{LP}(C_r)$ and is a selector by (a). A vertex is the unique solution of a nonsingular system of active constraints, so by Cramer's rule its bit length is polynomial in that of the data.
\end{proof}

Part (c) gives neither a polynomial way to find $C_r$ nor a bit bound in terms of the original network alone: before a stall, \cref{lem:bits} bounds the endpoint sizes only in terms of the number of rounds, which these results do not bound in terms of the network (\cref{prop:nostall} gives a run that never stalls). The rounded algorithm of \cref{sec:rounded} controls the endpoints but returns near-stalls, at which the identities used in \cref{prop:clipchar} need not hold.

\begin{example}[a selector that is not a projected fixed point]\label{ex:notfixed}
In \cref{ex:threecycle} at $E=4/25$, with stalled box $[12/25,21/25]^3$, the point $r=(7/10,7/10,7/10)$ has $\Psi(r)=(3/5,3/5,3/5)$ and residual $1/10\le E$, so it is a selector, but $G(r)=(3/5,3/5,3/5)\ne r$. In this example $\rho(N)=2$, while every proper principal minor of $I+N$ equals $1$ and $\det(I+N)=9$; so the expansion of \cref{prop:stall}(c) does not prevent $I+N$ from being a P-matrix. We make no claim about polynomial algorithms for linear complementarity problems with P-matrices.
\end{example}

\paragraph{A possibly failing search.} The following procedure, \Select{}, searches for a member of $\mathcal F$. Start with $C=\emptyset$ and repeat: if $\{r\in P:r_i=1-E\ (i\in C),\ (Ar)_i\le\tilde a_i+E\ (i\in U)\}$ is empty, stop with failure; otherwise solve the linear program
\[
\begin{aligned}
&\min\ \sum_{i\notin C}\xi_i\quad\text{subject to}\quad r\in P,\quad r_i=1-E\ (i\in C),\quad \xi\ge0,\\
&\qquad(Ar)_i\le\tilde a_i+E\ (i\in U),\qquad (Ar)_i\ge\tilde a_i-E-\xi_i\ (i\notin C).
\end{aligned}
\]
If the optimum is $0$, output the optimal $r$. Otherwise let $V=\{i\in\mathrm{Sat}\setminus C:\xi_i>0\}$ at the computed optimal solution; if $V=\emptyset$, stop with ``no certificate''; else replace $C$ by $C\cup V$.

\begin{proposition}\label{prop:select}
\Select{} runs at most $|\mathrm{Sat}|+1$ rounds, each with one feasibility test and at most one linear program, all of size polynomial in the configuration. Whenever it outputs a point $r$, $r\in\mathrm{LP}(C)$ for the final $C$, so $r$ is a selector, and this can be verified exactly by substitution.
\end{proposition}
\begin{proof}
$C$ grows strictly at every continuation and stays inside $\mathrm{Sat}$. An optimum $0$ means $\xi_i=0$ for every $i\notin C$, so the output lies in $\mathrm{LP}(C)$, and \cref{prop:clipsets}(a) applies.
\end{proof}

We do not know whether \Select{} always succeeds. If $\mathcal F$ were closed under intersections it would have a least element, which is a natural route to completeness; \cref{prop:nointersection} shows that this closure fails at an actually reached stall, while \Select{} still succeeds there.

\subsection{A reached stall without intersection closure}

\begin{proposition}\label{prop:nointersection}
Let $E=1/10$. Let $b_1,b_2,b_3$ have zero external assets and owe $1$ each to a sink $z$ with $e_z=1$ and no liabilities, and let one writer $s$, with no ordinary debt and $e_s=111/5=2L_s^{\max}+1$, write CDSs held by $b_i$ with reference $b_j$ of the notionals $N_{ij}$ given by
\[
N=\begin{pmatrix}0&13/5&3/2\\13/5&0&3/2\\6/5&6/5&0\end{pmatrix}.
\]
The network is non-degenerate and one-writer. $\BOX(1/10)$ reaches $P=[51/100,9/10]^2\times[17/50,9/10]$ in round $1$ and stalls in round $2$, with no settlement, all three coordinates saturated, and condition (H) of \cref{sec:stalls}. For this stall $\mathcal F=\{\{1\},\{2\},\{1,2\}\}$; in particular $\{1\},\{2\}\in\mathcal F$ but $\{1\}\cap\{2\}\notin\mathcal F$. \Select{} succeeds on this stall, whichever optimal solutions its linear programs return.
\end{proposition}
\begin{proof}
All roles are distinct, every $b_i$ owes a debt and $e_s,e_z>0$; the total notional is $53/5$, so $s\in T$. With $r_s=r_z=1$, each $b_i$ has liability $1$ and $\Psi=\tilde a-Nr$ with $\tilde a=N\mathbf1=(41/10,41/10,12/5)$. Initially all boxes are $[1/10,9/10]$. At the upper corner $\Psi=\tilde a/10=(41/100,41/100,6/25)$, and at the lower corner $\Psi=9\tilde a/10>\mathbf1$; so no bank settles, and the boxes become $P$. In round $2$ the upper corner is unchanged, and at the lower corner of $P$, $\Psi=(283/125,283/125,147/125)\ge\mathbf1$; so $P$ is unchanged and $\BOX$ stalls. Every $hi_i=9/10=1-E$.

Write $q_i=\Psi_i(r)-r_i$. Then $q_1+q_2=41/5-(18/5)(r_1+r_2)-3r_3=1+3q_3$ for every $r$. In $\mathrm{LP}(C)$ we have $q_i\ge-E$ for all $i$ and $q_i\le E$ for $i\notin C$.

$\emptyset\notin\mathcal F$: the identity would give $q_1+q_2\ge1-3/10=7/10$, while $q_1+q_2\le1/5$. $\{3\}\notin\mathcal F$: $q_1+q_2\le1/5$ forces $q_3\le-4/15<-E$. $\{1,3\}\notin\mathcal F$: with $r_1=r_3=9/10$, $q_3=21/50-(6/5)r_2\ge-1/10$ forces $r_2\le13/30<51/100$; $\{2,3\}$ is symmetric; and $\{1,2,3\}\notin\mathcal F$ because $r=(9/10,9/10,9/10)$ has $q_3=-33/50$. On the other hand $r^{(1)}=(9/10,51/100,4/5)\in P$ has $q=(337/500,1/20,-23/250)$, so $r^{(1)}\in\mathrm{LP}(\{1\})$; swapping the first two coordinates gives a point of $\mathrm{LP}(\{2\})$; and $(9/10,9/10,17/50)$ has $q=(7/20,7/20,-1/10)$, so it lies in $\mathrm{LP}(\{1,2\})$. At $r^{(1)}$ the actual residuals are $1/10$ (bank $1$ is clipped, $\Psi_1=787/500>1$), $1/20$ and $23/250$, so $r^{(1)}$ with $r_s=r_z=1$ is an (I)-$1/10$ vector.

\Select{}: the first feasibility test passes (for instance at $(3/4,3/4,7/10)$). In the first linear program $\xi_i\ge q_i-E$, so $\xi_1+\xi_2+\xi_3\ge1+3q_3-2E+\xi_3\ge1/2+\xi_3$, and $(3/4,3/4,7/10)$ with $\xi=(1/4,1/4,0)$ attains $1/2$. So every optimal solution has $\xi_3=0$ and $\xi_1+\xi_2=1/2$, and $V$ is $\{1\}$, $\{2\}$ or $\{1,2\}$. Each of these lies in $\mathcal F$, so in the next round the feasibility test passes and the optimum is $0$.
\end{proof}

The example comes from a family: for $c,d>1$, $b=2cd-1$ and $N=\begin{psmallmatrix}0&b&c\\b&0&c\\d&d&0\end{psmallmatrix}$ with $\tilde a=N\mathbf1$, the identity $q_1+q_2=2(c-1)+2cq_3$ holds, and the unconditional affine band $|q_i|\le E$ is infeasible whenever $c>(1+E)/(1-E)$. For every member of this family $\det(I+N)=0$. The error $1/10$ lies below $\alpha$, so the example says nothing about stalls inside the writer-solvent window.

\subsection{A run that never stalls exactly}

\begin{proposition}\label{prop:nostall}
Let $E=17/100$. Let $b_1,b_2$ have zero external assets, each owe the ordinary debt $1/100$ to the other and $99/100$ to a sink with external assets $1$ and no liabilities, and each hold a CDS of notional $23/10$ on the other, written by one writer $s$ with $e_s=2L_s^{\max}+1$. The network is non-degenerate and one-writer. $\BOX(17/100)$ never settles $b_1$ or $b_2$ and never stalls: after $t\ge0$ rounds both boxes are
\[
\Bigl[\frac{17}{30}-\frac{119}{300}\,100^{-t},\ \frac{83}{100}\Bigr].
\]
The vector with $r_{b_1}=r_{b_2}=230/329$ and $r_s=r_z=1$ is an exact clearing vector, and it lies in all these boxes.
\end{proposition}
\begin{proof}
Put $c=1/100$, $d=23/10$, $h=83/100$ and $l^\ast=17/30$. Each $b_i$ has liability $1$ and assets $cr_j+d(1-r_j)$, $j\ne i$. While the common lower endpoint satisfies $E\le l<l^\ast$ and the common upper endpoint is $h$, the raw enclosures are $cl+dE$ and $ch+d(1-l)\ge83/10000+(23/10)(13/30)=30149/30000>1$, so $\lambda=cl+dE$ and $\mu=1$. So $\lambda\le cl^\ast+dE=119/300=l^\ast-E$ and $\mu-\lambda\ge181/300>2E$: no bank settles, the upper endpoint stays $\mu-E=h$, and the lower endpoint becomes $cl+dE+E=l/100+561/1000$. This affine map has the fixed point $l^\ast$, and the initial lower endpoint is $E=l^\ast-119/300$, which gives the formula; it increases strictly with $t$ and stays below $l^\ast$, so no round leaves the boxes unchanged. At the symmetric point the asset ratio is $d-(d-c)r$, with fixed point $d/(1+d-c)=230/329$, which lies strictly between $17/30$ and $83/100$ and is below $1$; the writer and the sink have $\Phi\equiv1$.
\end{proof}

So a complete method cannot wait for an exact stall before invoking a global rule, even when the remaining equations are easy. The rounded algorithm of \cref{sec:rounded} always stops within $nM+1$ rounds, with every bank settled or at a near-stall.

\subsection{An obstruction to affine-band repair}

\begin{proposition}\label{prop:affineband}
Let $E=17/100$. Let $b_1,b_2,b_3$ have zero external assets and owe $1$ each to a sink $z$ with $e_z=1$ and no liabilities, and let one writer $s$, with no ordinary debt and $e_s=2L_s^{\max}+1=439/25$, write four CDSs: $b_1$ holds one on $b_2$ and $b_2$ one on $b_1$, of notional $11/5$, and $b_3$ holds one on $b_1$ and one on $b_2$, of notional $97/50$. The network is non-degenerate and one-writer. $\BOX(17/100)$ reaches $P=[68/125,83/100]^2\times[1037/1250,83/100]$ in round $1$ and stalls in round $2$ without settling any bank. The box $P$ contains an (I)-$E$ vector, but no $r\in P$ satisfies $|\Psi_i(r)-r_i|\le E$ for $i=1,2,3$. The same holds with $E+\eta$ in place of $E$ in these affine inequalities, for every $0\le\eta<17/5900$.
\end{proposition}
\begin{proof}
With $r_s=r_z=1$, $\Psi_1=(11/5)(1-r_2)$, $\Psi_2=(11/5)(1-r_1)$ and $\Psi_3=(97/50)(2-r_1-r_2)$. Initially all boxes are $[17/100,83/100]$; the lower enclosures are $187/500$, $187/500$ and $1649/2500$, and all upper enclosures are $1$. All widths exceed $2E=17/50$ (for $b_3$ by $1/2500$), so no bank settles and the boxes become $P$. In round $2$ the lower enclosures are unchanged, and at the lower corner of $P$, $\Psi_1=\Psi_2=627/625>1$ and $\Psi_3=5529/3125>1$; so every upper enclosure is $1$ and $P$ is unchanged.

Suppose $r\in P$ satisfies $|\Psi_i(r)-r_i|\le E+\eta$ for all $i$. From $\Psi_1-r_1\ge-E-\eta$ and $\Psi_2-r_2\ge-E-\eta$ we get $r_1+\frac{11}5r_2\le\frac{11}5+E+\eta$ and $\frac{11}5r_1+r_2\le\frac{11}5+E+\eta$, whose sum gives $r_1+r_2\le\frac{237}{160}+\frac58\eta$. From $\Psi_3\le r_3+E+\eta\le1+\eta$ we get $r_1+r_2\ge\frac{144}{97}-\frac{50}{97}\eta$. Since $\frac{144}{97}-\frac{237}{160}=\frac{51}{15520}$, these conflict for $0\le\eta<\frac{17}{5900}$.

The point $r=(11/16,11/16,83/100)$ lies in $P$, and $\Psi(r)=(11/16,11/16,97/80)$, so $\Phi(r)=(11/16,11/16,1)$ and the residuals are $0,0,17/100$; with $r_s=r_z=1$ it is an (I)-$E$ vector. (It is not a (D)-$E$ vector, since $\Psi_3=97/80\ge1+E$ while $r_3<1$.)
\end{proof}

\Cref{ex:stallfail} shows that the exact affine solution of a stall can lie outside the box; \cref{prop:affineband} shows that even the weaker requirement of an affine residual at most $E$ can be infeasible on a reached stalled box, inside the writer-solvent window. In both cases the box contains an (I)-$E$ vector (\cref{rem:exactbox}, with $\delta=0$); in the case of \cref{prop:affineband} every such vector has a coordinate whose affine residual exceeds $E$, so that the cap at $1$ in $\Phi_i=\min(1,\Psi_i)$ is active there. This does not prevent a rule that identifies clipped coordinates, as in \cref{prop:clipchar}, or another global method.

\section{Selection at near-stalls: complete results}\label[appendix]{app:selection}

This appendix contains the normal form of a terminal configuration of $\BOX(E_0,M)$ and the selectors outlined in \cref{sec:selection}, with their proofs.

\subsection{Normal form and expansion at a terminal configuration}

\begin{lemma}[normal form at a terminal configuration]\label{lem:nearnormal}
Suppose every CDS writer lies in $F$, and let $\BOX(E_0,M)$ stop with a nonempty unsettled set $U$ and final boxes $[l_i,h_i]$, $i\in U$. Every $i\in U$ writes no CDS and has a positive constant liability $D_i$, and, with the settled banks at their points, its asset ratio is
\[
f_i(r_U)=a_i+\sum_{j\in U}P_{ij}r_j+\sum_{j\in U}N_{ij}(1-r_j),\qquad a_i,P_{ij},N_{ij}\ge0,\quad P_{ii}=N_{ii}=0,
\]
and $\Phi_i=\min(1,f_i)$, where $P_{ij}=c_{ji}/D_i$ and $N_{ij}$ is $1/D_i$ times the total notional of the CDSs held by $i$ with reference $j$. The enclosures computed on the final product are, coordinatewise on $U$, $\lambda=\underline f:=a+Pl+N(\mathbf1-h)$ and $\mu=\min(\mathbf1,\overline f)$ with $\overline f:=a+Ph+N(\mathbf1-l)$.
\end{lemma}
\begin{proof}
A bank of $U$ lies outside $F$, so it writes no CDS, and its liability is the constant $D_i=L_i^{\max}>e_i\ge0$. Its assets are $e_i$, the debts owed to it, and the payments of CDSs written by banks of $F$ at recovery $1$; fixing the settled banks gives the stated form, with $P_{ii}=0$ because $c_{ii}=0$ and $N_{ii}=0$ because the three roles of a CDS are distinct. The interval arithmetic of $\BOX$ evaluates each term at an endpoint, so the raw enclosures are $\underline f$ and $\overline f$, with $\lambda=\min(\mathbf1,\underline f)$ and $\mu=\min(\mathbf1,\overline f)$ coordinatewise, even if debt and CDS terms in the same variable would cancel in $f_i$. By \cref{thm:roundedbox}(4), $\mu_i-\lambda_i>2E_0$ and $\mu_i\le1$, so $\lambda_i<1$ and $\lambda_i=\underline f_i$.
\end{proof}

\begin{proposition}[expansion at a near-stall]\label{prop:nearexpansion}
In the setting of \cref{lem:nearnormal}, let $w=h-l>0$. Then
\begin{equation}\label{eq:nearexp}
(P+N)w>w+2(E_0-\delta)\mathbf1 ,
\end{equation}
and if $\delta\le E_0$, then $\rho(P+N)>1$, while always $\rho(P)\le1$. So if $\delta\le E_0$, then in the directed graph on $U$ with an edge $j\to i$ whenever $P_{ij}+N_{ij}>0$, some strongly connected component contains an edge with $N_{ij}>0$, which lies on a directed cycle of that component. In particular ordinary debts alone cannot sustain a near-stall with $\delta\le E_0$.
\end{proposition}
\begin{proof}
By \cref{lem:nearnormal} and~\eqref{eq:nearstall}, $\overline f_i\ge\mu_i$, $\lambda_i=\underline f_i$ and
\[
\overline f_i-\underline f_i\ge\mu_i-\lambda_i>(h_i+E_0-\delta)-(l_i-E_0+\delta)=w_i+2(E_0-\delta).
\]
But $\overline f-\underline f=(P+N)w$, which proves~\eqref{eq:nearexp}. If $\delta\le E_0$, then $(P+N)w>w$, and the Collatz--Wielandt bound $\rho(P+N)\ge\min_i((P+N)w)_i/w_i>1$ follows. For the positive vector $(D_i)_{i\in U}$ we have $\sum_{i\in U}D_iP_{ij}=\sum_{i\in U}c_{ji}\le D_j$, so $(P^\top D)_j\le D_j$ for all $j$ and $\rho(P)=\rho(P^\top)\le1$. The spectral radius of a nonnegative matrix is the largest spectral radius of its diagonal blocks on strongly connected components. A block with spectral radius above $1$ cannot have $N_{BB}=0$, since then it would be a principal submatrix of $P$, of spectral radius at most $\rho(P)\le1$. So some component contains an edge with $N_{ij}>0$, and every edge inside a strongly connected component lies on a directed cycle of it.
\end{proof}

\Cref{prop:nearexpansion} is a necessary condition, not a certificate of hardness. It extends \cref{prop:stall}(c) from exact inverter stalls to rounded near-stalls with ordinary debts between unsettled banks, with the explicit defect $2(E_0-\delta)$, which is positive when $\delta<E_0$.

\subsection{Normalization of a near-stall}

Throughout this appendix every CDS writer lies in $F$, and $\BOX(E_0,M)$ of \cref{sec:rounded}, with rational working accuracy $0<E_0<1/2$ and grid step $\delta=(1-2E_0)/M$, has stopped with a nonempty unsettled set $U$, product $\mathcal B=\prod_{i\in U}[l_i,h_i]$ and settled banks fixed at their points. By \cref{lem:nearnormal}, every $i\in U$ has a positive constant liability and the asset ratio
\[
f_i(r)=a_i+\sum_{j\in U}P_{ij}r_j+\sum_{j\in U}N_{ij}(1-r_j),\qquad a_i,P_{ij},N_{ij}\ge0,\quad P_{ii}=N_{ii}=0,
\]
and $\Phi_i=\min(1,f_i)$. Put $b=a+N\mathbf1$ and
\[
A=I-P+N,\qquad g(r)=r-f(r)=Ar-b .
\]
The matrix $A$ is signed: ordinary debts between unsettled banks enter with a negative sign. The coefficients $a,P,N$ are nonnegative before this combination, and this is used below; an arbitrary affine map on a box would not do.

\begin{lemma}[coefficient bound]\label{lem:coefbound}
Put $C=1/E_0-1>1$. Then, for every $i\in U$,
\[
\sum_{j}(P_{ij}+N_{ij})<\frac1{E_0}-2,\qquad \sum_j|A_{ij}|\le C,\qquad |g_i(r)|\le C\ \text{ for every } r\in\mathcal B .
\]
These bounds remain valid for every principal submatrix of $A$ and every vector of $\mathcal B$, in particular after the coordinates of some components have been fixed at values in their intervals.
\end{lemma}
\begin{proof}
By \cref{thm:roundedbox}(4), $\mu_i-\lambda_i>2E_0$, and $\mu_i\le1$, so $\lambda_i<1-2E_0$; hence the raw lower enclosure $\underline f_i=a_i+\sum_jP_{ij}l_j+\sum_jN_{ij}(1-h_j)$ equals $\lambda_i$ and is below $1-2E_0$ (\cref{lem:nearnormal}). Since $l_j\ge E_0$, $1-h_j\ge E_0$ and all coefficients are nonnegative, $E_0\sum_j(P_{ij}+N_{ij})\le\underline f_i<1-2E_0$, which is the first bound. As $A_{ii}=1$ and $|N_{ij}-P_{ij}|\le N_{ij}+P_{ij}$, the second follows. Also $a_i\le\underline f_i<1-2E_0$, so for $r\in\mathcal B\subseteq[0,1]^U$ we get $0\le f_i(r)\le a_i+\sum_j(P_{ij}+N_{ij})<C$, and with $0\le r_i\le1<C$ this gives $|g_i(r)|=|r_i-f_i(r)|\le C$. Row sums of a principal submatrix are at most those of $A$, and the last bound holds at every vector of $\mathcal B$.
\end{proof}

Let $\Pi$ be the coordinatewise projection onto $\mathcal B$, and $G(r)=\Pi(f(r))=\Pi(r-g(r))$ for $r\in\mathcal B$.

\begin{lemma}[projection transfer]\label{lem:projtransfer}
For every $r\in\mathcal B$ and $i\in U$, $|G_i(r)-\Phi_i(r)|\le E_0$. Consequently, if $\|r-G(r)\|_\infty\le\eta$ for some $r\in\mathcal B$, then every unsettled bank has residual at most $E_0+\eta$ at $r$, every settled bank has residual at most $E_0+\delta/2$, and the banks of $F$ are at $1$; so $r$, completed by the settled points and by $1$ on $F$, is an (I)-$\max(E_0+\eta,E_0+\delta/2)$ vector, and hence a (W)-vector with the same error.
\end{lemma}
\begin{proof}
Since $h_i\le1-E_0<1$ and $f_i\ge0$ on $\mathcal B$, projecting $f_i(r)$ or $\Phi_i(r)=\min(1,f_i(r))$ onto $[l_i,h_i]$ gives the same point. By~\eqref{eq:roundinv} and \cref{thm:roundedbox}(1), $\Phi_i(\mathcal B)\subseteq[\lambda_i,\mu_i]\subseteq[l_i-E_0,h_i+E_0]$, and the projection of a number of $[l_i-E_0,h_i+E_0]$ onto $[l_i,h_i]$ moves it by at most $E_0$. The rest follows from \cref{thm:roundedbox}(2), because $r$ lies in the same product, and from $T\subseteq F$.
\end{proof}

No exact-stall identity is used: \cref{lem:coefbound,lem:projtransfer} hold at every terminal configuration of $\BOX(E_0,M)$, and settled banks are never reopened. A selector at target $E$ is therefore obtained from any method that finds $r\in\mathcal B$ with $\|r-G(r)\|_\infty\le E-E_0$, provided $E_0+\delta/2\le E$.

\subsection{A potential selector}

Let $\Omega=\mathrm{diag}(\omega_i)$ be a supplied diagonal matrix with positive rational entries; its scale is immaterial. Put
\[
M_{\rm s}=\tfrac12(\Omega A+A^\top\Omega),\qquad M_{\rm k}=\tfrac12(\Omega A-A^\top\Omega),\qquad m=\tfrac12(l+h),\qquad w=h-l,
\]
\[
\chi_i=\frac1{4\omega_i}\sum_j|\omega_iA_{ij}-\omega_jA_{ji}|\,w_j,\qquad \chi=\max_i\chi_i,\qquad R=\frac{\sum_i\omega_i}{\min_i\omega_i}.
\]
The \emph{skew allowance} $\chi$ is $0$ if $\Omega A$ is symmetric. No positive semidefiniteness is assumed. For $x$ in the centered box $\mathcal X=\prod_i[l_i-m_i,h_i-m_i]$ define the potential and its normalized gradient
\[
V(x)=\tfrac12x^\top M_{\rm s}x+(\Omega g(m))^\top x,\qquad u_i(x)=\frac{(M_{\rm s}x+\Omega g(m))_i}{\omega_i}.
\]
The coordinate curvatures are $(M_{\rm s})_{ii}=\omega_iA_{ii}=\omega_i>0$. Since $g(m+x)=g(m)+Ax$ and $\Omega A=M_{\rm s}+M_{\rm k}$,
\begin{equation}\label{eq:potgrad}
u(x)=g(m+x)-\Omega^{-1}M_{\rm k}x,\qquad |u_i(x)-g_i(m+x)|\le\chi_i ,
\end{equation}
where the inequality uses $|x_j|\le w_j/2$ on $\mathcal X$.

\paragraph{Algorithm $\mathrm{POTENTIAL}(\tau)$.} Start at $x=0$. At each step compute, for all $i$,
\[
y_i=\mathrm{clamp}_{[l_i-m_i,\,h_i-m_i]}\bigl(x_i-u_i(x)\bigr).
\]
If $\max_i|y_i-x_i|\le\tau$, return $r=m+x$. Otherwise select an index $i$ with $|y_i-x_i|>\tau$ and replace $x_i$ by $y_i$, leaving the other coordinates unchanged.

\begin{theorem}[potential selector]\label{thm:potential}
For rational $\tau>0$, $\mathrm{POTENTIAL}(\tau)$ stops after at most $\lceil5CR/(2\tau^2)\rceil+1$ steps and returns a rational $r\in\mathcal B$ with $\|r-G(r)\|_\infty\le\chi+\tau$, so the actual residual of every unsettled bank is at most $E_0+\chi+\tau$. Its running time and bit complexity are polynomial in the encoded input and in the numbers $C$, $R$ and $1/\tau$. For $\Omega=I$, $R=|U|$.
\end{theorem}
\begin{proof}
Along coordinate $i$, $V(x+se_i)=V(x)+\omega_iu_i(x)s+\tfrac{\omega_i}2s^2$, a convex quadratic in $s$, so $y_i$ minimizes $V$ along coordinate $i$ over $\mathcal X$. With $s=y_i-x_i$, the variational inequality of the projection, tested at $x_i$, gives $(x_i-u_i(x)-y_i)(x_i-y_i)\le0$, that is, $u_i(x)s\le-s^2$; hence
\[
V(x+se_i)-V(x)=\omega_iu_i(x)s+\tfrac{\omega_i}2s^2\le-\tfrac{\omega_i}2s^2 ,
\]
and every replacement decreases $V$ by more than $(\min_i\omega_i)\tau^2/2$. On $\mathcal X$ we have $|x_i|\le w_i/2<1/2$. Moreover $\sum_{ij}|(M_{\rm s})_{ij}|\le\sum_{ij}\omega_i|A_{ij}|\le C\sum_i\omega_i$ and $\sum_i\omega_i|g_i(m)|\le C\sum_i\omega_i$ by \cref{lem:coefbound}, so $|V(x)|\le C(\sum_i\omega_i)/8+C(\sum_i\omega_i)/2=5C(\sum_i\omega_i)/8$, and the total decrease is at most $5C(\sum_i\omega_i)/4$. This proves the bound on the number of steps. At termination, $|r_i-\Pi_i(r_i-u_i(x))|\le\tau$ for all $i$; projection onto an interval is $1$-Lipschitz, and $|u_i-g_i|\le\chi_i$ by~\eqref{eq:potgrad}, so $\|r-\Pi(r-g(r))\|_\infty\le\tau+\chi$. \Cref{lem:projtransfer} gives the actual residual.

All constants, the midpoint and the affine update coefficients are rational. Clear their denominators once by a common integer $q$, which also clears the box endpoints. After $k$ coordinate updates every coordinate has a denominator dividing $q^{k+1}$, because each new coordinate is either an endpoint or an affine combination of old coordinates with fixed coefficients whose denominators divide $q$; numerators are bounded accordingly, since all iterates stay in a bounded box. So bit lengths grow linearly in the number of updates, which is polynomial, and every comparison and arithmetic operation is exact and polynomial.
\end{proof}

If $\Omega A$ is symmetric, then $\chi=0$, even when $\Omega A$ is indefinite or $A$ is not a P-matrix. With a supplied diagonal symmetrizer the bound is polynomial only when the numerical ratio $R$ is polynomially bounded; symmetrizability by itself is not a polynomial guarantee, because $R$ can be exponentially large in the encoding length. The skew allowance $\chi$ is computed on the reached product, so a component that satisfies $\chi\le E-E_0-\tau$ for a supplied $\Omega$ yields a certified vector, and the routine reports that it has no certificate otherwise; this is an a-posteriori certificate, not a promise about all networks. For a fixed budget $\kappa_0$ and a fixed bound $R_0$, a suitable $\Omega$ can be found by linear programming if one exists: $\chi_i\le\kappa_0$ is the linear condition $\sum_jt_{ij}w_j\le4\kappa_0\omega_i$ with auxiliary variables $t_{ij}\ge\pm(\omega_iA_{ij}-\omega_jA_{ji})$, and since $\chi$ and $R$ do not change when $\Omega$ is scaled, $R\le R_0$ can be written as $\omega_i\ge1$ for all $i$ and $\sum_i\omega_i\le R_0$.

\subsection{A monotone selector}

Suppose now that the symmetric part $\tfrac12(A+A^\top)$ is positive semidefinite; $A$ need not be symmetric, and its skew part need not be small. Then $g$ is monotone: $(g(x)-g(y))^\top(x-y)\ge0$. Let $n=|U|$ and $\varsigma=1/(2nC)$.

\paragraph{Algorithm $\mathrm{MONOTONE}(\tau)$.} Start from any rational $x\in\mathcal B$. Repeat: put $y=\Pi(x-\varsigma g(x))$; if $\|x-y\|_2^2\le\varsigma^2\tau^2$, return $x$; otherwise replace $x$ by $\Pi(x-\varsigma g(y))$.

\begin{theorem}[monotone selector]\label{thm:monotone}
For rational $\tau>0$, $\mathrm{MONOTONE}(\tau)$ stops after at most $\lceil16n^3C^2/(3\tau^2)\rceil+1$ iterations and returns a rational $r\in\mathcal B$ with $\|r-G(r)\|_\infty\le\tau$, so the actual residual of every unsettled bank is at most $E_0+\tau$. Its running time and bit complexity are polynomial in the input size, $1/E_0$ and $1/\tau$.
\end{theorem}
\begin{proof}
$G$ is a continuous self-map of $\mathcal B$, so by Brouwer's theorem it has a fixed point $z=\Pi(z-g(z))$. The characterization of the projection gives $g(z)^\top(v-z)\ge0$ for every $v\in\mathcal B$, and monotonicity gives $g(y)^\top(y-z)\ge g(z)^\top(y-z)\ge0$ for every $y\in\mathcal B$. By \cref{lem:coefbound} the Euclidean Lipschitz constant of $g$ is at most $\|A\|_F\le\sqrt n\,C\le nC=:L$, so $\varsigma L=1/2$.

Write $x^+=\Pi(x-\varsigma g(y))$, $d=x-y$ and $q=x^+-y$. The projection characterization of $y$, tested at $x^+$, gives $d^\top q\le\varsigma g(x)^\top q$. The projection characterization of $x^+$, tested at $z$, gives
\[
\|x^+-z\|_2^2\le\|x-z\|_2^2-\|x-x^+\|_2^2-2\varsigma g(y)^\top(x^+-z).
\]
Write $\|x-x^+\|^2=\|d\|^2-2d^\top q+\|q\|^2$ and $x^+-z=q+(y-z)$. Then $2d^\top q-2\varsigma g(y)^\top q\le2\varsigma(g(x)-g(y))^\top q\le2\varsigma L\|d\|\,\|q\|\le\varsigma^2L^2\|d\|^2+\|q\|^2$, and therefore
\[
\|x^+-z\|_2^2\le\|x-z\|_2^2-(1-\varsigma^2L^2)\|x-y\|_2^2-2\varsigma g(y)^\top(y-z)\le\|x-z\|_2^2-\tfrac34\|x-y\|_2^2 .
\]
The squared diameter of $\mathcal B$ is at most $n$, so there are at most $4n/(3\varsigma^2\tau^2)=16n^3C^2/(3\tau^2)$ iterations with $\|x-y\|_2>\varsigma\tau$.

For an interval $[l,h]$, a point $x$ in it, a real $v$ and $0<\varsigma\le1$, we have $|x-\mathrm{clamp}(x-v)|\le\varsigma^{-1}|x-\mathrm{clamp}(x-\varsigma v)|$: for $v\ge0$ the two displacements are $\min(v,x-l)$ and $\min(\varsigma v,x-l)$, and the second is at least $\varsigma$ times the first; $v\le0$ is symmetric. Since $\varsigma\le1$, the stopping test gives $\|x-\Pi(x-g(x))\|_\infty\le\varsigma^{-1}\|x-y\|_2\le\tau$, and \cref{lem:projtransfer} applies. Each iteration applies two affine maps with fixed rational coefficients and a coordinatewise projection, and the test uses squared norms, so no square roots are needed; as in \cref{thm:potential}, the denominators grow by a fixed factor per iteration. Positive semidefiniteness of $\tfrac12(A+A^\top)$ can be decided once in polynomial time by exact rational symmetric elimination.
\end{proof}

\Cref{thm:monotone} is not the best possible dependence on $\tau$: a box-constrained variational inequality with a monotone affine operator can be written as a monotone linear complementarity problem, for which interior-point methods run in time polynomial in $\log(1/\tau)$~\cite{KMY89}. We include the elementary method because its proof is self-contained and uses only the bounds of \cref{lem:coefbound}.

\begin{example}[a directed three-cycle]\label{ex:threecycle}
Let three receivers with zero external assets owe $1$ each to a sink with $e_z=1$, and let each hold a CDS of notional $2$ on the next, written by one ex-ante solvent writer. Then $P=0$, $N=2S$ for the cyclic shift $S$, and $\tfrac12(A+A^\top)=I+S+S^\top$ is the all-ones matrix, which is positive semidefinite, although $\rho(N)=2$. At $E=4/25$ the exact algorithm $\BOX(4/25)$ reaches $[12/25,21/25]^3$ in one round and stalls in the next, and the rounded algorithm $\BOX(4/25,17)$, whose grid step is $1/25$, reaches the same product; so \cref{lem:coefbound,lem:projtransfer} apply with $E_0=4/25$, and \cref{thm:monotone} applies. The same construction on a directed five-cycle gives a symmetric part with the eigenvalue $1+2\cos(4\pi/5)<0$, so monotonicity is a genuine restriction.
\end{example}

\subsection{Composition}

Form the directed graph on $U$ with an edge $j\to i$ whenever the actual signed coefficient $(P-N)_{ij}$ is nonzero; exact cancellations between debt and CDS terms remove edges. Process its strongly connected components in topological order. When a component $B$ is processed, the coordinates of earlier components are fixed at the values already selected, which lie in their intervals, and later components do not occur in the asset ratios of $B$. The asset ratios of $B$ are those of the original network with the upstream values substituted, so $A_{BB}$ is a principal submatrix of $A$, only the affine offset changes, and \cref{lem:coefbound,lem:projtransfer} hold on $B$ because every full vector stays in $\mathcal B$.

\begin{theorem}[composition]\label{thm:composition}
In the setting of this section let $E$ be rational with $E_0+\delta/2\le E$. Suppose every strongly connected component $B$ is handled in one of the following ways: by \cref{thm:potential} on $A_{BB}$, with a supplied positive diagonal $\Omega_B$ whose ratio $R_B$ is bounded by a polynomial in the input size and with $\chi_B+\tau_B\le E-E_0$ on the reached product; or by \cref{thm:monotone}, if $\tfrac12(A_{BB}+A_{BB}^\top)$ is positive semidefinite, with $\tau_B\le E-E_0$. Then the components can be combined in topological order into a vector that, with the settled points and $1$ on $F$, is an (I)-$E$ vector of the whole network, computed in time polynomial in the input size, $M$, $1/E_0$, $\max_B1/\tau_B$ and the encoding lengths of the supplied $\Omega_B$ and $\tau_B$.
\end{theorem}
\begin{proof}
For $i\in B$, $G_i$ depends only on coordinates of $B$ and of earlier components. The selector for $B$ makes $|r_i-G_i(r)|\le E-E_0$ for $i\in B$, and later selections do not change $G_i$. So the combined vector satisfies $\|r-G(r)\|_\infty\le E-E_0$, and \cref{lem:projtransfer} with $\eta=E-E_0$ and $E_0+\delta/2\le E$ gives the claim. The time bound is the sum of those of \cref{thm:roundedbox} and of the selectors.
\end{proof}

\section{Sources, dependencies and verification}\label[appendix]{app:verification}

This appendix identifies the mathematical dependencies of the results on the research records, the reproducibility limits, and the scope of the literature checks. The computational evidence, with a claim-to-check index and the commands that reproduce it, is documented in the support bundle~\cite{Supp}. Finite computations do not replace universal proofs. A separate checker is an implementation that does not import the producer's mathematical code and operates on exported contracts or explicitly stated parameters. This technical separation is not independent human authorship or peer review. Internal research records are cited through the public support bundle~\cite{Supp}, with their record identifiers and original statement or section locators. Its \texttt{PROVENANCE.md}, record indexes and file maps resolve these identifiers to the archived scientific materials and retain detailed research-record provenance.

\paragraph{Available materials.} The public support bundle of this version~\cite{Supp} contains a scientific-only derivative of the curated reproducibility package of the earlier version of this paper (supporting materials v1.1.2), with its mathematics preserved and historical machine paths normalized, with the publication-edited records~L1--Rev1 in~\cite{Supp} under its directory \texttt{historical/} and its exact-arithmetic verification programs with their expected outputs. It also contains the scientific notes of the later records~\cite[records~S1, S2, S3, S4, S5]{Supp}, each either a byte-identical copy of a file of those records or a curated copy from which coordination material has been removed, with withdrawn or superseded claims marked in its indexes and a map from every file to the SHA-256 hash of its source; the exact-arithmetic programs of the later series and of this version with their expected outputs, with documented minimal path adaptations; a replay script that runs them from the bundle alone; and SHA-256 manifests. The curated copies are marked as such and are not byte-identical to their sources; the original records are preserved separately and are not needed for the replay. The paths of the records~L1--Rev1 in~\cite{Supp} listed in the support bundle's provenance records refer to \texttt{historical/} inside the earlier package, and program paths of the form \texttt{code/\dots} cited in the paper refer to that package as well. The rest of this paragraph and the next describe that package. It contains the manuscript sources of the earlier version, six exact-arithmetic verification programs, build and cross-reference checks, and publication-edited scientific records. It is a curated derivative of the historical Supplement v2, rather than the original immutable archive. Private correspondence and coordination material are excluded; three nested verification packages are unpacked. Its \texttt{MANIFEST.sha256} authenticates the files it contains. Unpacked verification materials occur under \texttt{historical/nested/}. Its README gives runnable commands, dependency requirements and scope limits.

The original Supplement v2 contained $480$ files, including its manifest, and had SHA-256
\begin{center}\small\texttt{ca373ddd2d2a8f597f6cf5fd140c703ec65c223b2b8206edb24ccaf7f9fdbe4d}.\end{center}
That digest identifies the historical archive, not this curated package. Unavailable historical material includes records $2$ and $4$, the programs of records $6$--$18$ produced separately from the odd-numbered record implementations, the programs underlying the $28$ stalls in~\cite[record~Lsolo]{Supp}, and the separate note \texttt{hardness-019.md} and its programs. Missing material is not reconstructed or claimed to be reproducible here.

\paragraph{Mathematical dependencies.} \Cref{tab:pointers} distinguishes external auxiliary proof details from arguments reproduced in the paper, alternatives and source references. It does not certify the cited reports as independent peer review.

{\small
\setlength{\tabcolsep}{4pt}
\begin{longtable}{@{}R{3.0cm}R{3.5cm}R{9.1cm}@{}}
\caption{Research-record sources, their role and their checks.}\label{tab:pointers}\\
\toprule
statement & source & check\\
\midrule
\endfirsthead
\toprule
statement & source & check\\
\midrule
\endhead
\cref{rem:EIendpoint}: endpoint of the design family & \cite[record~L13; Theorem~5.4]{Supp}; \cite[record~L14; Theorem~7.2]{Supp} with \cite[record~L15; Theorem~2.3]{Supp}; \cite[record~L15; Proposition~4.1]{Supp}; \cite[record~L16; Theorem~B]{Supp} & Proofs reproduced in \cref{app:hardnessaux}; the family qualifications in the cited statements apply.\\
paragraph after \cref{prop:basin}: basin test & proof of \cite[record~L15; Proposition~2.1]{Supp} & Only the ``if'' direction is used, and it is proved in this paper.\\
\cref{rem:leak}: slack $(15-\sqrt{153})/12$ & \cite[record~L16; Section~7.1]{Supp} & Algebra is given in the remark; the identity is checked by \texttt{constants.py}.\\
\cref{rem:leak}: direct (W) construction up to $0.14053956\ldots$ & \cite[record~L3; Theorem~4.3]{Supp} & Construction and exact root isolation reproduced in \cref{app:hardnessaux}; the root is also recomputed by \texttt{constants.py}.\\
\cref{rem:obstruction}: one-writer obstruction & \cite[record~L13; Section~3]{Supp} & The argument is given in the remark.\\
\cref{cor:Fendpoint}: second gadget & \cite[record~L16; Section~7.1]{Supp}, \cite[record~L17; Proposition~7.8]{Supp} & Optional alternative; the proof here uses the gadget of \cref{thm:wsI}.\\
\cref{thm:locallower}(a) & \cite[record~L7; Theorem~6.1]{Supp}, hypothesis corrected in \cref{rem:L7} & The correction is recorded in~\cite[record~L19; Remark~4.8]{Supp}. The proof is reproduced here with the corrected hypothesis; the cases in \cref{rem:L7} are recomputed by \texttt{constants.py}.\\
\cref{rem:rounds}: $1$-local threshold $(3-\sqrt5)/2$ & \cite[record~L3; Theorem~7.1(a)]{Supp} & Lower-bound proof reproduced in \cref{app:localaux}; the upper guarantee follows from \cref{sec:box}.\\
\cref{sec:local}: bound $0.1609$ for writer-solvent (W) & \cite[record~L11; Proposition~5.1]{Supp} & Proof reproduced in \cref{app:localaux}; the sharing of writers is restricted to consecutive stages.\\
one-pass algorithm (\cref{sec:box}) & \cite[record~L1; Section~2]{Supp} & Origin and earlier analysis; the upper guarantees also follow from the first-round bounds here.\\
\cref{rem:nonuniformscope}: obstruction for $\mathfrak F$ & \cite[record~L18; Theorem~8.1]{Supp} & Related earlier result; not used in the proof of \cref{thm:nonuniform}.\\
\cref{sec:routes}: two-receiver template & \cite[record~L18; Theorem~9.1]{Supp} & Proof reproduced in \cref{prop:tworeceiver}; the scope is one helper that reads only the main receiver and exact internal solution sets.\\
\cref{sec:literature}: abstract of the journal version of~\cite{IKV23} & \cite[record~L5]{Supp} & Literature provenance based on the abstract, not an audit of the full journal paper.\\
\cref{thm:interfaces}: separate saturated rail bounds & \cite[record~L19]{Supp} & The least-bound argument for this special case appears there; the proof here is complete and covers non-saturated intervals.\\
\cref{thm:Pk,thm:A2,thm:A3,thm:weightedD,thm:weightedPk}: supported propagation & \cite[Lemma~F.8, Theorem~F.9]{KM} & Mechanism credited; all proofs are given here.\\
\cref{sec:Rclass,sec:degree,app:chains,app:limits} & \cite[record~S4]{Supp}, with the refined rule of \cref{thm:Pk} from~\cite[record~S3]{Supp}, corrected in~\cite[record~S5]{Supp} & All proofs are reproduced here, with the clarifications of~\cite[record~S5]{Supp} (creditor holdings, distinct chain banks, the domain $E<1$, complexity model).\\
\cref{sec:selection,app:stalls} & \cite[records~S1, S2, S3]{Supp}, scope in~\cite[record~S5]{Supp} & All proofs are reproduced here; claims of~\cite[record~S2]{Supp} that were withdrawn are not used.\\
\cref{sec:rounded,sec:interfaces,app:weighted} & \cite[records~S1, S3]{Supp}, scope in~\cite[record~S5]{Supp} & All proofs are reproduced here.\\
\bottomrule
\end{longtable}
}

\paragraph{Computational evidence.} Statements marked \CO\ report finite exact computations or numerical diagnostics. For each of them the check index of the support bundle~\cite{Supp} names the producing record, the separate checks that were made, and whether the producing programs were archived; historical checks are not presented as new executions. The programs of the later series and of this version, and the six programs of the earlier package, are rerun by the replay script of the support bundle, which compares their outputs with the recorded ones. All these are finite checks of designed instances, boundary cases and case analyses; the theorems rest on the proofs given in the paper.

\paragraph{Literature scope.} Theorem~5, Definition~3 and the existence statement of Section~2.1 of~\cite{IKV22} were checked in the official conference PDF; Theorems~1 and~3 and Definition~3 of~\cite{IKV21} were checked in the cited arXiv version. For~\cite{IKV26b}, Definition~4, Theorem~2 and Corollary~2 were checked at statement level in indexed publisher text; the typeset article and its proofs were not reviewed. For~\cite{IKV26a}, only publication data were checked. The abstract of~\cite{RV13} and the cited arXiv statements of~\cite{DHK} were consulted. Other historical literature comparisons based on web excerpts remain limited to those excerpts, as recorded in~\cite[record~L17; Section~9]{Supp} and~\cite[record~L18; Section~11]{Supp}. For~\cite{KM}, Definitions~F.5 and~F.6, Lemma~F.8, Theorems~F.7 and~F.9 and Theorem~3.4 were read in arXiv version~1; for~\cite{DFHM24}, Definition~1, Theorem~2.1 and Corollary~2.2 in the complete accepted manuscript; for~\cite{DHK}, Definition~3, Theorem~1 and Section~3.2 in the arXiv version and, for the conference version, only the publication data, abstract and footnote on its public publisher page (retrieved 28 October 2026), its full text and proofs not being available; for~\cite{IKV26b}, Section~5 in indexed publisher text, while the full journal proof was not available; for~\cite{PP21}, Section~5.1, Theorem~5.1, Section~4 and Appendix~A in arXiv version~2; and for~\cite{DDH24}, the published abstract and highlights. Detailed locators and retrieval records are in~\cite[record~S5]{Supp}; the checks of the public page of the conference version of~\cite{DHK}, of footnote~1 of its arXiv version and of the numbering of the full version of~\cite{DFHM22}, made on 28 October 2026 for this version, are recorded in the literature notes of the support bundle~\cite{Supp}. These comparisons are bounded and version-specific; they are not an exhaustive search for prior work.

\end{document}